\documentclass[a4paper,11pt,oneside,openany]{book}

\usepackage[english]{babel}

\usepackage[utf8]{inputenc} 
\usepackage[T1]{fontenc} 

\usepackage{fancyhdr} 

\usepackage{graphicx, xcolor} 

\usepackage{tabularx} 
\usepackage{longtable} 

\usepackage[titletoc]{appendix} 

\usepackage{float} 

\usepackage{fancyvrb} 

\usepackage{amsmath, amsfonts} 
\numberwithin{equation}{section}

\usepackage{hyperref} 

\usepackage{lipsum} 

\usepackage{caption}
\usepackage{subcaption}
\usepackage{lscape}
\usepackage{booktabs}

\usepackage[square,sort,comma,numbers]{natbib}

\usepackage{bbm}

\usepackage{tikz}

\usepackage{multirow}

\usepackage{pdfpages}

\hypersetup{ 
    colorlinks,
    citecolor=black,
    filecolor=black,
    linkcolor=black,
    urlcolor=black
}

\usepackage[some]{background}
\SetBgScale{1}
\SetBgContents{
\includegraphics[scale=1]{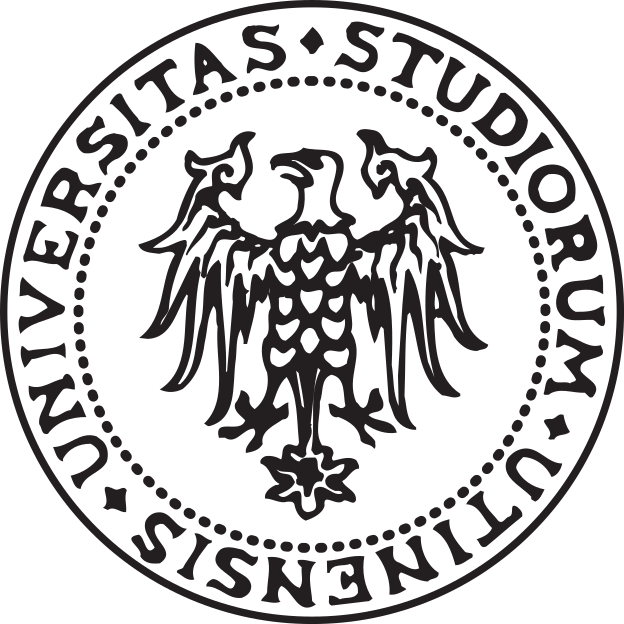}}
\SetBgColor{gray}
\SetBgAngle{0}
\SetBgOpacity{0.07}

\newtheorem{theorem}{Theorem}
\title{Deep Learning for Gamma-Ray Bursts: A data driven event framework for X/Gamma-Ray analysis in space telescopes}
\author{Riccardo Crupi}
\begin{document}


\pagenumbering{roman}
\begin{titlepage}
    \begin{center}
    \BgThispage
    {\LARGE {\bfseries UNIVERSIT\`A DEGLI STUDI DI UDINE \\}}
    \vspace{.5cm}
    {\Large {\bfseries Dipartimento di Scienze Matematiche, Informatiche e Fisiche \\}}
    \vspace{1cm}
    \includegraphics[width=6cm,height=6cm]{images/uniud_logo.png}\\[1.0cm]

    Corso di dottorato di ricerca in Informatica e Scienze Matematiche e Fisiche\\
        $36^\circ$ ciclo\\
    \vspace{.5cm}
    {\LARGE
         Tesi di dottorato \\
    }
    \vspace{0.75cm}
    {\LARGE 
        {\bfseries Deep Learning for Gamma-Ray Bursts: \\ A data driven event framework for X/Gamma-Ray analysis in space telescopes}
    }
    \vspace{1cm}


    \vfill
    \begin{table}[h]
        {\large
            \begin{tabular}{c c c c r c c | c c l}
                & & & & Dottorando & & & & & Supervisore \\
                & & & & \bfseries Riccardo \textsc{Crupi} & & & & & \bfseries ~Prof.ssa~Barbara~\textsc{De Lotto} \\ 
                & & & & & & & & & \\
                & & & & & & & & & Co-supervisori \\ 
                & & & & & & & & & \bfseries ~Prof.~Andrea~\textsc{Vacchi} \\
                 & & & & & & & & &  \bfseries ~Dr.~Fabrizio~\textsc{Fiore} \\
            \end{tabular}
        }
    \end{table}
    Anno 2024
    \end{center}
\end{titlepage}

\phantomsection
\thispagestyle{empty}
\vspace*{3cm}

\begin{center}
In matematica non si capiscono le cose.\\
Semplicemente ci si abitua ad esse. \\
--- John von Neuman ---
\end{center}


	
\frontmatter

\chapter{Abstract}
\markboth{Abstract}{Abstract}

The HERMES (High Energy Rapid Modular Ensemble of Satellites) Pathfinder mission serves as an in-orbit demonstration of a constellation of nanosatellites whose primary scientific purpose is to discover intense high-energy transients, such as gamma-ray bursts, across a broad energy range (few keV to few MeV) with unparalleled temporal precision and exact localisation.
By 2024, the first constellation of six nanosatellites is expected to be launched. \\

To fully exploit satellite data and allow faint astronomical events to emerge, a precise estimation of satellite background count rates is required to determine whether the event is statistically valid or not. The dynamics of the background are related to the satellite's orbital information, which varies in the order of minutes, potentially hiding long transient events. \\

This work introduces two main contributions I have brought ahead; first a novel background estimator is presented that could potentially be fitted to any type of X/Gamma-ray satellite space telescope, capable of capturing long-term dynamics and accurate enough to detect faint transients. This estimator is built using a Neural Network and tested on data from the Fermi Gamma-ray Space Telescope's Gamma Burst Monitor (GBM). \\

As a second objective, it is employed a trigger algorithm, called FOCuS (Functional Online CUSUM), to extract events from the background using the background estimator. The resulting framework, DeepGRB, can identify astronomical events that are both present and absent from the Fermi-GBM catalog. The analysis of the discovered events reveals the strengths and weaknesses of the framework.

\chapter{Acknowledgement}
\vspace{1.5cm}
\phantomsection 
\markboth{Acknowledgement}{Acknowledgement}
\vspace{0.5cm}

There are so many people to thank, and I'll try to do it in an almost random order. This journey started with my desire to contribute to the field of astrophysics, despite my initial lack of knowledge in this domain, coming from a background primarily in computer science. As an experienced experimentalist leading the HERMES team in Udine, Professor Andrea Vacchi embraced this challenge with an open mindset and enthusiasm, strongly believing that the interdisciplinary approach could generate interesting and valuable results. He guided me until now, and I can't thank him enough.

One thing is certain: I wouldn't have had a chance without Giuseppe Dilillo. He supported and "sopported" (which in Italian sounds like "tolerated") me in the main work, providing an essential step toward an end-to-end solution for high-energy transient detection. Related to this, Kester Ward deserves thanks for the Poisson-FOCuS implementation.

I'd like to express my gratitude to the entire HERMES team at Udine for their valuable advice and discussions during our recurrent Tuesday meetings: Giovanni Della Casa, Nicola Zampa, Daniela Cirrincione, Simone Monzani, and Marco Citossi. From Udine, I want to express my gratitude to Professor Barbara De Lotto, who accompanied me in the final stages of my PhD. I'd also like to thank Irene Burelli, with whom I shared some PhD lessons and anxiously completed bureaucratic tasks.

The principal investigator of HERMES and director of the Osservatorio Astronomico di Trieste, Fabrizio Fiore, made it possible and set the direction for this work to leverage the future data collection of HERMES. He also provided access to the powerful Nebula Server in Trieste with the assistance of Gianmarco Maggio, Chiara Feruglio, and Manuela Bischetti.

I'm grateful to all the professors and teachers I had, who patiently taught me, answered my questions and provided me the tools to build my education and passion for the research topics I studied.
Especially for the insightful comments and suggestions of the thesis referees, Professor Massimo Brescia, a leading advocate of AI in astrophysics, and Professor Maria Dainotti, a renowned expert in GRBs and AI. Their expertise and guidance significantly enhanced the quality of my thesis.

From the Intesa Sanpaolo side (yes, I'm a Data Scientist working in a bank, and I applied for a PhD in Astrophysics; after the discussion, I'll sleep at nights), a huge thank you goes to my boss, the head of the Data Science \& Artificial Intelligence office, Andrea Cosentini. He embarked on a similar journey as mine and encouraged me to do the same (without mentioning the nights) because he strongly believes that PhD training and research are necessary in our line of work. He's still the best boss I've ever had.

My colleague Daniele Regoli is the person I rely on the most for any kind of advice. He's like an inspiring guardian angel who patiently read this thesis and provided me with useful comments to make the main text clearer and the results more robust.

\begin{figure}
    \centering
    \includegraphics[width=1.\textwidth]{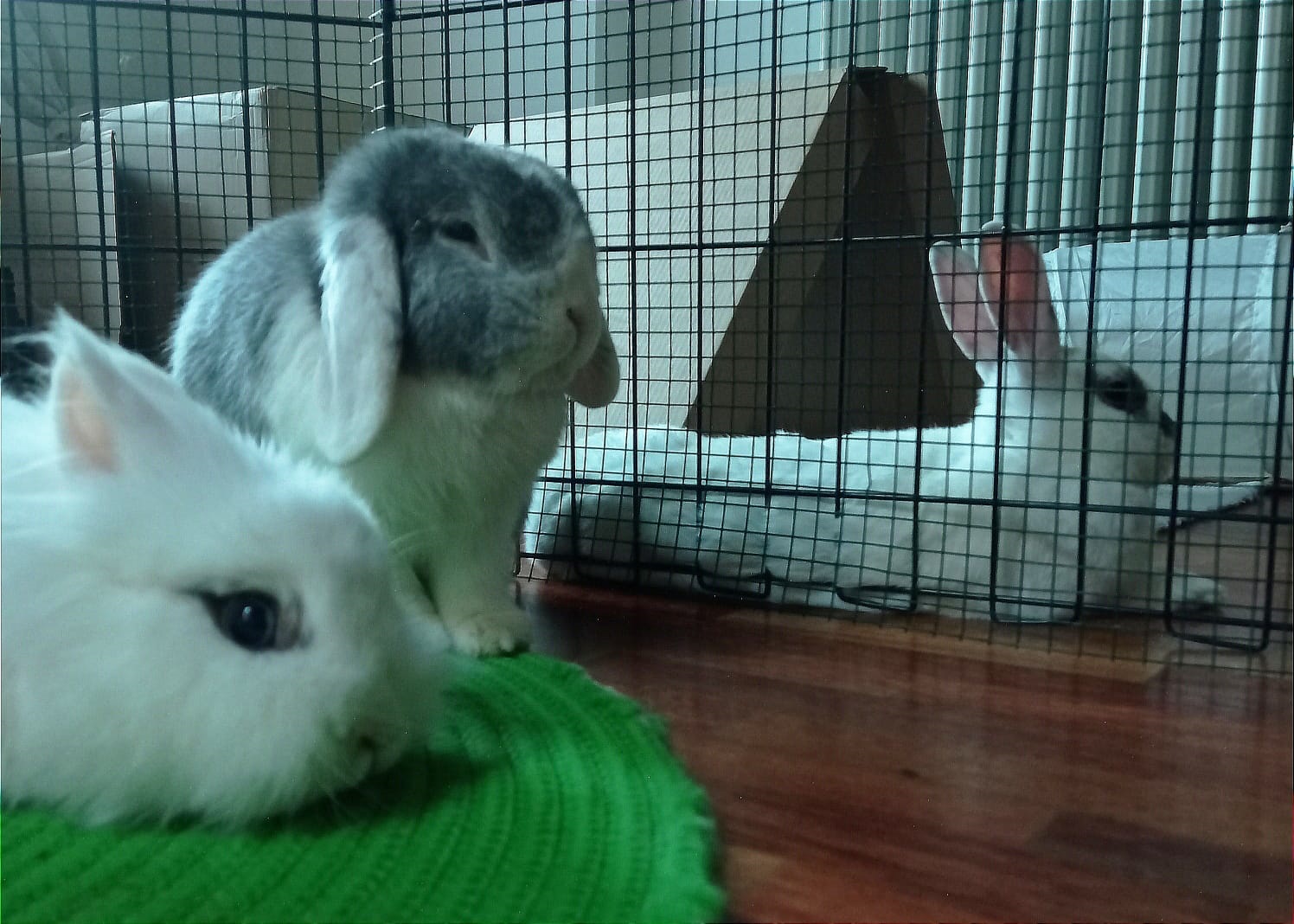}
    \caption{My three rabbits, from left to right: Nuvola, Olaf and Arturo.}
    \label{fig:rabbits}
\end{figure}

Back at home, I want to thank my three rabbits (see Figure \ref{fig:rabbits}), especially Olaf, who has exceptionally soft fur and relaxed me during sad moments, such as when the Neural Networks didn't converge properly (see Section \ref{sec:hyperparam}).

Finally, my last thanks go to my wife. Even in the search for the secret to Life, the Universe, and Everything\footnote{42, see Adams, Douglas, 1952-2001, "The Hitchhiker's Guide to the Galaxy".}, it wouldn't be worth it without her, neither if I found it for real\footnote{The author has a doubt it is already her the answer but it is left for future works.}.

\tableofcontents

\renewcommand{\chaptermark}[1]{\markboth{\MakeUppercase{\ #1}}{}}


\chapter{Introduction}
\markboth{Introduction}{Introduction}

HERMES (High Energy Rapid Modular Ensemble of Satellites) Pathfinder is an in-orbit demonstration consisting of a constellation of six 3U nano-satellites hosting simple but innovative detectors for the monitoring of cosmic high-energy transients. Actually, HERMES Pathfinder is not an observatory, and the main objective is to 1) validate the concept; study the uncertainty in detection/localization to validate the design for full constellation proposal, 2) demonstrate precise timing (300ns) achievable with small detectors, surpassing Fermi/GBM, within a nano-satellite experiment, 3) investigate uncertainties in combining signals from multiple detectors to enhance statistics in high-resolution time series.
The transient position is obtained by studying the delay time of arrival of the signal to detectors hosted by different nano-satellites on low Earth orbits. To this purpose, particular attention is placed on optimizing the detector's signal time accuracy, with the goal of reaching an overall accuracy of a fraction of a microsecond. In this context, we need to develop novel tools to fully exploit the future scientific data output
of HERMES Pathfinder. 

In this thesis, Chapter \ref{chp:1} provides an overview of the state of the art of GRBs, including their properties and potential progenitors. It also discusses the telescopes and instruments used for detecting GRBs, as well as the challenges in X/Gamma-ray photon background estimation providing the motivation for the work described in Chapters \ref{chap:bkg} and \ref{chap:frm}.

After an introduction of AI state of the art in Chapter \ref{chp:ml}, the Chapter \ref{chp:grb_ml} delves into AI's application in the context of GRBs, offering a comprehensive overview of AI techniques applied in GRB research.

Chapters \ref{chap:bkg} and \ref{chap:frm}, the major contribution of this thesis, is dedicated to introduce a new framework, \texttt{DeepGRB}, to assess the background count rate of a space-born, high-energy detector; a key step towards the identification of long/faint astrophysical transients. 
Chapter \ref{chap:bkg} introduces a Neural Network (NN) for background count rate estimation using satellite data. This problem is framed as a multi-regression task, and the NN's performance is analyzed during high and low solar activity periods, as well as during a period with an ultra-long GRB. The subsequent sections focus on hyperparameter selection and the application of eXplainable Artificial Intelligence (XAI) to understand the most important features both globally and for specific estimation aimed at debugging the NN. 
Subsequently, in Chapter \ref{chap:frm}, it is employed a fast change-point and anomaly detection technique, FOCuS-Poisson, to identify segments in the observations where there are statistically significant excesses in the observed count rate compared to the background estimate. It is tested over a period of about 9 months of data retrieved by the new software from archival data from the NASA Fermi Gamma-ray Burst Monitor (GBM), in which every single detector has a collecting area and background level of the same order of magnitude as HERMES Pathfinder.
Finally, the focus shifts to the events identified with \texttt{DeepGRB}. The framework is able to elaborate the daily Fermi/GBM data products and confirm events in the Fermi/GBM catalog, both Solar Flares and GRBs, but also identifies events, not present in Fermi/GBM catalog, that could be attributed to Solar Flares, Terrestrial Gamma-ray Flashes, GRBs, Galactic X-ray flash. For the latter events it is provided a catalog with an estimation of localisation, duration, detectors triggered, significance and a tentative classification (see Appendix \ref{chp:appendix}). Furthermore, seven of them are thoroughly examined, discussing the reasons for the classification based on the lightcurve's event plot, the event's localization, and the position of the Fermi GBM satellite. The last section describes a method that attempts automated classification of the transients using Machine Learning techniques, transparent by design or building on top a XAI technique to provide explanations alongside predictions.

\mainmatter

\renewcommand{\chaptermark}[1]{\markboth{\MakeUppercase{\chaptername\ \thechapter.\ #1}}{}}


\chapter{State of the art - GRB}\label{chp:1}


GRBs come from the most energetic and rich of information events in the universe. These bursts run from a few milliseconds to many minutes and release strong bursts of gamma-rays, the most energetic type of electromagnetic radiation. They were detected by US military satellites seeking for evidence of nuclear weapons testing in the late 1960s by the Vela satellites \citep{klebesadel1973observations}. However, it was not until the 1990s that astronomers began to carefully examine these occurrences and understand their astrophysical importance.

Over the years, significant progress has been made in understanding the underlying physical processes that lead to these explosive events. The following section will provide an overview of GRBs, divided into two main phases: the prompt emission and the afterglow. Based on these two phases, several potential progenitors have been proposed to explain their physical origin. However, despite the advancements, there are still many unanswered questions surrounding GRBs, and the open question section also highlights some of the ongoing research inquiries in this field.

\section{GRB overview}

GRBs have two different phases. The first is called \textit{prompt} and it is distinguished by their high energy emission, with gamma-ray energies typically ranging from a few keV to several hundred GeV. Then followed by the \textit{afterglow}, a longer-lasting emissions across the electromagnetic spectrum such as X/Gamma-rays, optical waves, and radio waves, allowing for multi-wavelength observations and analysis \cite{kumar2015physics}. 
Telescopes search for the corresponding afterglow after detecting the prompt emission. 
Prompt emission localization is typically coarse, and when telescopes/detectors dedicated to afterglow detection are alerted, they must search for the faint afterglow signal in a consistent region of space to precisely localize it. So it is unclear whether the GRB always has an afterglow component, but we sometimes miss it because it is too weak to be detected by our telescopes or simply does not exist.
The evolution of the GRB carries basic information of the evolution of the generating event and for this reason it is important to have the capability to obtain the most complete vision of each detected event.

GRBs are often divided into two categories depending on their duration: short and long. Short GRBs last less than two seconds, but long GRBs might last many minutes. This property has an high correlation with the physical origins of these two types of bursts. 
The physical origin of GRBs is still being studied, but the most commonly accepted theory is that they are caused by the stars evolutions phases or the merging of compact objects like neutron stars.

A big star runs out of fuel and falls under its own gravity, becoming a black hole in the collapsar model. The infalling material forms a disk around the black hole, emitting high-energy radiation as it accretes onto it. Because the disk may release radiation for several minutes, this process results in a long-duration GRB.

Two neutron stars or a neutron star and a black hole combine in the compact object merger model, resulting in a short-duration GRB. The merging produces a very intense outflow of matter that releases gamma rays as it collides with surrounding matter.

\begin{figure}[!htb]
\centering
  \includegraphics[width=1\textwidth]{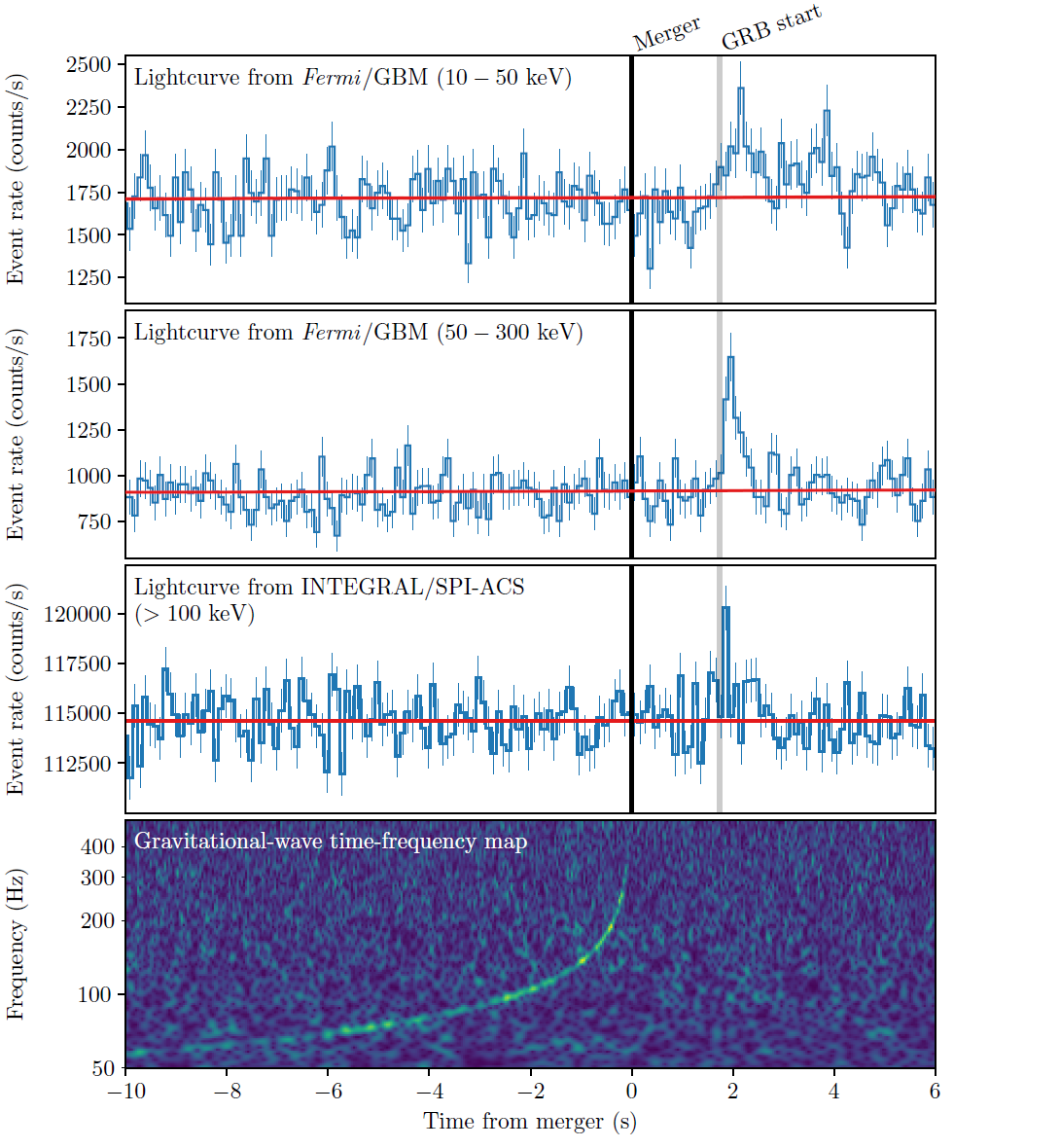}
\caption{\label{fig:ligovirgo} 
The discovery of GW170817 and GRB 170817A using data from Fermi-GBM and INTEGRAL, along with a time-frequency map from LIGO detectors. \cite{abbott2017gravitational} \copyright AAS. Reproduced with permission.}
\end{figure}

The discovery of GRBs has resulted in countless astrophysical discoveries. Astronomers, for example, have been able to analyze the features of the interstellar medium and the host galaxies of the bursts by observing afterglows. Furthermore, in 2017, the joint detection of the gravitational waves GW170817 and GRB 170817 \cite{abbott2017gw170817} (Figure \ref{fig:ligovirgo}) support the relation between GRB and compact object merger scenario.

\subsection{Prompt phase}\label{sec:grb_prompt}

This phase has distinct temporal properties, such as the presence of peaks, duration, burst shape and spectral properties.

\begin{figure}[!htb] 
\centering
\begin{subfigure}{.5\textwidth}
  \centering
  \includegraphics[width=1\linewidth]{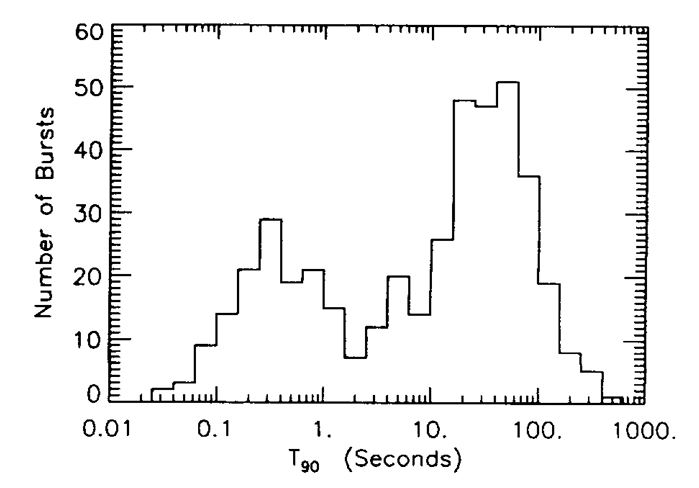}
  \caption{}
\end{subfigure}
\qquad
\begin{subfigure}{.5\textwidth}
  \centering
  \includegraphics[width=1\textwidth]{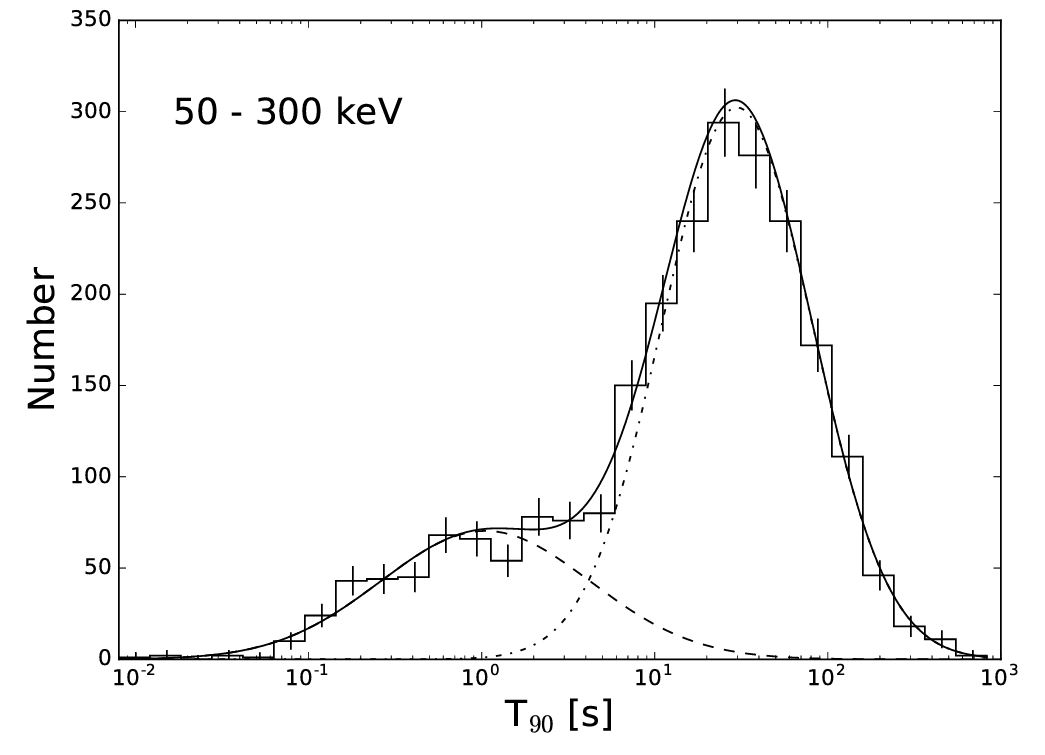}
  \caption{}
\end{subfigure}
\caption{\label{fig:t90batse}
T90 distribution for GRBs detected with a) BATSE \cite{meegan1996third} b) Fermi-GBM \cite{von2020fourth} \copyright AAS. T90 is the time at which the cumulative GRB counts increase from 5\% to 95\% of the total detected counts. Reproduced with permission.}
\end{figure}

A burst's duration is typically defined by the time interval between the start and end of the emission, known as the \textit{T90}. T90 is the time it takes for the cumulative counts in the burst to reach 90\% of the total observed counts. It measures the duration of the burst and is commonly used to classify GRBs as short-duration (T90 < 2 seconds) or long-duration (T90 > 2 seconds). The reason why the duration threshold is used 2s lays historically in the fact that in the distribution of T90, Figure \ref{fig:t90batse}, has two broad peaks centred at about 0.3s and 20s and the bimodal distribution can be separates GRBs in two broad categories cutting at 2s \cite{costa2011gamma}. This quantity can slightly vary among different satellites because of the different sensitivity of the detectors.

\begin{figure}[!htb]
\centering
  \includegraphics[width=1\textwidth]{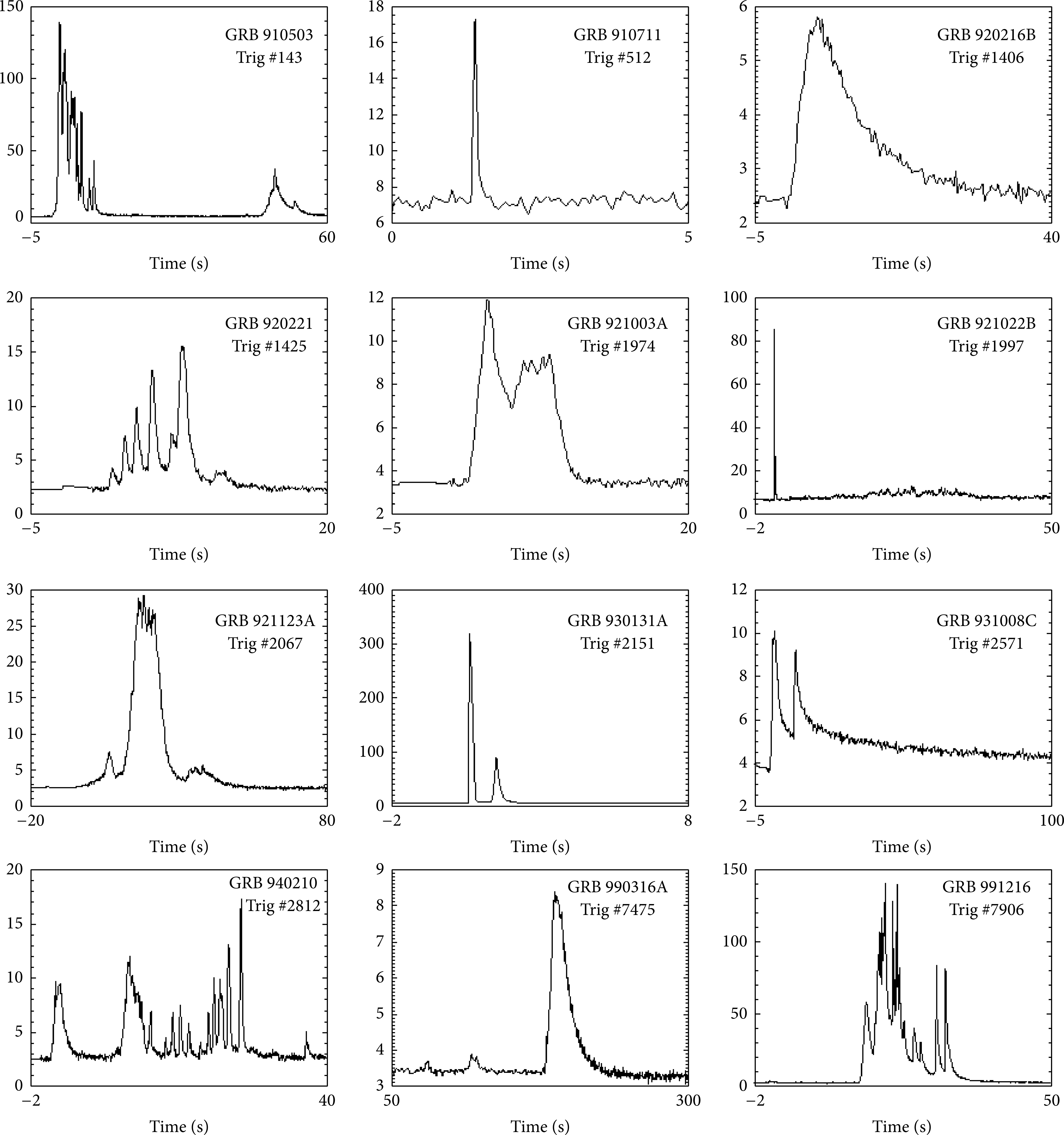}
\caption{\label{fig:grbex}
Examples of GRBs lightcurves \cite{pe2015physics}. The lightcurves of GRB are extremely diverse, with few recognizable patterns. This sample includes short and long events, single peaks or multiple peaks, noisy or very smooth, symmetric asymmetric profiles. Data from the public BATSE archive (\href{http://gammaray.msfc.nasa.gov/batse/grb/catalog/}{http://gammaray.msfc.nasa.gov/batse/grb/catalog/}), credict to Daniel Perley.}
\end{figure}

Aside from duration, a GRB may be distinguished by the temporal structure. Some GRBs have a single pulse, while others have a spiky structure with durations as short as ms. The multiple peaks sometimes can be separated by a quiet phase. Other temporal properties important for understanding the underlying physical processes include the rise time, decay time, and duration of individual peaks. In general, there is a quick raising phase followed by a gradual decreasing phase. Figure \ref{fig:grbex} shows some examples of lightcurve GRB emission. The complex profiles may include precursor emission, followed by the main peak and subsequent post-cursor emissions.

GRBs are also distinguished by spectral features related to the energy of the released gamma-ray photons in each phase.
The Hardness Ratio (HR) is a simple measure that consists of a ratio between photon counts in different energy ranges; a common measure can be found in $HR = (H - S) / (H + S)$, where H and S represent counts in the hard and soft energy bands, respectively. 

\begin{figure}[!htb] 
\centering
  \includegraphics[width=1\textwidth]{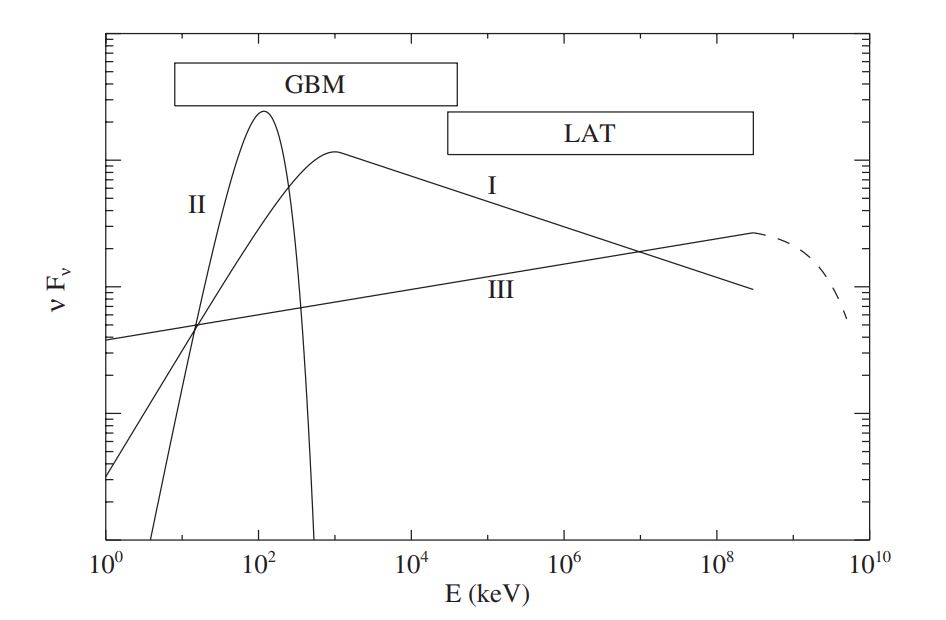}
\caption{\label{fig:spectral}
Three elemental spectral components that shape GRB prompt emission spectra: (I) a non-thermal Band-function component, (II) a black-body (quasi-thermal) component, and (III) an additional power-law component. The Fermi satellite utilizes two instruments, GBM (Gamma Burst Monitor) and LAT (Large Area Telescope), each with different energy sensitivities. \cite{zhang2011comprehensive} \copyright AAS. Reproduced with permission.}
\end{figure}

The spectra can be thought of as the result of various components, and Figure \ref{fig:spectral} depicts three of these components.
The most common spectra is a smoothly broken power-law function, called Band function \citep{band1993batse}:
\begin{equation}
F(E) = 
\begin{cases}
A \left(\frac{E}{100 \, \mathrm{keV}}\right)^{\alpha} \exp\left(-\frac{E}{E_{\rm peak}}\right) & \text{for } E \leq (\alpha - \beta)E_{\rm peak} \\
A \left(\frac{(\alpha - \beta)E_{\rm peak}}{100 \, \mathrm{keV}}\right)^{\alpha - \beta} \exp(\beta - \alpha) \left(\frac{E}{100 \, \mathrm{keV}}\right)^{\beta} & \text{for } E > (\alpha - \beta)E_{\rm peak}
\end{cases}
\end{equation}

Where $\alpha$ is low-energy (below $E_{\rm peak}$) power-law index, $\beta$ the high-energy (above $E_{\rm peak}$) power-law and $A$ is the normalization parameter. Usually, the values assumes the values $\alpha=1$, $\beta=-2.3$ and $E_p=150$ keV \citep{costa2011gamma}. 
Most spectral GRB emission shows a good fit with the Band function but other type of emission like black-body (quasi-thermal) or power-law emission is possible, see Figure \ref{fig:spectral}. The Band component must be due to a non-thermal mechanism because it may extend for six to seven orders of magnitude beyond the thermal emission. So even GRB should be generated by a non-thermal mechanism. The capabilities of the Fermi instruments made it possible to analyze the spectra of certain GRBs and reveal the power-law component in these spectra. 

Is natural to ask what type of process can generate those components. 
Highly relativistic outflows, such as jets moving at close to the speed of light, shocks, and turbulence, can efficiently accelerate particles to very high energies. These energetic particles interact with the surrounding magnetic fields and radiation, resulting in synchrotron radiation. This radiation has a wide range of energies, including gamma-ray frequencies, and the spectrum has a characteristic power-law shape. Another process that contributes to the non-thermal spectrum of GRBs is inverse Compton scattering. The high-energy electrons in the outflows interact with low-energy photons from the surrounding medium or the synchrotron radiation itself during this process. Low-energy photons gain energy as a result of this interaction, resulting in scattering to higher energies. This can cause the spectrum to broaden significantly and the appearance of a high-energy power-law tail. Finally, the photosphere, where photons are released for Compton scattering, is the major possibility for this emission in the Black Body or quasi-thermal component, and it was anticipated in one of the earliest models used to explain the core engine, the fireball model \cite{zhang2011comprehensive}. The thermal emission could arise from the interaction of radiation with matter, such as shocks or expanding fireballs, resulting in thermal equilibrium. 

Regarding the localization of GRB prompt emissions, it is known that they are distributed isotropically in the sky. This significant result came from BATSE \citep{costa2011gamma}. 

\subsection{Afterglow}\label{sec:afterglow}

The first afterglow was detected in 1997 with BeppoSAX, after 8 hours from trigger time of GRB 970228 \citep{costa1997discovery}. In particular, this afterglow lightcurve lasted 9 hours and was decaying according to a power-law, consistent with the average flux of the prompt phase. 

\begin{figure}[!htb] 
\centering
  \includegraphics[width=0.8\textwidth]{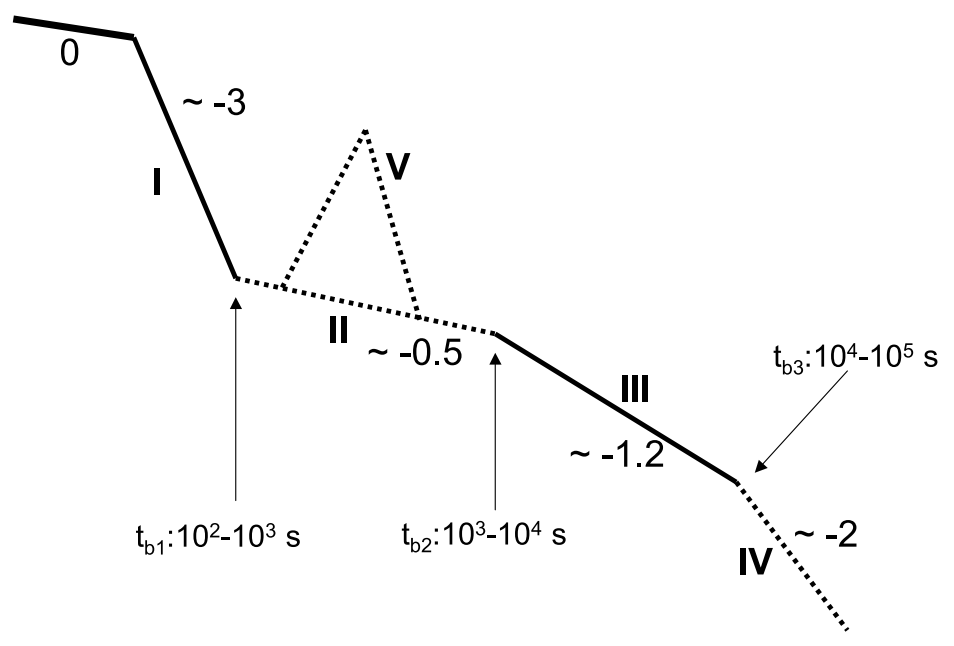}
\caption{\label{fig:afterglow}
The canonical X-ray afterglow lightcurve, which shows 5 different components: I. steep decay phase following the prompt emission, II. the plateau phase, III the normal decay phase, IV. the late steep phase, V. X-ray flares. \cite{zhang2006physical} \copyright AAS. Reproduced with permission.} 
\end{figure}

In general, the early afterglow phase, occurring within the first 10 hours, exhibits distinct characteristics, see Figure \ref{fig:afterglow}. It starts with a steep decay phase, which is an extension of the prompt emission, followed by a plateau phase with a slope greater than -0.5. This is succeeded by a normal decay phase with a power-law index of -1, and finally, a late steep decay phase with an index of -2 or steeper. X-flares may also occur during these phases, resulting from the central engine of the GRB experiencing a restart. It is important to note that not all GRB afterglows exhibit these specific features.

In contrast, the late afterglow phase, occurring after 10 hours, generally follows a broken power-law behaviour, with different characteristics in different observed bands. The optical emission initially follows a power-law decay with an exponent of approximately -1, which steepens to -2 after one day. On the other hand, the radio emission initially grows and then starts to decline after approximately ten days. The late-time afterglow radiation from radio to X-ray frequencies can be well-described by the synchrotron radiation mechanism in external shocks. However, the early afterglow phase poses additional complexities and requires more intricate models for a comprehensive understanding.

These findings highlight the temporal evolution and diverse behaviours observed in the afterglow phase of GRBs, shedding light on the physical processes involved in the emission at different timescales and across various wavelengths \cite{kumar2015physics}. 

The importance of the afterglow is crucial for determining the distance of the progenitor. The first GRB detected with redshift was GRB 970508 \citep{reichart1998redshift}. Three days later the detection an optical afterglow was still relatively bright and various absorption lines could be identified and a redshift $z=0.835$ established.

The discovery of the X-ray afterglow played a vital role in indirectly measuring the distance of the GRB. The growing number of measured redshifts solidified the extragalactic origin of GRBs, resolving the debate about their distance that culminated in the Great Debate of 1995, which pitted D. Lamb, advocating for a galactic origin \cite{lamb1995distance}, against B. Paczynski, supporting an extragalactic origin \cite{paczynski1991cosmological}. 

Thanks to the satellite Swift (Swift Gamma-Ray Burst Explorer) it has been identified the two high-redshift GRBs up to date, GRB 090423 and 090429B, with an estimated redshift respectively of $8.2$ and $9.4$ \citep{campana2022finding, gomboc2012unveiling}. Swift was also instrumental in observing the afterglow of a short GRB for the first time. Prior to 2005, the X-ray and optical counterparts for several long GRBs had been discovered, but never for a short-duration GRB. The observation of the first short GRB afterglow enabled the first redshift measurement for this type of GRB \cite{gehrels2005short}.

Since then, Swift has played a vital role in obtaining evidence about the progenitors of GRBs. It has identified the host galaxies of both long and short GRBs, allowing for the study of the interaction of the blast with its surroundings. It has been observed that long GRBs are typically found in regions with high star formation rates, while short GRBs are localized in galaxies with low star formation rates. These findings have given rise to the two progenitor models discussed in the next section.

\subsection{Progenitors}\label{sec:progenitors}

The identification of the progenitors, or stellar systems that give rise to GRBs, has been the subject of extensive research and investigation. The release of a massive amount of energy must be a common denominator. The duration of the burst is critical in understanding the potential progenitor scenarios. 

The first link between GRB and supernova was made in 1998 with GRB 980425, which was found in the same location as the supernova SN1998bw. 
Within a day of the GRB detection, the SN exploded. 
Long-duration GRBs, which typically last several seconds to minutes, are commonly associated with massive star core collapse. When the nuclear fuel in a massive star runs out, it collapses gravitationally, forming a central black hole or a rapidly spinning neutron star. The GRB is powered by the energy released during the collapse, and the associated supernova explosion may also be seen. In particular, SN Ic can produce GRBs since these are often seen in strongly star-forming, irregular or spiral galaxies, and their spectra show no indication of hydrogen or helium emission. But this is not always the case, for example, long GRBs 060614 and 060505 cannot be attributed to a massive star collapse. Moreover, in GRB 060614 was not found any SN emission but it is similar to a short GRB under many observational aspect \citep{valle2006enigmatic}.

Short-duration GRBs are frequently attributed to binary systems and compact object mergers. Two compact objects, such as neutron stars or a neutron star and a black hole, are in a binary orbit in this scenario. The binary system's gradual inspiral due to gravitational wave emission culminates in a merger event, releasing a burst of energy. The observed short duration of these GRBs is consistent with the timescales of inspiral and merger. On August 17, 2017, compelling observational evidence was obtained through the simultaneous detection of a short gamma-ray burst (SGRB) and a gravitational signal that bore the signature of a neutron star merger \cite{abbott2017gw170817}.

Magnetars, which are highly magnetized neutron stars, have also been proposed as potential GRB progenitors. Magnetars' extreme magnetic fields can release enormous amounts of energy, potentially driving a GRB. Magnetar-powered burst durations can range from short to long depending on the specific mechanisms involved. Magnetar flares or magnetic field instabilities can cause short-duration bursts, while continuous energy release processes can cause long-duration bursts.

\paragraph{GRB model}

\begin{figure}[!htb]
\centering
  \includegraphics[width=1\textwidth]{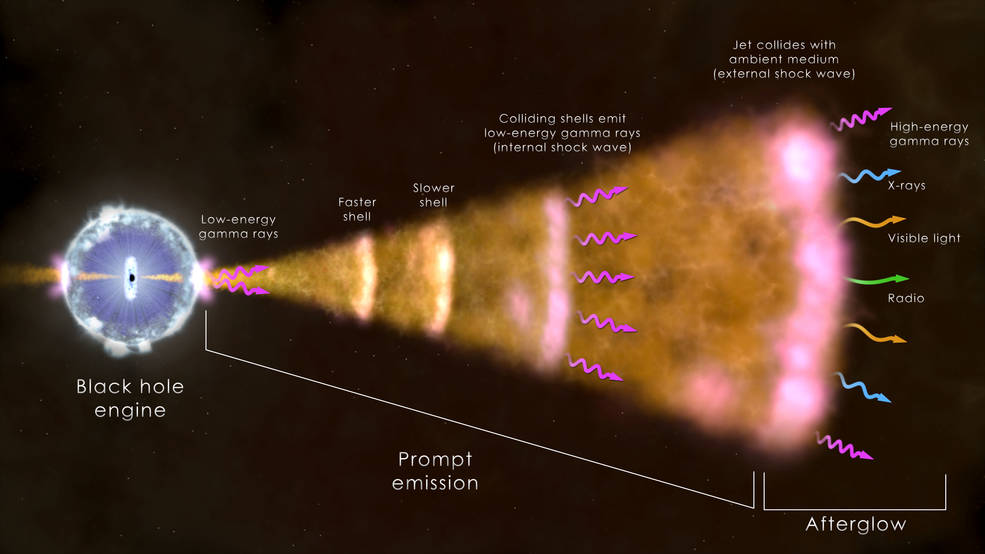}
\caption{\label{fig:grb_model}
The components of the most common model of gamma-ray burst. The core of a big star (left) has imploded, generating a black hole that blasts a jet out into space at near the speed of light. Radiation from hot ionized gas in the region of the newborn black hole, collisions between shells of fast-moving gas inside the jet, and the leading edge of the jet as it sweeps up and interacts with its surroundings all produce radiation across the spectrum. Credits: NASA's Goddard Space Flight Center.}
\end{figure}

GRB modeling seeks to comprehend the complex processes that occur during these high-energy astrophysical events. Investigating the phenomenon of internal shock absorbers is a prominent aspect of GRB modeling \citep{piran1999gamma}. These shock absorbers are thought to form as a result of relativistic outflows produced by the central engine of a collapsing massive star or merging compact objects. Variations in velocity or energy content of these outflows as they travel through the surrounding medium can cause collisions within the flow itself. Internal shocks caused by these collisions release massive amounts of energy in the form of gamma-ray radiation. Internal shock absorber research reveals insights into the dynamics of outflows, Figure \ref{fig:grb_model}, shedding light on the mechanisms behind the intense burst of gamma-ray emission.

Furthermore, the variability in the emitted gamma-ray radiation is critical in understanding the operation of the central engine that powers GRBs \citep{piran1997variability}. The rapid fluctuations and temporal patterns in the emitted radiation provide important information about the physical processes at the heart of these cataclysmic events. Variability in gamma-ray emission can reveal the behavior of the compact object, the geometry of the emission region, and the accretion processes that occur near the central engine. Astronomers can infer information about the nature of the progenitor star, the magnetic fields involved, and the efficiency of energy conversion within the central engine by analyzing these temporal features.

\begin{figure}[!htb] 
\centering
  \includegraphics[width=1\textwidth]{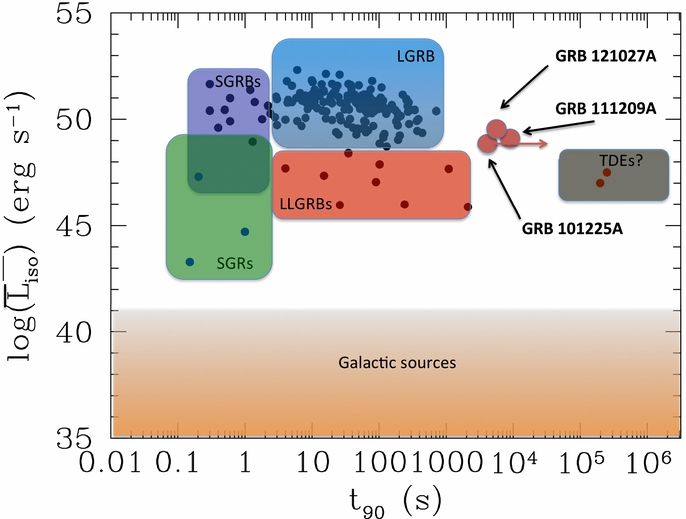}
\caption{\label{fig:ultralong_grb}
 The parameter space for transients in the gamma-ray sky is depicted, illustrating the burst duration and the approximate average luminosity throughout that duration. At lower luminosities, there exist numerous Galactic sources. At higher luminosities, the chart displays outbursts from soft-gamma repeaters (SGRs) within our own Galaxy, along with extragalactic transients such as long and short-duration gamma-ray bursts (LGRBs and SGRBs), and the probable population of low-luminosity gamma-ray bursts (LLGRBs). Additionally, are indicated very long transients believed to originate from tidal disruption events (TDEs). Notably, the bursts GRB 101225A, GRB 111209A, and GRB 121027A clearly stand out as outliers within any of the aforementioned categories. \cite{levan2013new} \copyright AAS. reproduced with permission.}
\end{figure}

\paragraph{Ultra long-GRB}
In the literature, there are studies focusing on faint events, such as the Low-Luminosity GRB (LLGRB) \cite{levan2013new}, as well as events with duration from hundreds to thousands of seconds, thus comparable to the duration of the Fermi orbit, the so-called ultra-long GRBs \cite{levan2013new, gendre2019can, dagoneau2020ultra, boer2015ultra}. See Figure \ref{fig:ultralong_grb}.

Currently, there is no consensus on a clear distinction between long and ultra-long GRBs, although the latter may have different progenitors, such as blue supergiants with a low metallicity (GRB 111209A \cite{gendre2013ultra, stratta2013ultra}) or magnetars \cite{zou2019magnetar, gompertz2017magnetars}. 
For GRB 101225A, also known as \emph{Christmas burst}, it has been proposed that the emission might be originated by the tidal shredding of an asteroid by a neutron star, or a burst in coincidence with a supernova inside a dense envelope. For GRB 110328A it has been proposed that the emission might be originated by tidal disruption event caused by a star falling in a supermassive black hole \cite{levan2013new}.
Estimating the burst duration using classical methods like T90 is challenging because the duration of the burst depends on the observing band and the prompt phase could spread across thousands of seconds and therefore including gaps in signal, due to, for example, passages of the satellites through the South Atlantic Anomaly or around the Poles, where the particle background is too high to allow normal operation of X-ray and gamma-ray instruments, or due to reorientation of the satellite because of download of the data over a ground station. These factors make the estimation of burst duration more complex and require careful consideration in the analysis \cite{gendre2013ultra}, in particular in the estimate of the background. As an example, we can refer to the estimated duration of three ultra-long GRBs discussed in Levan et al. (2013) \cite{levan2013new}: GRB 101225A, with an estimated prompt emission duration exceeding 7000s, GRB 111209A about 10000s and GRB 121027A about 6000s. In \cite{gruber2011fermi} the ultra-long GRB 091024 has an estimated prompt duration of about 1020s.

\subsection{Open questions}\label{sec:open_quest}

There are several open questions in the field of GRB research that continue to intrigue scientists and drive ongoing investigations \cite{zhang2011open, gomboc2012unveiling}. Some of the key open questions include:
\begin{itemize}
    \item[1] Progenitor Systems and Classification: The nature of GRB progenitor systems remains a topic of intense study. Identifying the precise stellar systems or compact objects responsible for GRBs is crucial for understanding their origins. There are competing theories suggesting progenitors such as massive star core collapse, compact binary mergers, or other exotic scenarios. This is in line with the GRB classification task based on spectral and temporal features. Traditionally the analysis of GRBs was performed in the T90–Hardness Ratio (HR) space that suggested that there are two classes of GRBs (short and long): are they associated to SN and coalescence of compact object? Are there more than two classes? Is there any parameter (feature) space in which the GRBs are completely separeted?    
    \item[2] Prompt Emission Mechanisms: The exact processes responsible for the intense prompt emission observed in GRBs are still not well established. Understanding how high-energy particles are accelerated and produce the observed radiation is a fundamental question. Models involving synchrotron radiation, inverse Compton scattering, and other emission mechanisms are being explored, but the details of these processes remain to be fully elucidated. Is it possible to involve the form of the prompt lightcurve to get new insight? Can the GRBs be analysed considering together temporal and spectral properties?
    \item[3] Afterglow Physics: The afterglow phase following the prompt emission provides valuable information about the environment surrounding the GRB and the mechanisms driving the long-lasting emission across various wavelengths. Investigating the emission properties, energy injection processes, and the role of magnetic fields in the afterglow phase is an ongoing research endeavor.
    \item[4] Multi-messenger Connections: Exploring the connections between GRBs and other astrophysical messengers, such as gravitational waves, neutrinos, and cosmic rays, is an active area of investigation. Detecting and analyzing coincident signals from multiple messengers can provide unique insights into the physical processes and environments associated with GRBs.
    \item[5] Cosmological setting: GRBs are cosmological events. Can GRBs be used to constrain cosmological parameters? The observed time dilation in GRB light curves, where distant bursts (high redshift) appear to have longer durations, is consistent with the stretching of time due to the expansion of the universe. By analyzing the time dilation effect, researchers can refine our understanding of the Hubble constant, which quantifies the rate of expansion of the universe.
\end{itemize}

These open questions highlight the evolving nature of GRB research. Progress in these areas requires the concerted efforts of observational studies, theoretical modeling, and advancements in instrumentation and data analysis techniques. For instance, in \cite{dainotti2016fundamental}, an analysis of 176 Swift GRBs with afterglow plateaus unveils a new three-parameter correlation and identifies a GRB "fundamental plane". This study provides insights into the underlying physical processes of long GRBs, connecting prompt and afterglows parameters. 
 
Continued exploration of these questions promises to deepen our understanding of the enigmatic phenomenon of gamma-ray bursts.

\section{Telescopes for GRB}\label{sec:telescope} 

The first (accidental!) detection of GRBs, specifically GRB 670702, took place in 1967 through the Vela satellite network. Originally designed for monitoring nuclear activity from the USSR during the Cold War, this network of American military satellites observed a burst of gamma rays from a location outside the field of view of the Earth, ruling out nuclear tests as the cause \cite{klebesadel1973observations}.

Despite several missions conducted by different countries and the establishment of the first Inter-Planetary Network (IPN) over the following three decades, no significant breakthroughs were achieved in understanding GRBs. The challenges included the rapid fading of GRB sources, unpredictability due to the absence of recurrent temporal patterns, and difficulties in gamma ray focusing, leading to unclear signals and imprecise positions, all complicating the study of these high-energy phenomena.

\subsection{BATSE}

\begin{figure}[!htb]
\centering
  \includegraphics[width=0.75\textwidth]{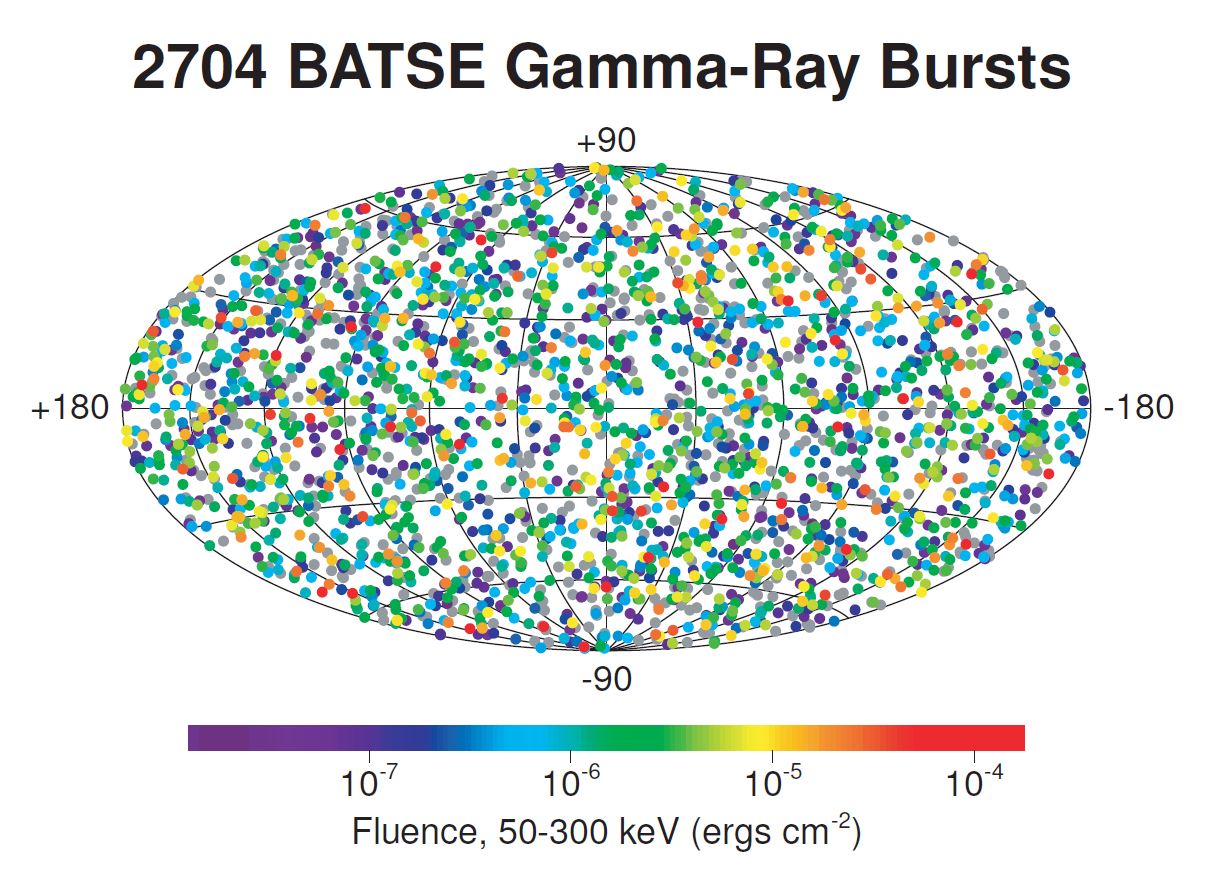}
\caption{\label{fig:loc_batse_grb} The locations of 2704 gamma-ray bursts identified by the BATSE instrument over a nine-year period. The distribution of sources is isotropic. Image taken from \cite{BatseLoc}.}
\end{figure}

The BATSE (Burst And Transient Source Experiment) telescope \cite{band1993batse}, operating from 1991 to 2000 aboard the Compton Gamma-Ray Observatory (CGRO), marked a significant milestone in understanding GRBs. Before BATSE, the origin of GRBs was speculated to be galactic; however, the telescope's observation of thousands of GRBs showcased an isotropic distribution (see Figure \ref{fig:loc_batse_grb}). 
This isotropic distribution contradicted the galactic origin hypothesis, as it was expected that GRBs would be more concentrated along the Milky Way's disk rather than distributed uniformly in all directions.
With its eight detector modules covering different directions and advanced technologies, BATSE detected over 2700 GRBs during its operational period (Figure \ref{fig:loc_batse_grb}), enabling a comprehensive study of these phenomena.
Another significant discovery made by BATSE was the differentiation between short and long-duration GRBs (Figure \ref{fig:t90batse}). This breakthrough became possible due to the abundance of GRBs detected by BATSE, allowing for a robust statistical analysis of these phenomena.

Utilizing crystal scintillators, made of sodium iodide (NaI), coupled with photomultiplier tubes, BATSE covered an energy range of 20 keV to 8 MeV to study the GRB emission spectrum. 
GRBs were detected by the computer on board Compton-BATSE through the monitoring of count rates from each of the eight detectors. The primary energy range used for forming count rates was 50-300 keV, but additional energy channels were considered for setting alternative trigger criteria, including 25-50 keV, 50-100 keV, 100-300 keV, and >300 keV \citep{paciesas1999fourth}. Background count rates were evaluated and updated every 17.408 s. The count rates were primarily monitored across the 1024 ms timescale, with the 64 ms and 256 ms timescales being enabled only during specific periods specified in the history of triggers in the fourth BATSE catalog \cite{paciesas1999fourth}. To trigger an event, the on-board computer issued a signal when at least two detectors simultaneously recorded a count excess that exceeded an adjustable threshold. During BATSE operations, the most commonly utilized threshold value was 5.5$\sigma$, although various significance thresholds were tested. In certain cases, higher threshold values (up to 26$\sigma$) were also tested, particularly for count rates with a time scale of 64 ms. This approach allowed for a flexible and adaptive detection strategy, optimizing the sensitivity to GRBs while minimizing false detections.

\subsection{BeppoSAX}\label{sec:bepposax}
Another significant advancement occurred with the launch of BeppoSAX (Satellite italiano per Astronomia X, Beppo in honor of
Giuseppe Occhialini) \cite{boella1997bepposax} in 1996, a collaborative effort between the Netherlands and Italy. 

The scientific payload of the satellite satellite is composed by four narrow field X-ray telescopes (NFI, 0.1 - 300 keV) and two Wide Field Cameras (WFC, 2 - 26 keV). More specifically the NFI are: MECS (Medium Energy Concentrator Spectrometers) 1.3 - 10 keV, LECS (Low Energy Concentrator Spectrometer) 0.1 - 10 keV, HPGSPC (High Pressure Gas Scintillation Proportional Counter) 4 - 120 keV, PDS (Phoswich Detector System) 15 - 300 keV. The Beppo-SAX Gamma-Ray Burst Monitor (GRBM) is formed by the four lateral active shields of the PDS which has an energy range 40 - 700 keV, with a temporal resolution of about 1 ms.

This satellite played a pivotal role in the field of GRB research. Apart from its ability to observe a broad energy range (0.1-700 keV), BeppoSAX made a crucial contribution by enabling observations in the X-ray energy range. Its two low-energy telescopes allowed for the first-ever measurement of the X-ray counterpart of a GRB and the first redshift measurement a few months later as mentioned in Section \ref{sec:afterglow}. Moreover, it found the first GRB (GRB980425) which was associated with a Supernova (SN1998bw) due to the same location of the emission in the SN's optical and radio bands \citep{costa2011gamma}.

The trigger for GRBM operates on the signals detected between the Lower Level Threshold (LLT), nominally ranging 20 - 90 keV, and Upper Level Threshold (ULT), nominally ranging 200 - 700 keV \cite{feroci1997flight}. The time resolution is 7.8125 ms and the estimated background is continuously computed on a Long Integration Time (LIT), which can be adjusted within the range of 8 to 128 s. Meanwhile, the counts within a Short Integration Time (SIT), adjustable between 7.8125 ms and 4 seconds, are compared to the moving average. If the counts exceed a certain threshold, defined by $n$ times the Poissonian standard deviation (where $n$ can be 4, 8, or 16), the trigger condition is satisfied for the corresponding shield.
For the GRBM trigger to be activated, the same trigger condition must be simultaneously met for at least two shields. This requirement ensures that the trigger is only activated when multiple shields detect signals above the specified threshold, providing a more robust indication of a potential GRB event. At the end of the first year of operation the trigger parameters were set to LLT$=32.8$ keV, ULT$=604$ keV, LTI$=128$ s, SIT$=1$ s and $n=4\sigma$.

\subsection{Swift}\label{sec:swift}

The Swift Gamma-Ray Burst Explorer \cite{gehrels2004swift}, operational since 2004, is a significant mission for observing GRBs. Comprising three main instriments, namely the Burst Alert Telescope (BAT, 15-150 keV), the X-Ray Telescope (XRT, 0.2-10 keV), and the UltraViolet/Optical Telescope (UVOT, 170-600 nm). Swift's goal is to quickly identify GRBs and undertake multi-wavelength follow-up studies of both the bursts and their afterglows by combining the capabilities of all three instruments. First, BAT to initially detect and locate a burst and then the other two instruments can effective observe the GRB afterglows from their early stages. The spacecraft swiftly (as the name telescope promise) reposition itself in the correct direction within a minute. Additionally, the spacecraft can be repositioned based on position information received from other satellites via the Tracking and Data Relay Satellite System, an American network of communication satellites and ground-based stations. This approach ensures timely and comprehensive data collection for detailed study and analysis.

BAT on Swift employs two main types of triggers: short and long rate triggers. Short triggers have timescales ranging from 4 ms to 64 ms, while long triggers span from 64 ms to 64 s. Both short and long triggers can utilize criteria with a single background sample, following the traditional "one-sided" trigger approach. However, long triggers can also incorporate criteria with two background samples, known as "bracketed" triggers, which a polynomial fit (constant, linear, quadratic) is performed to remove background trends \cite{mclean2004setting}. Both short and long triggers can be configured to select energy ranges of 15-25 keV, 15-50 keV, 25-100 keV, and 50-350 keV. The BAT system includes over 180 short triggering criteria and around 500 long triggering criteria that are used to detect gamma-ray bursts. 

Count rates are evaluated over timescales of 4, 8, 16, 32, and 64 ms, compared against the corresponding estimated background rate determined using the long trigger algorithm. The significance of the short trigger is computed using the following formula:

$$S = (C_{i,k} - B_{i}2^{k-10})^2 / (C_{i,k} + \sigma_{min}^2)$$

where, the label $i$ corresponds to one of the 36 region-energy combinations, $B_i$ represents the expected background counts over 1024 ms, and $C_{i,k}$ denotes the maximum count observed over the timescale previously mentioned. The variable $\sigma^2_{min}$ prevents the significance value from becoming too large when counts are low. A trigger is declared whenever $S$ exceeds a threshold value. This approach allows for the definition of 180 different trigger criteria \cite{fenimore2003trigger}.

For long trigger the equation is:

$$S = (C_{fore} - C_{back})^2 / (\sigma_{fore}^2 + \sigma_{back}^2 + \sigma_{min}^2 + \beta^2\sigma_{back}^2)$$

where $C_{fore}$ and $\sigma_{fore}$ are respectively the count rates and variance in the foreground period. $C_{back}$ and $\sigma_{back}$ are respectively the count rates and variance in the background period which can precede the foreground section (extrapolation, the polynomial estimate the next count rates) or "bracket" it (interpolation, the polynomial estimate the count rates in the foreground period that is between the background period).

With high count rates, $C_{back}$ will be greater than the $\sigma^2$ terms and than the above equation reduce to:

$$S = \frac{1}{\beta^2} \Biggl(\frac{C_{fore} - C_{back}}{C_{back}}\Biggl)^2.$$

Each long rate trigger in the BAT system is controlled by approximately 30 commandable parameters. These parameters define factors such as background timing, foreground phase, polynomial fitting, variance thresholds, systematic noise levels, CPU usage control, and enable/disable settings. Adjusting these parameters allows for precise customization of the trigger operation, optimizing its performance for different scenarios and data characteristics.

\subsection{Fermi}
\begin{figure}[!htb] 
\centering
  \includegraphics[width=0.75\textwidth]{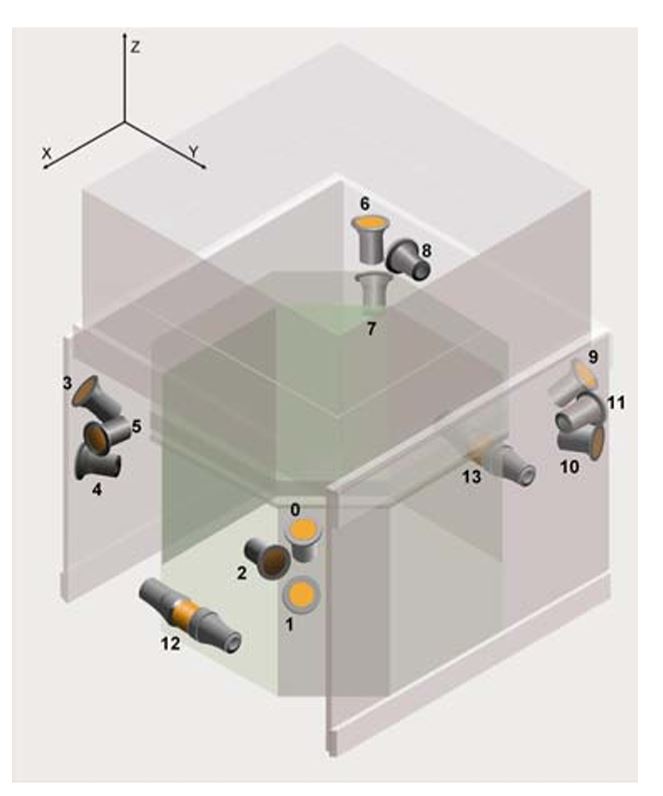}
\caption{\label{fig:loc_det_gbm}
 Locations and orientations of the GBM detectors. \cite{meegan2009fermi} \copyright AAS. Reproduced with permission.}
\end{figure}

The Fermi Gamma-Ray Space Telescope, launched in 2008, is a satellite observatory that covers a wide energy range from 8 keV to 300 GeV. It achieves this by employing two telescopes on board: the Large Area Telescope (LAT) and the Gamma-ray Burst Monitor (GBM). The LAT operates based on pair production, while the GBM utilizes 12 scintillators made of 12 NaI (sodium iodide) and 2 BGM (Bismuth Germanate), see Figure \ref{fig:loc_det_gbm}. Consequently, the GBM covers the lower energy range from 8 keV to approximately 40 MeV, while the LAT can detect gamma-rays within the interval of 20 MeV to over 300 GeV. 
Leveraging the high-energy capabilities of LAT, it became feasible to analyze the spectra of specific GRBs and unveil the presence of a power-law component in these spectra, as depicted in Figure \ref{fig:spectral}.

The detection process of transients by GBM involves a combination of count rates, energy ranges, and timescales. The on-board computer compares count rates from NaI and BGO scintillation detectors to an average background rate and its standard deviation, estimated using a sliding window approach over a parametrizable window, typically 17 seconds. To prevent transients from influencing the background rate, the most recent 4 seconds of data are excluded from the window.
GBM employs four distinct energy ranges (25-50 keV, 50-300 keV, $>$100 keV, and $>$300 keV) and can detect GRBs on timescales ranging from 16 ms to about 16 s (specifically $\{0.016 \times 2^k \mid k=0:10\}$). The are supported timescales for each energy range vary; for $>$100 keV, timescales go up to 4.096 s, while the $>$300 keV range supports shorter timescales, up to 128 ms.
The GBM trigger algorithm is designed to be highly sensitive, capable of detecting various GRBs, from short and hard bursts to long and soft bursts. By utilizing multiple energy ranges and timescales, the GBM can effectively detect GRBs emitting different amounts of radiation at various intervals.

In the catalogs of Fermi \cite{von2014second, bhat2016third, von2020fourth} it is reported the chronology of GBM's trigger operations on-board Fermi, showing that most algorithms working in the energy range of 25-50 keV and $>300$~keV were "disabled during most of the mission" to reduce the number of false positive and mitigate the effects of high solar activity on the Fermi-GBM false trigger rates and to prevent the on-board data storage devices from becoming saturated. To identify Terrestrial Gamma-ray Flashes (TGFs) four extra high-energy (2-40 MeV) trigger criteria with a timescale of 16 ms are specifically designed for the BGO detectors. To issue a trigger, there must be a simultaneous exceedance of the trigger threshold in at least two detectors. The total number of trigger criterion is 119, and the threshold ranges from $4.5$ to $8.0\sigma$ depending on the trigger criteria \citep{von2020fourth}.

Fermi/GBM data are collected \emph{daily}, continuously recording and sending data to the ground, \emph{trigger}, data products whenever a trigger has been detected, and \emph{burst}, for data triggers classified as gamma-ray bursts.
These data can be found in different data products to meet various research needs and scenarios:
\begin{itemize}
    \item Time-Tagged Event (TTE) Data. Event data (the arrival time and energy of individual gamma-ray photons) in 128 energy channels for each detector with a time precision of 2 $\mu s$. 
    \item CTIME. It provides count information accumulated at regular intervals, either every 0.256 seconds for daily data or every 0.064 seconds for trigger or burst data. It's segmented into eight energy channels for each of the 14 detectors.
    \item CSPEC. These data contain counts aggregated over longer intervals, specifically every 4.096 seconds for daily data and every 1.024 seconds for trigger or burst data. Like TTE, it comprises 128 energy channels for each of the 14 detectors.
    \item POSHIST. Fermi's position and attitude have been recorded in POSHIST data. These data are required for calculating Detector Response Matrices (DRMs). This file is only available for daily data.
    \item HEALPix. HEALPix files provide comprehensive details about the localization of GBM-localized GRBs, presented as HEALPix arrays. These files also include information about the Geocenter's equatorial location, the position of the sun, and the equatorial pointing directions of each GBM detector at the time of the GRB. Additionally, a PNG skymap is included. These files are only available for burst data.
    \item RSP. DRMs are provided in both 8 and 128 energy channel versions for all 14 detectors. This file is available for all burst data but may not be generated for trigger data.
\end{itemize}

In particular, CSPEC and POSHIST files are employed in the background estimation described in Chapter \ref{chap:bkg}.

\subsection{AGILE}
AGILE (Astro-rivelatore Gamma an Immagini LEggero) \citep{tavani2009agile} is an Italian Space Agency mission. It was launched in 2007 in an equatorial orbit with very low particle background to observe and investigate astronomical gamma-ray sources. Its instrumentation includes a gamma-ray imaging detector (30 MeV - 50 GeV), a hard X-ray imager (18-60 keV), a calorimeter (350 keV - 100 MeV) and an anticoincidence system. Notably, the AGILE mission stands out among astronomy space missions due to its ability to provide very good imaging capabilities simultaneously in the 30 MeV-50 GeV and 18-60 keV energy ranges, offering a large wide field of view in high-energy astrophysics space missions. Besides its power and cost-effectiveness, AGILE's distinguishing feature is its real-time identification and pinpointing of transient gamma-ray sources such as GRBs and AGN flares. Its agile pointing enables quick responses to alarms from other space and ground-based observatories, facilitating multi-wavelength follow-up investigations of gamma-ray sources.

\subsection{INTEGRAL}\label{sec:integral}

The INTEGRAL observatory \citep{winkler2003integral}, managed by ESA, is designed for detailed gamma-ray source analysis in the energy range of 15 keV to 10 MeV. It employs two primary gamma-ray instruments: the SPI spectrometer optimized for precise spectroscopy and the IBIS imager for high-resolution imaging. Complementary X-ray and optical monitors (JEM-X and OMC) enhance the observatory's capabilities. Launched in October 2002, INTEGRAL has been actively observing the sky, accumulating substantial data for high-energy studies. Although INTEGRAL was not primarily designed for GRB observation, its broad field of view enables detection of a GRB within its view roughly every 1-2 months \citep{foley2008global}. Additionally, the SPI instrument's Anticoincidence Shield (SPI-ACS) efficiently detects around one GRB per day, albeit without spatial and spectral details \citep{ESA2023INTEGRAL}.

\subsection{HETE-2}

The High Energy Transient Explorer Mission (HETE-2, \cite{ricker2003high}) is an international collaboration that uses soft and medium X/gamma-rays to detect and localize GRBs. The HETE-2 mission  is equipped with three distinct scientific instruments: FREGATE is a collection of wide-field gamma-ray spectrometers (6-400 keV), WXM is a wide-field X-ray monitor (2-25 keV), and SXC is a collection of soft X-ray cameras (1.3-14 keV). HETE-2 was launched in October 2000.

\subsection{Ground based observatory}
So far, the telescopes shown have been satellite space missions, although terrestrial tests and proposals were carried out.

MAGIC (Major Atmospheric Gamma Imaging Cherenkov) \citep{baixeras2003magic} is another ground-based gamma-ray observatory on La Palma, Canary Islands. It employs two telescopes, each with a large mirror (17 m in diameter) and a high-speed camera, to detect very high-energy gamma rays via Cherenkov radiation. MAGIC has been used to detect and study the topology (spatial) and spectral characteristics of gamma-ray sources such as AGN, pulsars, and GRBs. Its technology and sensitivity have allowed for in-depth investigations into the physical processes underlying these high-energy (> 50 GeV) phenomena.

HESS (High Energy Stereoscopic System) \citep{vasileiadis2005hess, aharonian2023hess, hess2023page} is a ground-based gamma-ray observatory located in Namibia. It consists of an array of five telescopes, four with mirror diameters of 12 m and one with a mirror diameter of 28 m, designed to detect very high-energy gamma rays (from 10 GeV to 10 of TeV) from celestial sources. HESS has made significant contributions to the study of gamma-ray sources, including the discovery of numerous galactic and extragalactic gamma-ray emitters. It has provided valuable insights into the nature of cosmic accelerators, such as supernova remnants and active galactic nuclei.

The VERITAS (Very Energetic Radiation Imaging Telescope Array System) Observatory \citep{holder2008status, veritas2022deep}, located at the Fred Lawrence Whipple Observatory in southern Arizona, USA, is composed of four imaging atmospheric Cherenkov telescopes with a diameter of 12 m each. These telescopes are designed to detect very high-energy gamma rays above 100 GeV, with a field of view (FoV) diameter of 3.5 degrees. The observatory has significantly contributed to our understanding of the universe by discovering new sources of gamma rays and conducting studies of active galactic nuclei.

The CTA (Cherenkov Telescope Array) \citep{cta2011design} is a large-scale future observatory that aims to revolutionize gamma-ray astronomy. It will be made up of multiple Cherenkov telescopes of various sizes and types that will be strategically placed in both the northern and southern hemispheres. CTA will have greater sensitivity and coverage than current observatories, allowing for unprecedented observations of gamma-ray sources across a wide energy range (expected to be less than 100 GeV and greater than 200 TeV). CTA hopes to solve mysteries surrounding cosmic particle acceleration, dark matter, and other fundamental astrophysical phenomena with its high-performance instruments.

\subsection{HERMES}\label{sec:HERMES_mission}

Present instrumentation dedicated to GRBs and cosmic transients has been launched during the 2010s. There is no guarantee that it will continue to operate beyond the mid-2020s. For this reason, several proposals to NASA and ESA have been already submitted to select the successors of these instruments. Some notable proposals include: THESEUS \citep{stratta2018theseus}, SVOM \citep{bernardini2021svom}, e-ASTROGAM \citep{e2017astrogam}, AMEGO-X \citep{caputo2022amego} and several CubeSats missions \citep{bloser2022cubesats, fiore2021distributed} (BlackCAT \cite{falcone2022soft},
BurstCUBE \cite{perkins2020burstcube},
COMCUBE \cite{laviron2021comcube},
COMPOL \cite{yang2020feasibility},
CUSP \cite{fabiani2022cusp},
EIRSAT-1 \cite{murphy2018eirsat},
GRID \cite{wen2019grid},
 GRBAlpha \cite{pal2020grbalpha},
 GTM \cite{chang2022gamma},
 IGOSat \cite{phan2018igosat},
IMPRESS \cite{setterberg2022geant4},
LECX \cite{braga2020lecx},
LIGHT-1 \cite{almazrouei2021complete},
MAMBO \cite{bloser2022mini},
MeVCube \cite{lucchetta2022introducing},
Min-XSS1 \cite{mason2016miniature},
Min-XSS2 \cite{mason2020minxss},
SOCRATES \cite{delange2016sensor},
VZLUSAT-2 \cite{ripa2022early}).

The HERMES (High Energy Rapid Modular Ensemble of Satellites) Pathfinder project aims to validate the HERMES concepts through an in-orbit demonstration, involving the detection and localization of GRBs using six satellite units (3U nano-satellites) \cite{fiore2020hermes,fiore2021}. This initiative is driven by the need to provide a fast-track, cost-effective solution bridging the gap between current X-ray monitors and the next generation. This experiment's success is pivotal in shaping the final design of the HERMES Full Constellation, which aims to monitor the entire sky and provide precise localization for most GRBs, in general for cosmic high-energy transients. The HERMES Pathfinder will serve the following key purposes:
	
\begin{enumerate}
	\item To validate the overall concept, conducting a comprehensive analysis of both the statistical and systematic uncertainties associated with detection and localization. This assessment will help validate and enhance the design of the payload and service modules, ensuring greater reliability for the proposed full constellation.
		
	\item To demonstrate the feasibility of achieving highly accurate timing within the unexplored temporal window between fractions of a microsecond and 1 millisecond, using detectors with relatively small collecting areas. The timing goal is to achieve a resolution of approximately 300 nanoseconds, which is about seven times better than Fermi/GBM, all while hosting the experiment on a nano-satellite.
		
	\item To investigate the uncertainties related to combining signals from different detectors and improving statistics for high-resolution time series.
\end{enumerate}
	
In pursuit of these objectives, the HERMES Pathfinder is required to meet specific mission requirements:
	
\begin{itemize}
	\item MIS-REQ-1: Consistently detect GRBs with a peak flux of $\ge$0.4-0.5 ph/s/$cm^2$ in the 50-300 keV band.
		
	\item MIS-REQ-2: Simultaneously detect 40 or more long GRBs and 8 or more short GRBs in at least 3 units, with an efficiency of $\ge$40-50\% in each unit. This will enable the assessment of GRB positions through the analysis of delay times in signal arrival on different detectors.
\end{itemize}
	
The accuracy of position determination is based on the study of the delay time of arrival of the signal to different detectors on the nano-satellites in low Earth orbits. The focus here is on optimizing the time accuracy to achieve an overall accuracy of a fraction of a microsecond. The accuracy depends on several factors, including the average baseline, uncertainty in signal delay times obtained through methods like cross-correlation functions, temporal characteristics of GRBs, their statistics, and any potential systematic errors. Meeting the specified GRB numbers in MIS-REQ-2 is essential for studying the statistical and systematic uncertainties related to detection and localization, as well as understanding the challenges tied to combining signals from diverse detectors to improve statistical analyses.

\begin{figure}[!htb]
\centering
  \includegraphics[width=0.85\textwidth]{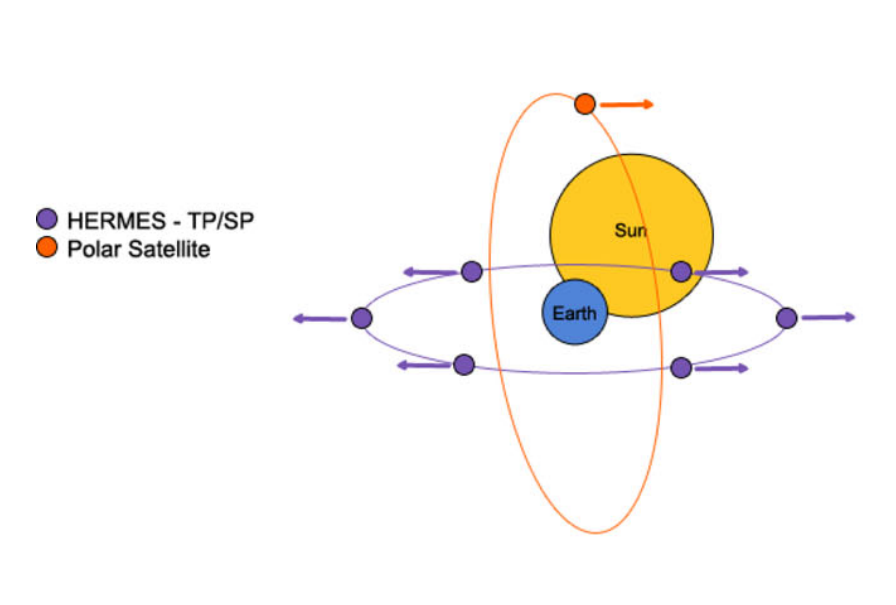}
\caption{\label{fig:loc_det_hermes}
A possible orbital configuration of six HERMES-SP CubeSats and SpIRIT. Image source: \cite{thomas2023localisation}.}
\end{figure}

To further this mission, a technological and scientific pathfinder known as HERMES-TP (funded by ASI) and HERMES-SP (funded by the European Commission) is in preparation. The primary aim is to prove the concept of detecting and localizing GRBs using miniaturized instrumentation hosted by nano-satellites. The initial phase is expected to involve launching the first six HERMES Pathfinder spacecraft into low-Earth, near-equatorial orbits during 2024, as illustrated in Figure \ref{fig:loc_det_hermes}.



A seventh payload unit identical to those hosted by HERMES Pathfinder will be hosted by SpIRIT \cite{auchettl2022spirit}, an Australia-Italy nano-satellite mission developed by a consortium led by the University of Melbourne and scheduled to launch in late 2023. SpIRIT will be the only satellite among the HERMES Pathfinder constellation to be launched into a polar orbit, improving the localization capability of the whole constellation \citep{thomas2023localisation}.
The HERMES Pathfinder and SpIRIT payload is a small yet innovative "siswich" detector providing broad-band energy coverage (few keV - $1$~MeV) and very good temporal resolution (a few hundreds ns) \cite{fuschino2019hermes, evangelista2020scientific,fiore2022,evangelista2022}.

\begin{figure}[!htb]
\centering
\begin{subfigure}{.5\textwidth}
  \centering
  \includegraphics[width=1\linewidth]{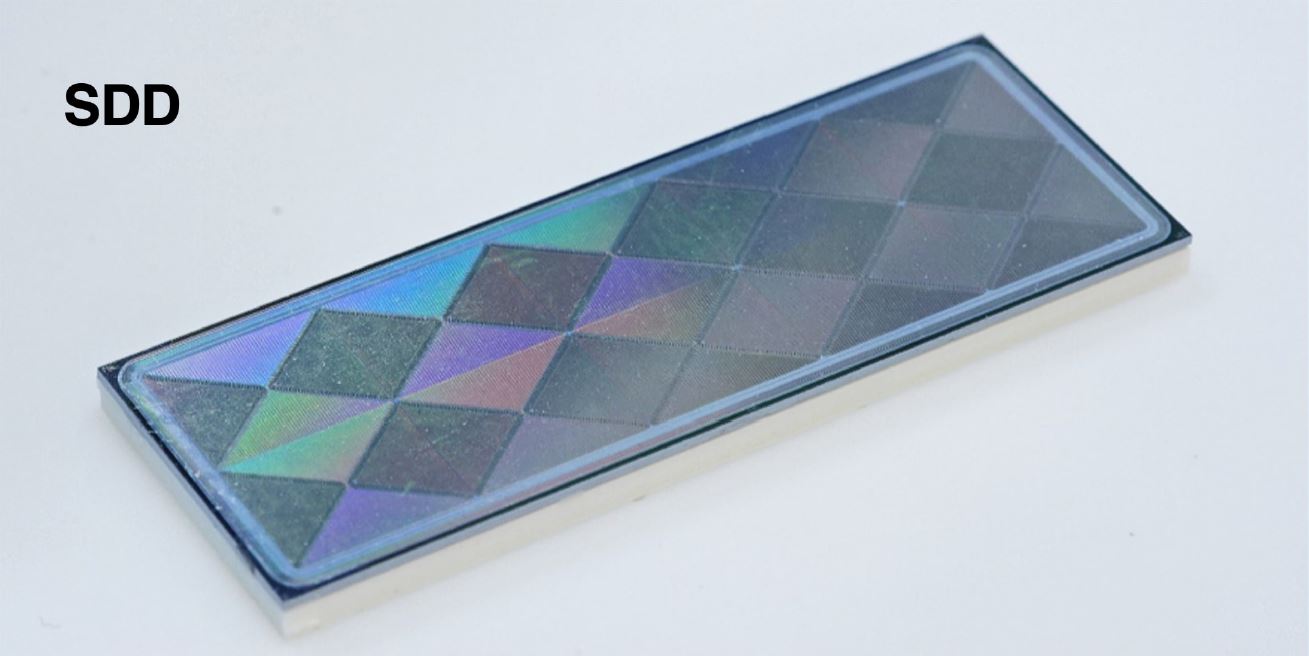}
  \caption{Silicon Drift Detectors (SDD).}
\end{subfigure}
\qquad
\begin{subfigure}{.5\textwidth}
  \centering
  \includegraphics[width=1\textwidth]{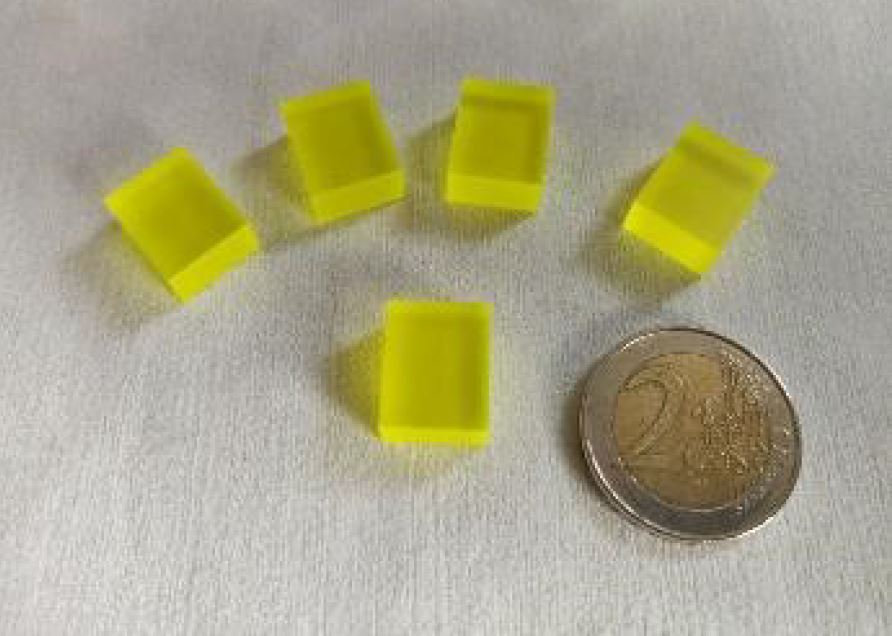}
  \caption{Cerium-doped Gadolinium Aluminium Gallium Garnet (GAGG) crystal.}
\end{subfigure}
\caption{\label{fig:hermes_detector}
The SDD is formed by the n-side, shown in (a), which is made up of $2 \times 5$ cells, and the p-side, which is made up of 5 cells. The latter will be coupled with the GAGG crystal, shown in (b), with the remaining faces wrapped in a reflective material. Images taken from the PhD thesis \cite{phddellacasa}.}
\end{figure}

\begin{figure}[!htb]
\centering
\begin{subfigure}{.5\textwidth}
  \centering
  \includegraphics[width=1\linewidth]{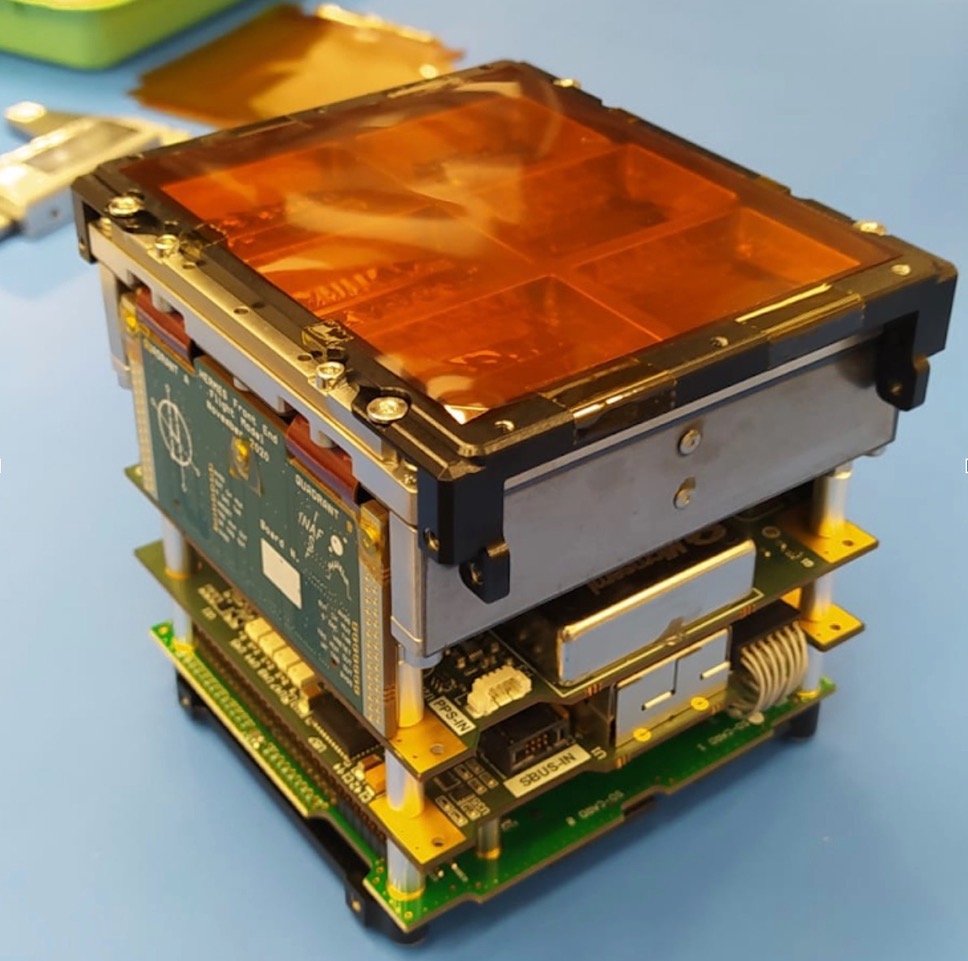}
  \caption{The HERMES Pathfinder FM1/SpIRIT nanosatellite's detector unit.}
\end{subfigure}
\qquad
\begin{subfigure}{0.75\textwidth}
  \centering
  \includegraphics[width=1\textwidth]{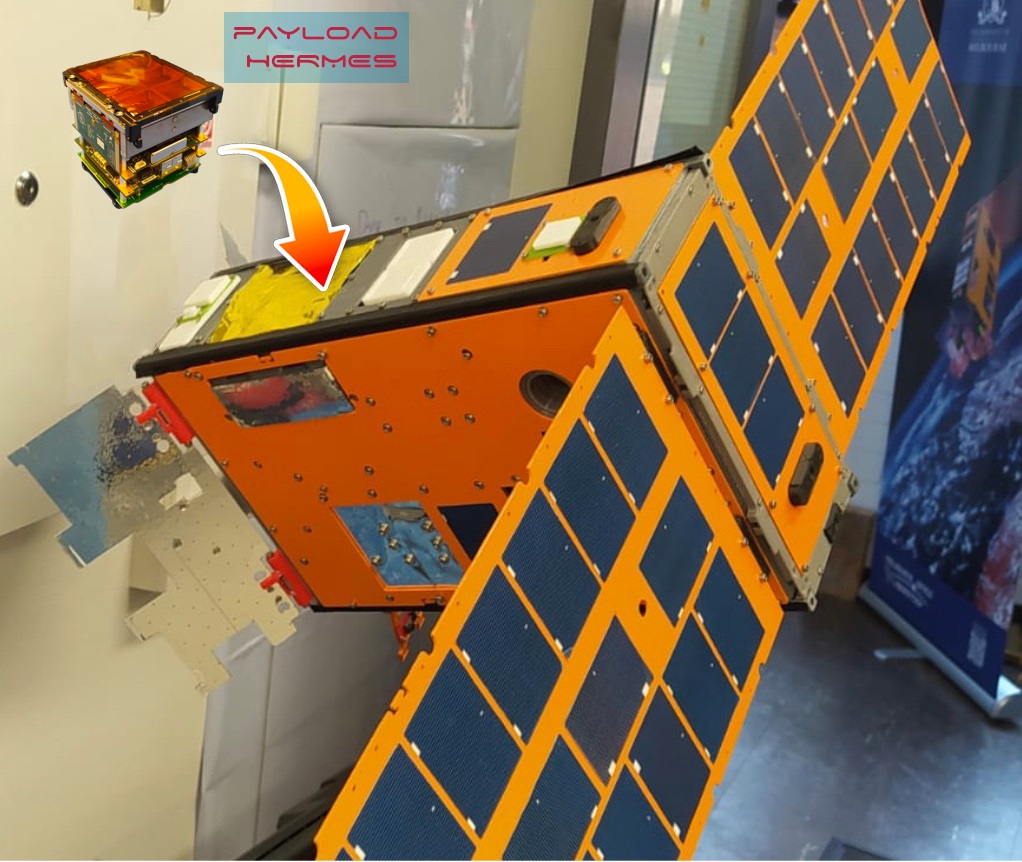}
  \caption{SpIRIT mockup.}
\end{subfigure}
\caption{\label{fig:hermes_detector2}
A detector unit is made up of 12 siswitch detectors. Credit source: \cite{HermesSite}.}
\end{figure}

In a siswich detector configuration, thin silicon detectors are connected to a scintillator crystal. When soft X-ray photons interact with the detector, they are absorbed within the silicon material. On the other hand, hard X-ray and gamma photons possess enough energy to penetrate the 450 $\mu m$ thick silicon detectors and reach the scintillator crystal, where they are subsequently absorbed. Inside the scintillator, these high-energy photons undergo a conversion process and emit visible light. This light is in turn absorbed by the active silicon slab,
resulting in an ionization signal proportional to the amount of light produced in the crystal. The discrimination between the two signals, soft X-rays and high-energy photons, is achieved through a segmented design. A single scintillator crystal is coupled to two separate silicon detectors. Events detected by only one silicon detector are more likely associated with soft X-rays, while events detected simultaneously in multiple adjacent detectors are attributed to the light produced in the scintillator by hard X-ray and gamma photons. The specific scintillation crystal chosen for the HERMES detector is GAGG:Ce (Gd3Al2Ga3O12:Ce), which is a Cerium-doped Gadolinium Aluminium Gallium Garnet crystal, Figure \ref{fig:hermes_detector}.
Twelve switches are mounted on one side of the detector unit before being integrated into the satellite case, as shown in Figure \ref{fig:hermes_detector2}.

\section{Background estimation for space satellites}
GRBs manifest as transient increases in the count rates of detectors. The activity of these phenomena appear as unexpected, and not explainable in terms of background or any other known sources. Any automated procedure for detecting GRBs is generally concerned with searching the time series of the observations for statistically significant excesses in photon counts, relative to a reference background estimate in the absence of $\gamma$/X-ray GRB related events. The on-orbit physical background observed by GRB monitor experiments is determined by factors inherent to the highly dynamical near-Earth radiation environment, to the spacecraft geographic position and attitude, as well as the spacecraft geometry, and the detector's pointing, design and response. 
Given the difficulty intrinsic to a real-time modelling of the expected scientific background, algorithms dedicated to the `online' search of GRBs often resort to extrapolate the background from recent observations. For example, as described in Section \ref{sec:telescope}, the trigger algorithms running on-board NASA Fermi/GBM assess a background estimate from an average of the photon count rates observed over the previous $17$~s excluding the most recent $4$~s of observations \cite{meegan2009fermi}; similar moving average approaches were used by Compton-BATSE \cite{paciesas1999fourth} and BeppoSAX-GRBM \cite{feroci1997flight}. 

\begin{figure}[!htb]
\centering
  \includegraphics[width=1\textwidth]{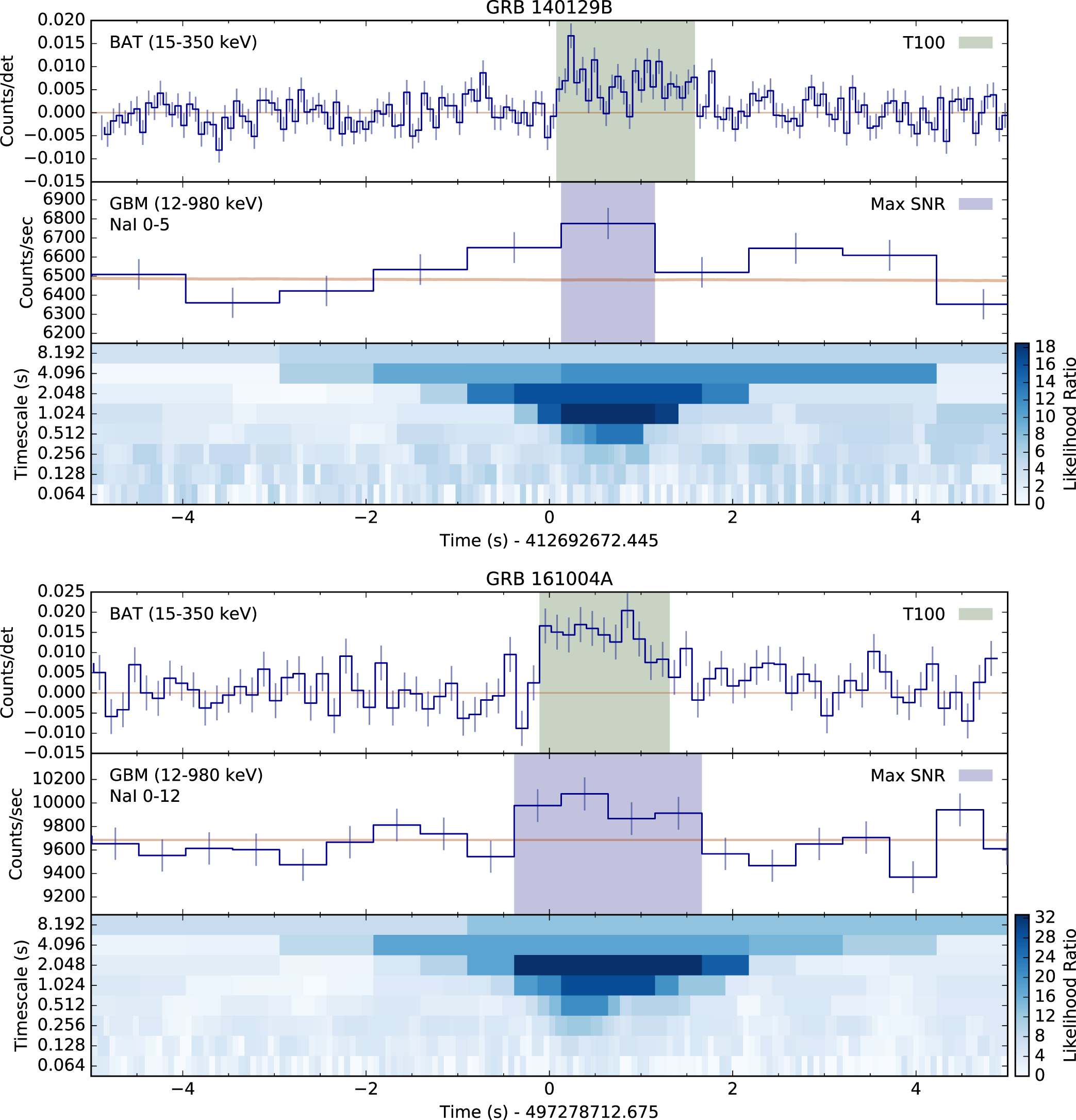}
\caption{\label{fig:polynomialGRB}
The top panels display the lightcurves from BAT for two GRBs. The middle panels show the corresponding lightcurves from GBM, indicating the detection of the GRBs. The lower panels present the event display, depicting the likelihood ratios or the significance of the signal above the local background. The event display shows a range of bin timescales and the blue shaded region represents the timescale that maximizes the signal significance in the GBM data. \cite{kocevski2018analysis} \copyright AAS. Reproduced with permission.}
\end{figure}

In `offline' analysis, archival data are searched for GRB events that the online and on-board algorithms may have missed. Examples of this approach can be found in \cite{kommers1999faint}, which uses the BATSE catalog, or in \cite{kocevski2018analysis} and \cite{hui2017finding} where they search for faint, short GRBs at times compatible with known gravitational wave events. Figure \ref{fig:polynomialGRB} shows an example of polynomial fitting as well as the significance of the event based on timescale binning.

In \cite{biltzinger2020physical} for example, an estimate is assessed starting from detailed models of the background expected for GBM, such as the detector response, the cosmic $\gamma$-ray background, the solar activity, the geomagnetic environment, the Earth albedo and the visibility of X and $\gamma$ point sources. 
The background description so achieved has been shown to reproduce very well the observations of Fermi/GBM and could potentially allow for the identification of otherwise hard to detect GRBs such as long-weak events with slow raising times. However, having been specifically tailored for the observations of Fermi/GBM, this technique is not immediately applicable to other experiments. 
In \cite{sadeh2019deep} a Recurrent Neural Network (RNN \cite{lecun2015deep}) is used to predict the background and, on top of it, classify or detect anomalies in the observations of a count rate detector. To recognize a GRB event, this RNN is trained onto existing catalogues of burst observations. We believe such an approach could inherit the detection biases of standard strategies for GRB detection, ultimately leading to missing events which already defied previous searches.

\subsection{New approach}
In Chapter \ref{chap:bkg} we introduce our approach to estimate the scientific background of a gamma-ray burst monitor experiment using a Neural Network (NN). In particular, we employ a Feed Forward Neural Network (\cite{bishop1995neural}) to estimate the count rate expected from background sources over the 12 NaI detectors of Fermi/GBM, in different energy bands and at regular time intervals. Our model is designed to learn the dynamics of the background over a timescale of months, enabling the detection of long-GRBs or even ultralong-GRBs \cite{gendre2019can}, as shown in an example in Section \ref{benchmark}.  Moreover, employing a robust loss function in the training phase, we are able to deal with outliers in count rate observations, such as transients due to astronomical events or brief period of detector high/low activity, see Section \ref{solarmaxmin}.
The choice of applying our framework to archival data from Fermi was motivated by the facts that (1) the HERMES Pathfinder spacecrafts are expected to be launched in a low inclination orbit with altitude $500-550$ km, an orbit where the background and its variations are expected to be smaller than those of Fermi/GBM \cite{meegan2009fermi}; and (2) the Fermi/GBM and HERMES Pathfinder detectors both rely onto scintillators and have similar effective areas \cite{bissaldi2009ground,campana2020hermes, dilillo2022space} resulting in background count rates of the same order of magnitude. 
To estimate the background observed by Fermi/GBM, we leverage on a large ensemble of information, including features both intrinsic to the satellite and its orbital setting such as the satellite attitude and geographic location in time, the Sun visibility and so on. These features are expected to be independent of events such as GRBs. This idea is consistent with \cite{fitzpatrick2012background}, which describes a method that estimates the background at the period of interest by using count rates from adjacent days when the satellite has similar geographical footprint. 
To retrieve these information's we use the Fermi/GBM Data Tools \cite{GbmDataTools} software package, an Application Programming Interface (API) allowing to download, analyse and visualise GBM data.
Being completely data-driven, we believe our approach to be in principle applicable to any GRB monitor experiment for which a similar dataset is available.

The background estimates produced by the NN are compared with the observations by mean of an efficient change-point detection technique called FOCuS-Poisson \cite{ward2022poisson}, aiming at the automatic identification of statistically significant astrophysical transients. We tested the combination of the NN background estimates and FOCuS-Poisson trigger on real Fermi/GBM data. We were able to confirm part of known events, but we also find events with no counterpart in the Fermi/GBM trigger catalog\footnote{https://heasarc.gsfc.nasa.gov/W3Browse/fermi/fermigtrig.html} \cite{von2020fourth}, yet with features resembling astronomical transients such as GRBs and solar flares and other galactic high-energy sources. 

Before delving into this novel approach, Chapters \ref{chp:ml} and \ref{chp:grb_ml} provide a contextual introduction. Chapter \ref{chp:ml} offers an overview of AI methods pertinent to this thesis, and Chapter \ref{chp:grb_ml} outlines the AI applications within the context of GRBs.

\chapter{State of the art - AI}\label{chp:ml}

This Chapter presents an overview of Artificial Intelligence tailored to the objectives of this thesis, while acknowledging that it does not aim to be exhaustive.

I attempt to provide brief definitions for the following topics:
\begin{enumerate}
    \item Artificial Intelligence (AI) is the collection of algorithms designed to mimic human cognition and other abilities.
    \item Machine Learning (ML), a subset of AI, consists of a range of statistical and optimization algorithms used to automatically get pattern and insights from data.
    \item Deep Learning (DL), a subset of ML, include a set of complex algorithms typically based on large neural networks.
\end{enumerate}

AI, ML and DL address a diverse range of challenges, including regression, time-series forecasting, classification, and clustering of similar objects. A notable ML subset is Reinforcement Learning (RL), although this will not be explored in detail within this chapter. RL provides solutions for problems demanding algorithmic interactions with dynamic environments to achieve predetermined goals, as seen in applications like robotic control, chess playing, and financial trading.

Despite the apparent diversity of these tasks, they share a common framework that entails defining input and target data, selecting the learning algorithm, and specifying the loss function. These algorithms exhibit versatility across a spectrum of data types, from structured data like tabular datasets to more complex data formats such as images, text, time series, videos, and graphs.

In the general setting the goal is to approximate an unknown phenomenon defined by:
\begin{equation}
    f: x \in \mathcal{X} \longrightarrow y \in \mathcal{Y},
\end{equation}
where $x$ and $y$ are respectively the input and the output data sample from the space $\mathcal{X}$ and $\mathcal{Y}$.
With a specific learner we can approximate $f$ with:
\begin{equation}\label{eq:generic_ml_map}
    F: x \in \mathcal{X} \longrightarrow \hat{y} \in \mathcal{Y},
\end{equation}
where $\hat{y} \approx y$ according to some evaluation metric. 

A dataset generally consists of $n$ samples denoted as pairs $(x, y)$ originating from the spaces $\mathcal{X}$ and $\mathcal{Y}$, respectively. Each input $x$ is associated with a corresponding target value $y$, belonging to the space $\mathcal{Y}$. When the target $y$ is known, the task is called \emph{supervised}. Conversely, when $y$ is either unknown or cannot be defined, the objective might be to map $x$ to an estimated value $\hat{y}$, thus giving insights into the input data distribution $\mathcal{D}_x$ (or $\mathbb{P}(x)$ in the probabilistic form). In such cases, it is referred to as an \emph{unsupervised} task. The spectrum between these two extremes is captured by the ``in the middle'' approach, known as the \emph{semi-supervised} approach, which includes various nuances and details.

\section{Supervised}

\subsection{Tabular data}

\begin{figure}[!htb]
\centering
  \includegraphics[width=1\textwidth]{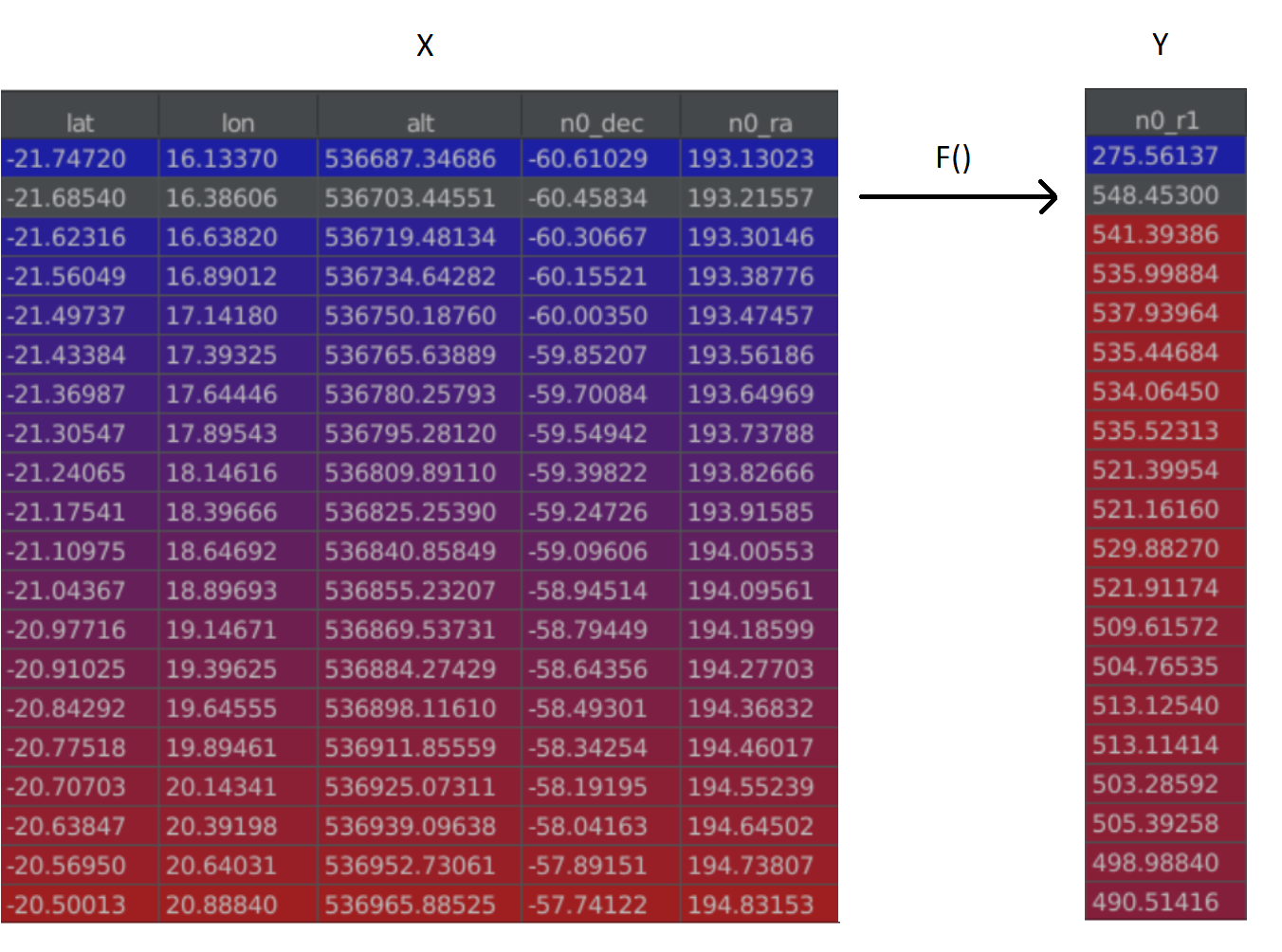}
\caption{\label{fig:tabular} 
Illustration of a tabular data task. The input consists of various features: latitude, longitude, altitude, detector's pointing in a particular timestep. The task is to map these features to one single target value, which represents the expected count rates of the detector.}
\end{figure}

In the most common settings, we refer to \emph{tabular data} when $x \in \mathbb{R}^k$ and $k$ denotes the number of \emph{features} within each sample, while $y \in \mathbb{R}^m$ (as shown in Figure \ref{fig:tabular}). The dataset is then represented as $X \in \mathbb{R}^{n \times k}$ and $Y \in \mathbb{R}^{n \times m}$. In its simplest form, when $m=1$, the problem can be framed as regression if $Y \in \mathbb{R}$ or as a classification task if $y \in \mathbb{L}=\{0, \dots, L-1\}$, involving $L$ different classes. 
When $m>1$, it is commonly referred to as multioutput regression. Within the context of this thesis, this formalization is employed to predict count rates from multiple detectors. In a basic regression scenario with $Y \in \mathbb{R}$, predictions can be employed to estimate attributes like the redshift of a GRB, while with $y \in \mathbb{L}$, classifications can be conducted to differentiate transient types (e.g., GRB, Solar Flare, etc.).

\begin{figure}[!htb]
\centering
  \includegraphics[width=1\textwidth]{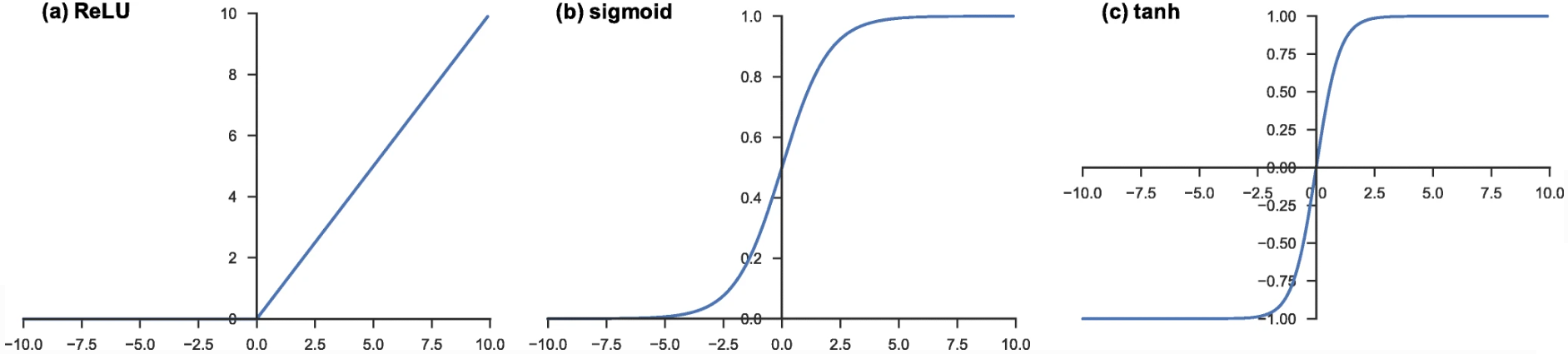}
\caption{\label{fig:act_fun}
Activation functions frequently utilized in neural networks: (a) Rectified Linear Unit (ReLU), (b) Sigmoid, and (c) Hyperbolic Tangent (tanh). \cite{yamashita2018convolutional} \copyright Yamashita et al. (2018), \href{http://creativecommons.org/licenses/by/4.0/}{CC BY 4.0}.}
\end{figure}

\begin{figure}[!htb]
\centering
  \includegraphics[width=1\textwidth]{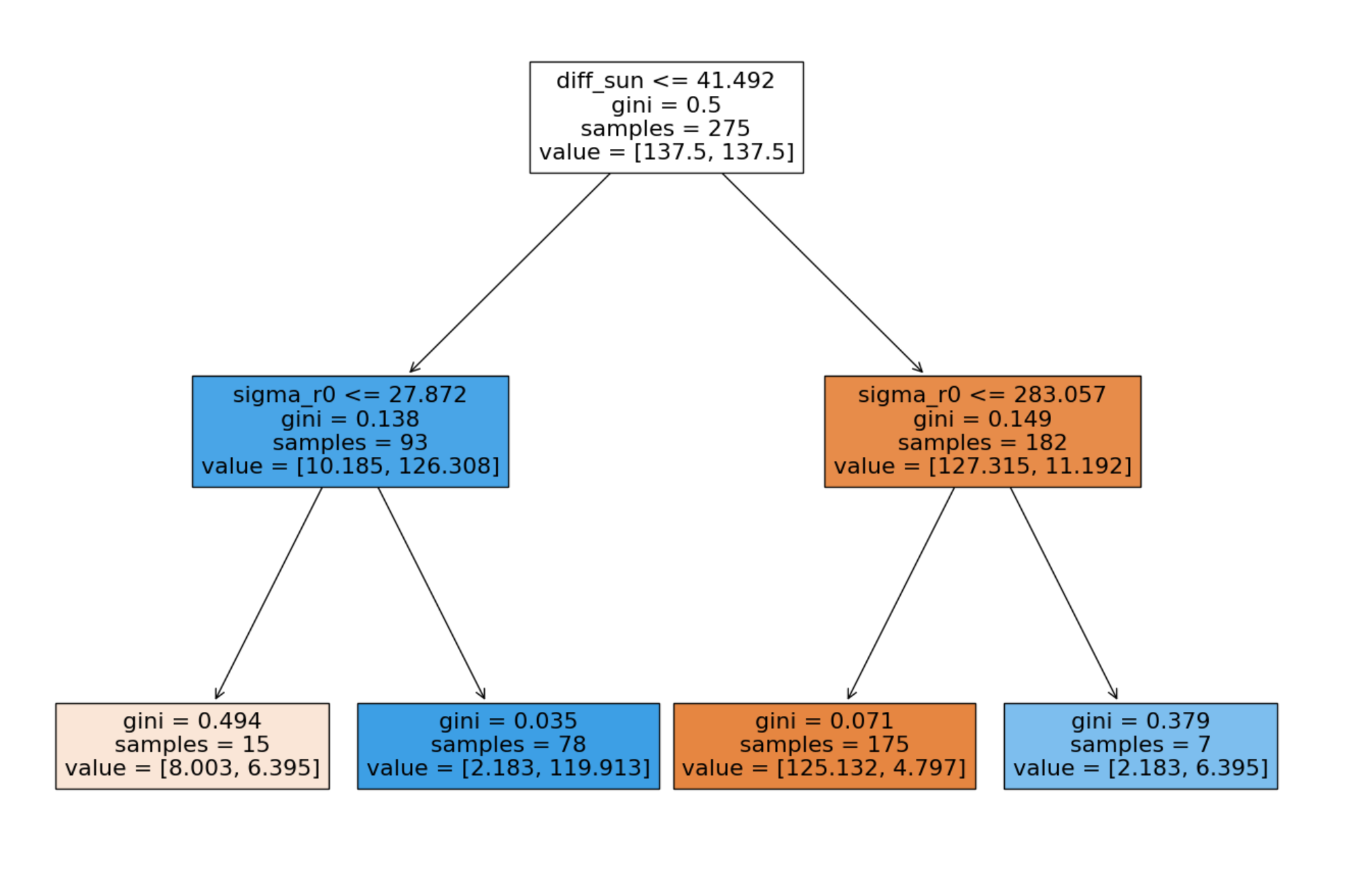}
\caption{\label{fig:decision_tree} 
The Decision Tree classifier involves an iterative process wherein the algorithm selects the optimal feature to split the data, aiming to minimize a specific criterion (e.g., Gini index, as described in Equation \ref{eq:gini}). At each node, the dataset is partitioned into subsets, and this recursive process continues until predefined stopping conditions are met, such as reaching a maximum tree depth or having a minimum number of samples per leaf. The resulting tree contains a set of \textbf{rules} that can classify new data by following the path from the root node to a leaf node, which determines the predicted class.}
\end{figure}

\begin{figure}
    \centering
    \includegraphics[width=1\textwidth]{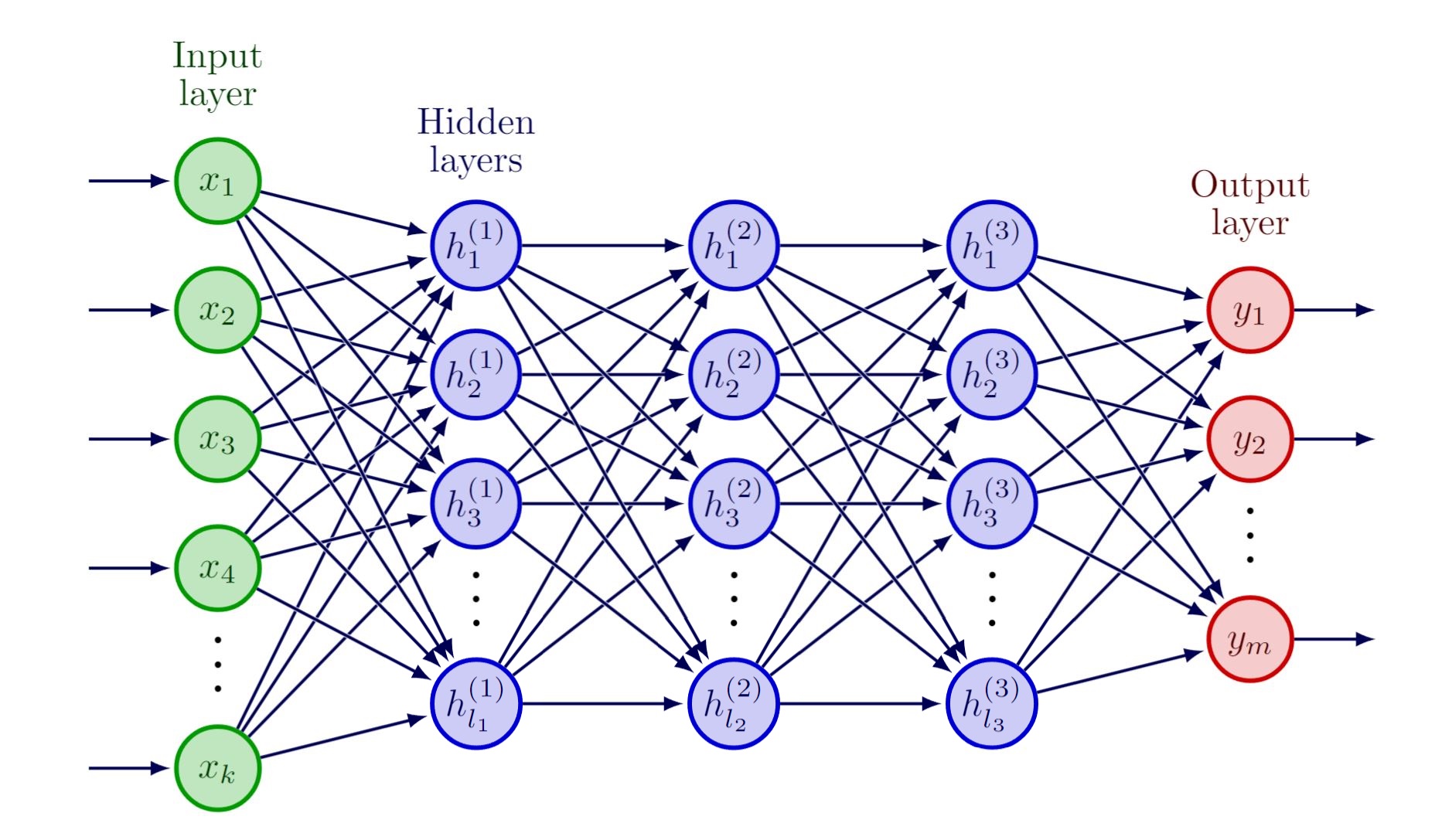}
    \caption{An example of the architecture of a feed forward neural network. The input has dimension $k$. The three hidden layers have $l_1$, $l_2$ and $l_3$ neurons, respectively. The output layer has dimension $m$. Each node in the network acts similarly to a logistic regression, with vector weights sized according to the preceding layer (plus an additional dimension for the bias term). For instance, the third node in the second hidden layer ($h_3^{(2)}$) possesses vector weights $w_3^{(2)} \in \mathbb{R}^{l_1+1}$. Similarly, the output node $y_m$ features vector weights $w_m^{(4)} \in \mathbb{R}^{l_3+1}$. This neural network architecture extends the concept of logistic regression, as can be framed into a logistic regression when the network consists solely of the input layer and a single output node with the sigmoid activation function.}
    \label{fig:nn_intro}
\end{figure}

In recent years, ML and DL have made significant advances, and several famous learners have emerged as powerful algorithms for a variety of tasks \citep{scikit-learn}. Some of the most well-known classical ML learners are:

\begin{enumerate}
    \item \emph{Linear regression} is used for predicting continuous output values based on a linear relationship between input features and the target variable, $\hat{y} = wx+c$ where $w \in \mathbb{R}^{1 \times k}$. The \emph{Logistic Regression} is a linear classifier algorithm designed to deal for binary classification tasks, it maps the output to a probability value $\sigma(x) = \frac{1}{1+e^{-wx-c}}$ and $\hat{y} = \mathbbm{1}_{\sigma(x)>T}$ where $T$ is a threshold for the output probability (e.g. 0.5) in order to obtain the binary output.
    \item Linear Discriminant Analysis (\emph{LDA}) assumes that the conditional distribution $\mathbb{P}(X \mid y=l)$ follows a normal distribution $\mathcal{N}(\mu_l,\,\Sigma)$ and aims to find the hyperplanes that maximize the separation among classes while preserving class-specific information. The discriminant criterion (one class versus all others) can be written as $w^Tx>c$, where $w$ is the vector defining the hyperplane and $c$ a constant. This is employed in classification tasks. In order to enhance data visualization and reduce the dimensionality of $x$ (see Section \ref{sec:dim_rec}), the dataset can be projected onto a lower-dimensional subspace defined by the eigenvectors corresponding to the highest eigenvalues of $\Sigma^{-1} \Sigma_b$, where $\Sigma_b = \frac{1}{L} \sum_{l=0}^{L-1} (\mu_l - \mu)(\mu_l - \mu)^T$.   
    \item Support Vector Machines \emph{SVM} is an algorithm that determines the best hyperplane to divide data into classes. SVM is known for its ability to separate data that is not linearly separable through the use of kernel functions to transform the data into a higher-dimensional space. The SVM estimates a quantity $\hat{y} = w\phi(x) + c$, where $\phi$ is the mapping of the data into a high dimensional space $\mathbb{R}^{k'}$, where $k \ll k'$. In practical computations, the actual mapping $\phi(x)$ is not performed; instead, only the scalar product $\mathcal{K}(x_i, x_j) = \langle \; \phi(x_i), \phi(x_j) \; \rangle$ is computed using a kernel function $\mathcal{K}$. Similarly, a regression version of SVM can also be implemented, taking advantage of the kernel.
    \item k-Nearest Neighbors (\emph{k-NN}) is a straightforward algorithm used for classification and regression tasks. It determines the class of a data point considering the majority class among its $k$-nearest neighbors, or by taking the mean if the target is continuous. These neighbors are identified based on a chosen distance metric, often employing the Euclidean distance.    
    \item \emph{Decision trees} are adaptable and interpretable models that recursively split data based on feature thresholds, making them useful for classification and regression tasks, as shown in Figure \ref{fig:decision_tree}.
    The optimal feature splits are determined by metrics that assess the heterogeneity of sub-samples, such as the Gini index:
    \begin{equation}\label{eq:gini}
    \text{Gini} = \frac{1}{c} \sum_{l=1}^c p_l (1-p_l)
    \end{equation}
    or the Log Loss:
    \begin{equation}\label{eq:log_loss}
    \text{Log Loss} = \frac{1}{c} \sum_{l=1}^c p_l \, \log(p_l),
    \end{equation}
    where $p_l$ represents the proportion of observations belonging to class $l$ within the subsample. A higher value of these metrics indicates greater concentration of a node within a particular class.
    \item \emph{Random Forest} is a method of ensemble learning that combines multiple decision trees to improve accuracy and reduce overfitting. It constructs several trees using random subsets of features and data and then averages their predictions for final classification or regression results.
    \item \emph{XGBoost} sequentially constructs an ensemble of weak learners (typically decision trees), with each tree correcting the errors of its predecessor, resulting in more accurate predictions.
    \item A \emph{Feed-forward neural network} (FFNN, \cite{bishop1994neural}) is a type of artificial neural network in which information flows in one direction, from input nodes through hidden layers to output nodes. It is a generalization of logistic regression, as it can learn more complex patterns and decision boundaries by incorporating multiple layers of neurons with non-linear activation functions $\sigma$ (see Figure \ref{fig:act_fun}). A common choice for $\sigma$ is the rectified linear unit $\text{ReLU}(x) = \max(0,x)$. Figure \ref{fig:nn_intro} shows an example with 3 hidden layers and the output can be represented as 
    $$\hat{y} = W_o(\sigma ( W_3 \sigma( W_2 \sigma (W_1 x + c_1) + c_2) + c_3) + c_4) $$ where $W_1 \in \mathbb{R}^{l_1 \times k}$, $W_2 \in \mathbb{R}^{l_2 \times l_1}$, $W_3 \in \mathbb{R}^{l_3 \times l_2}$, $W_o \in \mathbb{R}^{m \times l_3}$, $c_i \in \mathbb{R}^{l_i}$ are the bias terms. 
    When dealing with regression tasks, the non-linearity in the last layer is not not necessary: when dealing with classification, the final output is instead forced to represent a probability density, e.g. by using a softmax function $$\text{softmax}(y_i) = \dfrac{\exp(y_i)}{\sum_{j=1}^m \exp(y_j)}.$$
    
\end{enumerate}

To improve predictive performance and reduce overfitting (Figure \ref{fig:overfitting}), it is common practice to build \emph{ensemble} approaches which combine multiple individual models.
One example is Bagging, which builds multiple models independently and averages their predictions, as the Random Forest. Other examples are Boosting, which builds models sequentially, weighting misclassified instances more heavily (e.g. XGBoost), and Stacking, which combines predictions from multiple models as input to a meta-model for the final prediction. When compared to single models, ensemble methods frequently achieve better generalization and robustness.

When it comes to handling tabular data, tree-based algorithms have proven to be formidable competitors. This is partly due to the fact that Neural Networks have a high data requirement. Attempting to address this limitation, some researchers have explored the use of attention mechanisms with TabNet \citep{arik2021tabnet}. Yet, tuning the hyperparameters for Neural Networks remains challenging compared to the relatively straightforward process for Random Forest and XGBoost, which can often yield good results even with default hyperparameters. In contrast, Neural Networks offer unparalleled flexibility, allowing for easy implementation of multi-output tasks, while other Machine Learning techniques may require building separate models for each output dimension. Additionaly, the loss function in Neural Networks can also be highly customizable, offering flexibility to address a variety of challenges presented by the problem at hand. For instance, in the context of high-energy transients, where the presence of outliers is common, selecting a robust loss function is crucial.
The training phase and the loss function definition is described in Section \ref{sec:loss}.

\subsection{Image data}
As the number of features $k$ grows, we encounter the so-called ``curse of dimensionality'' \citep{verleysen2005curse}, where the the data points tend to become sparse in the $k$ dimensional space ($\mathbb{R}^k$), making it increasingly difficult for algorithms to identify meaningful patterns and relationships. This is evident in cases involving \emph{image data}, where $k$ can reach the order of thousands or even millions.
\emph{Image data} is often represented as a tensor with dimensions $p_1 \times p_2 \times ch$, where $p_1$ and $p_2$ represent the horizontal and vertical dimensions, and $ch$ represents the number of channels (e.g. RGB color channels). Such an image can be associated with a target value $y \in \mathbb{R}$ for regression tasks or $y \in \mathbb{L}=\{0, \ldots, L-1\}$ for classification tasks.

FFNN estimates outcomes by \emph{fully} connecting each node on a layer to all nodes in the surrounding layers. This structure is highly flexible, but at the same time is completely agnostic to any known regularity in the data. When one has prior knowledge of systematic symmetries in the data, the architecture of the NN can be constrained a priori. A positive side effect is potentially reducing the number of learning parameters required. To efficiently capture spatial patterns without relying in the hyperconnected FFNN, the convolutional layers are introduced. Convolutional Neural Networks (CNNs, \cite{lecun1995convolutional}) were designed to address the challenges of computer vision tasks. 
FFNNs, treating input data as a one-dimensional vector, were ill-suited for handling high-dimensional images due to their vast number of parameters and their inability to preserve spatial structures.

\begin{figure}
    \centering
    \includegraphics[width=1\textwidth]{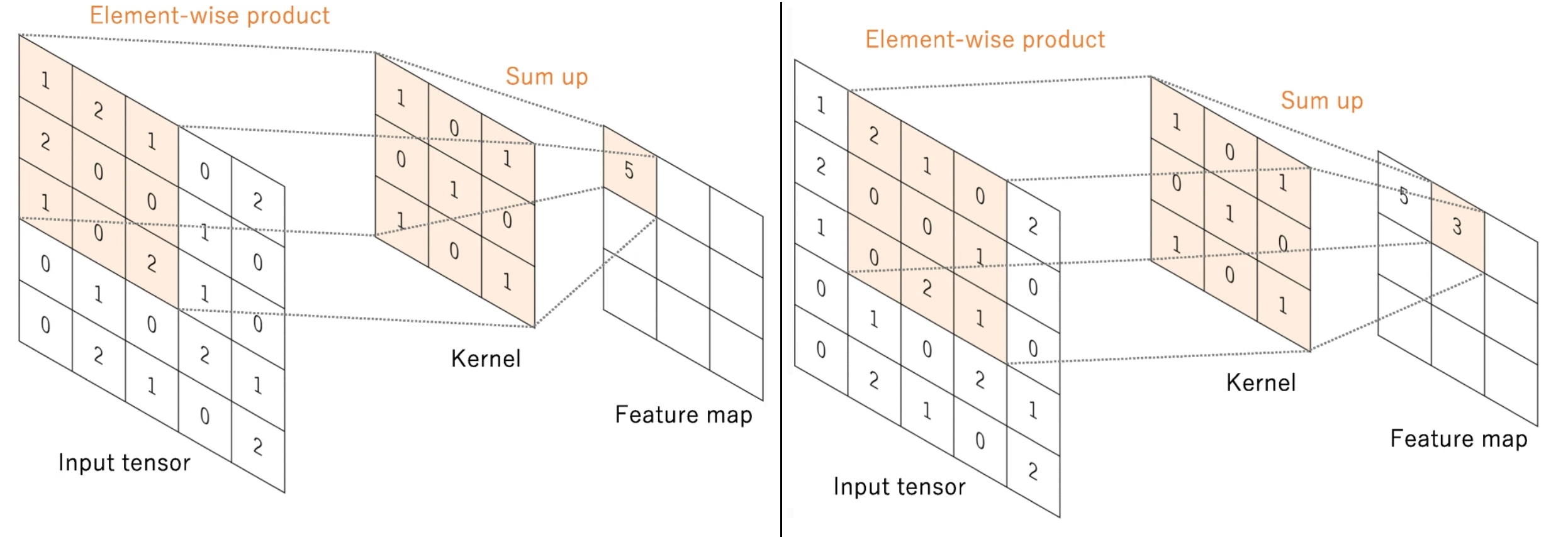}
    \caption{Example of a convolution operation with a kernel size of 3x3, no padding, and a stride of 1. The kernel is applied across the input tensor (image). At each location, an element-wise product is calculated between each element of the kernel and the input tensor. These products are then summed to obtain the output value in the corresponding position of the output tensor, which is referred to as a feature map. This process is repeated for all locations in the input tensor, producing the final feature map as the result of the convolution operation. The image has been adapted from \cite{yamashita2018convolutional} \copyright Yamashita et al. (2018), \href{http://creativecommons.org/licenses/by/4.0/}{CC BY 4.0}.}
    \label{fig:CNN_image_1}
\end{figure}

CNNs were designed to exploit the spatial relationships present in images. They use convolutional layers to apply filters (kernels) across the image, capturing local patterns and features, as shown in Figure \ref{fig:CNN_image_1}. By sharing weights across different parts of the image, CNNs significantly reduce the number of parameters compared to fully connected networks, making them computationally efficient and scalable.

\begin{figure}
    \centering
    \includegraphics[width=1\textwidth]{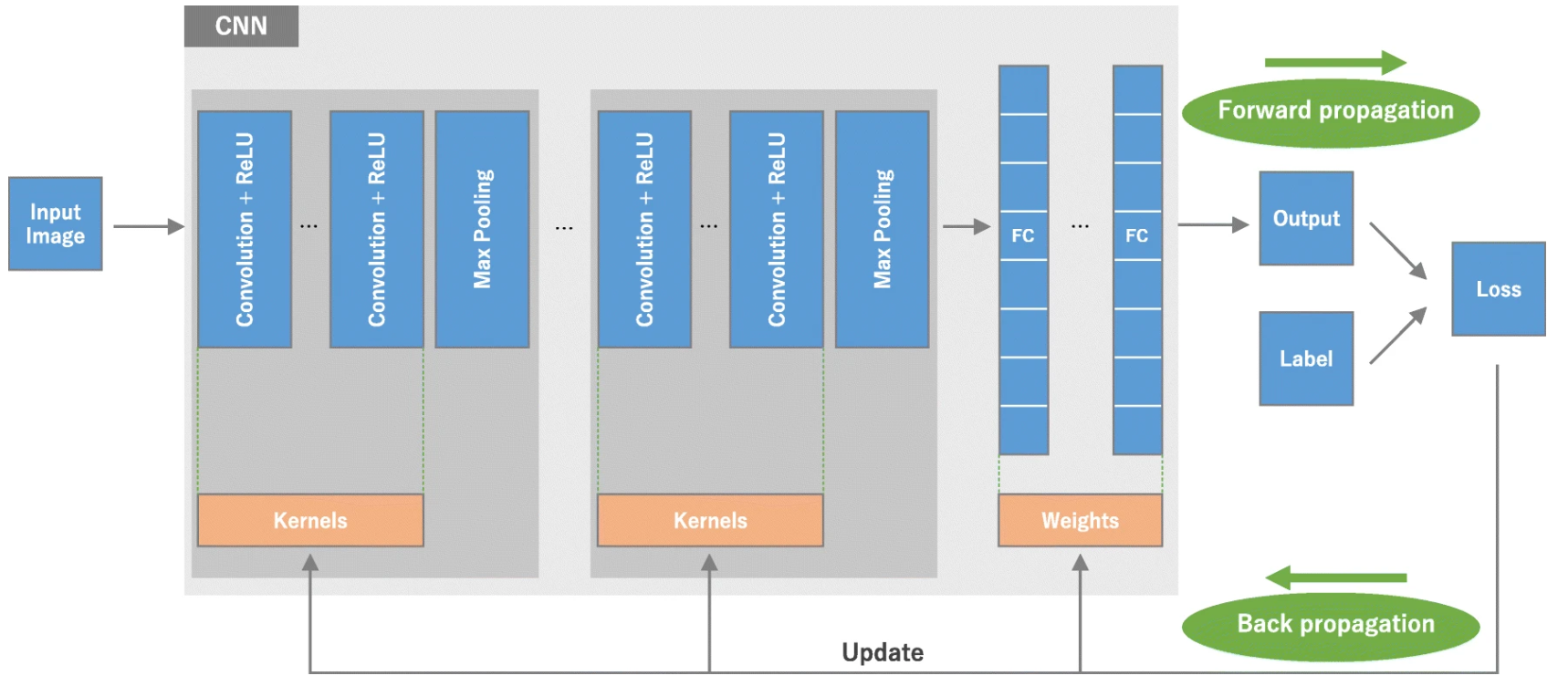}
    \caption{ 
    A CNN architecture comprises several building blocks, including convolution layers, pooling layers (e.g., max pooling), and fully connected (FC) layers. These layers are stacked together to form the network. During the training process, the model's performance is evaluated using a loss function (see Section \ref{sec:loss}), and forward propagation is performed on a training dataset to calculate the loss value under specific kernels and weights. Learnable parameters, such as kernels and weights, are then updated using the backpropagation algorithm with gradient descent optimization to minimize the loss value. \cite{yamashita2018convolutional} \copyright Yamashita et al. (2018), \href{http://creativecommons.org/licenses/by/4.0/}{CC BY 4.0}.}
    \label{fig:CNN_image_2}
\end{figure}

Following the convolutional stage, pooling layers are employed to downsample the spatial dimensions of the feature maps generated by the convolutional layers. Common pooling operations include max pooling, which retains the maximum value within a defined region, and average pooling, which calculates the average value. This process is illustrated in Figure \ref{fig:CNN_image_2}.

The ability of CNNs to automatically learn hierarchical representations from raw pixels enables them to recognize complex patterns and objects in images. As they learn to detect low-level features like edges, corners, and textures in early layers and combine them to form higher-level features in deeper layers, CNNs can effectively capture both local and global information from images. 
Even GRBs can be represented in the form of an image, with the x-dimension denoting time and the y-dimension representing the energy range. In this representation, the pixel value corresponds to the count rates within a specific time and energy range. Further details can be found in Chapter \ref{chp:grb_ml}.

\begin{figure}
    \centering
    \includegraphics[width=0.75\textwidth]{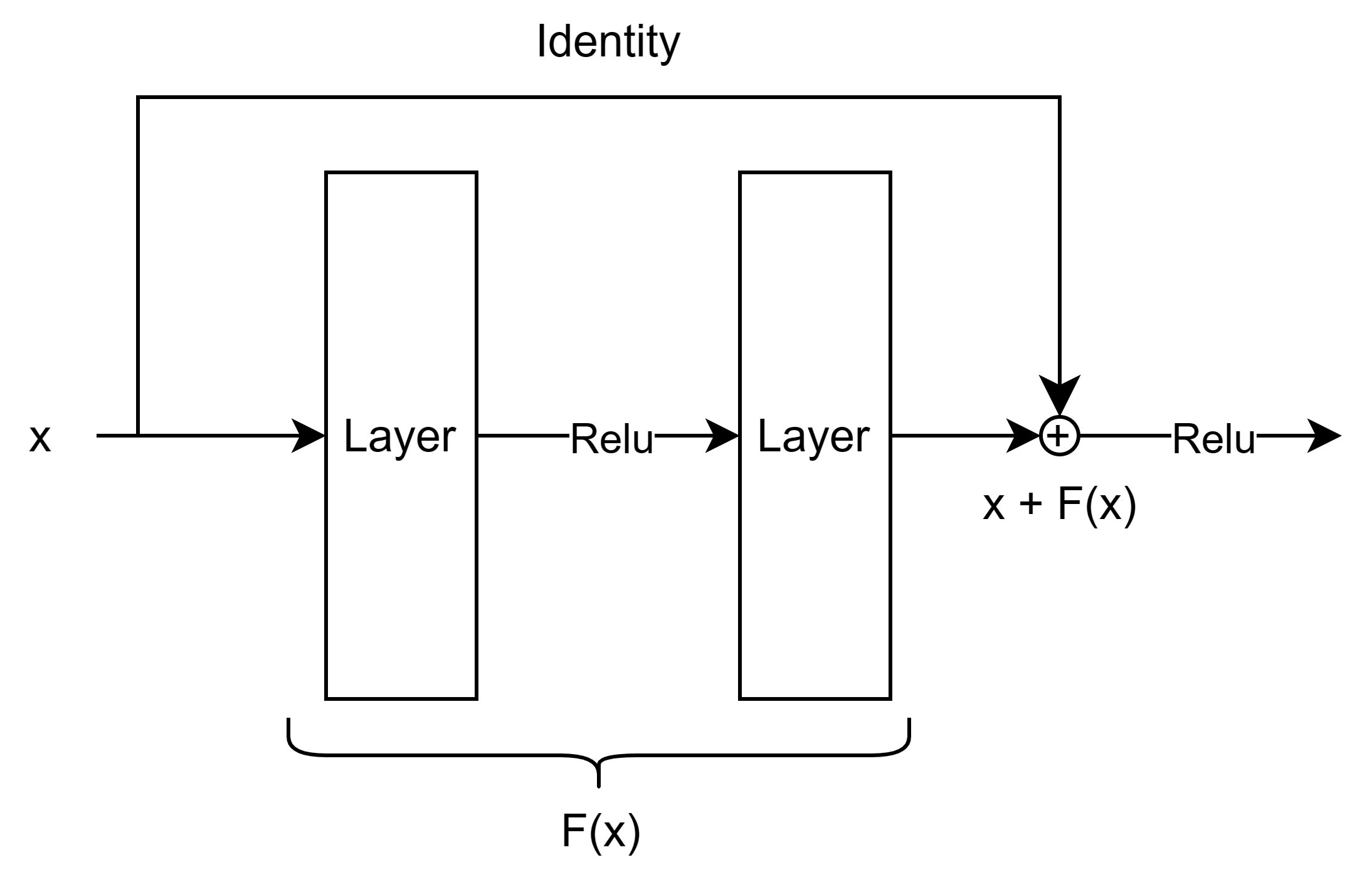}
    \caption{ 
    Residual building block. The values $x$ of a certain layer is added to the output of the application of two layers on $x$ itself.}
    \label{fig:res_net}
\end{figure}

More layers in the NN are useful because they can capture more complex patterns and representations in data, leading to higher model accuracy. However, one problem of deep networks is the vanishing gradient \citep{tan2019vanishing}. This issue arises when, for instance, the activation function are saturated, the errors cannot propagate properly on the entire NN and weights of early layers are updated minimally.
The ResNet \citep{he2016deep} architecture is introduced to address this problem in very deep networks. It utilizes residual blocks (Figure \ref{fig:res_net}) with skip connections to efficiently train deeper networks by allowing direct flow of gradients. 

An important advancement in neural networks is the Attention Mechanism \citep{vaswani2017attention}. This mechanism enables the network to focus on specific relevant parts of the input data, enhancing its ability to make accurate predictions. Initially developed for natural language processing tasks, where the model attends to different words in an input sentence for accurate translation, it has also been extended to computer vision, such as in CBAM (Convolutional Block Attention Module) \citep{woo2018cbam}.



\subsection{Time Series}

In the context of \emph{time series}, we denote with $x = (x_0, \ldots, x_t)$ a sequence of $t$ objects\footnote{It should be noted that the sequence length $t$ can vary from one sample to another.} (e.g. characters, words, or features represented as $x_i \in \mathbb{R}^k$). When $y \in \mathbb{R}$ represents values from time steps $> t+1$, the task is defined as \emph{forecasting}, even though the task can be formally expressed as classification or regression. If the objective is to classify $x$ based on certain attributes, such as shape, $y$ corresponds to a class number, thus constituting a formal classification task.
In Natural Language Processing (NLP), sequences of tokens are common, often denoted as integers corresponding to indices within a vocabulary. Tasks like \emph{translation} or \emph{Question-Answering} involve mapping a text $x$ to another text $y = (y_0, \ldots, y_t)$, representing either a translation in another language or an answer $y$ for a given question $x$.
Gamma-Ray Bursts (GRBs) are the context in which this representation aligns naturally. GRBs can be represented as a series of events, showing the times at which photons within a range of energies arrive, or as lightcurves, which are produced by counting photons over predetermined bin times. To enable comparison even when using different binning times, these counts are typically converted into count rates (counts per second).


In the context of ML, addressing time series data often involves reducing its dimensionality through feature extraction, which can include statistics (e.g., average, standard deviation, min, max) or applying transformations, such as utilizing the first components of the Discrete-Time Fourier Transform (DTFT). Alternatively, when the Machine Learning algorithm relies on distance measures between data points (sequences of numbers in this case), Dynamic Time Warping (DTW, \cite{muller2007dynamic}) becomes a suitable approach. DTW aligns two sequences by finding an optimal warping path that minimizes the total distance between corresponding elements, enabling it to handle sequences of different lengths and time variations. This adaptability makes DTW a valuable tool in diverse applications, including speech recognition.

\begin{figure}
    \centering
    \includegraphics[width=1\textwidth]{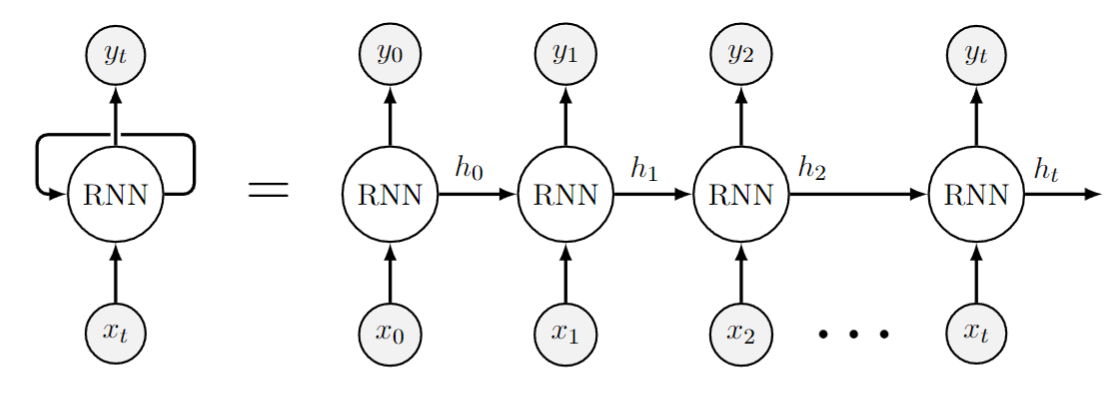}
    \caption{The RNN architecture consists of processing sequential data by passing the input $x_i$ through the RNN along with the hidden state $h_i$, which summarizes the history of the sequence up to the $i$-th step. At each step, the RNN updates the hidden state and produces an output $y_i$. For forecasting tasks, the output $y_i$ can be used as an estimate for the next step in the sequence, typically set as $x_{i+1}$. For classification tasks involving sequences, only the last output $y_t$ is considered to determine the final output, making use of the information learned from the entire sequence.}
    \label{fig:rnn}
\end{figure}

The introduction of specific DL architectures, such as Recurrent Neural Networks (RNNs) and Transformers \citep{vaswani2017attention}, has further enhanced time series data handling. In an RNN, each input $x_t$ in a sequence undergoes step-by-step processing, updating the network's hidden state $h_t$ at each iteration based on the present input and the prior hidden state, as depicted in Figure \ref{fig:rnn}. This recurrent nature allows RNNs the ability to retain a memory of previous inputs, making them well-suited for tasks involving time series data.

The general formula for a RNN is:
\begin{align*}
h_t &= \sigma(W_{ih} x_t + W_{hh} h_{t-1} + b_h), \\
y_t &= W_{hy} h_t + b_y.
\end{align*}
$x_t$, $h_t$, $y_t$ are the input, hidden state and output, respectively, at time $t$.
The weight matrix $W_{ih}$ connects input to hidden state, $W_{hh}$ connects hidden state at previous step $t-1$ to the current one (recurrent weights), $W_{hy}$ connects hidden state to output, $b_h$ and $b_y$ are bias terms, $\sigma$ is an activation function (e.g., $\tanh$ or ReLU). 
Notably, this architecture can be enhanced by introducing a ``cell state'', leading to the creation of the Long Short-Term Memory (LSTM, \cite{graves2012long}) which can store information for long periods of time. The LSTM tackles the vanishing gradient problem encountered in RNNs by enabling the flow of gradients.

\section{Loss function}\label{sec:loss}

Depending on the nature of the target variable $y$ (be it a class label, continuous value, image, etc.) and the specific learning algorithm employed for function $F$, the determination of the appropriate \emph{loss function} or \emph{cost function} arises, which measures the error between the predicted $\hat{y}$ and the actual $y$. This function plays a pivotal role during the training phase of the machine learning algorithm\footnote{It's important to note that the term ``Machine Learning algorithm'' refers to a general procedure that, when applied to training data, produces a fitted model $F$. For example, the Decision Tree Algorithm yields a Decision Tree model. This distinction is occasionally a source of confusion.}.

Decision Trees utilize the Gini index or Log Loss (Equations \ref{eq:gini} and \ref{eq:log_loss}) as criteria to minimize during each node split. This method, in particular, employs a greedy strategy, minimizing the loss locally within a subset of the split data. On the contrary, $k$-Nearest Neighbors ($k$-NN) does not require a specific loss function; its approach revolves around averaging the outputs of the $k$ neighboring points of $x$.

The loss function is critical in neural networks because it allows for error estimation per individual sample and gradient computation with respect to learnable parameters (weights). Iterative updates to the neural network are performed until convergence. For the successful application of gradient-based optimization techniques, the chosen loss function must be differentiable.

For an explanation of the evaluation metrics used to assess model performance after training, see Section \ref{sec:eval_metrics}.

\subsection{Supervised - Classification}\label{sec:classification_ML}
If the target variable $y$ takes binary values ${0, 1}$, the Machine Learning model $M$ typically estimates the probability of belonging to class $1$ given input $x$, formally represented as $\mathbb{P}(y=1 \mid x)$. This concept can be rephrased using equation \ref{eq:generic_ml_map} to define:
\begin{equation}\label{eq:generic_ml_classification_map}
F: \mathbbm{1}_{M(x)>T} \; ,
\end{equation}
where $T$ is a scalar threshold (usually set to 0.5).
A commonly used loss function is the cross-entropy:
\begin{equation}
L_{\text{bce}} = \frac{1}{n} \sum_{i=1}^n - (y_i \log(M(x_i)) + (1 - y_i) \log( 1 - M(x_i)) ).
\end{equation}

For the more general scenario with $c$ classes where $y \in K=\{0, \dots, c-1\}$, the ML model $M(x) = [M(x)^0, \dots, M(x)^l, \dots, M(x)^{c-1}]$ aims to learn the probabilities $\mathbb{P}(y=l \mid x)$ for each class $l \in K$.
The extension of the cross-entropy to multiple classes is the multiclass cross-entropy:
\begin{equation}\label{eq:multiclass_ce}
L_{\text{mce}} = \frac{1}{n} \sum_{i=1}^n \sum_{l=0}^{c-1} - y_i^l \log\left(\frac{e^{M(x_i)^l}}{\sum_{l'=0}^{c-1} e^{M(x_i)^{l'}}}\right),
\end{equation}
inside the $\log$ the prediction $M(x)^l$ is normalized through the softmax function.

These are two examples of a loss function for classification tasks, and alternative formulations can be employed to address issues like class imbalances (e.g. balanced cross-entropy or focal loss \cite{lin2017focal}). A common characteristic of these two losses is their differentiability, which enables the computation of gradients required by many optimization algorithms.

\subsection{Supervised - Regression}
A widely used loss function in regression tasks is the Mean Squared Error (MSE):
\begin{equation}
    L_{\text{MSE}} = \sum_{i=1}^n (y_i - \hat{y}_i)^2.
\end{equation}
This loss function possesses a well-defined derivative and offers itself to straightforward optimization. Another significant loss function is the Mean Absolute Error (MAE):
\begin{equation}\label{eq:mae_loss}
    L_{\text{MAE}} = \sum_{i=1}^n \lvert y_i - \hat{y}_i \rvert.
\end{equation}
This can be extended to the quantile regression loss:
\begin{equation}\label{eq:quantile_loss}
    L_{\text{quantile}} =  (q-1) \sum_{y_i < \hat{y}_i} (y_i - \hat{y}_i) + q \sum_{y_i \ge \hat{y}_i} (y_i - \hat{y}_i).
\end{equation}
Here, $q \in [0, 1]$ serves as the quantile parameter. When $q=0.5$, Equation \ref{eq:quantile_loss} is equivalent to Equation \ref{eq:mae_loss}. For $q>0.5$, positive residuals hold less weight than negative ones.

\begin{figure}
    \centering
    \includegraphics[width=1\textwidth]{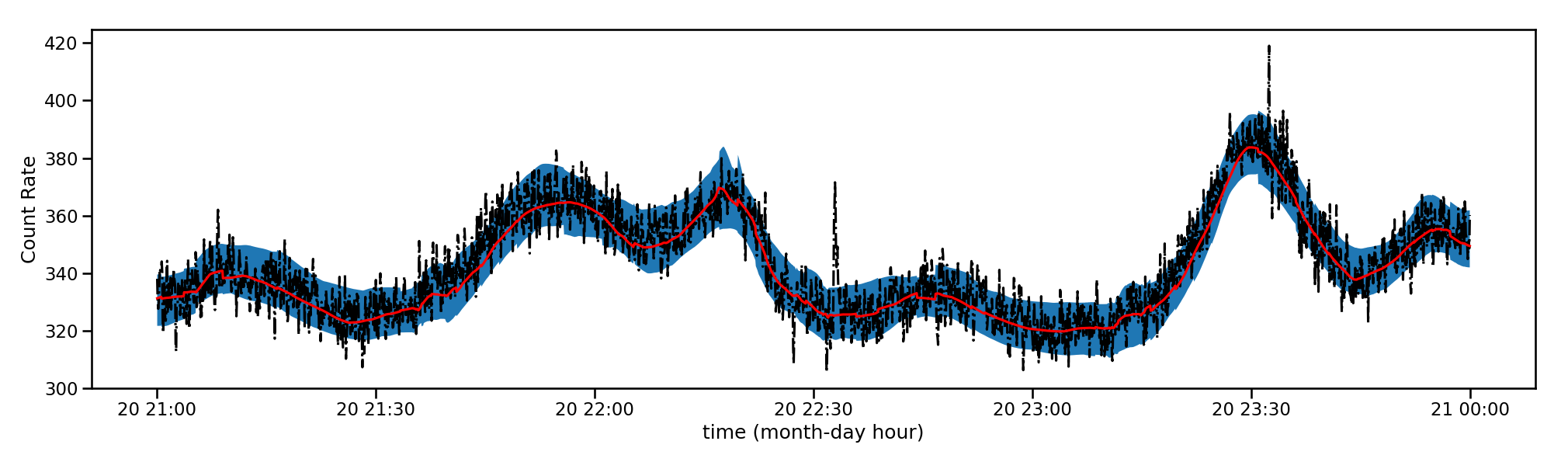}
    \caption{Quantile regression for count rates as the target variable. The red line represents the median estimate, while the blue shadow indicates the range between the upper value at quantile 0.9 and the lower value at quantile 0.1.}
    \label{fig:quantile_regression}
\end{figure}

By training a model with three different quantiles, such as the median, $q=0.1$, and $q=0.9$, it becomes possible to obtain predictions with uncertainty estimations. As shown in Figure \ref{fig:quantile_regression}, the red line represents the median estimate, while the blue shadow  represents the range between the upper values at $q=0.9$ and the lower value at $q=0.1$.

\subsection{Gradient}
\begin{figure}
    \centering
    \includegraphics[width=0.75\textwidth]{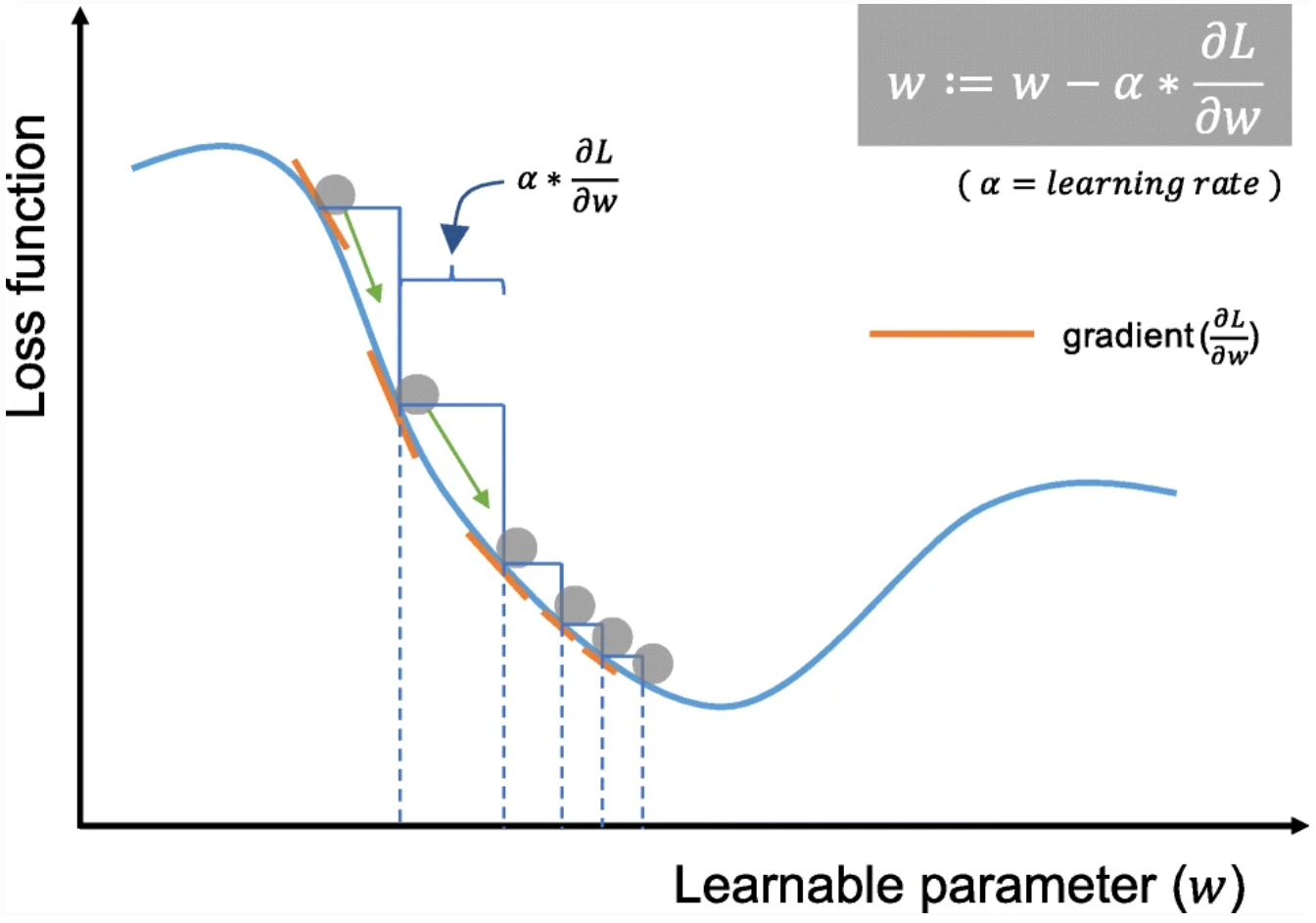}
    \caption{Gradient descent is an optimization algorithm used to minimize the loss function in machine learning. It works by iteratively updating the learnable parameters in the direction opposite to the gradient of the loss function. The step size of each update is controlled the learning rate $\alpha$. \cite{yamashita2018convolutional} \copyright Yamashita et al. (2018), \href{http://creativecommons.org/licenses/by/4.0/}{CC BY 4.0}.}
    \label{fig:gradient}
\end{figure}

Regardless of the specific neural network architecture, they all share a common optimization process using gradient descent (Figure \ref{fig:gradient}). There are various implementations of optimization algorithms for gradient descent \citep{ruder2016overview}, some examples include Stochastic Gradient Descent (SGD), Momentum, Nesterov Accelerated Gradient (NAG), Adagrad (Adaptive Gradient Algorithm), RMSprop (Root Mean Square Propagation), Adam (Adaptive Moment Estimation), and Nadam (Nesterov-accelerated Adaptive Moment Estimation). There is no optimizer that is always the best in every aspect. Although Adam is a common choice, Nadam was preferred in the context of multioutput regression for background estimation covered in Chapter \ref{chap:bkg}.

The weights of the neural network are updated using the equation:
\begin{equation}\label{eq:update_weights}
w_{il} \leftarrow w_{il} - \alpha \frac{\partial L}{\partial w_{il}};
\end{equation}
where $\alpha$ is the \textbf{learning rate}, $\frac{\partial L}{\partial w_{il}}$ is the derivative of the loss function with respect to vector weights $w_{il}$, $i$ the index of the node in the layer $l$.
The derivative of the loss function is calculated using the formula:
\begin{equation}\label{eq:derivative}
\frac{\partial L}{\partial w_{il}} = \sum_{j} \frac{\partial L}{\partial z_j} \frac{\partial z_j}{\partial w_{ij}},
\end{equation}
where $z_j$ represents the vector weights in the $j$-th node of the $(l+1)$-th layer. This approach is efficient since $w_{ij}$ directly affects only the nodes in layer $l+1$. Computing the derivative for a node in the last layer $\Bar{l}$ is straightforward because the output $\hat{y}=\sigma(W_{\Bar{l}}  h + c_{\Bar{l}})$ in the loss $L$ is explicit to each output node $i$: $\frac{\partial L}{\partial w_{i\Bar{l}}}$. By storing these quantities as $\frac{\partial L}{\partial z_i} = \frac{\partial L}{\partial w_{i \Bar{l}}}$, we can compute $\frac{\partial w_{i{\Bar{l}}}}{\partial w_{i,\Bar{l}-1}}$ = $\frac{\partial z_i}{\partial w_{i,\Bar{l}-1}}$, and finally compute $\frac{\partial L}{\partial w_{i,\Bar{l}-1}}$ through Equation \ref{eq:derivative}. This iterative process continues until the first layer is reached, and it is known as ``backpropagation'' due to its movement from the calculation of derivatives in the last layer back to the first one (see Figure \ref{fig:CNN_image_2}).

It's important to highlight that the application of the gradient step occurs not on all the samples in the dataset but rather on a batch of samples, where the \textbf{batch size} is a hyperparameter determined for the training of the neural network. Utilizing batch training enables more efficient computation and provides more consistent weight updates.

In Stochastic Gradient Descent (SGD), the term ``stochastic'' originates from the random sampling of the batch. An \textbf{epoch} is completed when all dataset samples are used at least once. Variations of Equation \ref{eq:update_weights} have been introduced to enhance gradient stability and faster convergence. The incorporation of momentum aims to mitigate oscillations by updating weights with an exponentially smoothed gradient:
\begin{align*}
V_{i+1} &= \gamma V_i + \alpha \nabla L(W_i) \\
W_{i+1} &= W_t - V_{i+1} \\
\end{align*}
Here, $W_i$ denotes the condensed representation of all the weights (or parameters) in the neural network at iteration step $i$. $\nabla L(W_i)$ is the gradient, and $V_{i+1}$ is the smoothed version, involving the hyperparameter $\gamma$, which updates $W_i$ to $W_{i+1}$.

In NAG (Nesterov Accelerated Gradient), momentum is calculated on the gradients computed not at the current parameters but at a corrected version $\nabla L(W_i - \gamma V_i)$, a step that aims to anticipate the updated parameters in advance.

To prevent the gradient norm from exploding, normalization can be employed. In Adagrad, the square of the gradients is computed and used for parameter updates:
\begin{align*}
G_{i+1} &= G_i + (\nabla L(W_i))^2 \\
W_{i+1} &= W_t - \frac{\alpha}{\sqrt{G_{i+1} + \epsilon}} \nabla L(W_i) \\
\end{align*}
RMSprop follows a similar concept, employing an exponential moving average for $G$.

Adam combines both momentum and squared gradient normalization, while Nadam extends this further by including Nesterov's trick \citep{dozat2016incorporating}. The update for Nadam goes as follows:
\begin{align*}
    M_{i+1} &= \beta_1 M_i + (1 - \beta_1) \nabla L(W_i) \\
    V_{i+1} &= \beta_2 V_i + (1 - \beta_2) (\nabla L(W_i))^2 \\
    \hat{M}_{i+1} &= \frac{M_{i+1}}{1 - \beta_1^{i+1}} \\
    \hat{V}_{i+1} &= \frac{V_{i+1}}{1 - \beta_2^{i}} \\
    W_{i+1} &= W_i - \frac{\alpha}{\sqrt{\hat{V}_{i+1}} + \epsilon} (\beta_1 \hat{M}_{i+1} + \frac{(1 - \beta_1)\nabla L(W_i)}{1 - \beta_1^{i}}) \\
\end{align*}

While various forms of gradient descent have been explored, one might consider computing the second-order derivative through the Hessian matrix and updating the step as in the Newton optimization process, as shown in \cite{cavuoti2012photometric}.

\subsection{Regularizer}
Regularizers are essential tools to improve the loss function and mitigate overfitting in ML algorithms, and they can vary across different learners. In decision tree-based algorithms, one can limit the maximum depth of the trees or set a minimum number of samples required for node splitting.
For linear regression, the loss function can be augmented with regularization terms to achieve Ridge Regression, where the Euclidean norm ($L_2$) of the weights is added:
\begin{equation}
L_{\text{Ridge}} = \parallel y - wx - c \parallel_2^2 + \lambda \parallel w \parallel_2^2
\end{equation}
Alternatively, by introducing the absolute norm ($L_1$), we obtain Lasso Regression:
\begin{equation}
L_{\text{Lasso}} = \parallel y - wx - c \parallel_2^2 + \lambda \parallel w {\parallel_1}
\end{equation}
The impact of these regularizers is to shrink the magnitude of the learnable weights. However, Lasso Regression has the added effect of promoting sparsity in the weights, leading to some coefficients being set to zero. This characteristic allows Lasso Regression to act as a feature selector, which can be advantageous in certain scenarios.
The same identical approaches can be adopted to the Neural Network, even adding both the regularizers controlled by some weighting parameters. 

\begin{figure}[ht]
\centering
\includegraphics[width=0.8\textwidth]{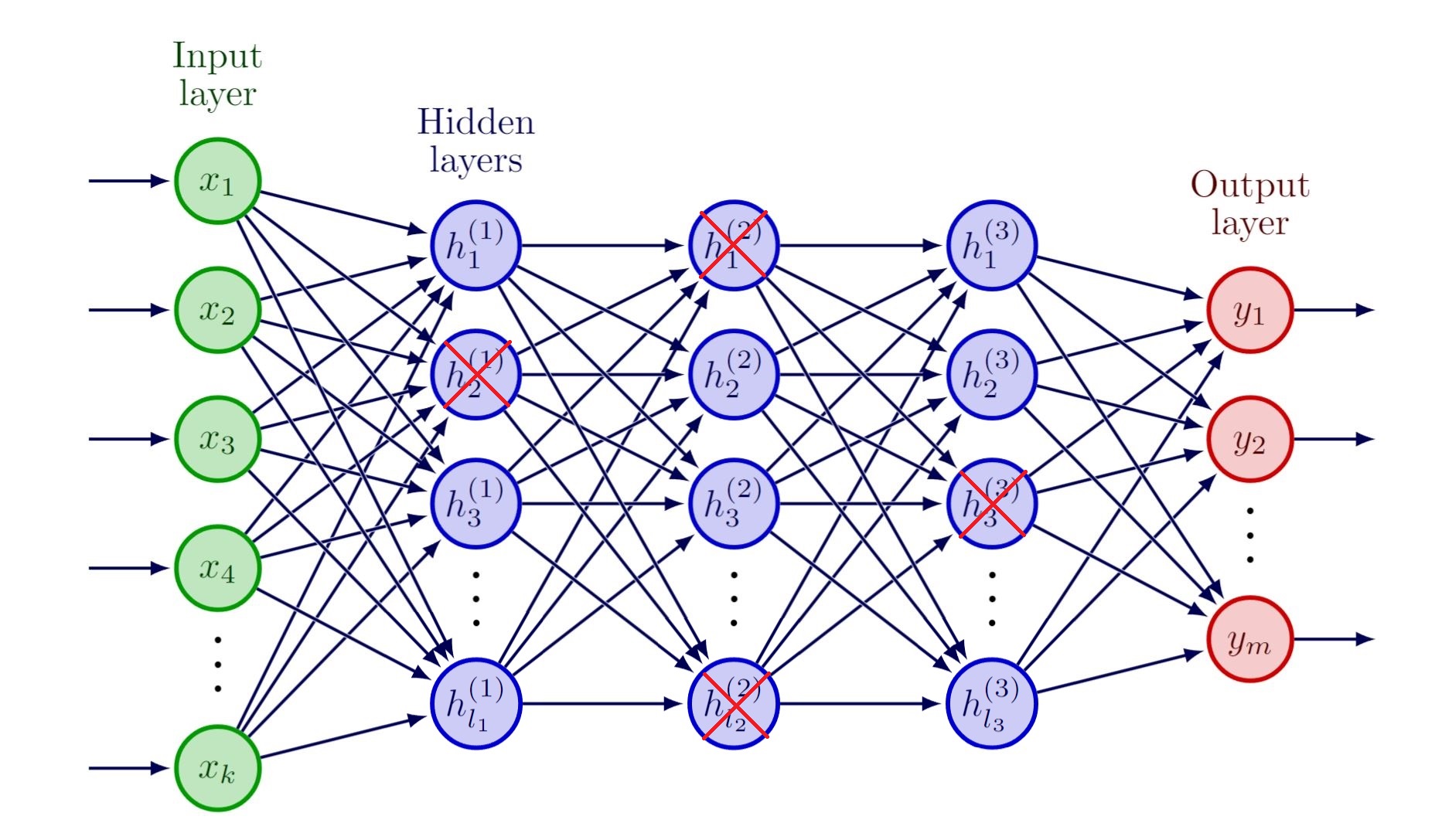}
\caption{With Dropout, some nodes in the Neural Network are deactivated with probability $p$ during the training phase, leading to changes in the architecture with each gradient step. In the prediction phase, all nodes are usually active (the weights have to be multiplied by $1-p$ to maintain the same expected values from training), contributing to better generalization.}
\label{fig:dropout_intro}
\end{figure}

Another type of regularizer specific to Neural Networks is Dropout. Dropout involves randomly ``switching off'' some neurons during the training phase, with the probability of dropout being an adjustable parameter. This process introduces an element of randomness, as the architecture is changed with each gradient step during training. However, in the prediction phase, typically, all nodes are active, simulating an ensamble of different neural networks (with deactivated nodes) and thus resulting in better generalization.
Figure \ref{fig:dropout_intro} provides a visual representation of Dropout in a Neural Network, where some nodes are deactivated with a probability of $p$ during the training phase, contributing to increased robustness and preventing overfitting.
The use of Dropout regularization was essential for the convergence of the Neural Network across three different years of data for the background estimation described in Chapter \ref{chap:bkg}.

In \cite{gal2016dropout} it is demonstrated that the Dropout process can be interpreted as a way to obtain an approximate distribution over the weights of a neural network, equivalent to the concept of a Bayesian Neural Network (BNN). BNN is a type of neural network that treats the weights as random variables and places a prior distribution over them, enabling probabilistic predictions, which is particularly valuable when dealing with small or noisy datasets.

In the context of BNN, for each sample, predictions can be obtained by sampling from the posterior distribution of the weights. Alternatively, in the Dropout setting, nodes can be deactivated even during the prediction phase. By taking enough output samples, one can compute the average to obtain the classic prediction. Moreover, it becomes possible to estimate the quantile-90 and quantile-10 for uncertainty estimation, providing a range of possible outcomes and quantifying the model's uncertainty, similar to Figure \ref{fig:quantile_regression}.

\section{Unsupervised}

\subsection{Clustering}\label{sec:clustering}
In the context of unsupervised learning, when the target variable $y$ is either absent or undefined, one objective could be to identify sets of data points sharing similar characteristics. Machine learning employs clustering techniques to achieve this goal, by grouping data points into clusters based on their feature space $X$ similarities or distances.
Diverse clustering algorithms exist, each accompanied by its specific loss function and distance metric tailored to the task. Some clustering methods allocate a single cluster to each data point, while others provide probabilities indicating membership in various clusters. These probabilities guide the final cluster assignment by choosing the cluster with the highest probability. The latter approach is often labeled as ``fuzzy'' or ``soft'' clustering.
These methods are used to find subgroups of GRBs that share characteristics, such as differentiating between short and long GRBs attributed to different progenitors, in the upcoming chapter, specifically in Section \ref{sec:GRB_cluster}. It raises the intriguing question of whether it is possible to find additional statistically significant divisions within these groups.

\begin{figure}
    \centering
    \includegraphics[width=1\textwidth]{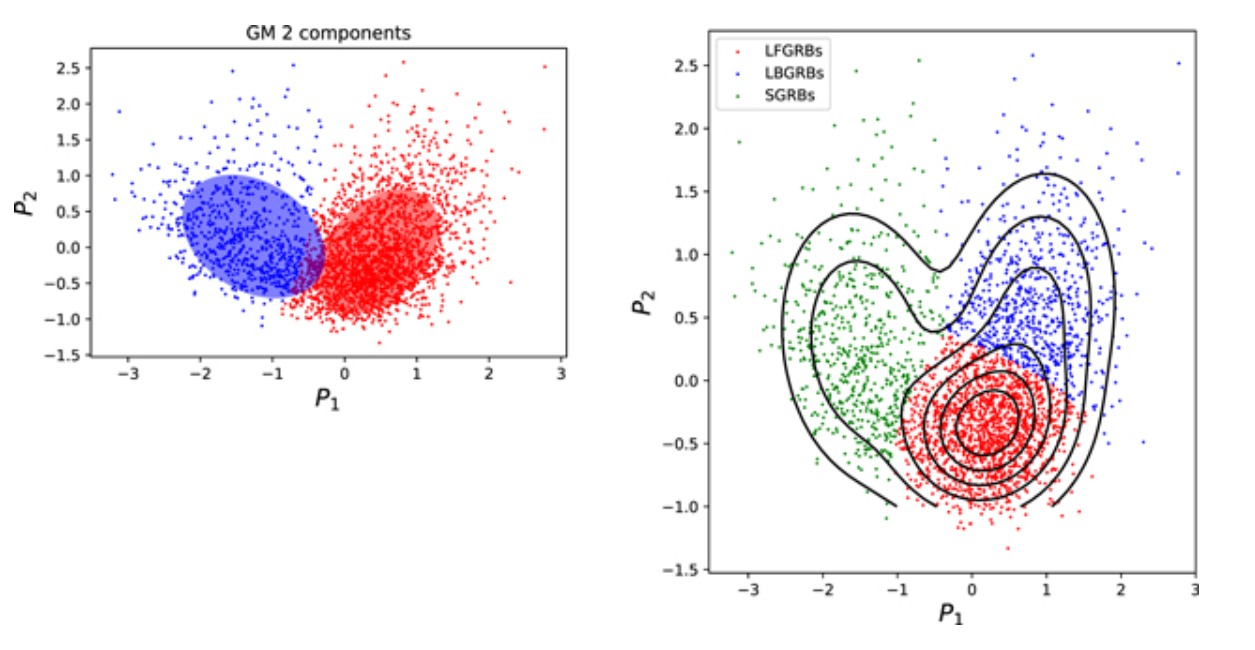}
    \caption{On the left the application of GMM with two cluster components. Blue dots represent short GRBs and red dots represent long GRBs, which are not strictly separated by T90. The covariance of two Gaussian cluster components is represented by the blue and red ellipses, respectively. On the right GRB clustering based on three GMM components. The contours represent lines of equal probability. The green cluster roughly corresponds to SGRBs (on the left figure), while the red and blue clusters are roughly separated from LGRBs. \cite{zhang2022tight} \copyright AAS. Reproduced with permission.}
    \label{fig:gmm}
\end{figure}

Here, we provide a brief overview of some clustering techniques:

\begin{enumerate}
    \item Hierarchical Clustering: This approach creates a hierarchical representation of clusters by either merging similar clusters (agglomerative) or dividing clusters into smaller subclusters (divisive). The key aspect here is defining a distance metric within the data space.
    \item Partitioning Clustering: In this method, data points are divided into non-overlapping clusters. An example is \emph{K-means}, where the goal is to find $k$ homogeneous clusters with similar samples, ensuring clear separation between clusters (minimizing within-cluster variance).
    \item Probabilistic model: The Gaussian mixture model (GMM) is a probabilistic model that assumes the data is distributed as a mixture of $k$ Gaussian distributions (as shown in Figure \ref{fig:gmm}). It's well-suited for density estimation and clustering tasks. The probabilistic nature of GMM allows for cluster overlap, and each sample can be assigned to a cluster based on the highest probability. In addition, the Dirichlet Process Mixture Model (DPMM) extends the GMM by incorporating a Dirichlet Process prior, allowing it to determine the number of clusters automatically. This is in contrast to GMM, which requires the number of clusters to be predefined.
    \item Conceptual Clustering: This category includes methods like CLASSIT and COBWEB \citep{al2006new}, which leverage decision trees where each node represents a concept. The objective is to form clusters based on similar concepts.
\end{enumerate}

A homogeneity loss function is defined in the context of the $k$-means algorithm to ensure the homogeneity of samples within a cluster:
\begin{equation}
L_{\text{Homogeneity}} = \frac{1}{k} \sum_{j=1}^k \sum_{x_i \in C_j} \parallel x_i - \mu(C_j) \parallel^2.
\end{equation}
Here, $\mu(C_j) = \frac{1}{\mid C_j \mid} \sum_{x_i \in C_j} x_i$ represents the centroid of cluster $C_j$.
In addition to the homogeneity loss, another important aspect is the separation among clusters, which can be quantified using the following loss:
\begin{equation}
L_{\text{Separation}} = \frac{2}{k(k-1)} \sum_{i<j} \parallel \mu(C_i) - \mu(C_j) \parallel^2.
\end{equation}

Furthermore, other measures like the contrastive loss can be incorporated, where each sample is compared with another sample of the same label and another of a different label. These losses can be combined to create more complex loss functions such as the triplet loss and the magnet loss \citep{dvedivc2020loss}, offering enhanced mechanisms for clustering evaluation and improvement.
Further research employing contrastive learning can be found in \cite{grill2020bootstrap, chen2021exploring}. Additionally, some methodologies involve a learning process where labels are replaced with ``pseudo labels'' that are estimated by the model itself. This concept, seen in works such as \cite{caron2018deep}, is often referred to as \emph{self-supervised} learning.

The GMM comprises multiple components, each characterized by a Gaussian distribution within the mixture. The GMM's loss function is defined as the negative log likelihood of the data. This is mathematically expressed as:
\begin{equation}
L_{\text{GMM}} = - \sum_{i=1}^n \log \sum_k p(k) \mathcal{N}(x_i; \mu_k, \Sigma_k)
\end{equation}
Here, $n$ represents the number of data samples, $x_i$ denotes the $i$-th data point, and $\mathcal{N}(x_i; \mu_k, \Sigma_k)$ is the probability density function of a Gaussian distribution with parameters $\mu_k$ and $\Sigma_k$, evaluated at the data point $x_i$. The summation runs across all components (clusters) indexed by $k$, while $p(k)$ denotes the probability of a given data sample belonging to cluster $k$. The goal of the GMM is to identify the optimal values of $\mu_k$, $\Sigma_k$, and $p(k)$ that maximize the likelihood of the data within the context of the GMM model.

\subsection{Dimensionality Reduction}\label{sec:dim_rec}

Dimensionality reduction involves the task of projecting the input data $x$ into a lower-dimensional space while retaining maximal information. Formally, dimensionality reduction can be represented as a function $F: x \in \mathbb{R}^k \longrightarrow z \in \mathbb{R}^d$, where $d < k$. When used for visualization purposes, $d$ is typically set to 2 or 3 in order to visualize the dataset in a reduced dimensionality ($Z$) space. In this type of task, the target variable $y$ is not strictly necessary, but if available, it can be treated as a special feature within the dataset $X$.
This visualization has the potential to reveal information about the division of clusters and the degree of similarity between various GRBs, as shown in some applications of Section \ref{sec:GRB_cluster}.

Several widely-used algorithms for dimensionality reduction include:
\begin{itemize}
\item PCA (Principal Component Analysis, \cite{pearson1901liii})
\item t-SNE (t-Distributed Stochastic Neighbor Embedding, \cite{van2008visualizing})
\item SOM (Self-Organizing Map, \cite{kohonen1990self})
\item LDA (previously discussed in the supervised learning section)
\end{itemize}

PCA is a linear dimensionality reduction approach that is used to convert high-dimensional data into a lower-dimensional space while retaining the greatest variance. This is accomplished by locating the primary components, which are orthogonal linear combinations of the original features. The first principal component captures the orthogonal direction with the largest variance, and each subsequent component catches the orthogonal direction with the lower variance. PCA's goal is to reduce the reconstruction error between the original data and the data rebuilt from the lower-dimensional representation. The mean squared reconstruction error, which measures the difference between the original data points and their projections into the lower-dimensional space, is the loss function in PCA.
\begin{equation}
L_\text{PCA} = \parallel X - Z \parallel_F^2 = \parallel X - WC^T \parallel_F^2 = \sum_{i=1}^n \sum_{j=1}^{d} (x_{ij} - \sum_{l=1}^k W_{jl} C_{jl})^2
\end{equation}
Where $X \in \mathbb{R}^{n \times k}$ is the dataset (standardized) and $Z \in \mathbb{R}^{n \times d}$ the dataset in the reducted space, $i$ the index of the sample, $k$ the number of initial features and $d$ the reduced feature space, therefore $d<k$, and $\parallel 
 \cdot \parallel_F$ the Frobenius norm. It can be demonstrated that $C$ are the eigenvector of the matrix $XX^T$ and $WC^T$ are the projection of each dataset $X$ onto the $d$ principal component. This can be generalized in Kernel-PCA \citep{scholkopf1997kernel}, an higher dimensional space $\phi(x)$ and then computing the eigenvector of $\phi(X)\phi(X)^T$ but leveraging only on the kernel $\mathcal{K}(x_i, x_j) = \langle \; \phi(x_i), \phi(x_j) \; \rangle$ (similar to the concept utilized in Support Vector Machines).

In t-SNE  the employed loss function can be expressed as follows:
\begin{equation}
L_\text{t-SNE} = \sum_{i,j} d_x(x_i, x_j) \log \left( \frac{d_x(x_i, x_j)}{d_z(z_i, z_j)} \right)
\end{equation}
where $d_x(x_i, x_j) = \frac{\exp\left(-\frac{\parallel x_i-x_j \parallel^2}{2\sigma_i^2}\right)}{\sum_{k \neq i}\exp\left(-\frac{\parallel x_i-x_k\parallel^2}{2\sigma_i^2}\right)}$ is the distance in the original space, and $d_z(z_i, z_j) = \frac{\exp(-\parallel z_i-z_j\parallel^2)}{\sum_{k \neq i} \exp(-\parallel z_i-z_j\parallel^2)}$ is the distance in the lower-dimensional space. The use of the Student-t distribution for $d_z$ helps preserve the local structure of the data, ensuring that nearby points in the original space remain close in the lower-dimensional embedding, which helps retain the intrinsic neighborhood relationships of the data.

An alternative approach can be found in UMAP \cite{mcinnes2018umap}, where it is claimed that the global structure are preserved. The UMAP loss function is formulated as:
\begin{equation}
L_\text{UMAP} = \sum_{i,j} d_x(x_i, x_j) \log \left( \frac{d_x(x_i, x_j)}{d_z(z_i, z_j)} \right) +
(1 - d_x(x_i, x_j)) \log \left( \frac{1 - d_x(x_i, x_j)}{1 - d_z(z_i, z_j)} \right).
\end{equation}
the similarity functions $d$ are equivalent to t-SNE but omitting the denominator terms (normalization terms).

A comparison of the equation loss of the different methods can be found in \cite{draganov2023unexplainable} and section C of  \cite{mcinnes2018umap}.

The SOM, also known as Kohonen maps, is an artificial neural network-based technique used for dimensionality reduction and visualization of high-dimensional data. It involves mapping the data onto a grid of nodes with each node having a random weight vector $w$ of the same dimension as the input $x$. The SOM is a type of neural network, but unlike typical neural networks, it does not use gradient descent for training. Instead, during the training phase, each input sample $x$ is passed to the network, and it is assigned to the node with the most similar weight vector $w$ (e.g., based on Euclidean distance).
The weight vectors of the winning node $v$ and its neighboring nodes are then updated based on the formula:
\begin{equation}
w_i(t + 1) = w_i(t) + \alpha(t) h_{v,i}(t) [x(t) - y_i(t)]
\end{equation}
where $w_i(t)$ is the weight vector of neuron $i$ at iteration step $t$, $x(t)$ is the input vector, 
$\alpha(t)$ is the learning rate, and $h_{v,i}(t)$ is the neighborhood function that determines the influence of neuron $i$ on the weight update based on its distance to the winning neuron $v$.
The training process continues for a fixed number of iterations or until convergence, resulting in a self-organizing map that represents the input data in a lower-dimensional grid space. Each input is assigned to one node, and the node weights can be further clustered using clustering methods to obtain a clustering in the original input space $X$. Alternatively, the number of nodes in the SOM can be set to the expected number of clusters, and each input can be assigned to the node in the SOM, directly obtaining a clustering result.

\subsection{Generative models}
When the target variable $y$ itself is an image, various tasks can be performed, such as image  \emph{denoising}, where the goal is to ``clean'' the image $x$ by referencing another image $y$, or learning a lower-dimensional representation of the image space to compress and subsequently reconstruct images, typically involving much fewer dimensions than $p_1 \times p_2 \times ch$, where $p_1$ and $p_2$ are the dimensions of the image and $ch$ is the number of channels. Generative models play a crucial role in addressing such tasks. In the GRB context, these techniques could be useful in reconstructing missing part of a signal or generate realistic lightcurve. This latter can be useful for evaluating transient new detection algorithms.

In contrast to the supervised task, such as the classification where the aim is to estimate $\mathbb{P}(y \mid X=x)$, with Generative models the interest is to estimate $\mathbb{P}(X)$. In general it is an heavy task because of the high dimensionality of $x$ and how to define the proper loss function that accurately quantifies the quality of the generated outputs.

Some more innovative applications involve mapping a textual description $x$ to an image $y$, which is achieved using diffusion models. This realm is known as \emph{Generative AI}.

Four prominent generative models are explored:
\begin{itemize}
    \item Autoencoder
    \item Variational Autoencoder
    \item GAN (Generative Adversarial Network)
    \item Diffusion Model
\end{itemize}

The Autoencoder is a model composed by a function encoder $E: x \in \mathcal{X} \longrightarrow z \in \mathcal{Z}$ and a decoder $D: z \longrightarrow x' \in \mathcal{X}$. $\mathcal{Z}$ is the latent space in which the input $x$ is encoded, usually in a lower dimension. Subsequently, $z$ is transformed back into $x'$ using the decoder $D$. The primary objective is to reconstruct $x'$ to closely resemble the original input $x$, and this is often achieved through a loss function, most commonly using Mean Squared Error (MSE):
\begin{equation}
L_\text{AE} = \sum_{i=1}^n \parallel x_i - {x'}_i \parallel^2 =  \sum_{i=1}^n \parallel x_i - {D(E(x_i))} \parallel^2.
\end{equation}
However, for datasets with categorical features, an improvement can be made by incorporating the multiclass cross-entropy loss (as shown in Equation \ref{eq:multiclass_ce}) to more effectively estimate categorical characteristics of $x$.

The Variational Autoencoder \citep{kingma2013auto} is a modification of the Autoencoder in which the latent space distribution is forced to follow a $d$ dimensional standard normal distribution $\mathcal{N}(0, \mathbb{I})$. In practice, for each $x$, the encoder maps it into $\mu_x$ and $\sigma_x$, the two parameters that define the posterior distribution in the latent space.
The VAE loss function is expressed as follows:
\begin{equation}\label{eq:vae}
L_\text{VAE} = \sum_{i=1}^n \parallel x_i - {D(E(x_i))} \parallel^2 + \text{KL}(\mathcal{N}(\mu_{X}, \sigma_{X}), \mathcal{N}(0, \mathbb{I})),
\end{equation}
where $X$ is the dataset $\{x_i \mid i=1:n\}$ and $\text{KL}$ is the Kullback-Leibler divergence, defined as:
\begin{equation}
D_{KL}(P||Q) = \sum_x P(x) \log \left( \frac{P(x)}{Q(x)} \right)
\end{equation}
In the context of the VAE, with Normal distributions, Equation \ref{eq:vae} simplifies to:
\begin{equation}
L_\text{VAE} = \sum_{i=1}^n \parallel x_i - {D(E(x))}_i \parallel^2 + \frac{1}{2} \left( \sum_{i=1}^n \sum_{l=1}^d (\mu_{l,x_i}^2 + \sigma_{l,x_i}^2 - 1 - \log(\sigma_{l,x_i}^2)) \right),
\end{equation} 
where $\mu_{l,x_i}$ and $\sigma_{l,x_i}$ are the mean and standard deviation of the posterior representation in the latent space for the sample $x_i$ across the $l$ component of the latent dimension $d$.

The GAN is a deep learning model consisting of two neural networks: a generator $G: z \longrightarrow x$ and a discriminator $D: x \longrightarrow \{0, 1\}$. The generator's role is to produce new data from a latent space $z$, while the discriminator's role is to determine whether the generated data is real or fake. The loss functions of GANs comprise two components: the generator loss and the discriminator loss. The generator loss quantifies how realistic the generated data is, while the discriminator loss evaluates how effectively the discriminator distinguishes between real and generated data.
The purpose of GANs is to reduce both the generator and discriminator losses. As the generator loss decreases, the generated data becomes more realistic. Similarly, as the discriminator loss decreases, the discriminator becomes more proficient at accurately identifying real and synthetic data.
The mathematical expression for the loss function of GANs is as follows:
\begin{equation}
L_{\text{GAN}}(D, G) = - \sum_{i=1}^n \log D(x_i) - \sum_{j=1}^n \log (1 - D(G(z_j))).
\end{equation}
This loss function aims to minimize the generator's loss while maximizing the discriminator's loss, searching for an equilibrium between the two networks:
\begin{equation}
    \min_G \max_D L_{\text{GAN}}(D, G).
\end{equation} 


The Diffusion Model is a recent generative model introduced by \cite{ho2020denoising} learns to transform a simple noise distribution into complex data samples, such as images, through a series of sequential $T$ diffusion steps. Each diffusion step applies a diffusion process to the noise, resulting in iteratively updated samples that gradually become more complex and closer to the target data distribution. 
A simplified version of this loss function is given by:
\begin{equation}
L_{\text{DM}} := \sum_{t, x_0, \epsilon} \left\| \epsilon - \epsilon_\theta \left( \sqrt{\overline{\alpha}_t} x_0 + \sqrt{1 -  \overline{\alpha}_t} \epsilon, t \right) \right\|^2
\end{equation}
where $\epsilon \in \mathcal{N}(0, I)$, $x_0$ represents an image in the input space $X$, $t$ the diffusion step ranging from 0 to $T$, $\theta$ refers to the parameter of an approximator $\epsilon_{\theta}$ that estimates $\epsilon$ based on $x_t$, and $\overline{\alpha}_t$ are parameters controlling the level of noise added during each diffusion step.

\section{Semi-supervised}

Semi-supervised learning is a machine learning paradigm that leverages both labeled and unlabeled data for training. This approach stands between supervised learning, which uses only labeled data, and unsupervised learning, which relies solely on unlabeled data. The goal of semi-supervised learning is to utilize the patterns and structure present in the unlabeled data to enhance the performance of a model. By incorporating unlabeled data, the model can be regularized to prevent overfitting on the labeled data. Semi-supervised learning becomes particularly powerful when labeled data is limited. GRB are a frequent example of this problem, where progenitor nature is unknown until afterglow analysis, host galaxy observation, or the detection of coincident gravitational waves (GWs) can be done.

One approach in the family of semi-supervised models involves a joint optimization of supervised and unsupervised loss functions:

\begin{equation}
L = L_{\text{supervised}} + \lambda L_{\text{unsupervised}},
\end{equation}

Here, $L_{\text{supervised}}$ represents the loss associated with the labeled data, $L_{\text{unsupervised}}$ corresponds to the loss derived from unlabeled data, and $\lambda$ controls the balance between these two components during training.

\begin{figure}[!htb]
\centering
\includegraphics[width=0.7\textwidth]{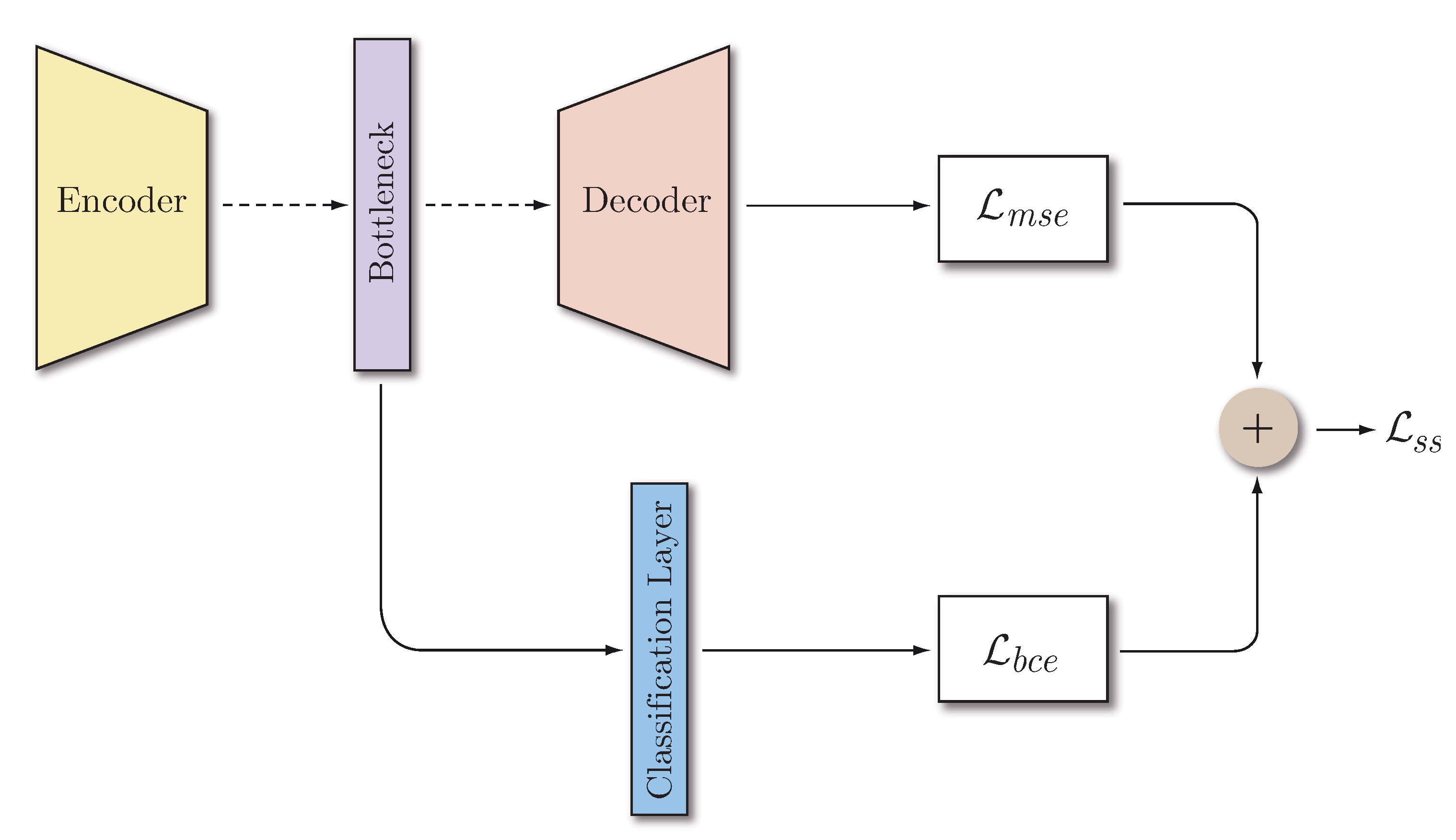}
\caption{\label{fig:semisup_autoenc}
In this semi-supervised neural network architecture the $x$ input is reconstructed by an autoencoder passing through the bottleneck layer. The unsupervised loss function is MSE, $L_{\text{unsupervised}} = L_\text{AE}$. Starting from the latent (bottleneck) vector it is added a classification layer (e.g. a fully connected layer) and the supervised task is performed. In particular in the image deal with binary label with the binary cross-entropy $L_{\text{supervised}} = L_{\text{bce}}$. \cite{naranjo2020open} \copyright Naranjo et al. (2020), \href{http://creativecommons.org/licenses/by/4.0/}{CC BY 4.0}.}
\end{figure}

For instance, consider a scenario where you have a dataset of images, with only a few of them labeled with corresponding targets. An autoencoder could be trained to capture the probability distribution of input images $x$. The encoder component, responsible for generating latent vectors $z$, can be utilized as a feature extractor. An additional dense layer can be added to discriminate the target variable $y$ for labeled samples. Then this process can be performed alternating the training phase between batches of labeled and unlabeled samples, aims to optimize both feature extraction and the supervised task (see Figure \ref{fig:semisup_autoenc}). 

\begin{figure}[!htb]
\centering
\includegraphics[width=.7\textwidth]{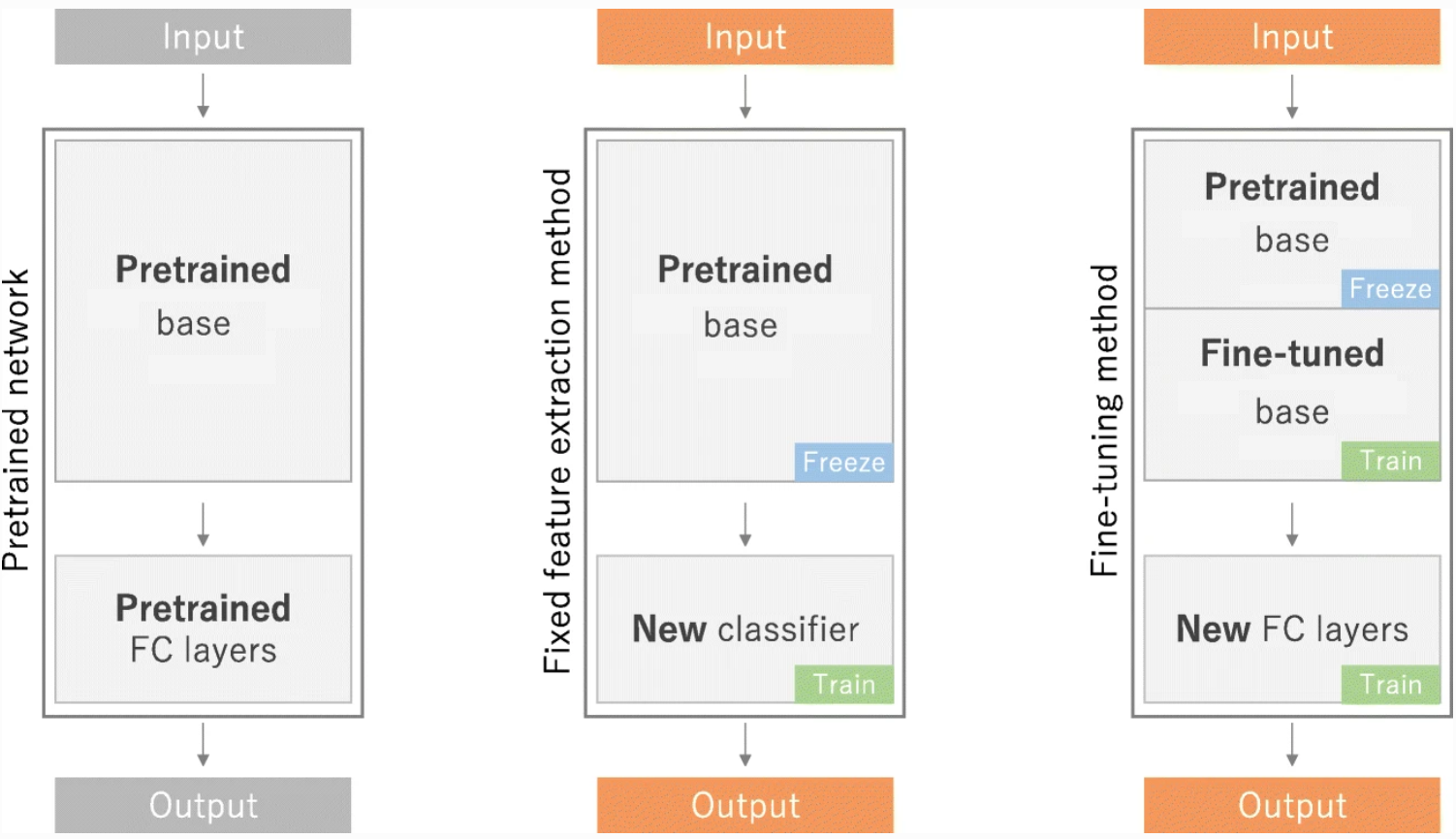}
\caption{\label{fig:transfer_learning}
Transfer learning is a widely used strategy for training networks on small datasets. The approach involves pretraining a network on a large dataset, and then applying it to a specific task. 
On the left, the network is employed directly for the specific task with no modification.
In the middle, the initial layers (base) works as a feature extractor and the fully connected (FC) layers are removed. This allows other classifiers such as Random Forest or SVM to be added on top. 
Alternatively, in the right, it is possible to fine-tune not only a new classifier but also a portion of the base layers through back-propagation to allow feature extraction personalized according to the specific data. The image has been slightly adapted from \cite{yamashita2018convolutional} \copyright Yamashita et al. (2018), \href{http://creativecommons.org/licenses/by/4.0/}{CC BY 4.0}.}
\end{figure}

Alternatively, the problem can be addressed in two separate steps. First, a neural network is trained to address a different task involving a large dataset where the target variable $\Tilde{y}$ is available. 
Depending on factors such as the dataset size, the amount of labeled samples, and the relevance of the input $x$ to the specific problem at hand, the parameters of the neural network might be ``freezed'' in the training phase. Second, perform the training starting from the pretrained network to estimate the target variable $y$, this process is known as \emph{fine-tuning}. This strategy for transferring knowledge from one domain to another is referred to as \emph{Transfer Learning} (Figure \ref{fig:transfer_learning}).

In a different approach, a supervised model could be trained on labeled data and then the labels are propagated to the unlabeled data, iteratively refining the model. This iterative process continues until the model is trained on all labelled data. This approach is similar to \emph{self-supervised} learning, with the distinction that in the initial step, some target labels are already available.

\begin{figure}[!htb]
	\centering
	\includegraphics[width=1\textwidth]{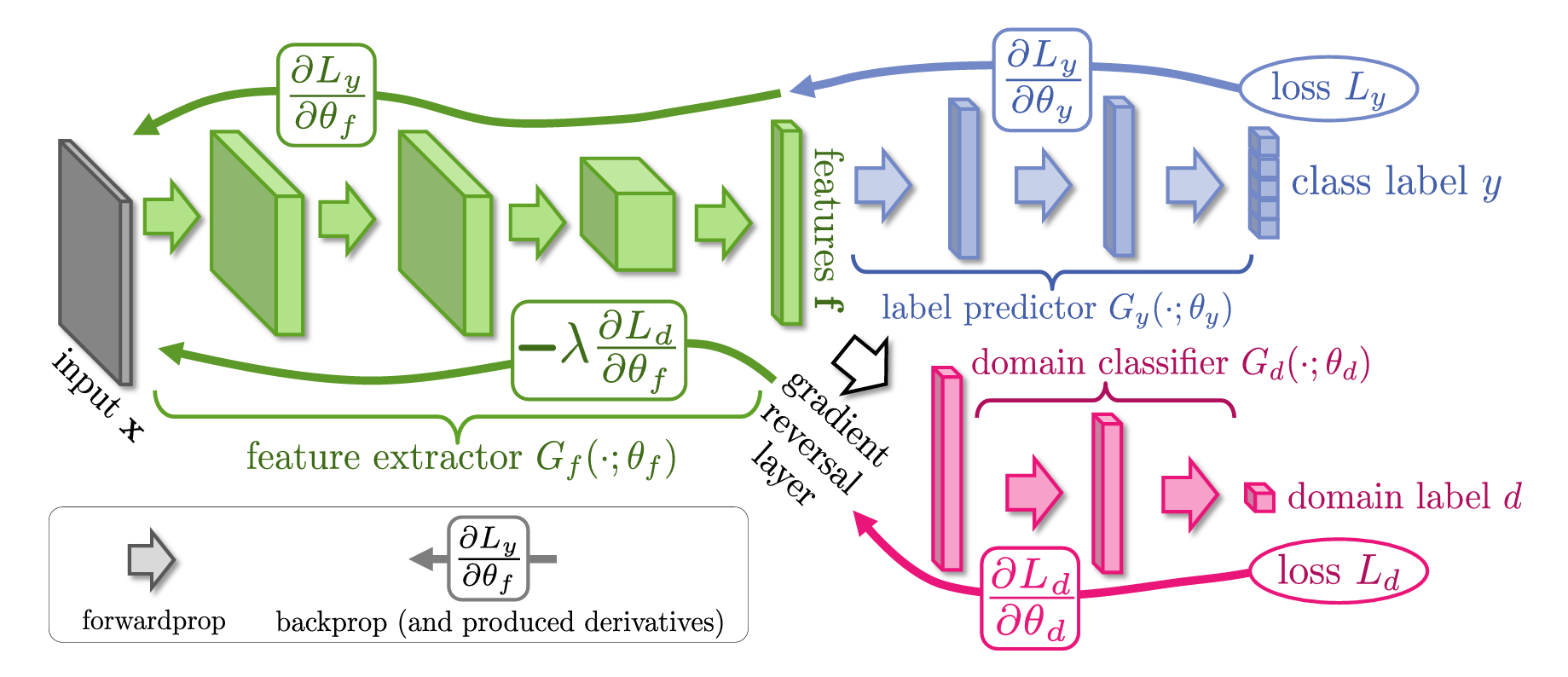}
	\caption{\label{fig:gradient_reversal} The Neural Network architecture includes a deep feature extractor (green) and a deep label predictor (blue). To induce domain-invariant features, a domain classifier is introduced during training, connected to the feature extractor through a gradient reversal layer (pink). \cite{ganin2015unsupervised} \copyright Ganin et al. (2015). Reproduced with permission.}
\end{figure}

Neural networks' flexibility in constructing custom loss functions allows them to address problems where the target domain differs from the source domain. Imagine a scenario where a model is trained to classify images from social media but needs to be applied to surveillance camera images, which lack labels. Is it possible to extract domain-independent features useful for classification? Techniques like the one proposed in \cite{ganin2015unsupervised} use gradient reversal layers to enforce domain-invariance in features, as illustrated in Figure \ref{fig:gradient_reversal}. Such methods can enable fair predictions and feature extraction that are independent of specific domains, as demonstrated in applications like predicting income while mitigating gender bias \cite{zhang2018mitigating}.
This kind of technique may be useful in the analysis of GRBs discovered by various satellites which have various sensitivities. A satellite-invariant analysis made possible by such a method would make it easier to examine GRBs coming from various sources in a coherent manner.


\section{Validation, test set and evaluation metrics}\label{sec:eval_metrics}
\begin{figure}
    \centering
    \includegraphics[width=0.75\textwidth]{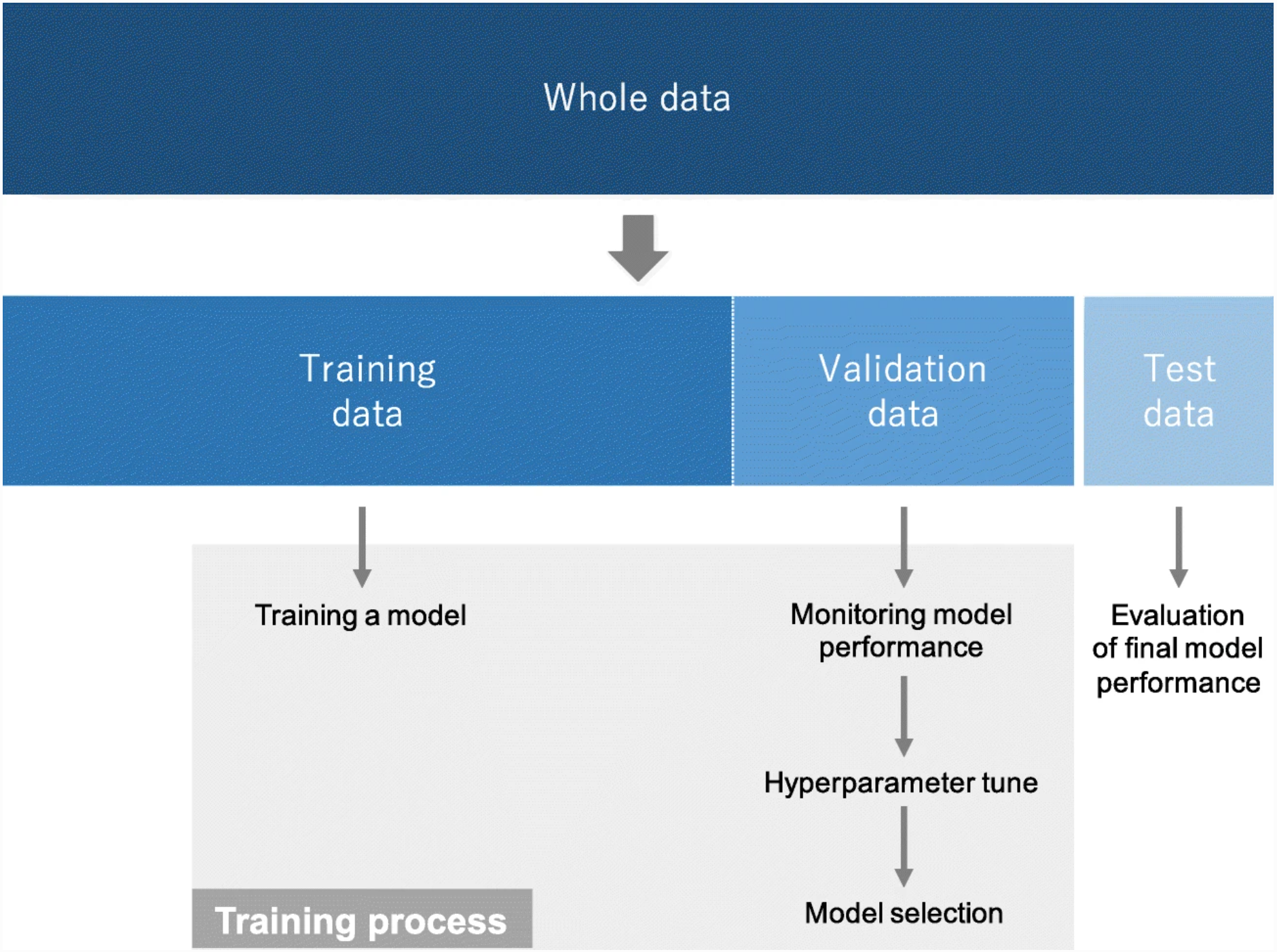}
    \caption{Train, validation and test set. \cite{yamashita2018convolutional} \copyright Yamashita et al. (2018), \href{http://creativecommons.org/licenses/by/4.0/}{CC BY 4.0}.
    }
    \label{fig:train_test_val}
\end{figure}

\begin{figure}
    \centering
    \includegraphics[width=0.75\textwidth]{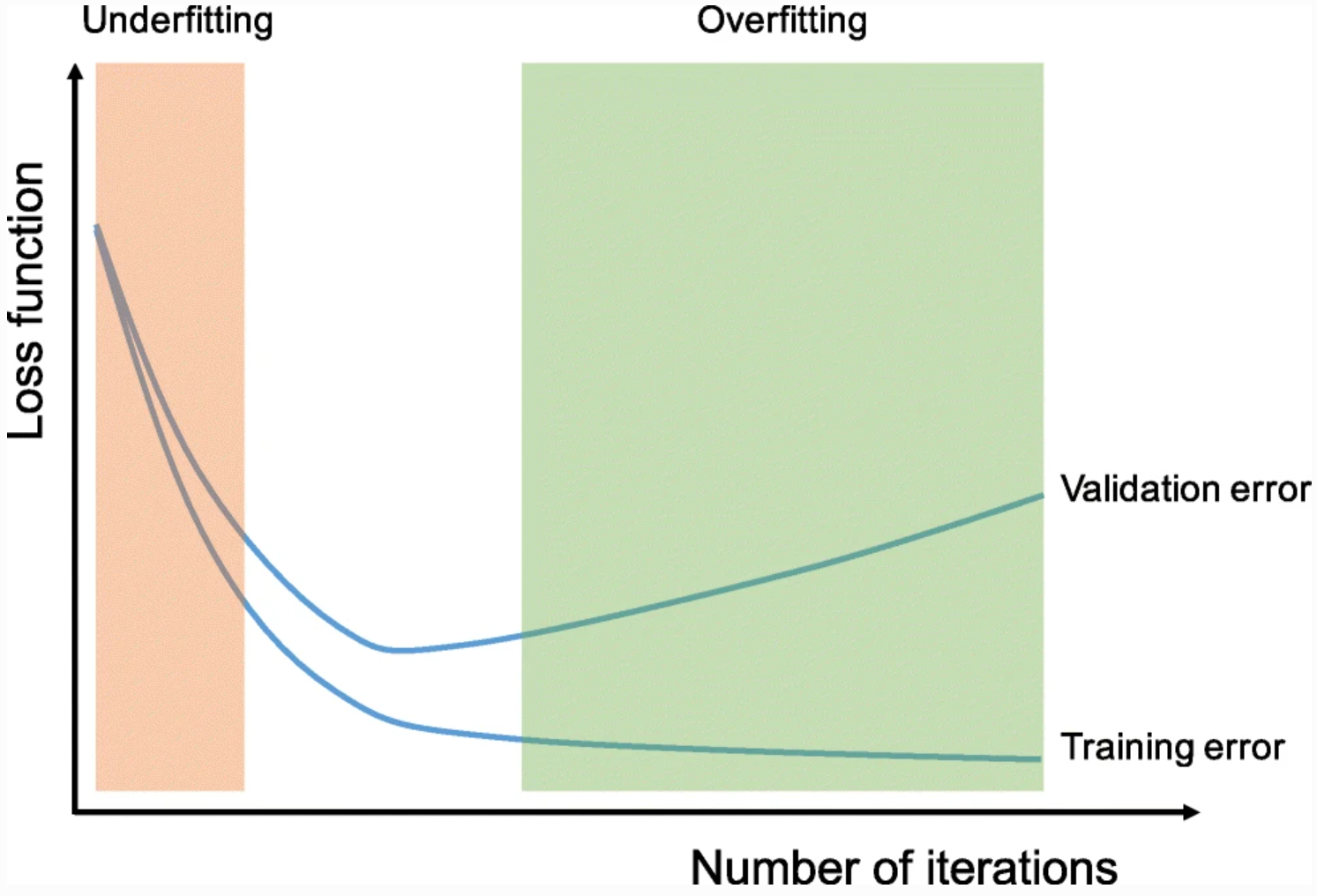}
    \caption{Overfitting occurs when the model becomes too specialized to the training data, learning noise or irrelevant patterns, which hurts its ability to perform well on new data. As a result, model's performance on the training set is significantly better than on the validation set. On the other hand if the model performs poorly on both the training and validation sets, it is likely that the model is underfitting the data and not capturing the underlying patterns. \cite{yamashita2018convolutional} \copyright Yamashita et al. (2018), \href{http://creativecommons.org/licenses/by/4.0/}{CC BY 4.0}.
    }
    \label{fig:overfitting}
\end{figure}

Datasets are typically divided into three subsets in machine learning: the training set, the validation set, and the test set (Figure \ref{fig:train_test_val}). The training set is used to train the model, which is then adjusted to minimize the loss function. During the training process, the validation set is used to tune hyperparameters and prevent overfitting (Figure \ref{fig:overfitting}) by monitoring the model's performance on unseen data. Finally, the test set is used to evaluate the model's generalization capabilities on data not belonging to train or validation set. Machine learning models can provide a reliable assessment of their performance on new and unseen data by using separate sets for training, validation, and testing, ensuring their suitability for real-world applications. 
However, the construction of these sets should be done with a thorough understanding of the problem being addressed. For forecasting tasks, it is essential to have the validation and test sets represent future periods compared to the training set. Additionally, when dealing with GRBs observed by multiple detectors, it is important to avoid separating correlated lightcurves into different sets, as this can lead to an underestimation of overfitting and compromise the validation results.

Evaluation metrics are critical in machine learning for assessing model performance and determining their effectiveness in solving specific tasks. Some common evaluation metrics for classification tasks include accuracy, precision, recall, balanced accuracy, F1-score, Matthews correlation coefficient and area under the receiver operating characteristic curve (AUC-ROC). 
Precision represents the ratio of true positive predictions to total predicted positive instances, highlighting the model's ability to avoid false positives. 
Recall, also known as sensitivity or true positive rate, is the ratio of true positive predictions to the total positive instances in the dataset, highlighting the model's ability to correctly capture all positive instances.
Accuracy measures the proportion of correctly classified instances in the dataset.
The F1-score is the harmonic mean of precision and recall, balancing the two metrics. It is particularly useful for imbalanced datasets where one label is more prevalent than the others. Additionally, to address imbalanced datasets, Balanced Accuracy and Matthews correlation coefficient (MCC) can be employed. Balanced Accuracy considers the proportions of positive and negative instances, and MCC is a normalized version of the confusion matrix $ \left[ \begin{matrix}
\text{TP} & \text{FP}\\
\text{FN} & \text{TN}
\end{matrix} \right]$ that includes true positives (TP), false positives (FP), false negatives (FN), and true negatives (TN).
TP is the number of data points correctly classified as positive. TN is the number of data points correctly classified as negative. FP is the number of data points that were incorrectly classified as positive. FN is the number of data points incorrectly classified as negative.
A comparison of these three metrics for imbalanced dataset can be found in \cite{chicco2020advantages}.
The AUC-ROC metric measures the model's ability to differentiate between classes, with a higher value indicating better discrimination. 

\begin{align*}
\text{Accuracy} &= \frac{\text{TP} + \text{TN}}{\text{Total}} = \frac{\text{TP} + \text{TN}}{\text{TP}+\text{FP}+\text{FN}+\text{TN}} \\
\text{Precision} &= \frac{\text{TP}}{\text{TP} + \text{FP}} \\
\text{Recall} &= \frac{\text{TP}}{\text{TP} + \text{FN}} \\
\text{F1-Score} &= 2 \cdot \frac{\text{Precision} \cdot \text{Recall}}{\text{Precision} + \text{Recall}} \\
\text{Balanced Accuracy} &= \frac{1}{2} \left( \frac{\text{TP}}{\text{TP} + \text{FN}} + \frac{\text{TN}}{\text{TN} + \text{FP}} \right) \\
\text{MCC} &= \frac{\text{TP} \cdot \text{TN} - \text{FP} \cdot \text{FN}}{\sqrt{(\text{TP} + \text{FP})(\text{TP} + \text{FN})(\text{TN} + \text{FP})(\text{TN} + \text{FN})}} \\
\text{AUC-ROC} &= \int_{0}^1 \text{TPR}(T) \, dT \\
\end{align*}

The ratio of TP to the sum of TP and FN is known as the TPR (True Positive Rate). The FPR (False Positive Rate) is defined as the ratio of FP to the sum of FP and TN. The $T$ value in the AUC-ROC is the threshold to determine the outcome 0 or 1 from the probability output of the ML model, see Equation \ref{eq:generic_ml_classification_map}.

MSE, root mean squared error (RMSE), MAE, and R-squared ($R^2$) coefficient of determination are common evaluation metrics for regression tasks. $R^2$ is a measure of the proportion of variance in the target variable that is explained by the model and ranges from 0 to 1, with a higher value indicating a better fit. For robust error estimation in the presence of outliers, the Median Absolute Error (MeAE) ($\text{median} \mid y - \hat{y} \mid$) can be employed. MeAE and MAE will be crucial metrics for the background estimator described in Chapter \ref{chap:bkg}, serving as evaluation and training metric, respectively.

Metrics such as MSE and MAE, which are used as loss functions, are directly minimized during the training process because they are differentiable\footnote{For the MAE, a special case is considered when $y - \hat{y} = 0$, which is non-differentiable. However, this is managed by the optimizers through the use of subgradients.}, allowing for the estimation of gradients and optimization. On the other hand, classification metrics and Median Absolute Error (MeAE) are not directly differentiable and cannot be easily minimized as loss functions. For classification predictions, a probability threshold with a value $T$ is often applied to obtain the final binary prediction (0 or 1), which further complicates the differentiability of the problem. Moreover, classification metrics and MeAE can only be computed for a group of consistent samples rather than individual data points. Therefore, they are more commonly used for evaluation purposes to assess the overall performance of the model rather than as loss functions in the training process. 

\section{XAI}\label{sec:xai}

As ML and DL models become more complex and are trained on large datasets with High Performance Computing infrastructure, there is an increasing risk of deploying decision systems that lack comprehensive human comprehension \citep{guidotti2018survey}.
The advancement of eXplainable AI (XAI) techniques in recent years has been motivated by the realization that the success of many AI applications has been challenged by ethical concerns and user distrust. The underlying premise driving these efforts is that by developing more transparent, interpretable, or explainable systems, users can gain a better understanding of intelligent agents, fostering greater trust and confidence in them \citep{miller2019explanation}. As a result, the pursuit of explainable AI serves as a critical response to these challenges, attempting to bridge the gap between AI's capabilities and user understanding while also enhancing the responsible and informed use of AI technologies.

There are several methods for providing explanations, each with its own set of advantages and disadvantages. One approach involves creating inherently interpretable models, often referred to as white box models, such as decision trees or linear regression. However, these models might not achieve the same performance levels as more complex black box models. 
One could rely on "global" information such as Feature Importance of a ensembles tree or a Sensitivity Analysis which provide a general overview of what features a model is using most and which has more effect. The Morris method \cite{morris1991factorial} is worth mentioning in the context of Sensitivity Analysis. This method performs a one-step-at-a-time global sensitivity analysis, which is suitable for quickly screening important inputs but does not account for nonlinear interactions. 

\begin{figure}[!htb]
\centering
\begin{subfigure}{1\textwidth}
  \centering
  \includegraphics[width=1\linewidth]{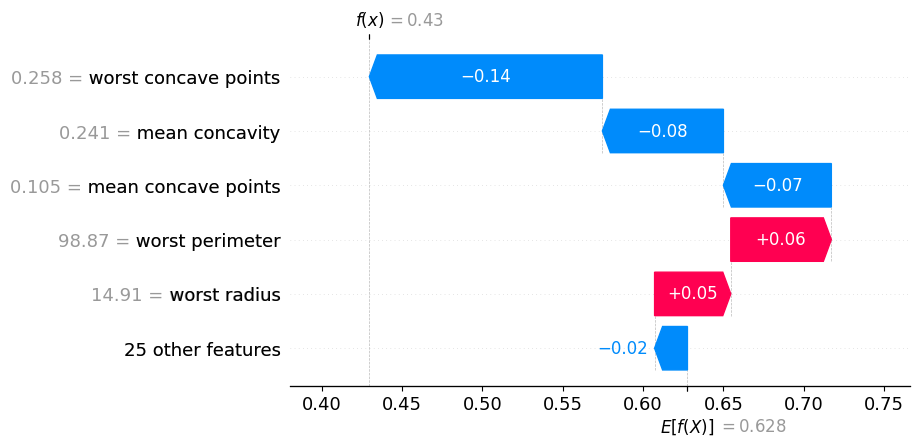}
  \caption{Illustration of feature importance attribution for a classification problem of the \href{https://archive.ics.uci.edu/dataset/17/breast+cancer+wisconsin+diagnostic}{breast cancer wisconsin dataset}, where features are sorted by their impact on determining the target label (0 and 1 correspond to malignant and benign cancer, respectively). This example employs SHAP, but similar results can be obtained with techniques like LIME, Integrated Gradients, and DeepLIFT. More details can be found in the notebook in this \href{https://colab.research.google.com/drive/1wZiz6GoPKMNxpP593SaYk76B2A_y55XS?usp=sharing}{hyperlink}.
  }\label{fig:lime_expl}
\end{subfigure}
\qquad
\begin{subfigure}{1\textwidth}
  \centering
  \includegraphics[width=0.75\textwidth]{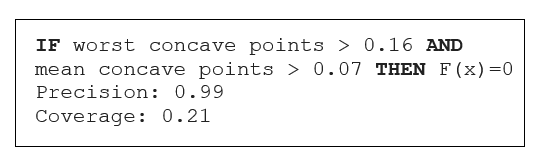}
  \caption{The Anchor offers a rule for explaining an instance, which is the same as the one in (a). Precision is an estimate of the percentage of points within the rule that satisfy the predicate, while coverage represents the percentage of points in the dataset that satisfy the condition rule.}\label{fig:anchor_expl}
\end{subfigure}
\qquad
\begin{subfigure}{1\textwidth}
  \centering
  \includegraphics[width=0.85\textwidth]{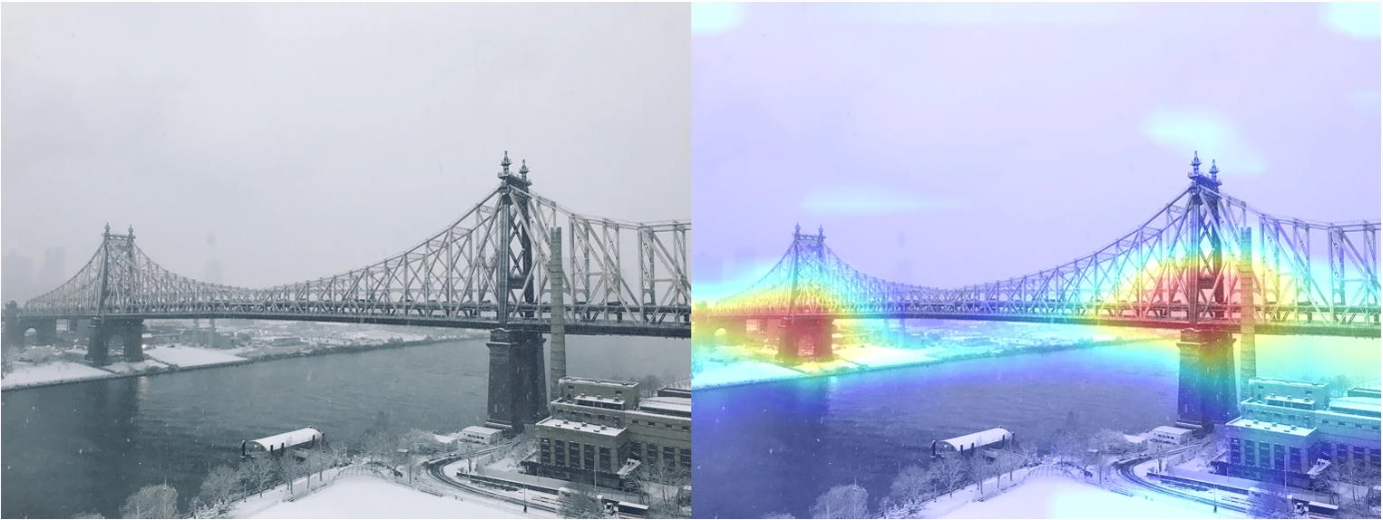}
  \caption{The left figure represents an image, and the right one depicts pixel activations for the label "bridge pier" employing Grad-CAM. \cite{yamashita2018convolutional} \copyright Yamashita et al. (2018), \href{http://creativecommons.org/licenses/by/4.0/}{CC BY 4.0}. Similar results could be obtain with DeepLIFT and Integrated Gradients.}\label{fig:gradcam_expl}
\end{subfigure}
\caption{\label{fig:xai_expl} Examples of common XAI techniques.}
\end{figure}

Given the limitations of global explanations, various techniques have been developed to provide more tailored explanations for individual predictions, emphasizing "locality" with respect to the input $x$. Among the most important XAI techniques are:

\begin{enumerate}
    \item LIME (Local Interpretable Model-agnostic Explanations): LIME generates linear local surrogate models around a specific instance to approximate the original model's decision-making process in an interpretable way \citep{ribeiro2016should}.
    \item SHAP (SHapley Additive exPlanations): SHAP values quantify each feature's contribution to the output of a model, providing a unified framework for feature importance analysis \citep{lundberg2017unified}. It's based on cooperative game theory concepts and calculates the average contribution of each feature to the prediction across various coalitions (groups of features). Figure \ref{fig:lime_expl} shows an example of the method's output.
    \item \emph{Anchors} are if-then rules that capture model behaviour for a specific instance and provide human-readable explanations \citep{ribeiro2018anchors}. Figure \ref{fig:anchor_expl} shows an example of the method's output.
    \item Given an input $x_0$ to explain, one may wonder what are the features to be changed in order to achieve a different prediction. A \emph{Counterfactual Explanation} \citep{wachter2017counterfactual, verma2021counterfactual} is an alternative input $x_{cf}$ ``near'' (according to a defined distance metric) that would result in different model outputs $F(x_0) \ne F(x_{cf})$. By generating one or more counterfactuals, one can provide insights on what features need to change in order to have a different outcome. 
    This approach can be valuable in the interaction with the end-user. In contexts such as credit lending, this technique would help answering the question ``What minimal changes in my application details could lead to loan approval?'' to an individual whose application was rejected, see Crupi et al. (2022) \citep{crupi2022counterfactual}.
    \item Gradient-weighted Class Activation Mapping (Grad-CAM, \cite{selvaraju2017grad}): is a technique that visually highlights important regions in an input by computing the gradient of the model's output with respect to the last convolutions layer. Grad-CAM offers a targeted explanation by focusing on class-specific image. Figure \ref{fig:gradcam_expl} shows an example of the method's output.
    \item Integrated Gradients \citep{sundararajan2017axiomatic} is a technique that attributes feature importance leveraging on the gradient of a model with respect to the input data $\nabla_x F(x)$. Given a baseline input $x_0$ and the input to be explained $x$ the feature importance are define by the integral $$\int_{\lambda \in [0,1]} (x_0 - x) \times \nabla_x F(x_0 + \lambda x) d\lambda.$$
    In the context of images, the baseline input $x_0$ is typically chosen as an input with all pixels black (e.g. with pixel values of zero) or a constant color. This serves as a reference point for determining the relevant pixels. When dealing with tabular data, however, $x_0$ may correspond to the average values of each feature.
    \item Unlike Integrated Gradients, DeepLIFT \citep{shrikumar2017learning} is a technique that directly uses the differences in output values, denoted as $\Delta_t = F(x) - F(x_0)$, to attribute feature importance to neural network nodes, ultimately reaching the input features. This distribution is achieved by utilizing the first-order approximation of gradients. This approach offers a way to assign importance scores to input features based on their contributions to the difference in model outputs.
\end{enumerate}

It's important to emphasize that in this context, the gradient is calculated with respect to the input space, denoted as $\nabla_x F(x)$, rather than the parameters space $\nabla_W F(x)$. This quantity helps to comprehend the local changes in the input that yield the most significant variation in the model's prediction $F(x)$, thus enhancing the interpretability of the analysis. 
Additional explainability techniques in computer vision, aimed at visualizing and interpreting deep neural network outputs, can be found in \cite{montavon2018methods}. However, their applicability and effectiveness in the context of astrophysics need to be demonstrated and evaluated. In Section \ref{sec:transient_classification} Grad-CAM is employed for identifying the pixel that contribute more in the transiet type classification (GRB, non-GRB) while SHAP can be employed for understand better and debug the NN background estimator in Chapter \ref{chap:bkg}.

\chapter{Applications of AI to GRB}\label{chp:grb_ml}

The importance of Data Science has raised in many fields, including astronomy. Astroinformatics arose as a promising sub-field of Data Science and Astrophysics around 2010. The increasing amount of available data \citep{borne2010astroinformatics} makes it quite difficult for the experts to actually analyse all of them and extract useful insights. AI, ML and DL are used in this area of research to automatise the analysis by learning from examples already classified, to forecast events or to give new scientific insights. Some applications in which ML is established are: discovering extrasolar planets, transient objects, quasars, and gravitationally-lensed systems; forecasting solar activity; and distinguishing between signals and instrumental effects in gravitational wave astronomy \citep{fluke2020surveying, ball2010data}.

In this thesis, the main astrophysical object of interest are X-gamma ray astronomical events, in particular GRBs. 
ML and DL algorithms can help scientists in multiple aspects of the research in GRBs to overcome the limitation of the classical approach. 
In this Chapter, I present a review of the possible applications of Artificial Intelligence, Machine Learning and/or Deep Learning techniques for the discovery or the analysis of GRBs.
The advantage of these algorithms is that they can deal with a lot of features and find relationships with interesting quantities or characteristics that are too complex to emerge using classical methodologies. Moreover, they do not rely on rigid rule-based algorithms, but rather learn patterns from data in a flexible manner. 

From the literature, the applications can be categorized into four line of research: 
\begin{enumerate}    
    \item \textbf{GRB clustering}. A GRB's prompt phase can be represented as a lightcurve from which temporal properties such as duration, fluence (integral of energy per unit area), peak flux (average of energy per unit area), and so on can be derived. 
    On the other hand, spectral properties can distinguish the GRBs, such as $E_{peak}$, $\alpha$ and $\beta$ of the Band function (see Section \ref{sec:grb_prompt}). With clustering techniques (Section \ref{sec:clustering}) GRBs can be clustered into subgroups based on similar features. The works described in Section \ref{sec:GRB_cluster} push in the direction of identifying progenitors (should similar GRBs have similar progenitors?) and distinguish GRBs with diverse prompt emission mechanism, in line with the Open Questions 1 and 2 in Section \ref{sec:open_quest}. Table \ref{tab:work_table} provides an overview of the different approaches described in the section.
    \item \textbf{Transient detection and classification}. The telescopes dedicated for detecting X/gamma-ray sources produce a vast amount of data that must be scanned by trigger algorithms in order to identify astronomical transients, which appear as a significant excess of photon count rates relative to background count rate noise. However, mere detection is not the only goal. Equally important is the subsequent classification of these transients, which allows for the identification of their nature or indicate situations in which a further analysis would be beneficial. This classification include the identification of transients such as GRB, solar flare, galactic X-ray flash, soft gamma repeater and others. See Section \ref{sec:transient_classification}.
    \item \textbf{Cosmological properties}. In order to answer Open Question 5 "Can GRBs be used to constrain cosmological parameters?", the works in \ref{sec:cosmological} aim to estimate the redshift leveraging from data coming from different detectors, including gravitational wave observations if available. 
    \item \textbf{GRB simulation}. The final Section \ref{sec:grb_simulation} is about GRB modeling and simulation. GRB prompt emissions are highly diverse (see Figure \ref{fig:grbex}), and while a canonical X-ray afterglow exists (see Figure \ref{fig:afterglow}), a single model to describe it remains difficult to identify. The section \ref{sec:GRB_cluster} explores AI approaches for reconstructing missing GRB afterglow or generating realistic prompt emissions while preserving their diversity. This method could be useful for testing detection algorithms for future missions to ensure sensitivity and robustness across different types of GRBs.
\end{enumerate}

\section{GRB subgroups}\label{sec:GRB_cluster}

Grouping GRBs in clusters can offer valuable insights into their progenitors. The association between GRBs and supernovae (SNe) was first observed in 1998 with the ``liberation burst'' detected by Beppo-SAX, suggesting that long GRBs originate from the core collapse of massive stars (Type II), see Sections \ref{sec:progenitors} and \ref{sec:bepposax}. In 2006, the proposal that short GRBs result from mergers of compact objects in binary systems (Type I) was introduced \citep{zhang2006burst}, which was later confirmed by the discovery of GW170817 and GRB 170817A \citep{abbott2017gravitational}.

However, distinguishing GRBs solely based on their duration is insufficient, and additional properties, such as spectral characteristics, are essential. For instance, the 1s duration GRB 200826A shares more properties with long GRBs, while the long GRB 060614, lacking an association with a SN, exhibits similarities to Type I events. Understanding GRB subclasses and their progenitors is crucial for cosmology, especially in studying the early universe and cosmological parameters in Type II events, and for multi-messenger astrophysics, as exemplified by the remarkable case of GW170817 - GRB 170817A \citep{amati2021short}.

In a recent study by \cite{chen2021novel}, a classification model was developed to discern whether a GRB emits high-energy photons (GeV) earlier at the source than during the prompt phase, leading to the identification of a new classification and providing novel insights into the dynamics of GRBs.

The following paragraphs are organized based on the data source from satellite observations and focus on the majority of works dedicated to subgrouping GRBs. While some of the cited papers refer to the ``classification'' of GRBs, the term ``clustering'' is more appropriate in the context of Data Science (see Sections \ref{sec:classification_ML} and \ref{sec:clustering}), as it includes grouping and analyzing GRBs based on their similarities and shared properties.

\subsection{BATSE}

\begin{figure}[!htb]
\centering
  \includegraphics[width=1\textwidth]{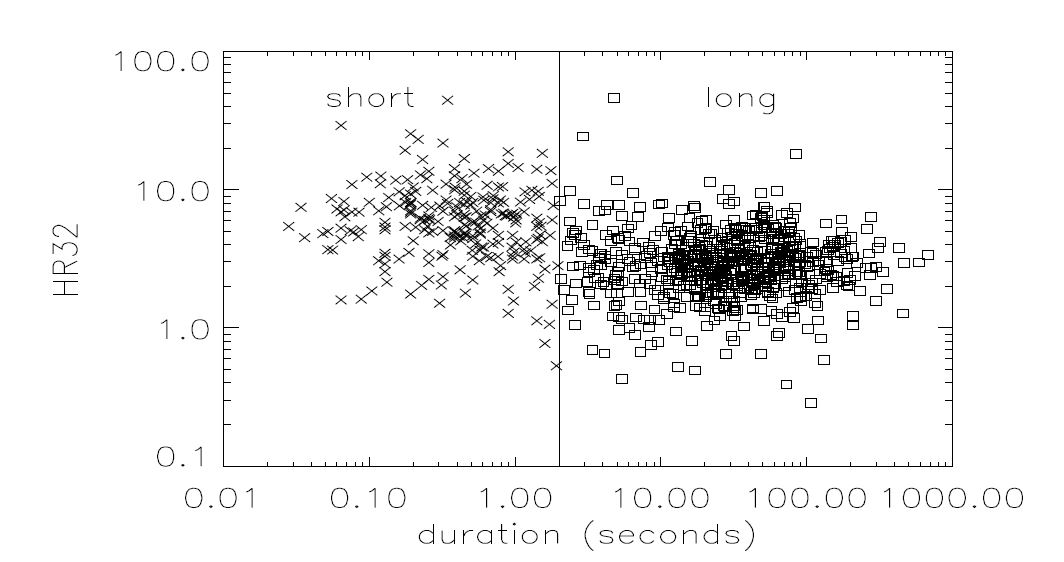}
\caption{\label{fig:grb_hr_t90}
HR32 vs. duration plot for BATSE GRBs. Reprinted from \cite{hakkila1998ai}, with the permission of AIP Publishing.}
\end{figure}

In the realm of GRB clustering, several studies have explored the use of AI techniques. In 1997/1998, \cite{hakkila1998ai} applied a conceptual clustering technique to six properties of BATSE GRBs, revealing that the long and short splitting is not the optimal way to subgroup the data. Instead, the division should rely on both hardness and duration (see Section \ref{sec:grb_prompt}), as shown in Figure \ref{fig:grb_hr_t90}. This is further confirmed by training a decision tree classifier only on hardness ratios and obtaining a balanced accuracy of 80\%. 

In 1998, \cite{mukherjee1998three} proposed two clustering algorithms (hierarchical and GMM), leading to the identification of three GRB classes: Class I with long/bright/ intermediate bursts, Class II with short/hard/faint bursts, and Class III with long/intermediate/soft bursts. Notably, Class III was introduced for the first time. 
\textit{The six features employed are: logarithm of total fluence $f_{tot}$, peak flux P256, durations (T90 and T50) and hardness ratios (HR321 and HR32)}.
Subsequently, \cite{hakkila2000gamma} employed a decision tree to classify the three classes, demonstrating statistical robustness. However, they argued that Class III might not necessarily represent a distinct source population, as some events could fall into Class I and II when analyzing systematic errors (the duration and fluence of some bursts could be underestimated) and the decision tree rules.

Different approaches yielded diverse results. \cite{balastegui2001reclassification} used an agglomerative hierarchical clustering, obtaining three classes, with the intermediate one having longer duration compared to that discussed in \cite{mukherjee1998three}. Later, \cite{balastegui2005neural} reanalysed the data using a Self-Organizing Map (SOM) incorporated nine quantities from the previous studies, further supporting the division into three classes.
\textit{The nine quantities are: four fluences (corresponding to the four energy channels: $Ch\#1$ 25–50 keV; $Ch\#2$ 50–100 keV; $Ch\#3$ 100–300 keV; $Ch\#4$ > 300 keV), three peak fluxes (corresponding to the three time-scales of integration: 64, 256 and 1024 ms) and two durations (T50 and T90). Additionally, \cite{balastegui2005neural} included two more parameters: the hardness ratio H32 and $\frac{V}{V_{max}}$.}

\cite{roiger2000unsupervised} suggested four classes, while \cite{pereira2010combining} found five of them (!), with one containing mostly short bursts and another mostly intermediate bursts. However, \cite{rajaniemi2002classifying} supported the hypothesis of two classes using a clustering implemented via a SOM, focusing on $\log(\text{T90})$, $\log(\text{H321})$ and $\log(\text{S})$ (logarithm of the sum of fluxes from channel 2 and 3) as features. It is worth to note that these are the same features employed by \cite{hakkila2000gamma}.

\begin{figure}[!htb] 
    \centering
      \includegraphics[width=0.75\textwidth]{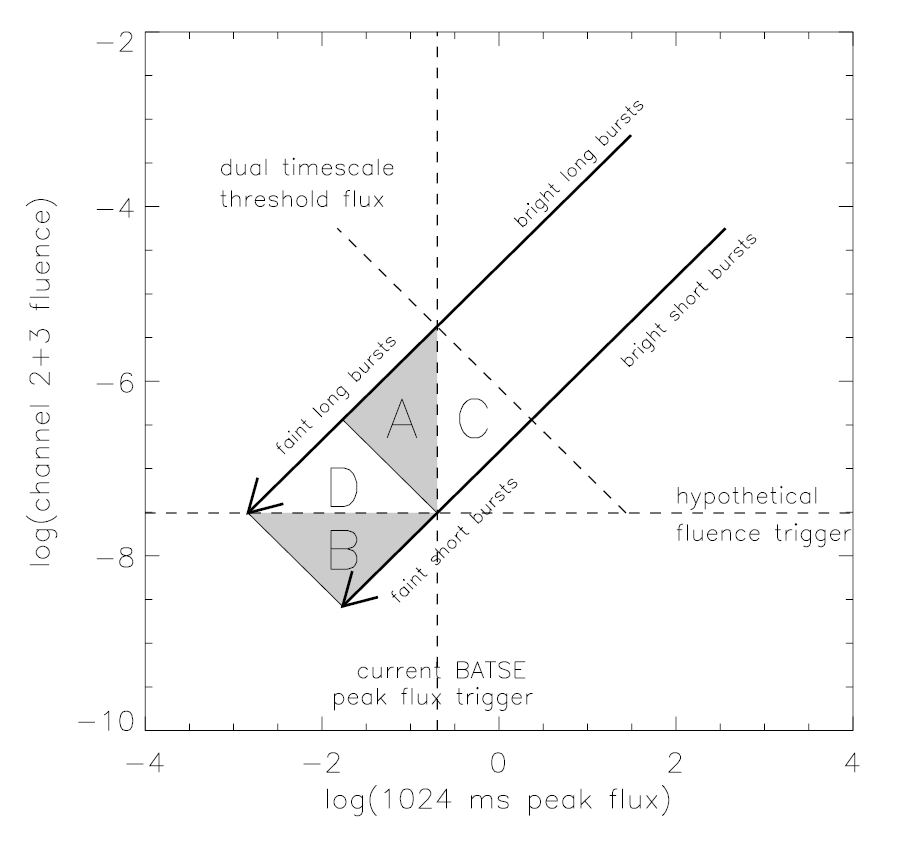}
    \caption{\label{fig:hakkila}
    BATSE 1024 ms trigger primarily detects short bursts near threshold (region C) while missing longer bursts (region A). A hypothetical fluence trigger would preferentially detect long bursts (region D) while missing shorter bursts (region B). To address this bias, a proposed dual-timescale threshold is suggested, which could be developed as an instrumental trigger on other experiments, effectively avoiding oversampling of long or short bursts. \cite{hakkila2003sample} \copyright AAS. Reproduced with permission.}
    \end{figure}

In their analysis, \cite{hakkila2003sample} integrated and expanded upon the findings from \cite{hakkila2000gamma} and arrived at the conclusion that the existence of Class III in the classification of GRBs is significantly influenced by sample incompleteness. They found that this incompleteness can be attributed to a trigger algorithm bias, wherein the algorithm exhibits a preference for detecting shorter bursts over longer ones. This bias is evident in the fluence/peak flux plane, where Class III is situated in a region where the trigger algorithm tends to favor the detection of shorter bursts. As a result, the apparent presence of Class III is more likely a consequence of this bias rather than representing a distinct and meaningful intermediate group of GRBs.
Moreover, it is proposed a trigger logic in which it is involved both peak flux and fluence to preserve the duration distribution of faint bursts, Figure \ref{fig:hakkila}.

In a subsequent work, \cite{hakkila2004subgroups} emphasized that the identification of the third group might depend on factors such as the size of the database, the specific clustering algorithm used, the choice of features considered, and the treatment of measurement errors of the specific instrument. However, they provided strong support for a clear distinction between short and long GRBs, with no compelling evidence for the existence of an intermediate group. In 2007, \cite{chattopadhyay2007statistical} used $k$-means and the Dirichlet process of Mixture Model, identifying two subgroups within long bursts based on total fluence and duration.
The latter model assumes the data distributed as a K mixture of multivariate normal, where K is determined automatically the Dirichlet process. Long bursts are further classified in this categorization into two classes, which are distinguished principally by their total fluence and duration, and are hence referred to as low and high fluence GRBs.
\textit{The features used are: log T50, log T90, log $f_{tot}$, log P256, log H321, log H32.}

Differentiating based on the ratio of $E_{peak}$ and fluence was proposed by \cite{goldstein2010new}, using a lognormal distribution with preference towards bimodal distribution.

In recent years, \cite{modak2018clustering} performed Kernel principal component analysis followed by $k$-means on various features \textit{(F1, F2, F3, F4, P64, P256, P1024, T50, T90)}, leading to the identification of three clusters. 
\textit{In particular: F1, F2, F3, F4 are time-integrated fluences in 20 - 50, 50 - 100, 100 - 300 and $>$ 300 keV spectral channels respectively; P64, P256, P1024 are peak fluxes measured in 64, 256 and 1024 ms bins respectively; T50, T90 are times within which $50\%$ and $90\%$ of the flux arrive.}
\cite{modak2021distinction} later employed a non-parametric fuzzy clustering method, incorporating hardness-ratio features H32 = F3/F2 and H321 = F3/(F2 + F1), which revealed the emergence of five groups redundantly from numerical separation of the established three groups. This supported the existence of at most three clusters in GRBs.

\subsection{BeppoSAX}

In the BeppoSAX catalog \cite{horvath2009classification}, a clustering based solely on duration using Maximum Likelihood is performed, revealing the best fit with three clusters. However, the authors acknowledge that the physically distinct class of intermediate GRBs requires caution, and future catalogs from new missions, like Fermi-GBM, are expected to shed more light on this issue.

\subsection{Swift}

In the context of the Swift satellite, changing the instrument dataset alters the sensitivity to GRBs, necessitating a new clustering approach, e.g., the BAT instrument of Swift satellite has different spectral sensitivity respect to CGRO of BATSE. In \cite{horvath2008classification}, a duration clustering is performed, resulting in three classes. Another study \cite{horvath2010detailed} employs features such as logarithmically normalized T90, HR, fluence 2 (25-50 keV), and fluence 3 (50-100 keV), defining hardness by the HR = F3/F2 ratio. The clusters obtained consist of long bursts dominating one class, a less populated short burst class (compared to BATSE), and a more numerous intermediate burst class (softer respect the short class). This distribution is reasonable considering that Swift's BAT is less sensitive to high energy and more sensitive to low energy compared to BATSE. Thus, Swift detects more intermediate bursts and fewer short bursts compared to BATSE.

In \cite{lu2010new}, a new parameter called $\epsilon$ is introduced, combining isotropic gamma-ray energy and the cosmic rest-frame spectral peak energy, along with T90 for clustering using Gaussian mixture models (KMM implementation). This analysis reveals a bimodal distribution.

\begin{figure}[!htb] 
\centering
  \includegraphics[width=1\textwidth]{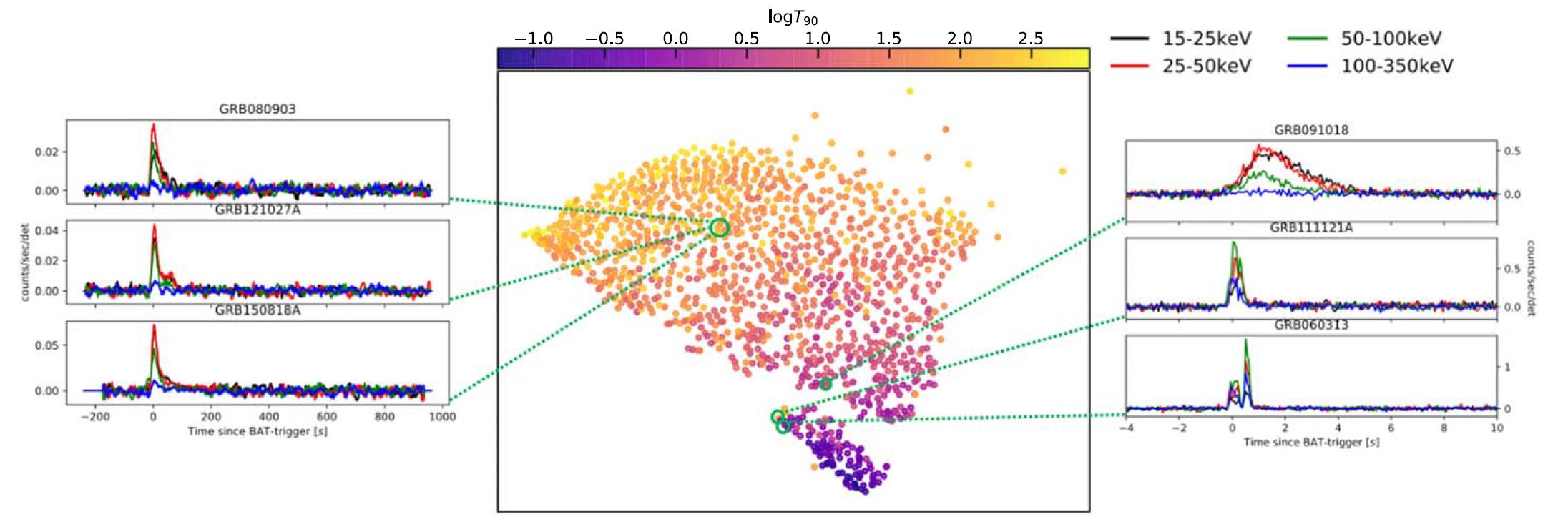}
\caption{\label{fig:GRB_tsne}
The color-coded logarithm of the time (T90) t-SNE mapping of Swift lightcurves indicates unique groupings. Lightcurves that are similar are placed closer together, while those that are distinct are placed farther away. The graphic clearly shows a division into two groups, with the smaller group at the bottom referred to as type-S, which has shorter durations in general. It is crucial to note that the axes produced by t-SNE do not have a direct physical meaning. Jespersen et al. (2020) \cite{jespersen2020unambiguous} \copyright AAS. Reproduced with permission.}
\end{figure}

In \cite{jespersen2020unambiguous} an interesting approach, probably the first on its own, utilizes the entire prompt lightcurves of GRBs without preprocessing into condensed quantities (e.g., fluence, hardness ratio, duration). They apply a discrete-time Fourier transform to the lightcurves and then perform dimensionality reduction using t-distributed Stochastic Neighbor Embedding (t-SNE). t-SNE shows two clear clusters (Figure \ref{fig:GRB_tsne}) without direct physical meaning in the axes. Additional explainability algorithms could help understand the meaningfulness of this division. 

Similar work is proposed in \cite{misra2023evidence} where five clusters are identified. Type I GRBs are in two separate clusters, indicating they may be due to subclasses of binary neutron star and/or NS–black hole mergers. 

Even in \cite{garcia2023identification} a t-SNE is employed showing how its hyperparameters learning rate and perplexity influence the results. In the end the method show a good separation in two clusters. 

In \cite{berretta2022comprehension} an analogous application of t-SNE is done in the prompt-emission detection by Swift, revealing two groups and relating those with the properties of the corresponding GRB afterglow. 

In \cite{bhardwaj2023grb} it is explored the possibility of clustering GRBs to match current observational subclasses (e.g. SGRB, LGRB, ULGRB, GRB with an SN associated, etc), based on a wide range of parameters, including plateau emission properties from X-rays and optical GRB lightcurve features. The researchers used the GMM to cluster the GRBs and analyze their properties. The results showed that traditional classifications of GRBs based on prompt emission properties may not be sufficient, and that a broader set of parameters may be necessary for a more accurate clustering. Small patterns emerged within the results, showcasing instances where GRBs of the same classification converged within the same cluster, both in terms of X-ray and optical features. The study provides insights into the physical processes behind GRBs and highlights the potential of unsupervised machine learning methods for future studies.

\subsection{Fermi GBM}
\begin{figure}[!htb] 
\centering
  \includegraphics[width=1\textwidth]{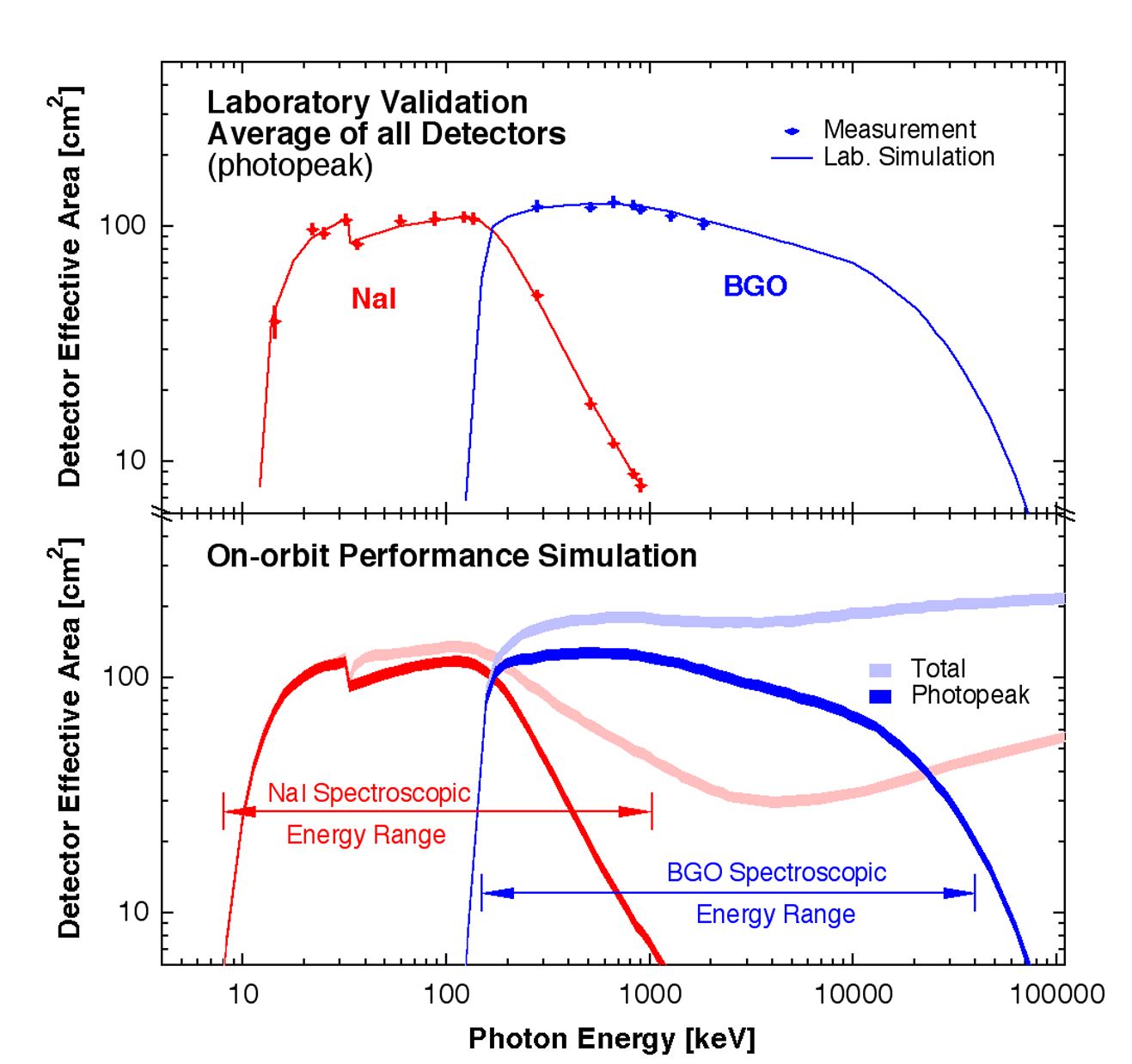}
\caption{\label{fig:Fermi_sensitivity}
The effective area at normal incidence is energy dependent for both detector types. The lower panel depicts the spacecraft's simulated impacts on a typical detector. \cite{meegan2009fermi} \copyright AAS. Reproduced with permission.}
\end{figure}

The Fermi GBM utilizes 12 NAI detectors with a good sensitivity energy range from 28 KeV up to 500 KeV \cite{meegan2009fermi}, as shown in Figure \ref{fig:Fermi_sensitivity}.
It's important to note that the clustering procedure may not produce identical cluster groups as observed in other satellite datasets due to different transient detection strategies and energy range sensitivities.

An analysis similar to \cite{horvath2009classification}, but focusing on the duration of GRBs detected by Fermi, resulted in a bimodal distribution \cite{tarnopolski2015analysis}. Instead of clustering, \cite{tarnopolski2015distinguishing} performed a classification (in the strict sense of Data Science) using Support Vector Machines (SVM) with the target variable being ``short'' or ``long'' GRBs. The features employed were Hurst Exponents (HEs), minimum variability time-scales (MVTS), and T90. The results demonstrated that HE and MVTS could improve the accuracy of the classification.
More details on MVTS can be found in \cite{walker2000gamma, maclachlan2012minimum} and for HE in \cite{maclachlan2013hurst}.

In \cite{horvath2015classification}, clustering was performed on the basis of T90, total fluence, hardness ratio, and peak flux256. After applying PCA for dimensionality reduction, Gaussian Mixture Models Clustering revealed three components as the best fit.

In \cite{horvath2019multidimensional} it is utilized 16 variables for clustering, which included parameters such as T90, fluence, peak flux, Compton amplitude, Compton Epeak, and Compton power law index from both fluence and peak flux spectral fits. \textit{The 16 variables comprehend: 
\begin{itemize}
	\item lgT90 (T90)
	\item lgT50 (T50)
	\item lgfluence (fluence)
	\item lgPflux64 (64 ms peak flux)
	\item lgPflux256 (256 ms peak flux)
	\item lgPflux1024 (1024 ms peak flux)
	\item lgfluenceBATSE (fluence in the BATSE energy channels)
	\item lgPflux64BATSE (64 ms peak flux in the BATSE energy channels)
	\item lgPflux256BATSE (256 ms peak flux in the BATSE energy channels)
	\item lgPflux1024BATSE (1024 ms peak flux in the BATSE energy channels)
	\item lgflncCompAmp (Compton amplitude from fluence spectral fit)
	\item lgflncCompEpeak (Compton Epeak from fluence spectral fit)
	\item flncCompIndex (Compton power law index from fluence spectral fit)
	\item lgpflxCompAmp (Compton amplitude from peak flux spectral fit)
	\item lgpPflxCompEpeak (Compton Epeak from peak flux spectral fit)
	\item pflxCompIndex (Compton power law index from fluence spectral fit)
\end{itemize}}
PCA was applied, and clustering identified three optimal classes. The known short class and two long classes with different peak flux distributions but similar hardness distributions were identified, whereas the previously identified Intermediate class did not match the obtained data due to weakly defined characteristics and different spectral fitting parameters.

In \cite{zhang2022tight}, the analysis focused on log T90, log f (fluence), and log F (flux) parameters, which were highly correlated and reduced to two dimensions using PCA. In this two-feature space, Gaussian Mixture Models (GMM) were used to find two classes, and Linear Discriminant Analysis (LDA) was employed to identify the optimal line that distinguishes the two classes. If more classes were fitted with GMM, only three result to minimize the BIC (Bayesian information criterion). The authors propose that, in addition to Short Gamma-Ray Bursts (SGRBs), Long Gamma-Ray Bursts (LGRBs) can be further divided into two subgroups: long-bright gamma-ray bursts (LBGRBs) and long-faint gamma-ray bursts (LFGRBs). Statistical analysis revealed that LBGRBs have greater "f" and "F" values than LFGRBs, and further research suggests that LBGRBs have a larger number of GRB pulses compared to LFGRBs.
The result of the clustering can be seen in Figure \ref{fig:gmm}.

\subsection{Analysis on multiple satellites}

In \cite{zhang2016classifying}, the duration distributions of GRBs from BATSE, Swift, and Fermi were studied using Gaussian Mixture Models. The results showed that GRB duration distributions are detector-dependent. Swift data sets supported three clusters (two if restricted only on the redshift-known), while BATSE and Fermi data sets strongly supported two clusters. The existing catalog of duration parameters may contain some noisy data that could affect the distribution, and durations need to be focused on prompt emission to be a distinguishing feature of the central engine. Similar results were presented in \cite{kulkarni2017classification}.

In \cite{tarnopolski2022graph}, a graph-based clustering approach, variant of the k-nearest neighbour graph, was used on several GRB samples from various satellites, including RHESSI, Konus, BATSE, Swift, Fermi, and Sazaku. Two features, T90 and H32, were employed, and the results showed that in most situations, two or three separate groups were possible. However, there was no compelling evidence for a third GRB class, and there were no signs of more than three classes being present.

In \cite{desai2022two}, the hardness ratio and duration T90 of GRBs from BATSE and Fermi GBM were analyzed. \textit{It’s worth to notice that the definition of hardness ratio is quite different due to the sensitivity of the detectors, indeed H = f(50-100 keV) / f(20-50 keV) for BATSE, f(50-300 keV)/ f(10-50 keV) for Fermi/GBM}. The features were completed with their errors, and an extended Gaussian Mixture Model (XDGMM) was used for clustering. For BATSE, two clusters were identified, while Fermi GBM showed a preference for three clusters.

\cite{salmon2022two} performed GMM clustering on samples of Fermi and Swift GRBs based on hardness ratio and T90. The results indicated two components on the HR-T90 plane, and further analysis using entropy criteria supported the presence of only two significant components, suggesting that the intermediate class might be a result of overfitting rather than a new GRB population.
Additionally, a table was presented in the paper, summarizing various works that used statistics and Machine Learning techniques to study different sub-populations of GRBs. Table \ref{tab:work_table} aligns with this table but serves a different purpose, providing an overview of different approaches used in the field, including works that explore the application of advanced ML techniques like t-SNE.

In \cite{salmon2022twoGMM}, the authors propose a study on Fermi, BATSE, and Swift datasets, where GRB lightcurves are preprocessed using a wavelet transform, followed by PCA and GMM clustering. Although the cluster separation is not distinct in all datasets, two groups consistently emerge.

In \cite{luo2022identifying}, the authors utilized an XGBoost classifier to perform supervised classification of GRBs into Type I and Type II. The classification was based on the target variable provided by the Greiner's GRB catalog\footnote{\href{https://www.mpe.mpg.de/~jcg/grbgen.html}{https://www.mpe.mpg.de/~jcg/grbgen.html}}. Features from prompt emission, afterglow, and the host galaxy (if present) were employed. The study found that T90 remains the most important feature for classification, but integrating it with fluence and hardness ratio improved classification. The correct classification of intermediate GRBs (i.e., short-duration Type II and long-duration Type I GRBs) supported the hypothesis that a third class does not exist.

In \cite{steinhardt2023classification}, the work of \cite{jespersen2020unambiguous} was extended to include GRBs from Fermi and BATSE, using both t-SNE and UMAP (Uniform Manifold Approximation and Projection) for dimensionality reduction. The embeddings showed a clear distinction between short and long bursts in each catalog, but a tiny proportion of bursts could not be consistently identified. Additionally, three of the 304 bursts seen by both Swift and Fermi had compelling but inconsistent classifications, suggesting the presence of other, unusual forms of bursts in addition to the prevalent short and long GRB classes. Multi-wavelength studies may be needed to distinguish these unusual bursts from the more common GRB types.

\paragraph{Summary and consideration}

Table \ref{tab:work_table} summarizes the works based on the satellite dataset used, techniques, and variables employed for obtaining GRB groups. Discordance in results is evident, which can be attributed to differences in instrument sensitivity and models used for feature extraction.
While T90 is an essential indicator, researchers aim to incorporate additional features to identify progenitor types like NS-NS, BH-NS, SNe I, SNe II, Magnetar, etc. However, relying solely on T90 can be misleading, especially given the universe's expansion. A short T90 in the rest frame of a high-redshift GRB, may be observed as a long transient on Earth.
Moreover, the existence of Class III (a sub-class of long GRBs) is questioned due to instrument biases, highlighting the need for improved trigger algorithms to detect long-faint events, such the ones described in Section \ref{sec:transient_classification}.

Several factors contribute to the variation in results:

\begin{enumerate}
    \item Dataset: Each satellite has unique instrument characteristics, such as energy range sensitivity and time resolution, leading to potential biases in the detection of certain GRB groups. For instance, long faint events may be underrepresented in BATSE due to the energy channel bias, as observed in \cite{hakkila2000gamma, hakkila2003sample}. The dataset size also plays a critical role in achieving robust analysis and results. On the other hand, a catalog enriched with GRB information from multiple datasets would contain multi-wavelength features, potentially facilitating clearer and more effective separation among GRB clusters.

    \item Features: The choice of features used in the clustering method is crucial for defining GRB groups. As highlighted in Table \ref{tab:work_table}, T90 combined with Hardness Ratio (HR) is often considered the most important feature. However, adding more features or the normalization (more generally the preprocessing step) can alter the similarity among GRBs. For instance, if the feature space includes (T90, HR, fluence), clustering tends to emphasize duration similarity due to the high correlation between T90 and fluence. Adding more features can complicate the interpretation, necessitating dimensionality reduction techniques like PCA, t-SNE, or UMAP to visualize the clustered data effectively. Existing works that rely on predefined features may not provide clear and satisfactory group definitions. Alternatively, leveraging direct lightcurve information in various energy bands, such as in \cite{jespersen2020unambiguous}, can preserve informativeness but may present challenges in providing a clear physical interpretation of the extracted Fourier Transform coefficients.
    
    \item Methods: The choice of clustering method is critical as each method defines similarity differently in the feature space. The distance metric used in the method is also crucial, typically relying on the Euclidean distance. A more nuanced difference consists in the cluster assigned, in either a "hard" or probabilistic manner. For example, $k$-means assigns samples to one cluster based on centroid distances, while SOM or GMM assigns probabilities of belonging to certain clusters based on the highest probability. 
    Some approaches employ dimensionality reduction techniques like PCA, t-SNE, or UMAP to visualize the data in 2D or 3D, enabling a clearer separation of clusters. Explainability is really needed in this context. Notably, \cite{hakkila1998ai} analyzed clustering results using decision rules fitted to a decision tree, now known as explanation with a global surrogate.
    In \cite{luo2022identifying}, a classification task is performed to distinguish confirmed GRBs Type I and II. The most important features in prompt emission are found to be T90, hardness ratio, and fluence, known from previous clustering works. A possible next step should involve using the SHAP technique to explain why individual GRBs are classified as certain types and to investigate the most uncertain classifications close to the decision boundary (e.g., with probabilities close to 0.5).

\end{enumerate}

For future research, there are several potential approaches to consider:
\begin{enumerate}
    \item[a.] \textbf{Extracting satellite-independent patterns by using all datasets together.} This can be achieved with techniques like \cite{ganin2015unsupervised}, which extract features from a Neural Network to be independent of specific labels, such as the satellite, see Figure \ref{fig:gradient_reversal}. \\
                Pros: access to a large amount of data and the identification of common patterns among datasets and instruments. \\
                Cons: this approach may not fully leverage the unique characteristics of individual instruments, potentially reducing effectiveness. For instance, if for Swift-BAT data is possible to distinguish type I and II but not with the Fermi-GBM due to the higher energy range, the features extracted will be limited for both. Additionally, designing a common input format for all GRB data would be necessary.
                   
    \item[b.] \textbf{Fitting one model per dataset.}\\
                Pros: have a standardized input based on the specific satellite dataset, enabling better utilization of instrument peculiarities. \\
                Cons: Different methods to fit and evaluate. the GRBs have to be clustered in a coherent way among the different methods. 
    \item[c.]
    \textbf{Creating a dataset consisting of confirmed Type I and II GRBs.} Since labels are only available for a few GRBs, a semi-supervised approach could be used, leveraging patterns extracted by models from previous points.
               Pros: a supervised model that simplifies result interpretation and clear measurements.\\
               Cons: however, labeled data may be limited compared to the total number of detected GRBs. Additionally, representing a GRB with multiple views from different satellites (e.g., Figure \ref{fig:cuoco2021multimodal}) can introduce missing information and potential selection bias in labelled GRBs.
    \item[d.] \textbf{Similar to a. but without the constrain of the satellite-independent patterns.} An implementation could be done training a Neural Network autoencoder on all the data input and cluster on the latent space embeddings but separately per each dataset to avoid the trivial separation on satellite source. 
                Pros: access to a large amount of data and enabling better utilization of instrument peculiarities. Manage only one model.
                Cons: designing a common input format for all GRB data would be necessary.
\end{enumerate}

In all cases, conducting research with the most informative data is crucial, which includes retaining both temporal and energy spectral information of GRBs (see Figures \ref{fig:zhang_input} and \ref{fig:grb_image}) and incorporating multiple satellite datasets. By keeping the data in its raw form, without feature engineering, the challenge lies in explaining the results. Explainability algorithms can aid in discovering new patterns or understanding the clustering using the previously mentioned features.

Validation of the clustering results should involve two sets: one with labelled GRBs as Type I and II (preferably excluded from the training set to avoid overfitting) and another set comprising GRBs detected by different satellites. Consistent clusters across different datasets, features, and methods indicate more reliable results.

As more data becomes available from missions like HERMES and upcoming missions, it will be possible to conduct a robust analysis on prompt and progenitor GRBs, similar to the study by \cite{luo2022identifying}. However, it should be noted that data collected by different instruments may be subject to selection biases, raising concerns about the classification of dark GRBs without proper analysis of their afterglow or due to a lack of faint GRB data.

{
\footnotesize

\begin{longtable}{@{}p{0.05\columnwidth}p{0.15\columnwidth}p{0.15\columnwidth}p{0.15\columnwidth}p{0.20\columnwidth}r@{}} 
\caption{\label{tab:work_table}\footnotesize Summary table works. $P$ is the peak flux and the suffix specify the ms time binning, $f$ the fluence and the suffix specify the energy channels, $HR_{xy}$ the hardness ratio $\frac{f_x}{f_y}$, $\frac{V}{V_{max}}$ is a measure of the maximum redshift, $\epsilon$ is based on the isotropic gamma-ray energy and the cosmic rest-frame spectral peak energy, MVTS is the minimum variability time-scales, HE the Hurst Exponents (a time series feature), z is the redshift when available. The \textit{Fermi parameters} are: lgPflux64 (64 ms peak flux), lgPflux256 (256 ms peak flux), lgPflux1024 (1024 ms peak flux), lgfluenceBATSE (fluence in the BATSE energy channels), lgPflux64BATSE (64 ms peak flux in the BATSE energy channels), lgPflux256BATSE, (256 ms peak flux in the BATSE energy channels), lgPflux1024BATSE(1024 ms peak flux in the BATSE energy channels), lgflncCompAmp (Compton amplitude from fluence spectral fit), lgflncCompEpeak (Compton Epeak from fluence spectral fit), flncCompIndex (Compton power law index from fluence spectral fit), lgpflxCompAmp (Compton amplitude from peak flux spectral fit), lgpPflxCompEpeak (Compton Epeak from peak flux spectral fit), and pflxCompIndex (Compton power law index from peak flux spectral fit). Notice that the channels depends on the instrument and sometimes it's different from work to work. DTFT stands for discrete-time Fourier transform.}\\
\toprule
Study & Telescope & Method unsup. & Method sup. & Parameters & Clusters \\
\midrule
\endfirsthead

\toprule
Study & Telescope & Method unsup. & Method sup. & Parameters & Clusters \\
\midrule
\endhead
\cite{hakkila1998ai} & BATSE & Conceptual clustering (CLASSIT) & Decision tree & T90, $P_{1024}$, $f_{23}$, $HR_{21}$, $HR_{32}$, $HR_{43}$ & 2 \\
\cite{mukherjee1998three} & BATSE & Hierarchical agglomerative clustering, GMM & - & T50, T90, $P_{1024}$, $P_{256}$, $P_{64}$, $f$, $f_1$, $f_2$, $f_3$, $f_4$, $H_{3-21}$ & 3 \\ 
\cite{balastegui2005neural} & BATSE & SOM & - & T90, T50, $P_{1024}$, $P_{256}$, $P_{64}$, $f_1$, $f_2$, $f_3$, $f_4$, $HR_{32}$, $\frac{V}{V_{max}}$ & 3 \\
\cite{hakkila2000gamma} & BATSE & - & Decision tree & T90, T50, $P_{1024}$, $P_{256}$, $P_{64}$, $f_{23}$, $f_1$, $f_2$, $f_3$, $f_4$, $HR_{32}$, $HR_{21}$, $HR_{3-21}$, $HR_{43}$ & 2 \\
\cite{balastegui2001reclassification} & BATSE & PCA, Hierarchical clustering, Neural Network & - & T90, $P_{1024}$, $P_{256}$, $P_{64}$, $f_1$, $f_2$, $f_3$, $f_4$ & 3 \\
\cite{roiger2000unsupervised} & BATSE & Hierarchical clustering & Decision tree & T90, $f$, $HR_{3-21}$ & 4 \\
\cite{pereira2010combining} & BATSE & Self-organizing map, PCA & - & T90, T50, $f_1$, $f_2$, $f_3$, $f_4$, $HR_{3-21}$ & 5 \\
\cite{rajaniemi2002classifying} & BATSE & SOM & - & T90, $f_{23}$, $HR_{3-21}$ & 2 \\
\cite{hakkila2003sample} & BATSE & K-means, SOM, GMM & - & T90, $f$, $HR_{3-21}$ & 2 \\
\cite{chattopadhyay2007statistical} & BATSE & K-means, Dirichlet mixture & - & T50, T90, $f$, $P_{256}$, $HR_{3-21}$, $HR_{32}$ & 3 \\
\cite{goldstein2010new} & BATSE & Lognormal & - & $E_{peak}$, $f$ & 2 \\
\cite{modak2018clustering} & BATSE & K-means & - &  T90, T50, $P_{1024}$, $P_{256}$, $P_{64}$, $f_1$, $f_2$, $f_3$, $f_4$ & 3 \\
\cite{modak2021distinction} & BATSE & Fuzzy cluster (FANNY) & - &  T90, T50, $P_{1024}$, $P_{256}$, $P_{64}$, $f_1$, $f_2$, $f_3$, $f_4$, $HR_{32}$, $HR_{3-21}$ & 3 \\
\cite{horvath2009classification} & BeppoSAX & Lognormal & - &  T90 & 3 \\
\cite{horvath2008classification} & Swift & Lognormal & - & T90 & 3 \\
\cite{horvath2010detailed} & Swift & GMM & - & T90, $HR_{32}$ & 2 \\
\cite{lu2010new} & Swift & GMM & - & T90, $\epsilon$ & 2 \\

\cite{jespersen2020unambiguous} & Swift & T-sne & - & DTFT on lightcurve & 2 \\
\cite{misra2023evidence} & Swift & T-sne & - & DTFT on lightcurve & 5 \\
\cite{garcia2023identification} & Swift & T-sne & - & DTFT on lightcurve & 2 \\
\cite{berretta2022comprehension} & Swift & T-sne & - & DTFT on lightcurve & 2 \\
\cite{bhardwaj2023grb} & Swift & GMM & - & XRT and optical parameters & $\ge$2 \\
\cite{tarnopolski2015analysis} & Fermi & Lognormal & - &  T90 & 2 \\
\cite{tarnopolski2015distinguishing} & Fermi & - & SVM & T90, HE, MVTS & 2 \\
\cite{horvath2015classification} & Fermi & PCA, GMM & - &  T90, $f$, HR, $P_{256}$ & 3 \\
\cite{horvath2019multidimensional} & Fermi & PCA, GMM & - &  T90, T50, $f$, \textit{Fermi parameters} & 3 \\
\cite{zhang2022tight} & Fermi & PCA, GMM, LDA & - &  T90, $f$, peak flux & 3 \\

\cite{zhang2016classifying} & BATSE & GMM & - &  T90 ($f2$, $f3$, $HR_{23}$ used only for filter) & 2 \\
\cite{zhang2016classifying} & Swift & GMM & - &  T90 ($HR_{31}$ used only for filter) & 3 \\
\cite{zhang2016classifying} & Swift & GMM & - &  T90 adjusted for z ($HR_{31}$ used only for filter) & 2 \\
\cite{zhang2016classifying} & Fermi & GMM & - &  T90 ($f$, HR, flncCompIndex, lgflncCompEpeak used only for filter) & 2 \\

\cite{kulkarni2017classification} & BATSE & Lognormal & - &  T90 & 2 \\
\cite{kulkarni2017classification} & Fermi & Lognormal & - &  T90 & 2 \\
\cite{kulkarni2017classification} & BeppoSAX & Lognormal & - &  T90 & 2 \\
\cite{kulkarni2017classification} & Swift & Lognormal & - &  T90 & 3, 2 \\

\cite{tarnopolski2022graph} & RHESSI & Graph-based algorithms & - &  T90, $HR_{32}$ & 2 \\
\cite{tarnopolski2022graph} & Konus-Wind & Graph-based algorithms & - &  T90, $HR_{32}$ & 2 \\
\cite{tarnopolski2022graph} & BATSE & Graph-based algorithms & - &  T90, $HR_{32}$ & 2, 3 \\
\cite{tarnopolski2022graph} & Swift & Graph-based algorithms & - &  T90, $HR_{32}$ & 2 \\
\cite{tarnopolski2022graph} & Fermi & Graph-based algorithms & - &  T90, $HR_{32}$ & 2, 3 \\
\cite{tarnopolski2022graph} & Sazaku & Graph-based algorithms & - &  T90, $HR_{32}$ & 2 \\
\cite{desai2022two} & BATSE & Extended GMM (XDGMM) & - &  T90, $HR$ & 2 \\ 
\cite{desai2022two} & Fermi & Extended GMM (XDGMM) & - &  T90, $HR$ & 3 \\ 
\cite{salmon2022two} & Fermi & GMM & - &  T90, $HR$ & 2 \\ 
\cite{salmon2022two} & Swift & GMM & - &  T90, $HR_{32}$ & 2 \\ 
\cite{salmon2022twoGMM} & BATSE & t-SNE, PCA, GMM & - &  Wavelet transform on lightcurve & 2 \\
\cite{salmon2022twoGMM} & Fermi & t-SNE, PCA, GMM & - &  Wavelet transform on lightcurve & 2 \\
\cite{salmon2022twoGMM} & Swift & t-SNE, PCA, GMM & - &  Wavelet transform on lightcurve & 2 \\
\cite{luo2022identifying} & - & - & XGBoost &  T90, $f$, $HR$, z, spectral parameters, afterglow features, host galaxy features & 2 \\
\cite{steinhardt2023classification} & Swift & T-sne, UMAP & - & DTFT on lightcurve & 2 \\
\cite{steinhardt2023classification} & Fermi & T-sne, UMAP & - & DTFT on lightcurve & 2 \\
\cite{steinhardt2023classification} & BATSE & T-sne, UMAP & - & DTFT on lightcurve & 2 \\

\bottomrule 
\end{longtable}
}

\section{Transient classification and detection}\label{sec:transient_classification}


Prompt identification of transient types (e.g., GRBs, solar flares, etc.) is crucial for alerting ground telescopes for pointing direction and afterglow observation, as well as for follow up studies. The deep learning approaches can lead to more accurate and efficient GRB identification against other transients, providing insights into the physics behind these events.
Additionally, this method's application extends to the classification of other astronomical objects and the analysis of large-scale surveys, offering potential implications and applications in astrophysics and beyond.

The journey of GRBs begins with their detection, which involves various algorithms depending on the type of telescope used (ground-based or space-orbiting) and the energy range to which its detectors are sensitive. This detection task can be formulated as either an anomaly detection problem, where the significance of the signal over background noise is assessed, or as a binary classification task of distinguishing between GRBs and non-GRBs.

\subsection{Transient classification}

\begin{figure}[!htb]
\centering
  \includegraphics[width=0.75\textwidth]{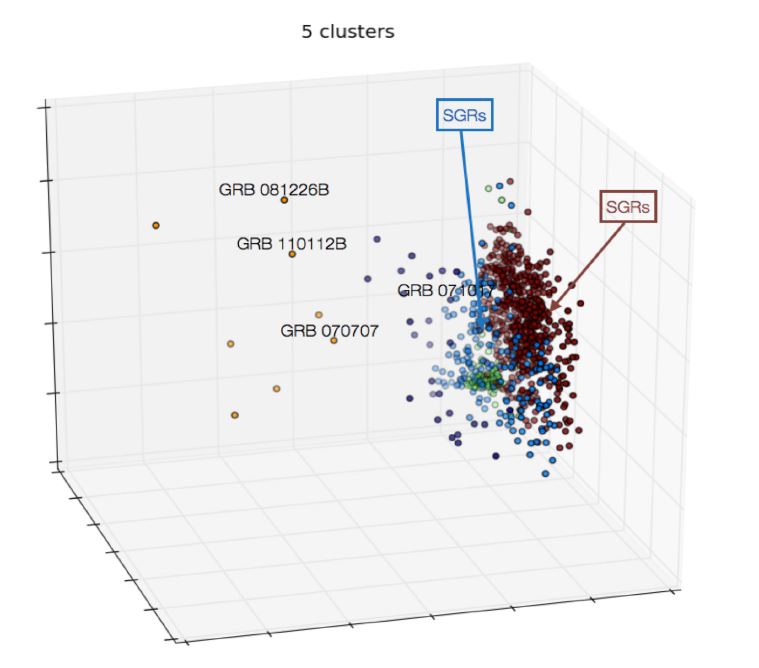}
\caption{\label{fig:topinka}
Three dimensional projection of the PCA flares feature space. The data points corresponding to known sGRBs and SGR flares identified by INTEGRAL are labelled. \cite{topinka2014search} \copyright Topinka et al. (2014), \href{https://creativecommons.org/licenses/by-nc-sa/4.0/}{CC BY-NC-SA}.}
\end{figure}

In \cite{topinka2014search}, the main objective is to search through the INTEGRAL (see Section \ref{sec:integral}) data archive to identify untriggered flares, short GRB or soft gamma repeater (SGR), on a millisecond time scale. The features used to characterize a flare include its duration, peak flux, fluence, spectral lag, hard/soft ratio, symmetry (based on rise and decay times), temporal variability, Galactic latitude of actual pointing, and the total number of flares in the same observation block. To analyze the data and group similar flares, PCA (see Section \ref{sec:dim_rec}) is first applied to reduce the feature space to three dimensions. Then, a $k$-means clustering algorithm is employed to cluster the flares into distinct groups. The analysis reveals five clusters, and, notably, GRBs are identified as an outlier group (Figure \ref{fig:topinka}). One specific GRB, GRB 071017, is suggested to be misclassified and should be considered a SGR instead. The ability to discriminate between different classes of objects, such as SGRs and GRBs, can have practical implications for adjusting strategies for real-time follow-up observations and for determining the fraction of SGRs among short GRBs.

\cite{topinka2016machine} focus on automatically classifying and distinguishing GRB afterglows from other sources in images. The goal is to enhance existing searches for GRB afterglows in wide-field images, even when temporal evolutionary data is missing, which is often the case in historical surveys. The study demonstrates that a minimal set of features, including at least three color indices g-r, r-i, and i-z, is sufficient to identify more than 90\% of GRB afterglows. To achieve this classification, various machine learning algorithms are utilized, including Support Vector Machine, Random Forest, and Neural Network.

\subsection{Classification GRB/non-GRB}

In the researches presented in \cite{hobson2014machine, graff2016modeling}, a feed-forward neural network (FFNN), Random Forest, AdaBoost and SVM were proposed as a detection classifier to replace the complex Swift trigger pipeline, which relies on about 500 triggering criteria to identify long GRBs (as discussed in Section \ref{sec:swift}). The FFNN offers a rapid and efficient way to determine whether a given GRB is detected, providing results in just a few microseconds. Each GRB is represented by 15 features that describe its properties and how it is observed by the detector. The FFNN's two outputs represent the probabilities of the GRB being detected or not being detected.

In the study \cite{abraham2021machine}, a Dynamic Time Warping (DTW) technique is used to measure distances between lightcurves of AstroSat Cadmium Zinc Telluride Imager (CZTI) data. Subsequently, hierarchical clustering is applied to generate 52 GRB templates. A lightcurve is classified as GRB if its DTW distance to at least one template is less than a certain threshold. At the time of the study, the pipeline was configured to alert the CZTI AstroSat support team about probable GRB candidates, and the team made the final decision on issuing GCN alerts.

\begin{figure}[!htb]
\centering
\begin{subfigure}{1\textwidth}
  \centering
  \includegraphics[width=1\linewidth]{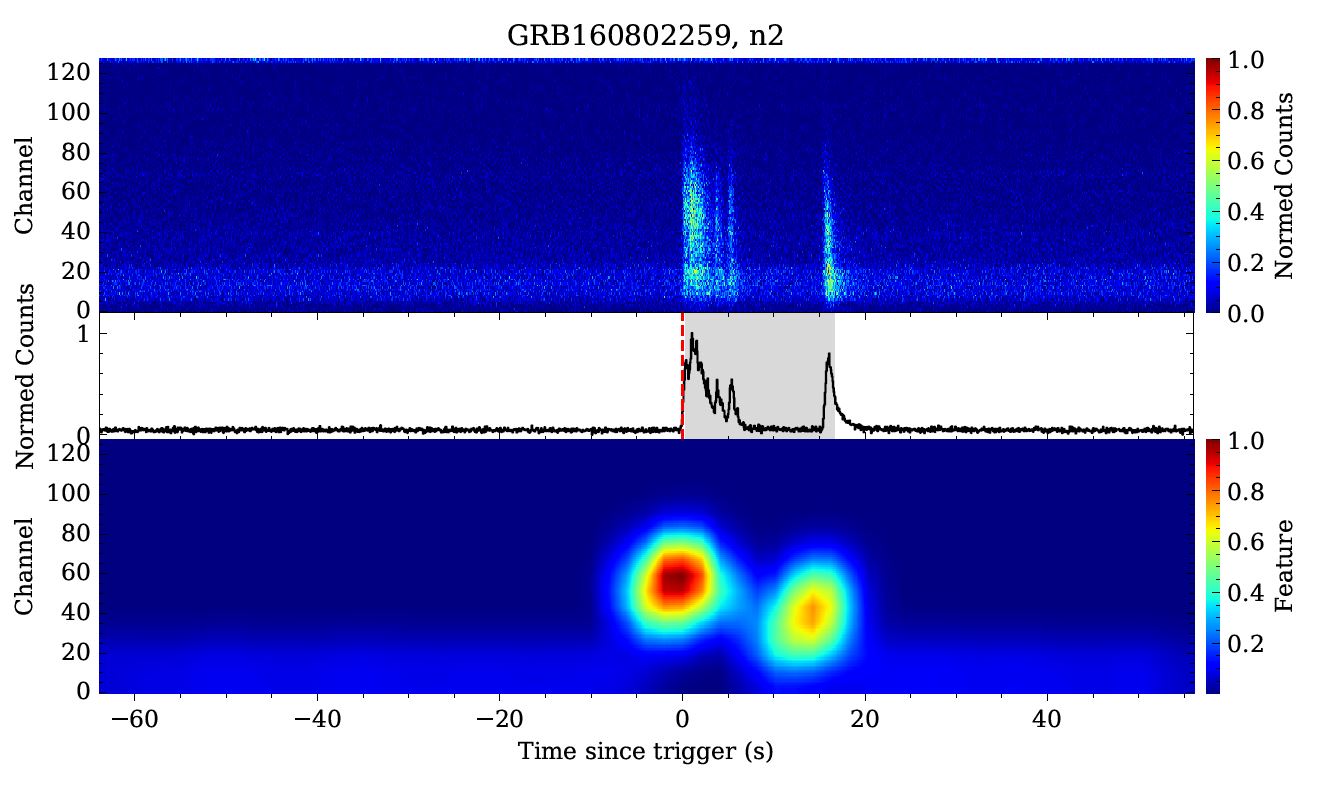}
  \caption{}\label{fig:zhang_input}
\end{subfigure}
\qquad
\begin{subfigure}{1\textwidth}
  \centering
  \includegraphics[width=1\textwidth]{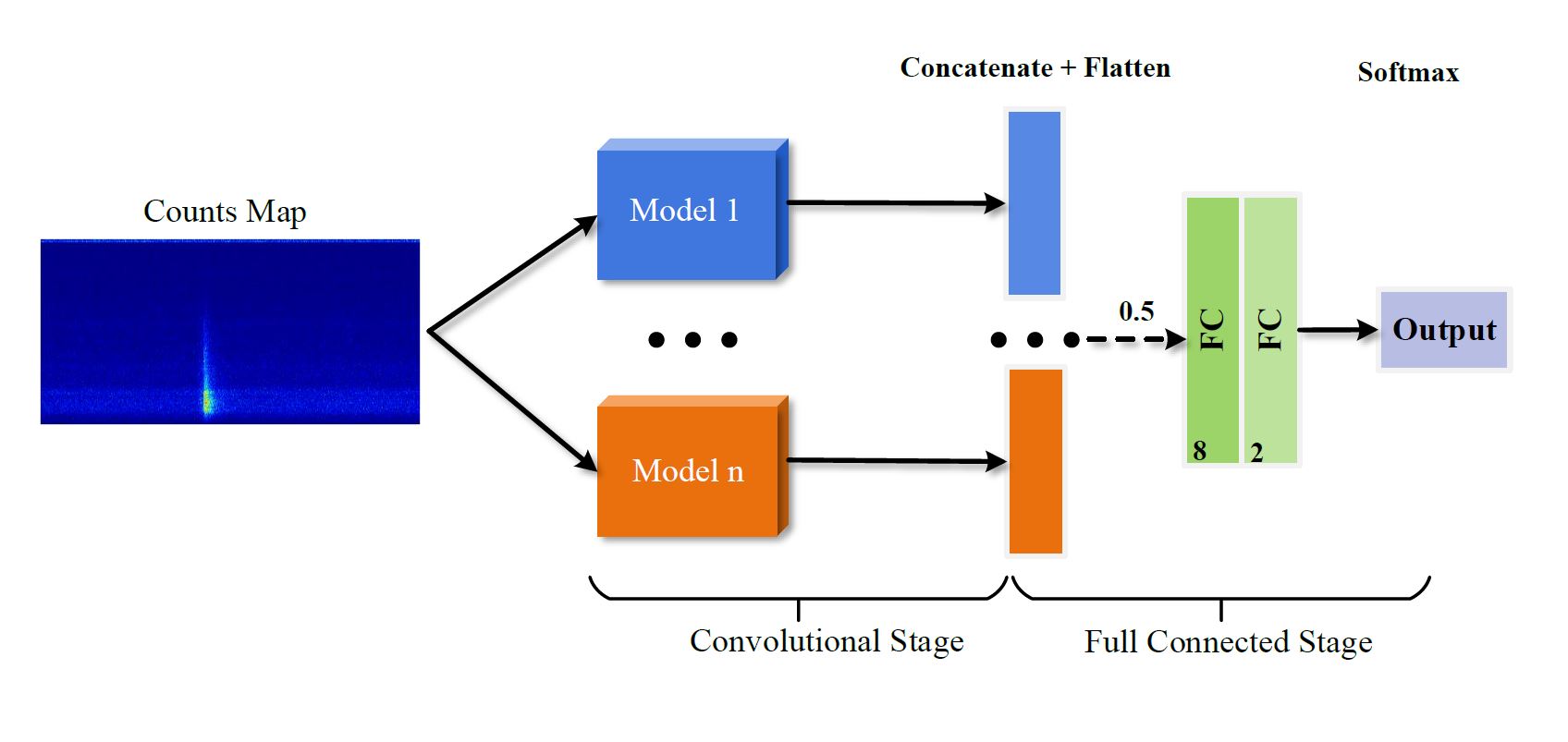}
  \caption{}\label{fig:zhang_cnn}
\end{subfigure}
\caption{\label{fig:zhang} In a) The top panel depicts the input model, normalized count map, the middle panel depicts the normalized lightcurve integrated over the entire energy band, and the bottom panel depicts the feature heat-map produced by the Grad-CAM approach. In b) the architecture of the complete model described in \cite{zhang2023application}. \cite{zhang2023application} \copyright Zhang et al. (2023). Reproduced with permission.}
\end{figure}

In \cite{zhang2023application}, Fermi-GBM data is treated as an image with time on the $x$-axis (binned at 64ms) and energy channels on the $y$-axis (128 channels, top panel of Figure \ref{fig:zhang_input}). Three Deep Learning models (CNN, ResNet, and ResNet-CBAM) are trained to classify whether the event is a GRB or not (Figure \ref{fig:zhang_cnn}). The explainability method, Gradient-weighted Class Activation Mapping (Grad-CAM), provides insight into the prediction and identifies crucial pixels (lower panel of Figure \ref{fig:zhang_input}).
The researchers employed the last convolutional layer to obtain the embedding values, which is an high-dimensional vector, and then apply a t-SNE in order to represent the dataset in 2D. These visualization approaches enabled the researchers to have a better understanding of the features that the model was learning and to pinpoint the physical qualities that were most relevant for distinguish GRB and non-GRB. 

The classification method exhibits excellent performance; however, a more in-depth examination of selection bias is needed. Training on a GRB sample aligned with Fermi's triggers may lead the deep learning algorithm to inadvertently ignore certain GRBs (e.g., faint or soft ones).
A future and promising application could involve using the embedding values of each GRB. By mapping each GRB into a embedding (output vector from one of the last layers on the Neural Network), clustering techniques can be applied to achieve results comparable to those described in Section \ref{sec:GRB_cluster}. The advantage of this approach is that it does not require explicit feature definitions and allows for a more comprehensive clustering analysis, leveraging both the temporal and spectral structures of the GRBs.

\subsection{Detection algorithm}

In \cite{parmiggiani2021deep} it is discussed a method based on CNN for classifying gamma-ray sky maps from the AGILE-GRID observatory as containing a GRB or not. The CNN processes arrays of maps, each with a size of 100 × 100 pixels, using a Convolution2D layer with 20 filters to detect features within the intensity maps. The final CNN architecture consists of 10 layers, chosen to achieve the best balance between training time and validation performance. The CNN is trained using a dataset comprising both simulated GRBs and background-only maps. Half of the data sets contain simulated GRBs with randomly generated fluxes, while the other half represents the background level obtained from the isotropic background distribution. This simulation approach ensures that the data sets closely resemble real data, enhancing the CNN's ability to learn from the simulated data and transfer that knowledge to analyze actual data. By including realistic background levels and varying GRB fluxes, the CNN is better prepared to handle the complexities of real-world data and improve its performance in classifying gamma-ray sky maps with GRBs. 
In \cite{parmiggiani2023deep}, a novel anomaly detection technique for multivariate time series (MTS) is presented. The technique involves using a CNN autoencoder to calculate the anomaly score based on the MTS reconstruction error. The MTS windows are composed of 140 bins of 1.024 seconds and five panels. However, the dataset used for training is restricted to include only background MTS, with periods containing idle mode or known gamma-ray bursts (GRBs) excluded.

\cite{sadeh2019deep, sadeh2020data} propose a pipeline employing an LSTM-based model coupled with a statistical test to calibrate the p-value during the inference phase. The algorithm is specifically designed to combine sequential datasets, enabling the derivation of realistic joint multimessenger datasets, such as gamma-ray bursts (GRBs) from CTA and neutrinos from IceCube. The model comprises an encoder and a decoder, and on top of the decoder, a classification layer can be attached for event or non-event classification, or an anomaly detection layer, where the actual values of the decoder are predicted and the anomaly score is determined by the reconstruction error. In \cite{veritas2022deep} a similar approach is employed.

In \cite{brill2019investigating}, a CNN combined with a LSTM architecture is utilized since the input data from VERITAS is represented as a sequence of images. The CNN-LSTM networks are trained using simulated events to effectively reject background signals.

The work presented by Crupi et al. \cite{crupi2023searching} introduces a feedforward neural network (FFNN) designed to estimate the X-gamma-ray background for space telescope detectors, including those used in HERMES. The NN is constructed using a robust loss function to account for outlier X/Gamma-ray count rates caused by transients. Following this, an efficient trigger algorithm is applied, using both observed and background count rates to detect transients. Notably, the trigger algorithm operates unsupervisedly, functioning as an anomaly detection method. This work establishes the groundwork for the developments discussed in the following chapters.

\paragraph{Summary and consideration}
The transient and GRB classification offer a straightforward and highly successful solution, providing significant relief from the burden of analyzing thousands of transients and identifying the most relevant ones from a researcher's perspective. Both Machine Learning, with feature engineering, and Deep Learning are viable options, as accuracy is crucial, and the black-box nature of the models does not pose a problem. While distinguishing a GRB from a non-GRB is of primary significance for practical applications, such as real-time alerting and follow-up observations, it may have less scientific impact on tasks like clustering GRBs or exploring their cosmological properties, as discussed in Sections \ref{sec:GRB_cluster} and \ref{sec:cosmological}.

The application of AI techniques for transient detection is less common in the literature, possibly because existing onboard satellite instruments' methods are considered efficient. However, these traditional methods might have inherent biases and limitations, leading to the potential omission of certain types of events, such as long or faint transients.

While \cite{sadeh2020data} and \cite{parmiggiani2021deep} focus on datasets free of transients (can this be verified?), Crupi et al. \cite{crupi2023searching} (the foundation for Chapters \ref{chap:bkg} and \ref{chap:frm}) introduces an alternative approach. This approach could avoid the exclusion of events within the dataset, offering a potentially more robust solution for detecting a broader range of transient events.

\section{Cosmological properties and progenitors identification}\label{sec:cosmological}

In the context of cosmology, GRBs have the potential to act as standard candles for distance estimation. These highly energetic events offer valuable insights into the early universe and its evolution, provided their redshifts can be determined. AI methods can be employed to estimate redshifts when traditional methods, which often involve time-consuming follow-up observations using multiple telescopes, are not feasible due to the inability to observe the afterglow. Utilizing AI allows to capture complex relationships among data, possibly leading to more accurate redshift estimations. The knowledge of redshift can also aid in identifying the host galaxy of the SGRB, providing further understanding of the progenitor system and the environment in which the event occurred.

\cite{shahmoradi2022bayesian} propose a bayesian methodology to estimate the redshift of SGRBs based on their observed features. This approach involves a semi-Bayesian data-driven technique, beginning with the segregation of SGRBs and LGRBs using the fuzzy clustering algorithm, applied to a larger sample of GRBs. Subsequently, the authors develop models for the rates of SGRBs, utilizing four key properties commonly reported in GRB catalogs: T90, time-resolved energy-resolved GRB light-curve, time-integrated spectrum, and energy-integrated light-curve of the GRB. Employing a Bayesian framework, the authors estimate the parameters of these models and infer the redshifts of individual SGRBs. This Bayesian approach accounts for all sources of uncertainty in the analysis, utilizing likelihood functions and posterior probability density functions to compare and evaluate different models effectively.

\cite{ukwatta2016machine} introduce an algorithm named ``machine-z'', enabling rapid redshift prediction of Swift GRBs by leveraging data accessible within the initial hours following the GRB trigger. This development is particularly valuable for identifying high-z candidates before their optical afterglow fades, offering significant implications for gamma-ray burst studies. The algorithm utilizes 25 features derived from readily available data from Swift's major instruments: the Burst Alert Telescope (BAT), X-ray Telescope (XRT), and UV Optical Telescope (UVOT). These features mainly capture the timing and spectral characteristics of both the prompt and afterglow emissions from the burst. The algorithm was trained on a dataset comprising 284 gamma-ray bursts with known redshifts, employing a Random Forest regressor and classifier. For the classification task, the algorithm predicts whether a GRB is high-z or low-z, while for the regression task, it estimates the redshift value.

\begin{figure}[!htb]
\centering
  \includegraphics[width=1\textwidth]{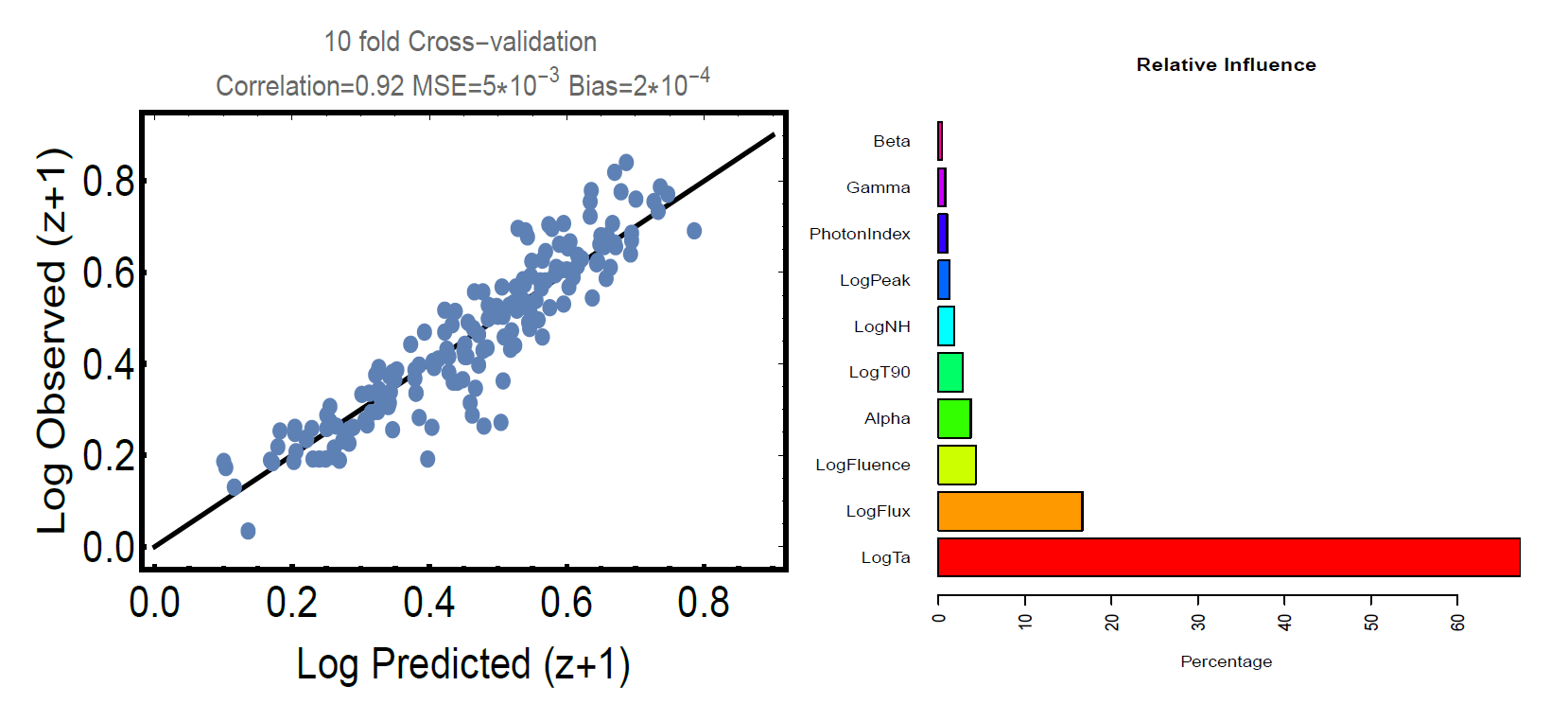}
\caption{\label{fig:dainotti}
In the left panel, we observe a comparison between predicted and observed redshifts. On the right panel, we can see the feature importance on the predicted outcomes. The afterglow's plateau flux and duration, along with the prompt fluence, stand out as the most influential predictors for the results. \cite{dainotti2019gamma} \copyright Dainotti et al. (2019).}
\end{figure}

\cite{dainotti2019gamma} utilizes SWIFT data for a regression task to estimate the redshift of GRBs using an ensemble of ML regressors. Notably, the three most important features (Figure \ref{fig:dainotti}) for this estimation are derived from the X-ray afterglow, rather than the prompt emission. This highlights the difficulty in establishing the distance of GRBs using count rates over a specific energy range, such as in the case of prompt emission data from Fermi-GBM. The study demonstrates that the GRB luminosity function and cumulative density rate evolutions, based on both predicted and observed redshifts, exhibit agreement, indicating that GRBs can be reliable distance indicators. To determine the feature importance, a local approach is used, where for each observation, a linear model is fitted to approximate the ML prediction with synthetic data perturbed by Gaussian noise from the original observation. 
 The feature importance for the feature $j$ is defined as $FI_j =  \frac{\mid B_j \mid}{\sum_{i=1}^{k} B_i}$, where $B_j$ is the coefficient of feature $j$ and $k$ the total number of features. 
 This method aligns with the principles of LIME \citep{ribeiro2016should}.
In \cite{dainotti2022optical}, GRB redshift estimates can improve $\Omega_M$ (matter content of the universe today) measurement, extending the distance ladder beyond Supernovae Type Ia (z=2.26), reducing errors in cosmological studies.

Instead of estimating the redshift from a GRB, \cite{escamilla2022neural} propose to calibrate the distance modulus-redshift relation of Type Ia supernovae (SNe Ia) and expand it to include the GRBs. The methodology involved the use of both a Recurrent Neural Network (RNN) and a Bayesian Neural Network (BNN) to achieve this task. An analogous work can be found in \cite{tang2022reconstructing}.

\begin{figure}[!htb]
\centering
  \includegraphics[width=1\textwidth]{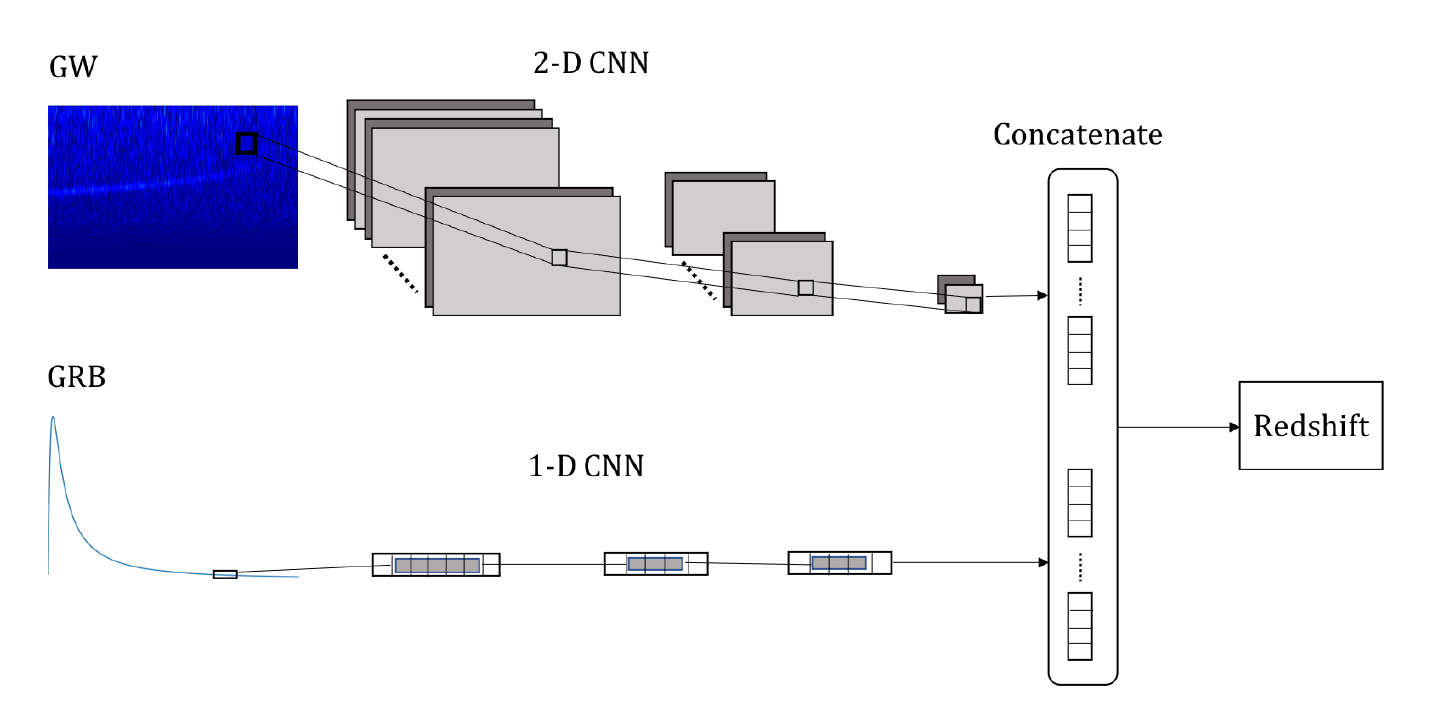}
\caption{\label{fig:cuoco2021multimodal}
To solve the regression problem, the network reads two types of data: images and time series. \cite{cuoco2021multimodal} \copyright Cuoco et al. (2021), \href{https://creativecommons.org/licenses/by/4.0/}{CC BY 4.0}.}
\end{figure}

The advent of third-generation gravitational wave detectors and modern astronomical facilities anticipates the occurrence of numerous multi-messenger events with similar characteristics. In \cite{cuoco2021multimodal}, the authors employ machine learning techniques to analyze multi-messenger data, particularly focusing on estimating the redshifts for joint GW and GRB sources, as shown in Figure \ref{fig:cuoco2021multimodal}.

In \cite{racz2017new}, the authors investigated GRBs and their afterglows using three instruments on board Swift, operating in gamma-ray (BAT), X-ray (XRT), ultraviolet, and optical (UVOT) wavebands. Instead of solely relying on the Swift GRB Table from the NASA website (as in \cite{ukwatta2016machine}), they combined this table with the Swift-XRT GRB Catalogue from the UK Swift Science Data Centre to create a new table that includes all X-ray spectral fitting data, offering higher precision compared to the data in the Swift GRB Table. With these enriched features, they applied both XGBoost and Random Forest regression models to estimate the redshift of the GRBs. 

Further statistical studies \cite{dainotti2023progenitors, dainotti2021cosmological} explores the formation rates of long and short GRBs in comparison to the cosmic star formation rate (SFR). It presents evidence for the presence of a distinct low redshift long GRB component and suggests that low redshift long GRBs could also have compact star mergers as progenitors, potentially increasing the expected rate of gravitational wave detections. The studies emphasize the importance of accurately determining the formation rates of GRBs to better estimate the occurrence of low mass merger gravitational wave sources and kilonovae, especially at low redshifts.

\paragraph{Summary and consideration}
Accurately determining GRBs redshift would give us a significant advantage, since it would provide us with information about their distance. The importance of alerting ground telescopes for follow-up grows with the distance, as the event could belong to a population of the first stars in the visible universe.
Moreover, when analysing a GRB it is very important to have the correct energy in the rest frame and this is something that can be done with redshift information. This can help in get fine and distinct clustering, i.e., improving the distinction between type I and type II GRBs.
The estimation of redshift remains a challenging task, and to date, the most successful solutions have been achieved using Swift data, which includes both prompt emission and X-ray afterglow emission. There have been no significant efforts to estimate redshift solely based on prompt emission, such as from Fermi-GBM, likely due to the critical role of afterglow information. However, a potential approach could involve leveraging the complete information of the GRB, considering both its temporal and spectral characteristics, as depicted in Figure \ref{fig:zhang_input}.

\section{GRB simulation}\label{sec:grb_simulation}

An additional type of application can be found in simulating GRB prompt or afterglow emission.
In the work presented by \cite{boersma2022deepglow}, a neural network called DeepGlow is trained to simulate a GRB afterglow lightcurve. This simulation is conditioned by pre-defined afterglow parameters, thus ensuring fidelity to the underlying physical properties. On the other hand, Gaussian processes can be employed as in \cite{dainotti2023stochastic} for the reconstruction of GRB X-ray afterglow emissions. 
These developments are fundamental for the investigation of theoretical models and enhancing the estimate the parameters related to GRB afterglows, including the redshift of GRBs.

Simulating the prompt emission of GRBs is critical for testing trigger algorithms and refining algorithms for estimating the localization of gamma-ray sources. By generating diverse shapes of GRB prompt emissions along with different background templates, the heterogeneity of input data is increased, thereby improving the robustness of these algorithms.

A synthetic light curve generator is available in the repository \url{https://github.com/peppedilillo/synthburst}. This tool generates synthetic lightcurves by starting with existing GRBs from the Fermi-GBM catalog and adding a background template as an option. This template can be customized to meet specific requirements, such as originating from a specific detector during a particular part of an orbit of Fermi.
Another useful resource is \texttt{cosmogrb}, which can be found at \url{https://github.com/grburgess/cosmogrb}. This tool enables the generation of GRB light curves by specifying various parameters describing the GRB, such as localization (ra, dec), peak flux, duration, spectral properties (e.g., $\alpha$ and $E_{peak}$), and more.
Furthermore, lightcurves and Time Tagged Events (TTE) generator is available within the Fermi Data Tools framework, which can be found at \url{https://fermi.gsfc.nasa.gov/ssc/data/analysis/gbm/gbm_data_tools/gdt-docs/notebooks/Simulations.html}. This feature allows the user to create lightcurves binned at specific time intervals or as TTE, all of which can be conditioned by a spectral model, various GRB shapes (time profiles), and the background model spectrum.

\begin{figure}[!htb]
\centering
  \includegraphics[width=1\textwidth]{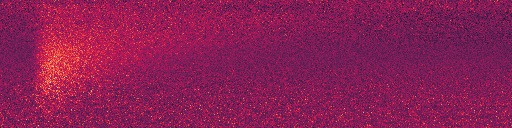}
  \includegraphics[width=1\textwidth]{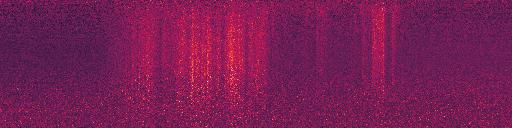}
  \includegraphics[width=1\textwidth]{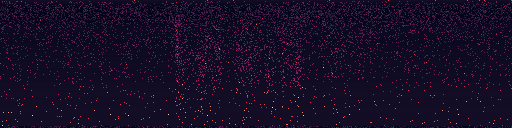}
\caption{\label{fig:grb_image}
The top, middle, and bottom panels shows GRB110920, GRB170214, and GRB170222, respectively, detected by Fermi-GBM and converted into a $512 \times 128$ image. The 512 time steps and 128 energy channels are normalized across the temporal period.}
\end{figure}

So far, these resources contribute significantly to the expanding toolkit for simulating and GRB prompt emissions, but a more data-driven version can be built employing ML and DL algorithms.
As already shown in \cite{zhang2023application}, a GRB can be represented as an image, preserving both temporal and energy spectral information. A GRB from the Fermi-GBM catalog has 128-energy channels and an estimation of duration, T90. To create a representation conserving the temporal representation, even for short GRBs, the time binning for count rates is determined by dividing the duration\footnote{Accounting for an additional 15 seconds before the event begins and 30 seconds after it ends.} by 512. This latter defines the image's $x$-axis, while the 128-energy channels the $y$-axis. Count rates are energy channel-normalized after background subtraction via linear interpolation.
Further implementation details are provided in the repository \cite{Crupi_ImageGRB_2023}, alongside illustrative examples shown in Figure \ref{fig:grb_image}.

\begin{figure}[H]
\centering
\includegraphics[width=0.95\textwidth]{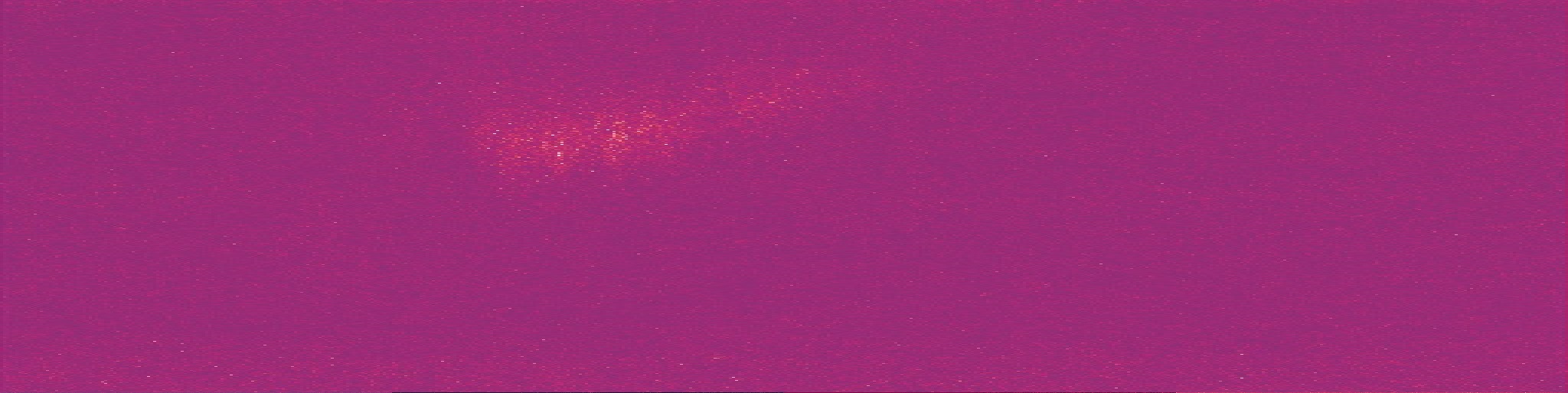}
\includegraphics[width=0.95\textwidth]{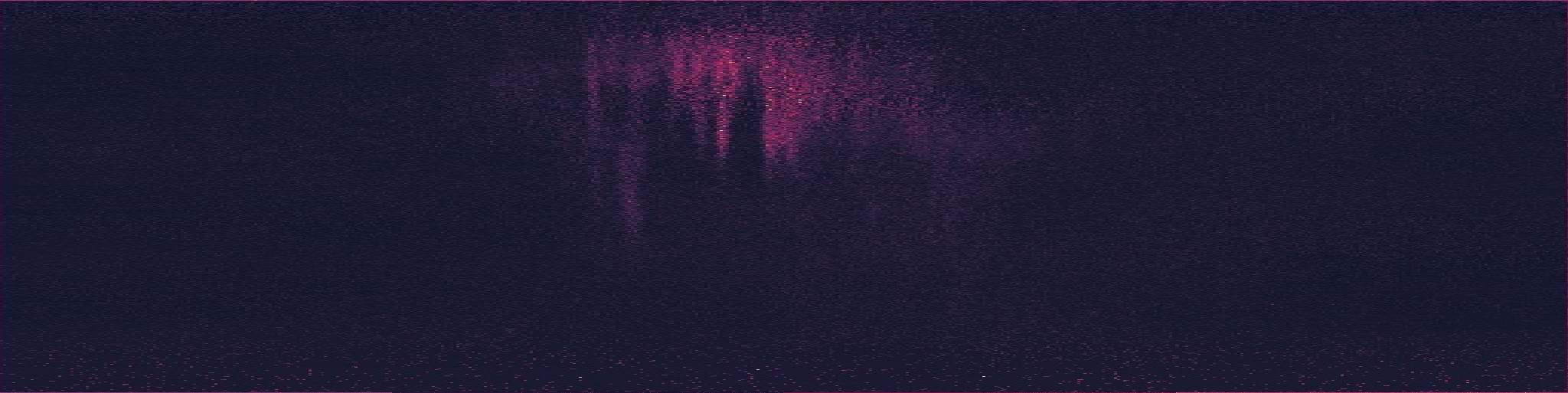}
\includegraphics[width=0.95\textwidth]{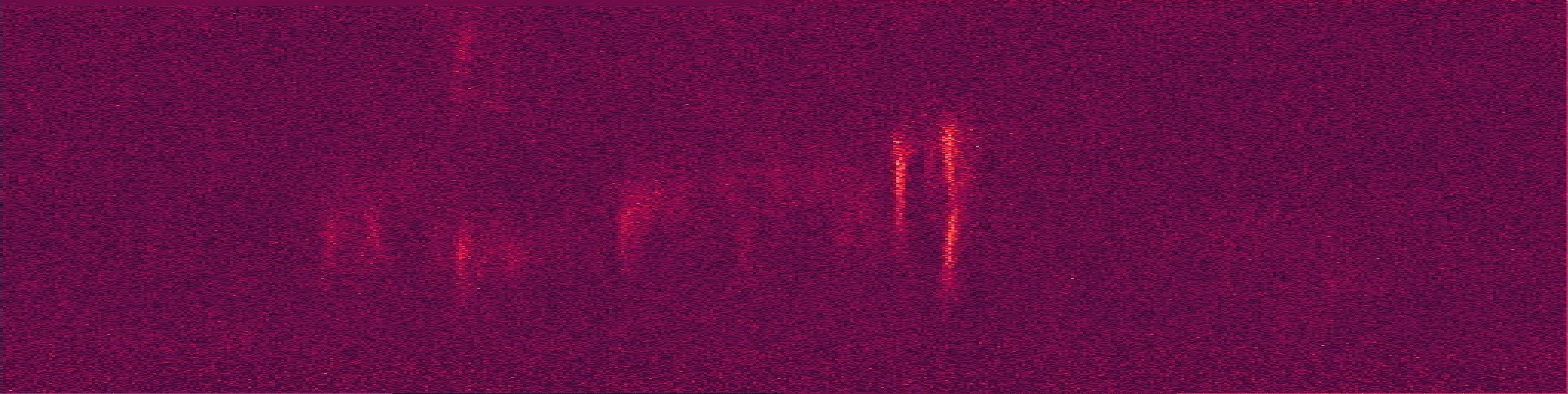}
\caption{\label{fig:grb_image_fake}
Three synthetic images generated by a Diffusion model after being trained over approximately 30000 images for 50 epochs.}
\end{figure}

\begin{figure}[H]
\centering
\hspace*{-0.5cm}
\includegraphics[width=1\textwidth]{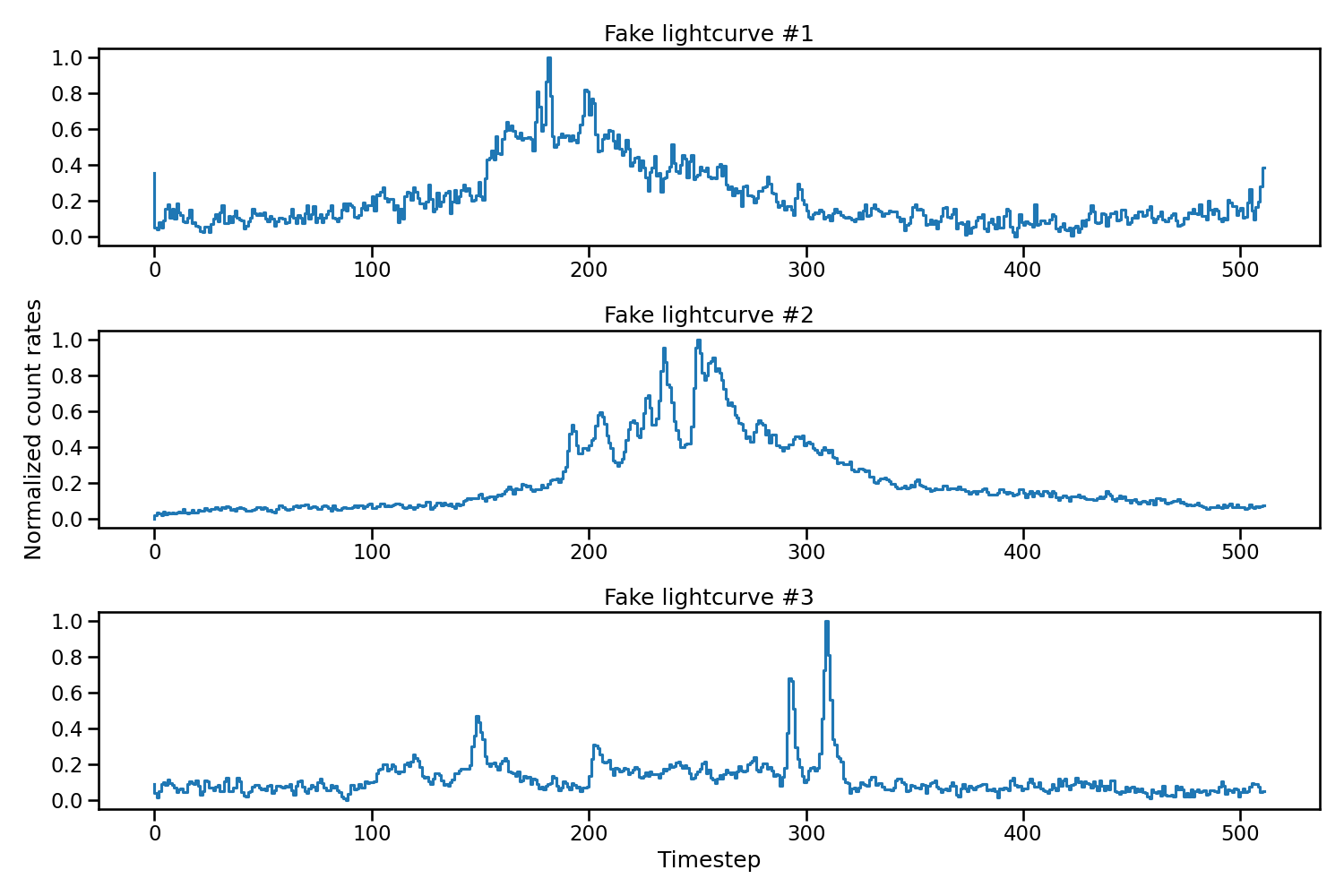}
\caption{\label{fig:grb_image_lc_fake}
Three synthetic light curves obtained by integrating over the energy ranges of the three synthetic GRBs from Figure \ref{fig:grb_image_fake}.}
\end{figure}

This representation can serve as input for generative deep learning algorithms that generate a distribution of GRB-like images resembling the input dataset. Such algorithms employ methodologies like GANs or diffusion models, avoiding the dependence on specific templates or predefined physics.
A first attempt to leverage on generative models for GRB prompt emission simulation is reported here, employing the Denoising Diffusion Probabilistic Models\footnote{Implementated by Hugging Face \url{https://huggingface.co/docs/diffusers/api/pipelines/ddpm}.} on a training dataset including around 1000 GRBs. Taking into account that each event is observed, on average, by 3 detectors and applying random shifts to the left and right, this dataset is further expanded to 10000 images. Following 50 training epochs, the diffusion model produces synthetic images, three of which are shown in Figure \ref{fig:grb_image_fake}. The corresponding light curves, integrated across all energy ranges (y-axis), are shown in Figure \ref{fig:grb_image_lc_fake}. 

Other potential approaches include exploiting the time series nature of GRBs, such as employing RNNs to learn the dynamic patterns or exploring the utilization of GANs \cite{yoon2019time, snow2020mtss} or Diffusion models for time series \cite{lin2023diffusion}.

\paragraph{Summary and consideration}

Exploring the use of AI techniques for GRB simulation could be a promising direction to pursue. By employing AI generative models, realistic GRB simulations can be created, providing a valuable stress test for on-board and offline trigger algorithms. If the generative models can accurately replicate the properties of real GRBs, such as duration, hardness ratio, and redshift, it would not only enhance our understanding of the GRB generation process but also offer insights into the underlying relationships between these properties. This approach has the potential to improve our knowledge of GRBs and their characteristics, aiding in the advancement of GRB research and trigger algorithms for future missions.

\chapter{Background Estimator}\label{chap:bkg}

In this chapter it is described a Machine Learning approach to estimate a background model for the GBM data. It is employed a Neural Network (NN) to estimate each detector background signal given the information of the satellite and its detectors. 
It is described the input data, the architecture of the NN, the training phase and finally some performance results. In the results are reported quantitative results based on measure like Mean Absolute Error and Median Absolute Error and a qualitative application estimating the background on GBM data for a GRB present in the GBM catalog, the ultra-long GRB 091024.
Solar maxima minima analysis is performed with the double aim for: 1) understand how the background estimation is affected by an high background noise due to solar activity and 2) estimate the expected count rates effect of the solar activity on the detectors for HERMES, which be launched in a period close to the maxima solar activity.
The estimated background can then be employed into a triggering algorithm to discover significant long/weak events that are not detected by other approaches.
The proposed approach is straightforwardly generalizable to estimate the background model of other satellites because all share the data described in the following section. 

\section{Data}\label{sec_data}

\begin{figure}
    \centering
    \includegraphics[width=1\textwidth]{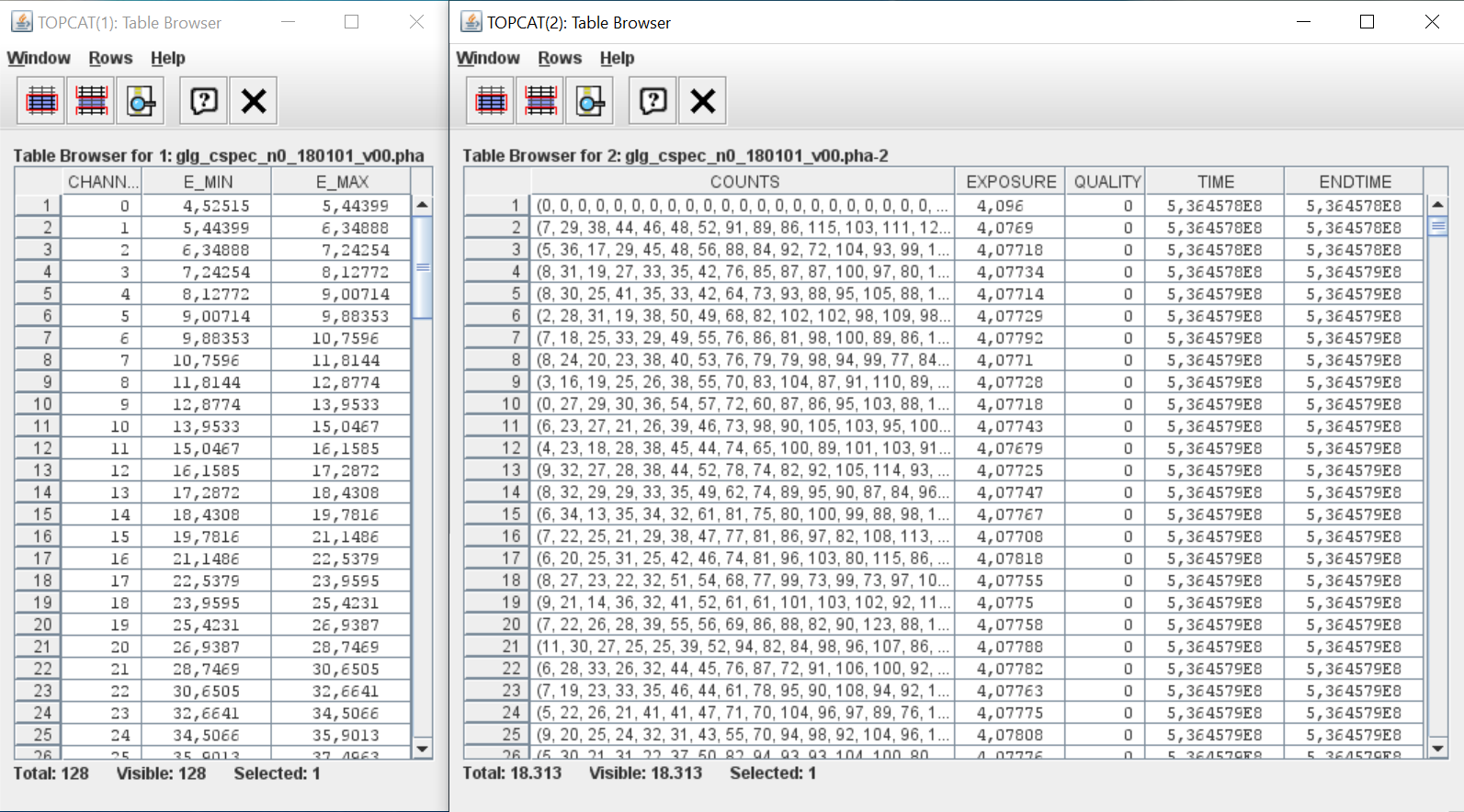}
    \caption{An example of CSPEC file opened with the TOPCAT software. On the left the association between energy channel label and its energy range. On the right the count rates data per each energy channel over the exposure time, around 4s.}
    \label{fig:cspec_topcat}
\end{figure}

The Fermi/GBM daily \verb|CSPEC| data products (see Figure \ref{fig:cspec_topcat}) were used for both the testing and the training of the neural network and for searching astrophysical transient events with FOCuS-Poisson. 
These data are photon count rates over a duration of $4.096$~s, binned over $128$ logarithmically spaced energy channels spanning from  $\approx 8$ keV to $\approx 900$ keV \cite{meegan2009fermi}. 
The time resolution provided by \verb|CSPEC| data is high enough to investigate long and ultra-long GRBs, yet it is too low to reliably identify short GRBs and other transients with characteristic duration shorter than a few seconds. 
This is unfortunate, yet justified for our use-case. Indeed, the variability of background over time intervals of duration comparable to the duration of short GRBs is negligible, hence our method provides little benefits relative to simpler approaches such as moving average or exponential smoothing.
On the other hand, an accurate description of the background become essential when searching for long, faint events, in particular events whose duration is comparable to that of the Fermi orbit such as ultra-long GRBs.
We consider data from all of the Fermi/GBM's twelve NaI detectors. Each detector is identified according to the standard GBM nomenclature (ten detectors labelled with integers ranging from $0$ to $9$, two detectors are identified by the letters $a$ and $b)$. 
In our analysis we disregard the Fermi/GBM bismuth-germanate detectors. These instruments are in fact sensible to energies much greater than the energies typically involved with GRBs prompt emission and are mainly used for the detection and observation of phenomena different from GRBs, such as Terrestrial Gamma-Ray Flashes (TGF) \cite{von2020fourth}.
To build the target variables $Y$, the input \verb|CSPEC| data from each detector are binned anew, this time over three coarser energy ranges (28-50 keV, 50-300 keV and 300-500 keV, see Table~\ref{tab:range}). The resulting dataset is arranged in a table with $36$ columns, one for each of the $36$ detector-energy combinations.


\begin{table}[!htb]
\centering
\begin{tabular}{l|r}
Range & Energy range (keV) \\\hline
r0 & 28-50 \\
r1 & 50-300 \\
r2 & 300-500
\end{tabular}
\caption{\label{tab:range} 
Energy ranges table. 
}
\end{table}

Beside the \verb|CSPEC| data product, the NN is trained using information on the satellite geographical location and the detectors pointing direction, as well as a number of auxiliary features such as the Earth occultation status and the visibility of the Sun for each detector at a given time. These informations are gathered from the Fermi/GBM \verb|POSHIST| data products. A detail of the orbital and detectors features used in the training of the NN is given in Table~\ref{tab:feature} and Table~\ref{tab:feature_det}. 

Sources of background for high-energy count detectors in low-Earth, near-equatorial orbits have been discussed thoroughly in literature (see \cite{campana2013background} and \cite{biltzinger2020physical} for discussions relevant to the present context). 
Our choice of the NN feature inputs was designed to provide a sufficient description of the different background components along the Fermi’s orbit. 
For example, the instantaneous rate of primary and secondary cosmic ray particles, a majour component of the background, will change depending on the the spacecraft geographical latitude, altitude, as well as the McIlwain parameter. Furthermore, the intensity of the cosmic photon background is influenced by the spacecraft's attitude and the position of the Earth in a detector's field of view. 
Leveraging information such as the pointing direction of different detectors and the spacecraft attitude can also aid the neural network in predicting the impact of point sources on the instantaneous background rate; as considering the presence of Earth in the field of view of a detector can potentially improve the neural network's ability to resolve the impact of components such as the albedo Gamma-ray and neutron background.
The rate of change of individual components is influenced by the spacecraft's velocity in the Earth's inertial system and the angular velocity of the spacecraft itself. Special flags were utilized to indicate the presence of the Sun in a detector's field of view, as well as transits through the high-radiation environment of the SAA.

To access both the \verb|CSPEC| and \verb|POSHIST| data products we use the Fermi Data Tools package, a python software API to the HEASARC Fermi archival database.
The resulting input datasets $X$ include a total of 60 different features, sampled with a step length of $4.096$ s. 

\begin{table}[H]
\centering
\begin{tabular}{c|l}
Feature label & Description \\\hline
$pos\_x$, $pos\_y$, $pos\_z$ & position of Fermi in Earth inertial coordinates \\
$a$, $b$, $c$, $d$ & Fermi attitude quaternions \\
$lat$ & Fermi geographical latitude \\
$lon$ & Fermi geographical longitude \\
$alt$ & Fermi orbital altitude \\
$vx$, $vy$, $vz$ & velocity of Fermi in Earth inertial coordinates \\
$w1$, $w2$, $w3$ & Fermi angular velocity \\
$sun\_vis$ & Sun's visibility boolean flag\\
$sun\_ra$ & Sun's right ascension\\
$sun\_dec$ & Sun's declination \\
$earth\_r$ & Earth's apparent radius \\
$earth\_ra$ & Earth center right ascension\\
$earth\_dec$ & Earth center declination \\
$saa$ & SAA transit boolean flag\\
$l$ & approximate McIlwain L value
\end{tabular}
\caption{\label{tab:feature}
A table of the 24 orbital features used to form the NN's input table. The features are obtained from the POSHIST files and processed by the library Fermi GBM Data Tools.
}
\end{table}

\begin{table}[H]
\centering
\begin{tabular}{c|l}
Feature label & Description \\\hline
$ni\_ra$ & $i$-labelled detector pointing right ascension \\
$ni\_dec$ & $i$-labelled detector pointing declination \\
$ni\_vis$ & $i$-labelled detector Earth occulation boolean flag\\
\end{tabular}
\caption{\label{tab:feature_det} 
A table of the 36 detector features used to form the NN's input table, where the detector label $i \in \{0,1,2,3,4,5,6,7,8,9,a,b\}$. The features are obtained from the POSHIST files and processed by the library Fermi GBM Data Tools. 
}
\end{table}

\begin{table}[H]
\centering
\begin{tabular}{c|l}
Target label & Description \\\hline
$ni\_r0$ & $i$-labelled detector count rates in range $r0$ \\
    $ni\_r1$ & $i$-labelled detector count rates in range $r1$ \\
    $ni\_r2$ & $i$-labelled detector count rates in range $r2$ \\
\end{tabular}
\caption{\label{tab:target_data} 
A table of the 36 detector features corresponding NN's target data table, where the detector label $i \in \{0,1,2,3,4,5,6,7,8,9,a,b\}$. The target data are the count rates binned at 4.096 s, obtained from the daily CSPEC files and processed by the library Fermi GBM Data Tools.
}
\end{table}

\section{Architecture and problem definition}
The background assessment problem is expressed as a supervised Machine Learning estimator, with the variables inherent to the satellite and its orbital position as inputs and the count-rate observed by each detector in three different energy bands as outputs.
The background estimates so obtained are compared against the actual observations using FOCuS-Poisson and described in Chapter \ref{chap:frm}.

\subsection{Neural Network background estimation}\label{sec_background}
\begin{figure}
    \centering
    \includegraphics[width=1\textwidth]{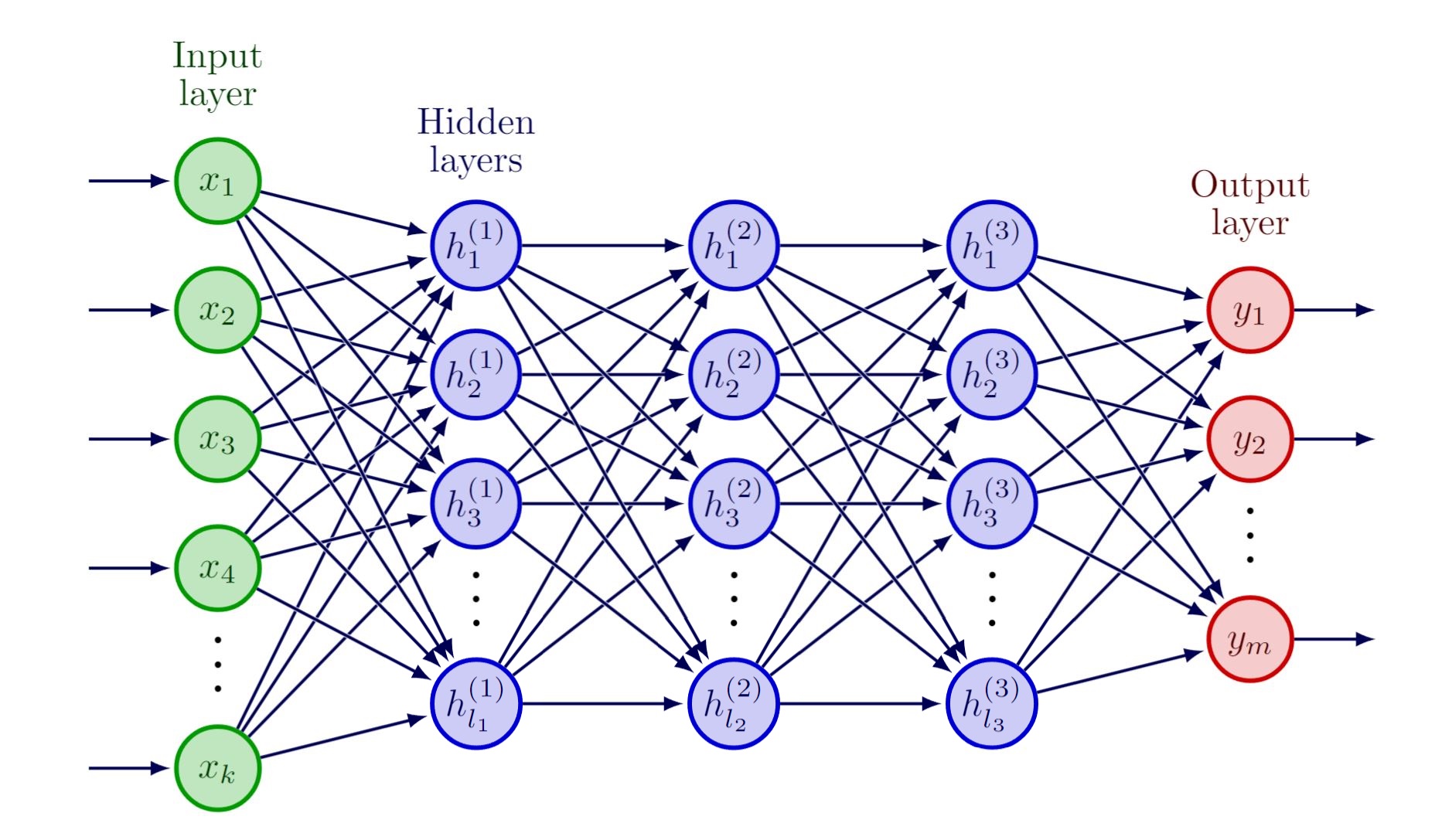}
    \caption{The architecture of a feed forward neural network. The input has dimension 60. The first two hidden layer have 2048 neurons, the third 1024. The output layer has dimension 36.}
    \label{fig:nn}
\end{figure}

 We define $X$ as the input variables, see features in Tables \ref{tab:feature} and \ref{tab:feature_det}, and $Y$ as the output variables, see variables in Table~\ref{tab:target_data}. 
We suppose that a function $f(X)$ exists which predict $Y$ given $X$, that is the solution that minimize $L(f(x), Y)$ ($\text{argmin}_f  L(f(x), Y)$) where $L$ is the loss function that quantify the error in the predictions. The model's goal is to estimate a quantity $F(x)$ such that $f(x) \approx F(x)$ \cite{hastie2009elements}. Here we are dealing with a multi-output regression: $F: X \in \mathbb{R}^{k} \longrightarrow Y \in \mathbb{R}^{m}$, where $k$ is the number of features into the model and $m$ the number of outputs.\\
The model employed is a feed forward neural network with 3 hidden dense layers (Figure~\ref{fig:nn}). Each hidden layer is followed by a batch normalization layer \cite{ioffe2015batch} and a dropout layer \cite{JMLR:v15:srivastava14a}. The NN is implemented in Tensorflow \cite{tensorflow2015-whitepaper}. 
The input layer has dimension $k=60$. Each of the first two hidden layers is composed of $2048$ neurons, while the third hidden layer hosts $1024$ neurons. The last (output) layer has $m=36$ neurons. Each of the output neurons is associated with a particular detector-energy combination. The probability parameter for the drouputs is $0.02$. The optimizer used is Nadam \cite{Ruder16} with learning rate $\eta$ varying accordingly to Equation \ref{lr_decay}, $\beta_1 = 0.9$, $\beta_2 = 0.99$ and $\epsilon = 10^{-7}$.
\begin{equation}\label{lr_decay}
  \eta =
    \begin{cases}
      0.01 & \text{if epoch $<4$}\\
      0.0016 & \text{if $4\ge$ epoch $<12$}\\
      0.0004 & \text{if epoch $\ge 12$}
    \end{cases}       
\end{equation}
We run the fitting for 64 epochs with a batch size of 2048. 

Other neural network architectures were considered during the design process. For instance, the background estimation could be approached by utilizing sequential count rates to predict future ones, i.e. employing a RNN. This approach has been discussed in the literature, such as in the work of \cite{sadeh2019deep}.
Training an RNN to predict background count rates, it is crucial to exclude periods that contain astrophysical transients from the training dataset. This prevents the RNN from learning the count rate dynamics in a way that would make it difficult to distinguish astrophysical transients from the actual background. 
To filter transients such as GRBs from the training dataset, the presence of these events should be known in advance. This implies relying on existing catalogs of transient astrophysical phenomena. We believe such an approach could result in the model inheriting the detection biases of standard strategies for GRB detection. This scenario would prove detrimental to the present work, as our goal is precisely to detect transients that may have evaded previous searches.
On the other hand, our approach differs in that we utilize input features related to the satellite/detector, which should be independent of events like GRBs, to estimate the expected count rates for each detector. This "mapping" from the satellite configuration to the expected count rates is currently accomplished through the previously described FFNN, but could in line of principle be extended by incorporating an RNN that considers the previous satellite configurations.

In a pre-processing step, the input training dataset is standardised and filtered. Data filtering takes place in two steps in which the following data subsets are removed:
\begin{itemize}
    \item data collected while Fermi is transiting through the high radiation environment of the South Atlantic Anomaly (SAA).
    \item data acquired at times in which an event of the Fermi/GBM trigger catalog occurred.
\end{itemize}
This latter choice isn't strictly necessary, yet it is useful to better understand the neural network performances over known events. 
The splitting procedure divides the dataset in 75\% train, 25\% test; 30\% of the training set is further kept as validation set. The resulting splitting is: 52\% train, 23\% validation and 25\% test. The instances inside these sets are not sequential but rather taken randomly. \\

The purpose of our framework is to evaluate the effectiveness of our model on a known dataset, hence the choice of a loss function $L$ which is robust against outliers is critical. The Mean Square Error loss function (MSE) is:
\begin{equation}\label{eq:mse}
\text{MSE}(z, y) = \frac{1}{n} \sum_{i=1}^{n}\left ( y_{i} - z_{i} \right )^2.
\end{equation}
where $n$ is the total number of samples in training set, $i$ refers to the specific sample, $y_i$ the target value (the observed count rate) and $z_i$ is the estimated value (the estimated count rate).
MSE is very sensitive to the discrepancy between the prediction and the target value, thus it is a bad choice when outliers are present in the training dataset. 
We remark that the filtering of catalog events is not enough to guarantee the optimization of the background estimator when using MSE. Indeed, anomalous events, which are not present in the GBM catalog, may be over-fitted when minimizing MSE; these events are the actual targets of our search. 

The Mean Absolute Error (MAE) loss function is less sensitive to residuals:
\begin{equation}
\text{MAE}(x, z) = \frac{1}{n} \sum_{i=1}^{n} \mid y_{i} - z_{i} \mid
\end{equation}
the term are the same as in Equation \ref{eq:mse}.

When anomalous events are included in the training dataset, the use of MAE instead of MSE can lead to a neural network less prone to overfitting, as discussed in Section \ref{sec:hyperparam_loss}.
In the settings of multi-output regression, the overall loss $\mathcal{L}$ is defined as the MAE average of the NN outputs:
\begin{equation}
\mathcal{L} =  \frac{1}{m} \sum_{j=1}^{m}( \text{MAE}(Z_j, Y_j) )
\end{equation}
where $j$ a specific detector/energy range, $m$ the total number of detector/energy range, $X \in \mathbb{R}^{n,k}$ the input feature matrix of the NN (samples times features), $Z=\{F(X_i), \, i=1:n\} \in \mathbb{R}^{n,m}$ is the Neural Network outputs (estimated count rates per each detector/energy range), $Y \in \mathbb{R}^{n,m}$ the observed count rates for each detector/energy range.

For evaluation purposes, the Median Absolute Error (MeAE) is employed because of its robustness against the outliers
\begin{equation}
     \text{MeAE}(z, y) = \text{median} ( \{ \mid y_{i} - z_{i} \mid \} ),
\end{equation}
where the terms are the same as in Equation \ref{eq:mse}.

\section{Performance results}
In this section we present the results of the background estimator and the trigger algorithm application.
The open source code implementation is available on \href{https://github.com/rcrupi/DeepGRB}{github.com/rcrupi/DeepGRB}.

\subsection{Background estimator performance}\label{sec_results_nn}
\begin{figure}[H]
	\centering
	\includegraphics[width=1.\textwidth]{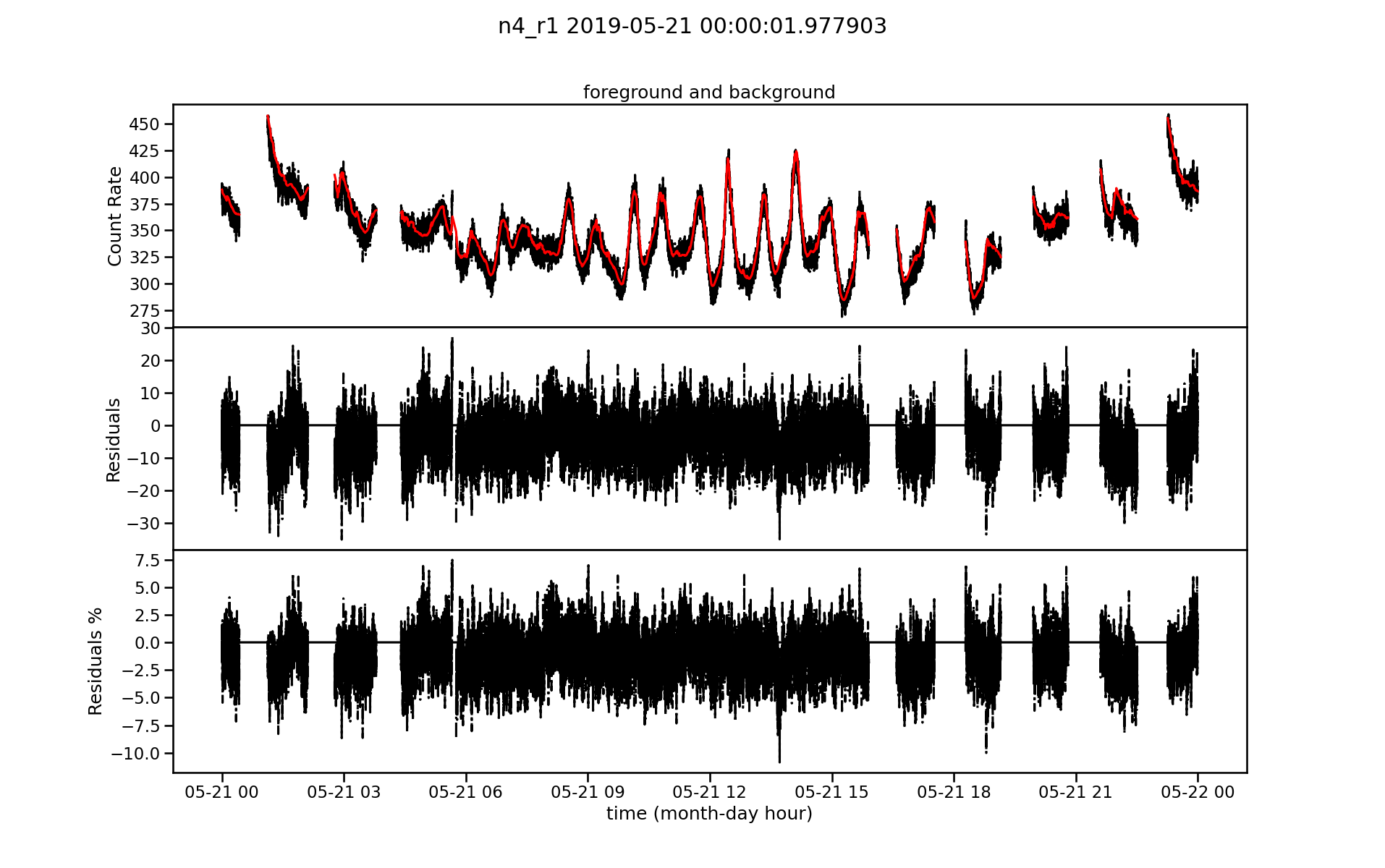}
	\caption{\label{fig:residual}
		The background estimation for the \texttt{n4} detector, in the energy range r1, during 21 May 2019. The Fermi-GBM count rate observations are represented over time as a black line, whereas the neural network estimation is plotted as a red solid line. The middle panel shows the residuals between the two quantities, with a black solid line denoting the reference of null residual. The lower panel shows the residuals as relative change percentage. Source: Crupi et al. \cite{crupi2023searching}.}
\end{figure}

To show the effectiveness of this approach, a NN is trained over 7 months of data from January to July 2019. An excerpt of the resulting background estimation is presented in Figure \ref{fig:residual} for one detector-range combination, during a day without any events in the Fermi-GBM trigger catalog. This provides both quantitative and qualitative insights into the background estimation.
 For a more concise and comprehensive quantitative performance evaluation, refer to the MAE values reported in Table~\ref{tab:mae_summary}. The energy range bins are the same as those used in Section \ref{sec_data} and are defined in Table \ref{tab:range}.

\begin{table}[H]
\centering
 \begin{tabular}{||c | c c ||} 
 \hline
 det range & MAE train (counts/s) & MAE test (counts/s) \\
 \hline\hline
 r0 & 4.942 $\pm$ 0.331 & 4.953 $\pm$ 0.328  \\
 r1 & 6.088 $\pm$ 0.167 &  6.098 $\pm$ 0.163 \\ 
 r2 & 1.790 $\pm$ 0.044 & 1.792 $\pm$ 0.045 \\
 \hline
 average & 4.273 & 4.281 \\ 
 \hline
 \end{tabular}
 \caption{The NN MAE loss function (within one standard deviation) per energy range, over the training and the testing datasets, averaged over the 12 Fermi's GBM NaI detectors.
 }\label{tab:mae_summary}
\end{table}

\begin{figure}[H]
	\hspace*{-1cm}   
	\centering
	\includegraphics[width=1.25\textwidth]{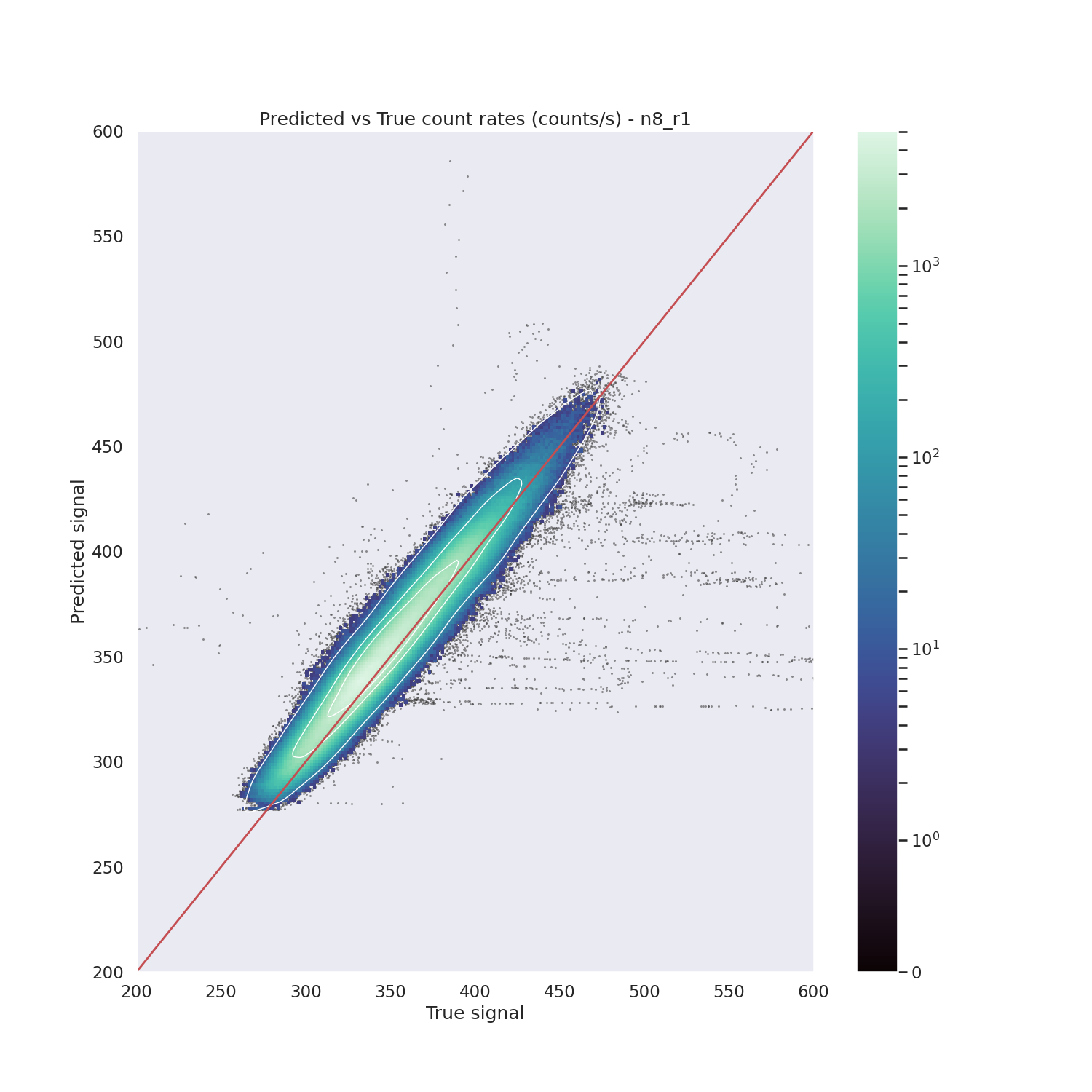}
	\caption{\label{fig:bkg_est} Fermi-GBM photons count rates from NaI-8 detector in the energy range $50$ - $300$ keV (r1) versus the respective prediction from the NN over the same combination of detector and energy range. Data spans from 1 January 2019 to 1 July 2019. The three white lines represent the contour plots at 1$\sigma$, 2$\sigma$ and 3$\sigma$. Source: Crupi et al. \cite{crupi2023searching}.}
\end{figure}

\begin{figure}[H]
	\hspace*{-1cm} 
	\centering
	\includegraphics[width=1.25\textwidth]{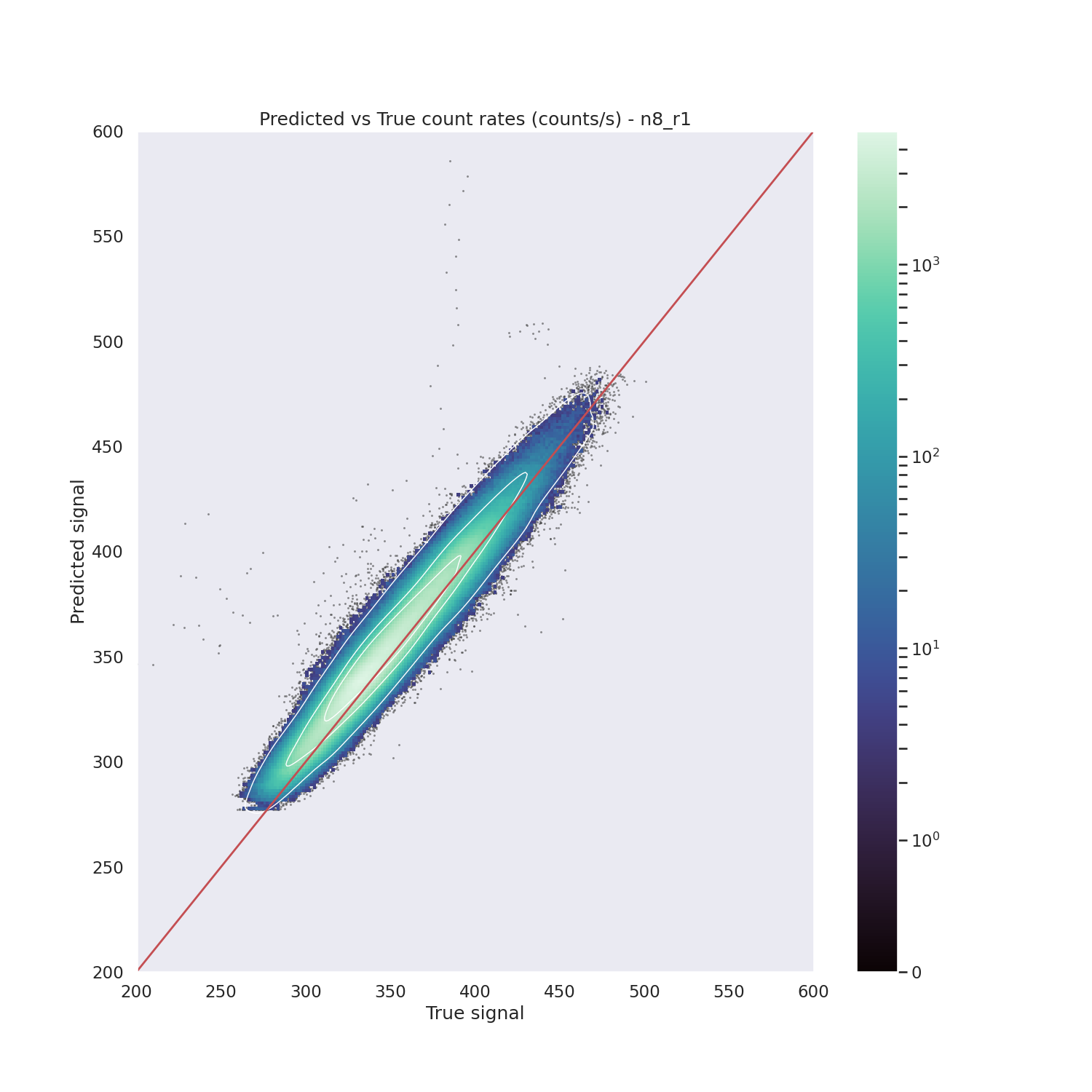}
	\caption{Figure equivalent to \ref{fig:bkg_est} but with the exclusion of count rates associated with transients detected by FOCuS (details in Chapter \ref{chap:frm}). While there are still some outliers present on the right side of the plot, it has been identified that these outliers correspond to the initial phases of triggered events. We can confirm that the horizontal line represents potential transients. \label{fig:bkg_est_no_event}}
\end{figure}

In Figure \ref{fig:bkg_est} it is plotted the NN predictions against the corresponding observed value, in particular we filtered out the data points 150 s before and after the SAA, specifically if the satellite remains in the SAA for at least 500s. 
The events of interest for this research should be found when the observed count rates exceed the NN prediction, that are the points below the bisector, for detectors-range combinations. Notice the horizontal lines in Figure \ref{fig:bkg_est}. The same plot, but excluding the identified transients with the method explained in the next chapter, can be seen in Figure \ref{fig:bkg_est_no_event}. 

\begin{figure}[H]
    \centering
    \includegraphics[width=1.\textwidth]{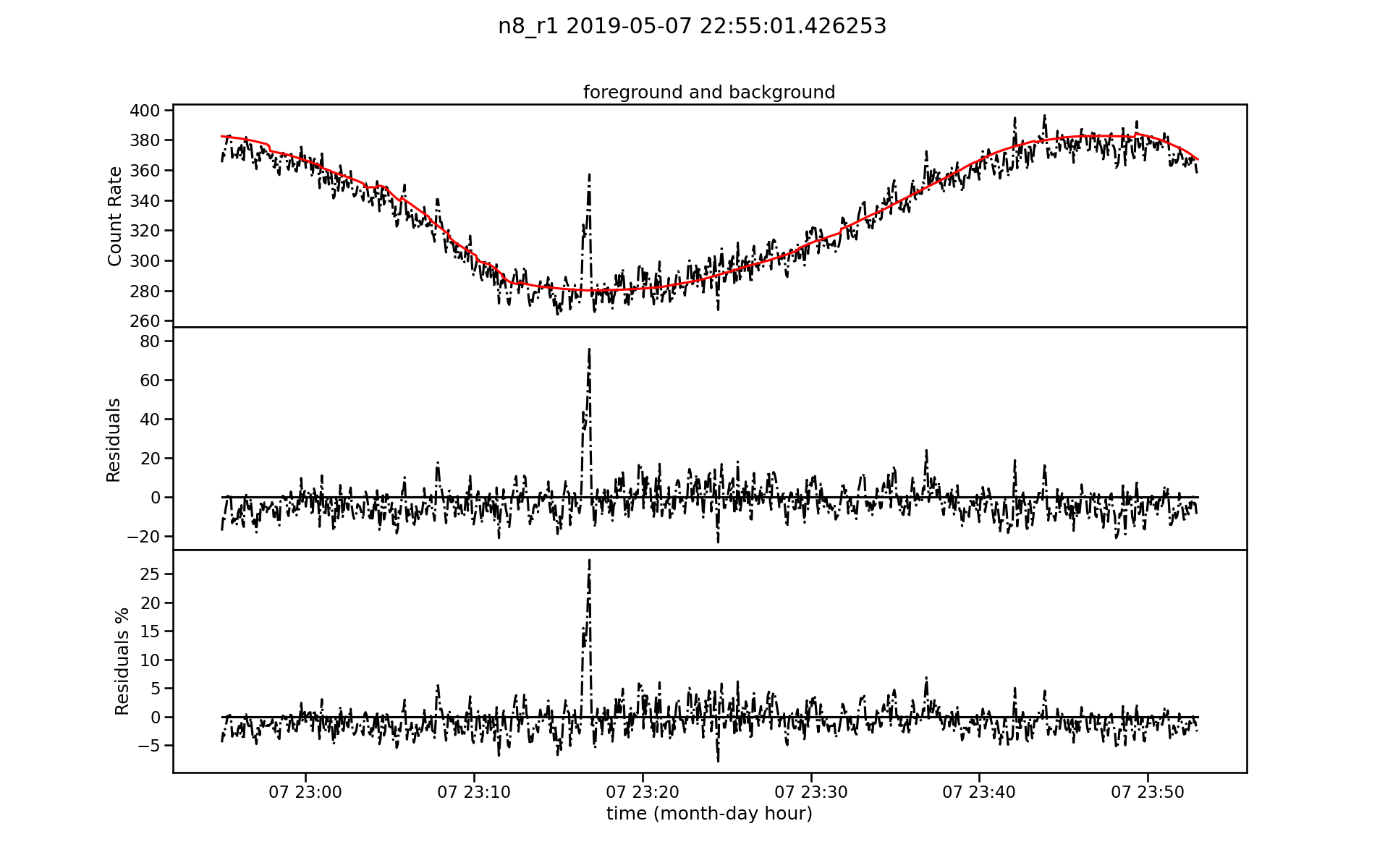}
    \caption{Observed and background estimation count rates for detector n8 energy range r1, around the GRB 190507970, with residual difference and residual as relative change percentage. This event is visible as the seven horizontal data points on the bottom right of the ellipse in Figure \ref{fig:bkg_est}. Source: Crupi et al. \cite{crupi2023searching}.
    \label{fig:res_GRB190507}}
\end{figure}

Transient, bright events such as GRBs may result in a temporary increase of the observed count rates (see Figure \ref{fig:res_GRB190507}) and, taking place at random times and directions, are not predictable from features intrinsic to the Fermi spacecraft motion and attitude, which are the actual inputs of the NN. 

\begin{figure}[H]
	\centering
	\includegraphics[width=.65\textwidth]{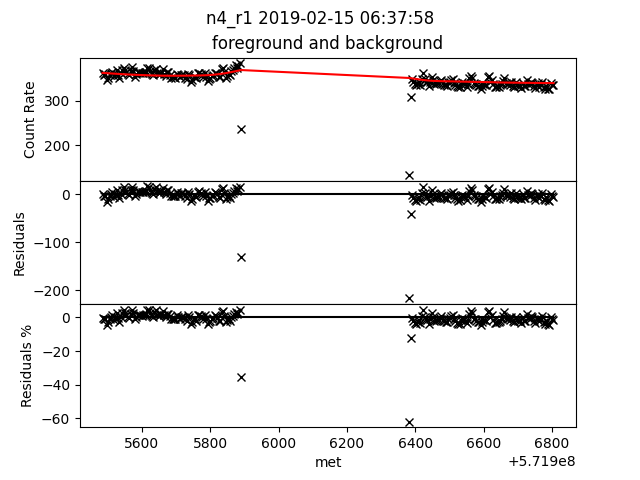} 
	\caption{\label{fig:switch}
		Fermi-GBM NaI-4 detector photon count rates (crosses) in the energy range $50$ - $300$ keV (r1) versus the respective prediction from the Neural Network (red solid lines). The lower panels show the residuals and relative change percentage between the two quantities, with a black solid line denoting the reference of null residual. Data span $1400$ s and one SAA crossing. The observed counts immediately before and after the instrument's switch-off during SAA crossings exhibit anomalously low values. Source: Crupi et al. \cite{crupi2023searching}.
	}
\end{figure}

Figure \ref{fig:switch} depicts low value outliers, background count rates greater than those observed, encountered in immediate proximity to SAA transit when the Fermi-GBM instruments are switched on or off.

The three time periods chosen for the application of the trigger algorithm spans:
\begin{itemize}
    \item 1 November 2010 to 19 February 2011,
    \item 1 January 2014 to 28 February 2014,
    \item 1 March 2019 to 9 July 2019.
\end{itemize}
For the sake of brevity, we will refer to these epochs as the `2010',  the `2014' and  the `2019' \emph{periods}.
These \emph{periods} are chosen to test the framework under a variety of conditions, including solar activity intensities and potential detector degradation.

A separate NN is trained and tested for each of these periods to account for variations in background count rates over long time scales (years), which may be caused by factors such as the solar activity and detector degradation. We report the performance metrics in Table \ref{tab:nn_perf_table}.

\begin{table}[!htb]
\centering
\begin{tabular}{ |c|c||c|c|} 
 \hline
 \multicolumn{4}{|c|}{NN performance metrics on test set} \\
 \hline
 Period & Energy range & MAE (counts/s) & MeAE (counts/s) \\
 \hline
 \hline
 2010 & r0 & 7.730 $\pm$ 4.842 & 3.963 $\pm$ 0.232 \\
 2010 & r1 & 6.469 $\pm$ 1.517 & 4.777 $\pm$ 0.100 \\
 2010 & r2 & 1.864 $\pm$ 0.033 & 1.563 $\pm$ 0.028 \\
 \hline
 2014 & r0 & 19.79 $\pm$ 18.92 & 4.545 $\pm$ 0.409 \\
 2014 & r1 & 13.29 $\pm$ 11.10 & 5.604 $\pm$ 0.196 \\
 2014 & r2 & 1.949 $\pm$ 0.099 & 1.598 $\pm$ 0.082 \\
 \hline
 2019 & r0 & 4.831 $\pm$ 0.300 & 3.938 $\pm$ 0.245 \\
 2019 & r1 & 5.640 $\pm$ 0.082 & 4.716 $\pm$ 0.070 \\
 2019 & r2 & 1.804 $\pm$ 0.038 & 1.510 $\pm$ 0.032 \\
 \hline
 \hline
\end{tabular}
 \caption{\label{tab:nn_perf_table} Mean Absolute Error (MAE) and Median Absolute Error (MeAE), on the test datasets, for each energy range and averaged for detectors within one standard deviation. The higher MAE in 2014 compared to 2019 and 2010
can be attributed to the effect of solar activity. However, the use of MeAE as
the evaluation metric shows comparable results among the three periods.}
\end{table}

Additional details on the neural network's performance during times of both high and low solar activity are provided in Section \ref{solarmaxmin}.

\subsection{Performance analysis and weakness}
\begin{figure}
\centering
\begin{subfigure}{0.45\textwidth}
\centering
\includegraphics[width=\linewidth]{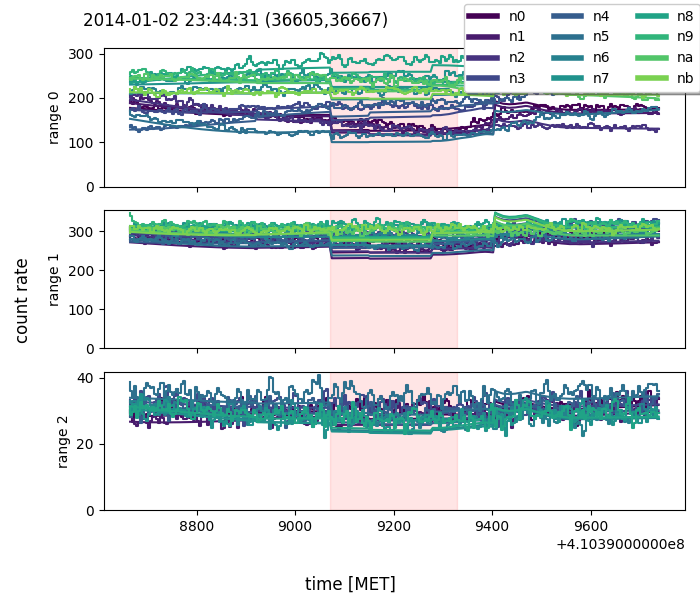}
\caption{\label{fig:erronn} 
}
\end{subfigure}
 \hspace*{\fill} 
\begin{subfigure}{0.45\textwidth}
\centering
\includegraphics[width=\linewidth]{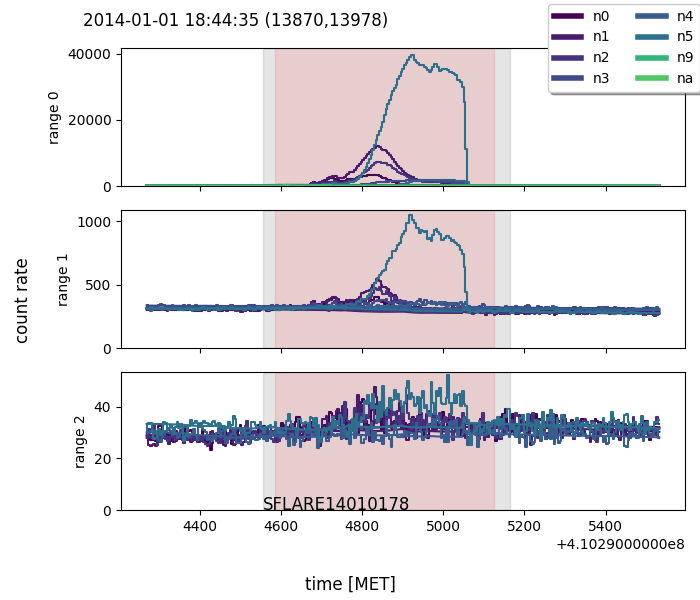}
\caption{\label{fig:bigsolarflare} 
}
\end{subfigure}

\caption{Photon count rates from each triggered detector are plotted with step lines, across three energy bands spanning $28-50$ keV, $50-300$ keV and $300-500$ keV (Table \ref{tab:range}), with a resolution of $4.096$~s. 
The neural network's prediction of background count rates is represented by solid lines.
Different detectors are identified using different colors.
 A red shaded area limits FOCuS-Poisson's best guess of the transient duration. Times are expressed in units of seconds according to Fermi's standard mission elapsed time (MET).
(a) Example of False Detection in which all the detector are triggered over an imprecision of the Neural Network estimation. (b) Example of a solar flare in the Fermi/GBM catalog detected by our approach. The event start and end MET time, as reported in the Fermi/GBM trigger catalog, is represented by a grey shaded area. Source: Crupi et al. \cite{crupi2023searching}.} \label{fig:2014errsf}
\end{figure}

According to Tables~\ref{tab:mae_summary} the test set and train set MAE values are similar up to $1 \%$ indicating no over-fitting and strong generalization across energy range and detector.
Table \ref{tab:nn_perf_table} shows that the neural network trained on data from the 2014 period has the highest (worst) MAE, which can be attributed to the presence of strong solar activity. This is understandable since, during an activity maximum, the background particle count rate is more unpredictable due to the influence of the Sun on the local radiation environment (see Figure \ref{fig:bigsolarflare}). Nonetheless, MeAE shows similar performance with the other two periods, thanks to its robustness against outliers. On the other hand, the 2019 period has the lowest MAE most likely due to low solar activity and low background variability.

Similar conclusions can be derived from the analysis presented in Section \ref{solarmaxmin}. The performance of the two neural networks trained on the complete data from 2014 and 2019 is nearly identical in terms of MeAE, see Table \ref{tab:mae_solar}. This suggests a comparable central tendency of the residuals in both periods. 

 It is important to note that the performance results are presented for both metrics, as they provide complementary information about the algorithm's performance. During the 2014 period, which had a solar maximum, the MAE was significantly larger than the MeAE due to the inclusion of 71 transient events not found in the Fermi-GBM trigger catalog (see Table \ref{tab:detections_stats}). These events, some of which are likely of solar origin, affected the MAE more than the MeAE. However, it is important to note that a low MeAE does not guarantee a perfect background estimation, as indicated by a high number of false detections in 2014 (Table \ref{tab:detections_stats}). Factors such as the inclusion of luminous solar transients and the reduced training dataset length can contribute to background estimation issues during this period.

In Figure \ref{fig:bkg_est}, most of the data points are distributed along the plot bisector $y = x$, indicating that most often the neural network estimate is in agreement with the actual observations.
Above the bisector, more count rates are expected than they are actually observed. From spot analysis, it is observed that outliers in this domain correspond to anomalously low values in the observed count rates. Many of these outliers are encountered in immediate proximity to SAA transit when the Fermi-GBM instruments are switched on or off\footnote{A possible explanation of this behaviour can be rooted in the Fermi/GBM experimental apparatus. Fermi/GBM is composed of multiple photomultipliers. These instruments require high voltages to operate. A "ramp-up" of a photomultiplier to operational voltage takes place over a time span of several seconds. During this time, the photomultiplier amplification factor (the gain) is hindered resulting in lower than nominal count rates. The same effect takes place when "ramping-down" before turning a photomultiplier off.}, as shown in Figure \ref{fig:switch}. 
In the lower part of the bisector the observed counts rate exceed the counts estimated. The trigger algorithm's goal is to determine whether this exceeding amount is part of an event or simply a random fluctuation.

In periods of high solar activity, Fermi/GBM data include a large number of soft transient events of solar origin; thus, the soft ($25$ - $50$ keV) trigger conditions have been disabled on multiple occasions (e.g. see Table 4 in \cite{bhat2016third} for 2014). Likewise, we required that at least one detector must be over threshold in the energy band spanning $50$ and $300$ keV in order for the trigger condition to be satisfied. 
Still, Table \ref{tab:detections_stats} shows a higher number of total events for the 2014 period. The majority of these events are most likely associated to solar flares; indeed, 50 of the 81 events in the GBM trigger catalog for this period are solar flares, and the majority of the events we find with no counterpart in the Fermi/GBM trigger catalog are triggered over Sun facing detectors (\texttt{n0}, \texttt{n1}, \texttt{n2}, \texttt{n3}, \texttt{n4}, \texttt{n5}).
False detections may be caused by artifacts in the background estimation. These are generally easy to identify; most of the time these artifacts take the form of sudden steps in the background estimate, simultaneously over all detector/range combination. One of these events is represented in Figure \ref{fig:erronn}. This behaviour is less frequently present in the other two periods analyzed, indicating that noisy background impacts on performance (see MAE) and therefore more false positive are detected. This issue should be investigated more, for instance integrating explainability techniques in the NN, see Section \ref{sec:xai_nn}, or implementing different and more capable architectures, such as RNNs.

\section{Background estimate for GRB 091024}\label{benchmark}

One qualitative approach for assessing the quality of a background estimator is to estimate the background during an event and then see whether the residuals can emerge clearly and if the dynamics estimated are coherent before, during, and after the event.

To demonstrate the potential of the NN background estimator in the presence of a long event, a background estimation is performed in a period containing the ultra-long GRB 091024 \cite{gruber2011fermi}, for which a similar evaluation is provided employing an established physical Fermi/GBM background model \cite{biltzinger2020physical}. 

\begin{figure}[H]
\bigskip
\bigskip
\centering
\hspace*{-2cm} 
\begin{subfigure}{.3\linewidth}
    \centering
    \includegraphics[width=1.5\textwidth]{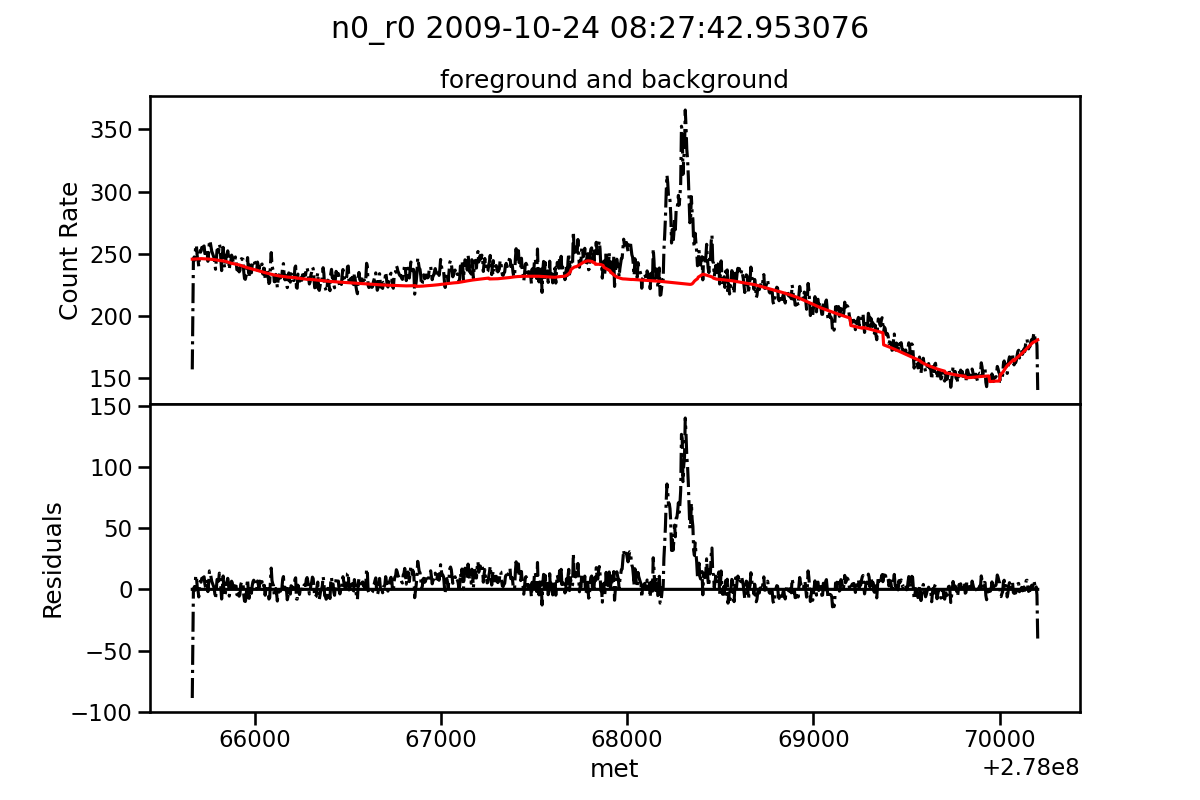}
\end{subfigure}
    \hfill
\begin{subfigure}{.3\linewidth}
    \centering
    \includegraphics[width=1.5\textwidth]{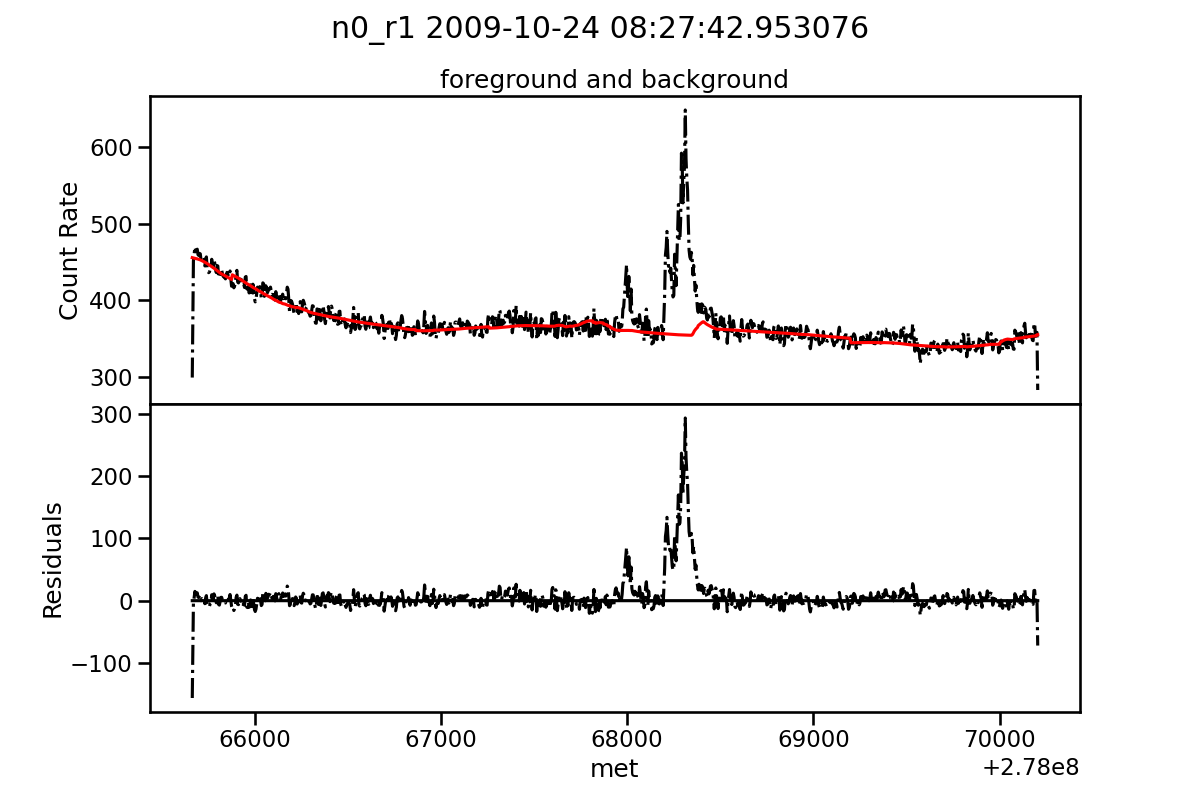}
\end{subfigure}
   \hfill
\begin{subfigure}{.3\linewidth}
    \centering
    \includegraphics[width=1.5\textwidth]{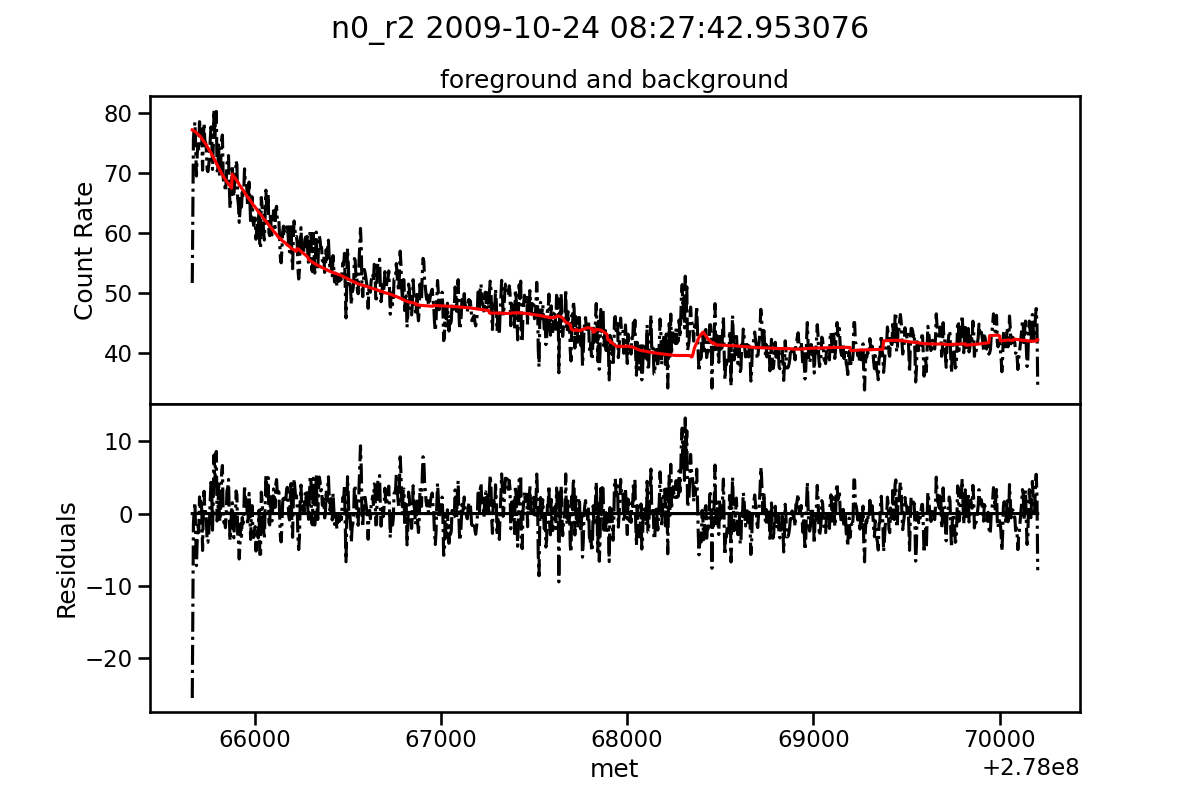}
\end{subfigure}

\bigskip
\bigskip
\hspace*{-2cm} 
\begin{subfigure}{.3\linewidth}
    \centering
    \includegraphics[width=1.5\textwidth]{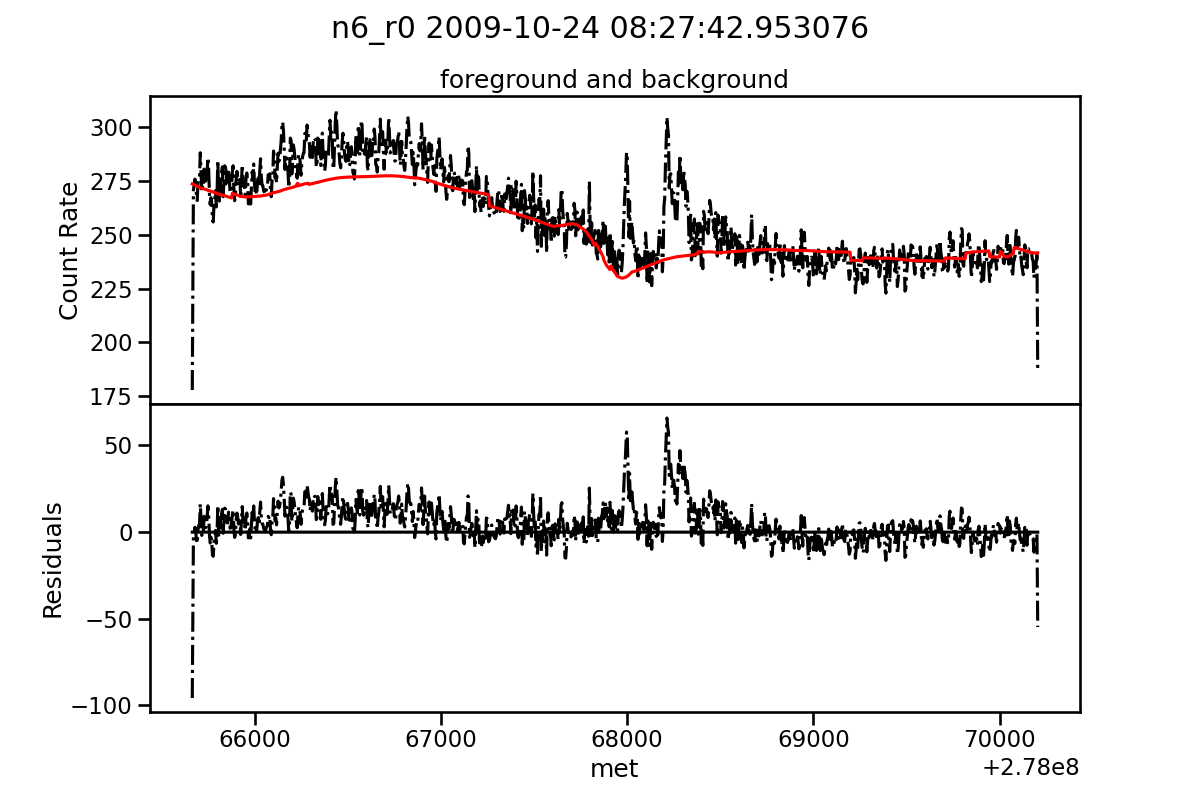}
\end{subfigure}
    \hfill
\begin{subfigure}{.3\linewidth}
    \centering
    \includegraphics[width=1.5\textwidth]{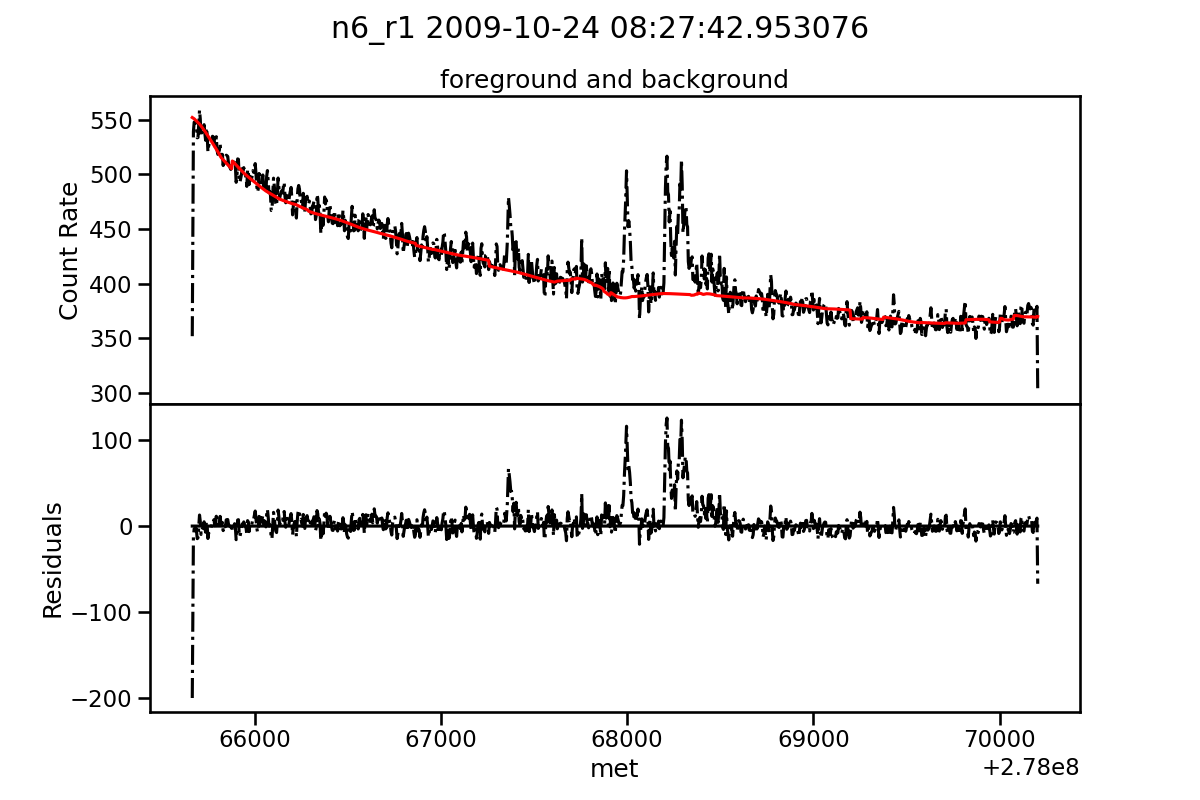}
\end{subfigure}
   \hfill
\begin{subfigure}{.3\linewidth}
    \centering
    \includegraphics[width=1.5\textwidth]{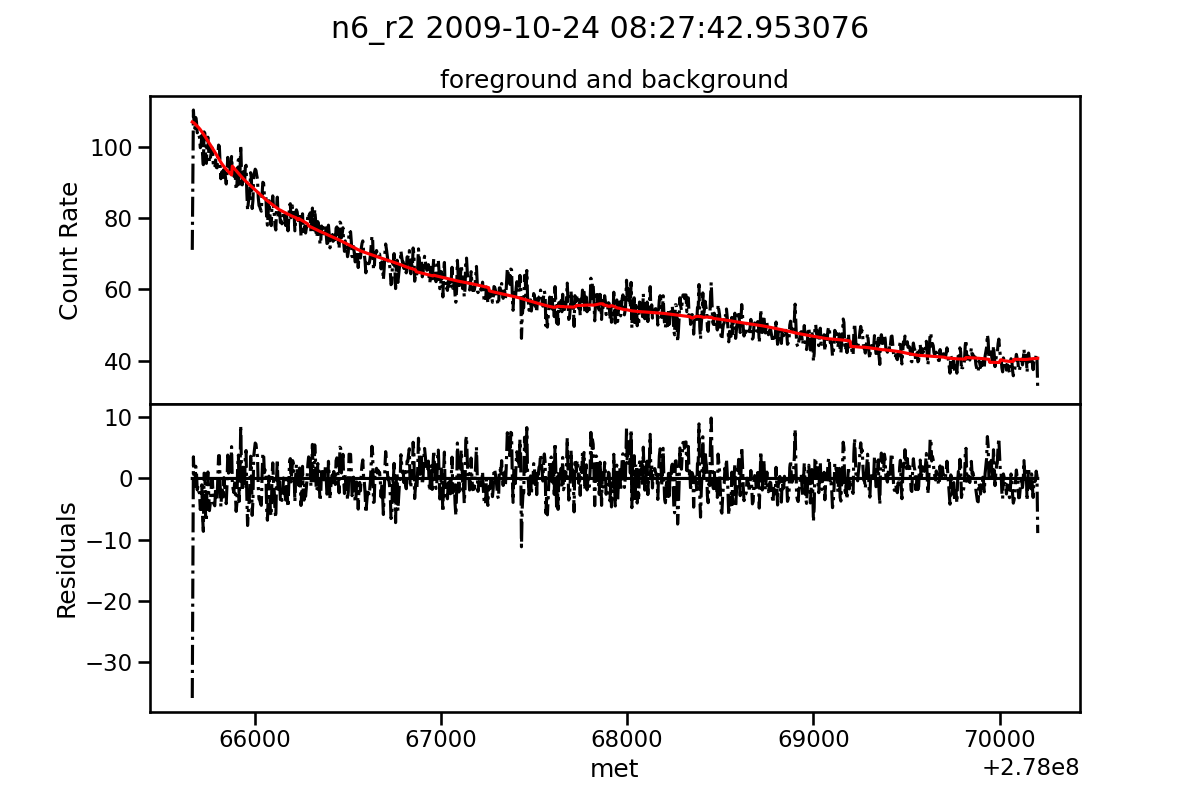}
\end{subfigure}

\bigskip
\bigskip
\hspace*{-2cm} 
\begin{subfigure}{.3\linewidth}
    \centering
    \includegraphics[width=1.5\textwidth]{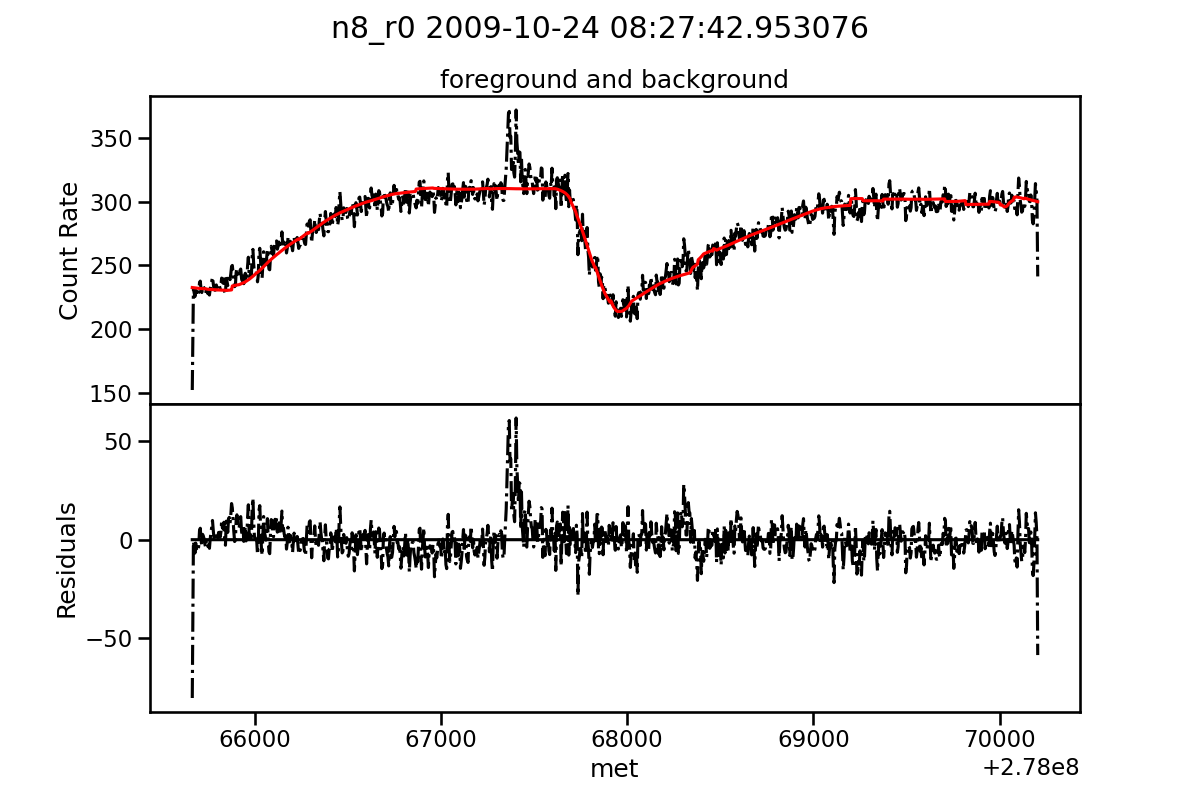}
\end{subfigure}
    \hfill
\begin{subfigure}{.3\linewidth}
    \centering
    \includegraphics[width=1.5\textwidth]{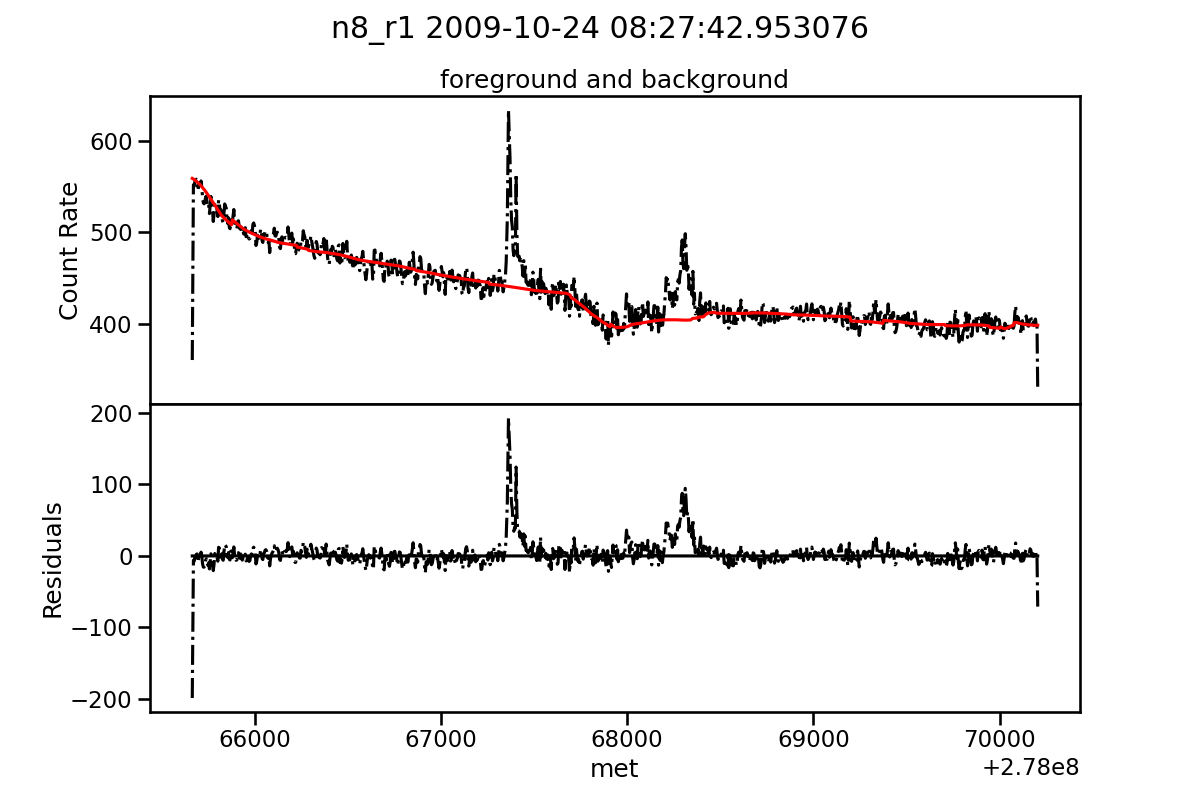}
\end{subfigure}
   \hfill
\begin{subfigure}{.3\linewidth}
    \centering
    \includegraphics[width=1.5\textwidth]{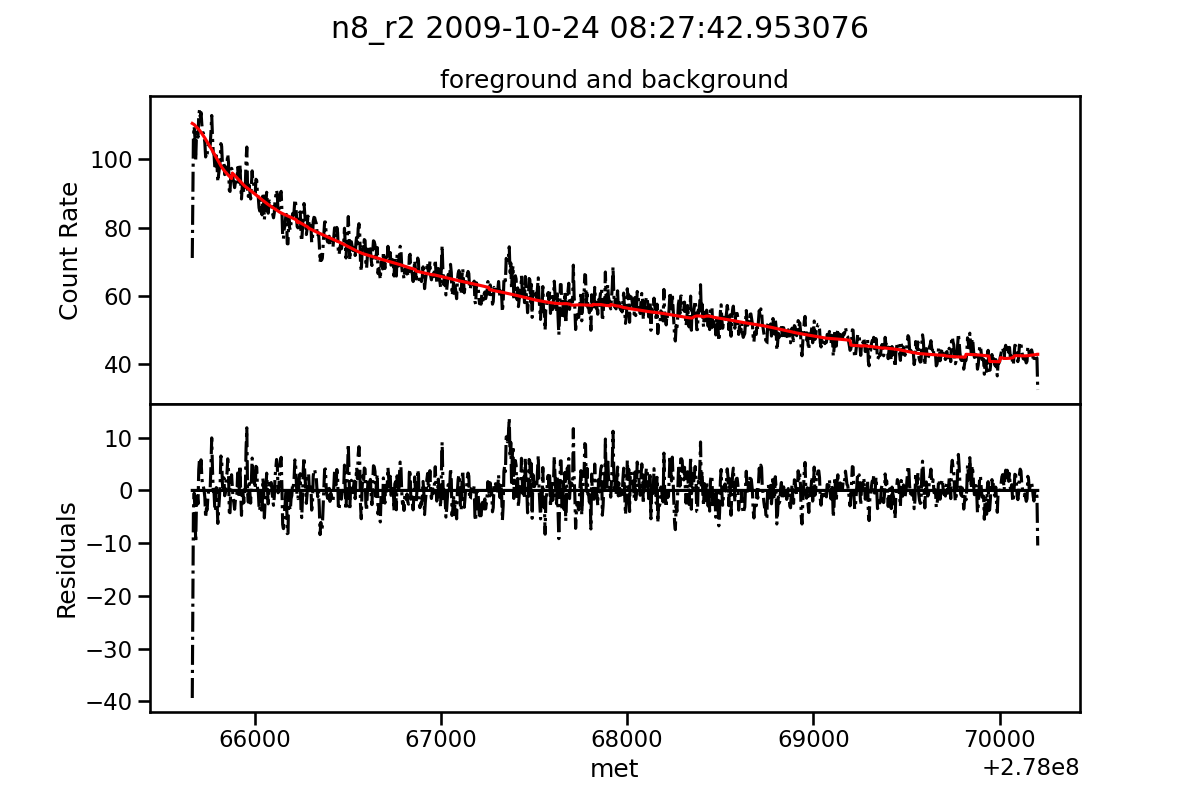}
\end{subfigure}

\caption{Observed and background estimate count rates around the event GRB 091024. From left to right the plots refer respectively to range r0, r1, r2, from top to bottom the plots refer respectively to detectors \texttt{n0}, \texttt{n6}, \texttt{n8}. The firsts two columns can be compared to Figure \ref{fig:blitz_091024}. Source: Crupi et al. \cite{crupi2023searching}.
\label{fig:091024}}
\end{figure}

\begin{figure}
    \centering
    \includegraphics[width=1.\textwidth]{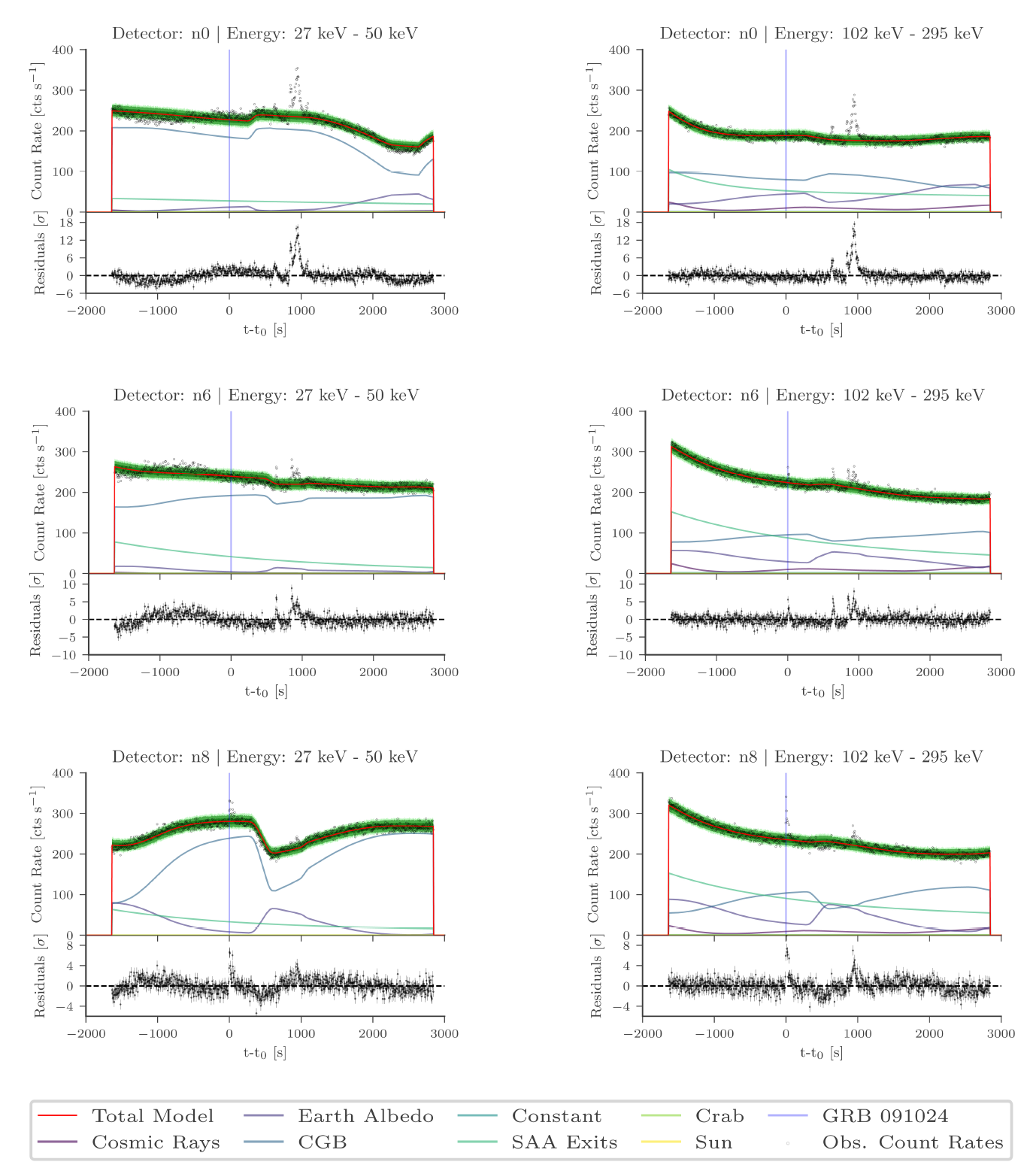}
    \caption{Data and physical background model estimation for the 5000 s around GRB 091024's GBM trigger time (t = 0) for detectors \texttt{n0} (top), \texttt{n6} (middle), \texttt{n8} (bottom), and two reconstructed energy ranges 27-50 keV (left, identical to r0) and 102-295 keV (right, similar to r1). \cite{biltzinger2020physical} \copyright Biltzinger et al. (2020), \href{https://creativecommons.org/licenses/by/4.0}{CC BY 4.0}.}
    \label{fig:blitz_091024}
\end{figure}

In Figure \ref{fig:091024} are shown detectors \texttt{n0}, \texttt{n6} and \texttt{n8} in the three energy band specified in Table~\ref{tab:range}. In Figure \ref{fig:blitz_091024} is depicted an equivalent plot taken from \cite{biltzinger2020physical}.
The data and background estimation of a Neural Network trained and tested during a three-month period, from September 1 to November 30, 2009, are presented in black and red, respectively.
The dataset consists of 1.63 million of samples and the hyperparameters are the same used in Section \ref{sec_background} except for the learning rate
\begin{equation*}
  \eta =
    \begin{cases}
      0.025 & \text{if epoch $<4$}\\
      0.004 & \text{if $4\ge$ epoch $<12$}\\
      0.001 & \text{if epoch $\ge 12$}
    \end{cases}.       
\end{equation*}
The event emerges clearly from the residuals of all the detectors in range r1 and r2, in detector \texttt{n0} and range r2 it is still visible a peak probably belonging to the end of it. In detector \texttt{n6} range r0, in the first part of the time series before the peaks of the event, the background estimation underestimates the foreground (data observed). This could be due to a too short period of training dataset, a non optimal parameter settings of the NN, a different event such as Local Particles or, more interestingly, the first part of the GRB, where photon count rates were too low to be detected due to background variability.

\newpage

\section{Solar minima and maxima}\label{solarmaxmin}
Hermes Pathfinder will be launched in 2024 that is near the next solar maximum forecast in 2025 \cite{space2019solar,biesecker2019solar}. This analysis is interesting because reveals what background is expected and how the NN background estimation performs in the two periods. 
The most sensitive detector for the solar activity is the Sun-facing \texttt{n5} \cite{meegan2009fermi}. In this analysis are considered background binned in a GBM period orbits (about 96m) and 16 GBM period orbits, for range 0, the most sensitive for solar flares, in the year of the last solar minima, 2014, and the local minima, 2020.

\begin{figure}[H]
\hspace*{-2cm} 
\centering
\includegraphics[width=1.25\textwidth]{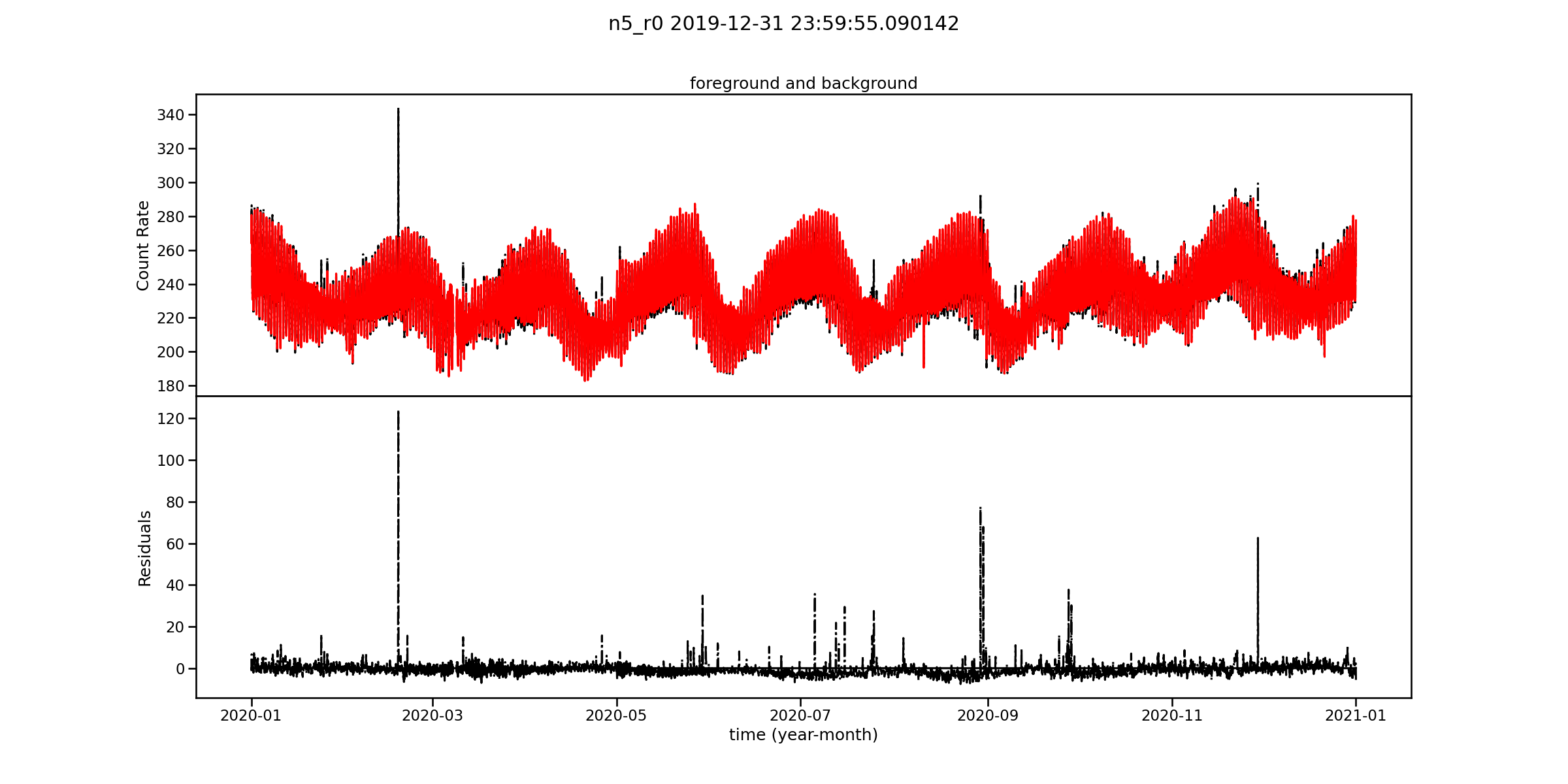}
\caption{\label{fig:solarmin2020_orbit1} The background estimation in year 2020 for detector \texttt{n5} (Sun-facing) in the energy range r0. The count rates are averaged over a bin time corresponding to 1 period orbit (96m). Source: Crupi et al. \cite{crupi2023searching}.}
\end{figure}

\begin{figure}[H]
\hspace*{-1.5cm} 
\centering
\includegraphics[width=1.2\textwidth]{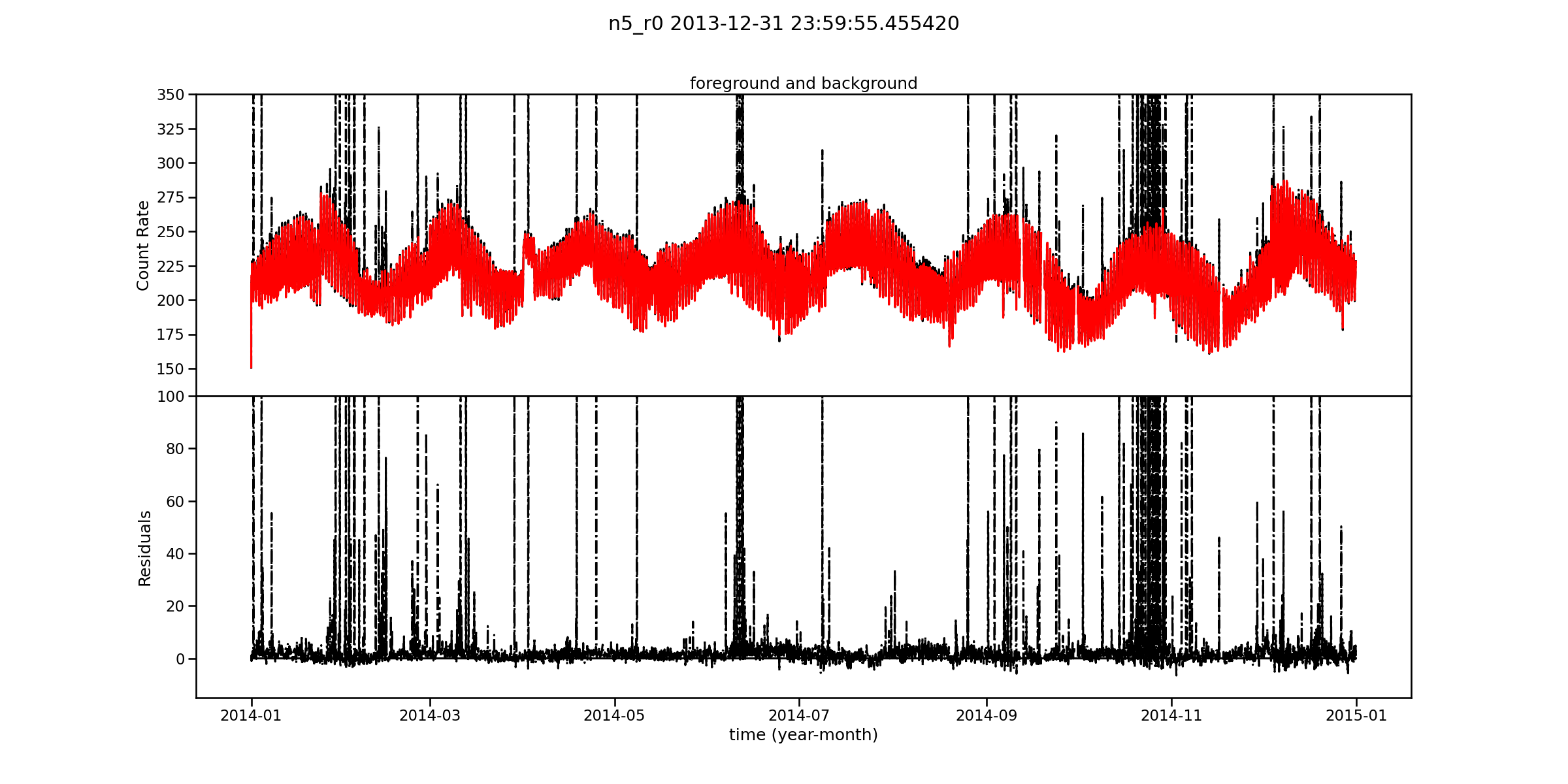}
\caption{\label{fig:solarmaxzoom2014_orbit1} The background estimation in year 2014 for detector \texttt{n5} (Sun-facing) in the energy range r0. The count rates are averaged over a bin time corresponding to 1 period orbit (96m). A zoom-in is applied to avoid the outliers shown in Figure \ref{fig:solarmax2014_orbit1}. Source: Crupi et al. \cite{crupi2023searching}.}
\end{figure}

\begin{figure}[H]
\hspace*{-1.5cm} 
\centering
\includegraphics[width=1.2\textwidth]{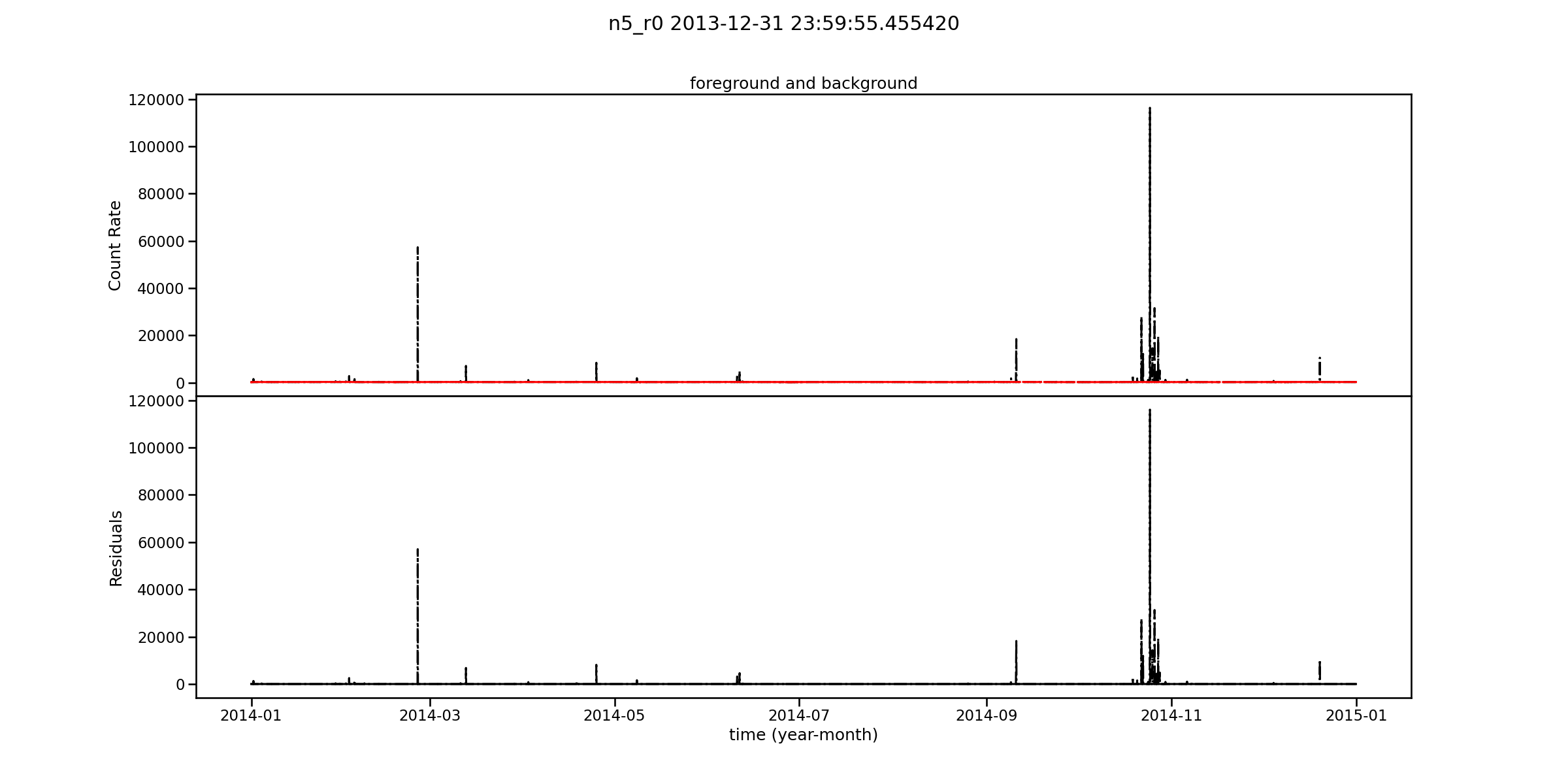}
\caption{\label{fig:solarmax2014_orbit1} The background estimation in year 2014 for detector \texttt{n5} (Sun-facing) in the energy range r0. The solar activity in this year is tremendously high. The count rates are averaged over a bin time corresponding to over 1 period orbit (96m). Source: Crupi et al. \cite{crupi2023searching}.}
\end{figure}

\begin{figure}[H]
\hspace*{-2cm} 
\centering
\includegraphics[width=1.25\textwidth]{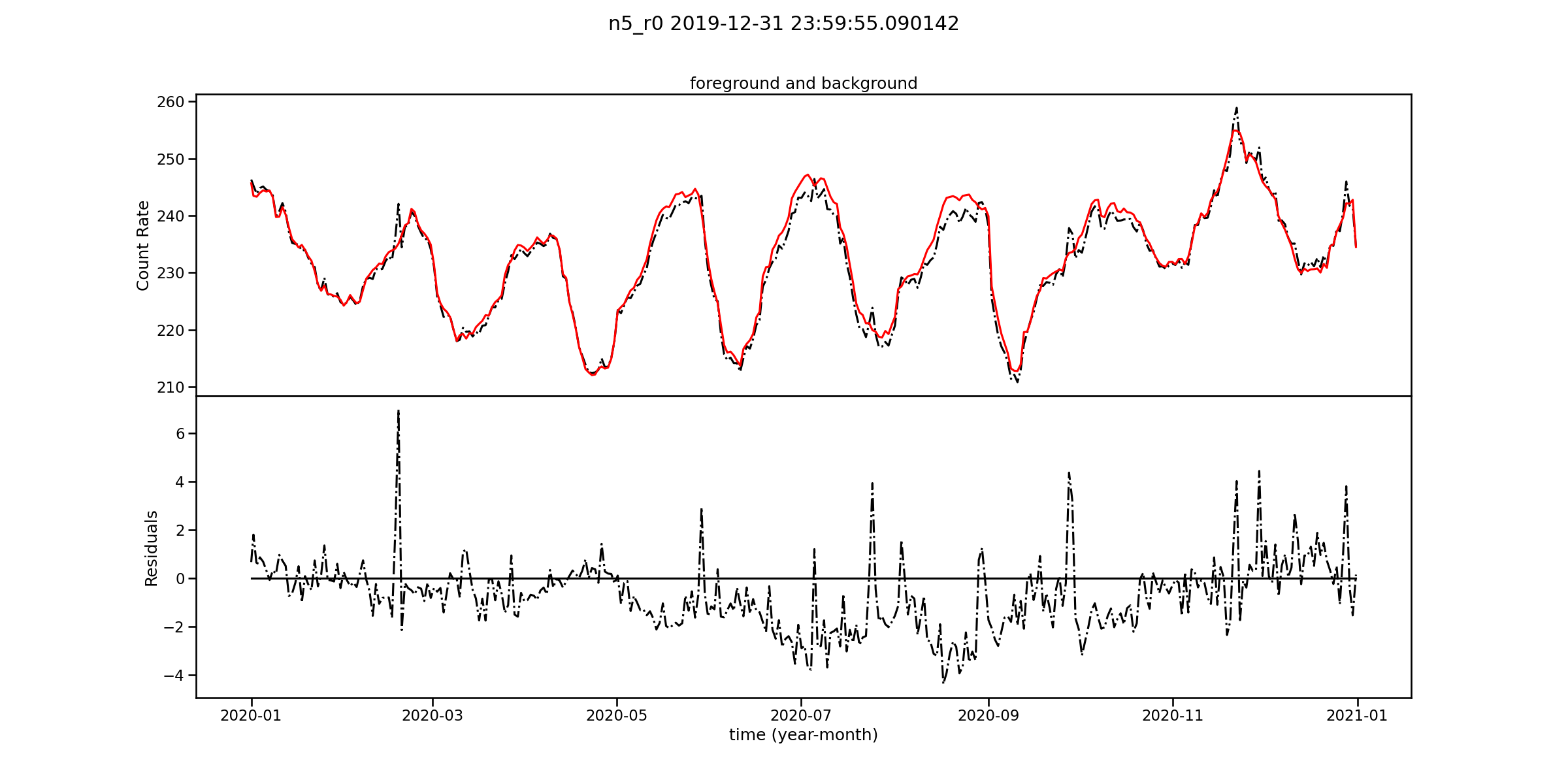}
\caption{\label{fig:solarmin2020_orbit16} The background estimation in year 2020 for detector \texttt{n5} (Sun-facing) in the energy range r0. The count rates are averaged over a bin time corresponding to 16 period orbits ($25.6$h). Source: Crupi et al. \cite{crupi2023searching}.}
\end{figure}

\begin{figure}[H]
\hspace*{-2cm} 
\centering
\includegraphics[width=1.25\textwidth]{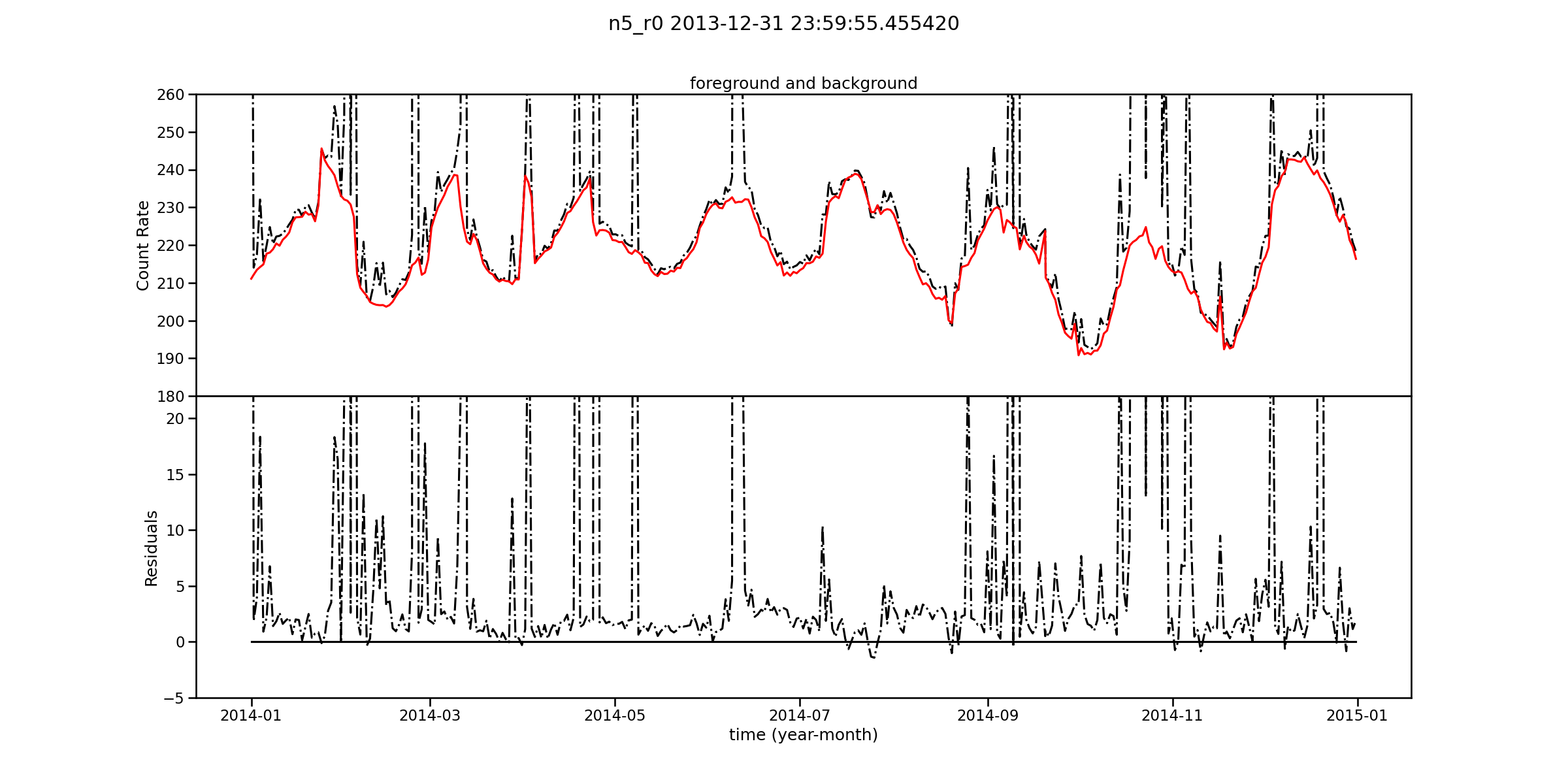}
\caption{\label{fig:solarmaxzoom2014_orbit16} The background estimation in year 2014 for detector \texttt{n5} (Sun-facing) in the energy range r0. The count rates are averaged over a bin time corresponding to 16 period orbits ($25.6$h). Source: Crupi et al. \cite{crupi2023searching}.}
\end{figure}

The Figures \ref{fig:solarmaxzoom2014_orbit1}, \ref{fig:solarmax2014_orbit1}, \ref{fig:solarmaxzoom2014_orbit16} and
\ref{fig:solarmin2020_orbit1}, \ref{fig:solarmin2020_orbit16} are obtained considering respectively years 2014 and 2020, a NN per each year is trained. One orbit time binning for 2020 \ref{fig:solarmin2020_orbit1}, around 240 counts/s, and 2014 \ref{fig:solarmax2014_orbit1} are not comparable due to the high values of the latter but if we zoom the estimated background part, Figure  \ref{fig:solarmaxzoom2014_orbit1}, we see count rates around 225 counts/s. The same reasoning applies for 16 orbits in 2020, Figure \ref{fig:solarmin2020_orbit16}, and 2014, Figure \ref{fig:solarmaxzoom2014_orbit16}. 
In Table \ref{tab:mae_solar} are presented the performance of the background estimation for the year 2014 and 2020.
The solar activity is known to follow a cycle of 11 years \cite{bhowmik2018prediction}. For periods consisting in few months we can assume the solar activity to be constant.

Some reference for the solar cycle prediction can be found in \cite{hathaway1994shape}, \cite{upton2018updated}, \cite{bhowmik2018prediction}. 

\begin{table}[H]
\centering
 \begin{tabular}{||c | p{2cm}p{1.9cm}p{2cm}p{1.9cm} ||} 
 \hline
 year & MAE train (counts/s) & MAE test (counts/s) & MeAE train (counts/s) & MeAE test (counts/s) \\ [0.5ex] 
 \hline\hline
  2014 & 10.601 & 10.467 & 3.940 & 3.944 \\
  2020 & 4.897 & 4.914 & 3.901 & 3.909 \\ [1ex] 
 \hline
 \end{tabular}
 \caption{For each year a neural network is trained and are presented MAE and
MeAE metrics averaged per detector and range. The comparison between the training and test results demonstrates that the neural network does not suffer from overfitting. The higher MAE in 2014 compared to 2020 can be attributed to the effect of solar activity. However, the use of MeAE as the evaluation metric shows comparable results because it is robust against outliers in the data.}\label{tab:mae_solar}
\end{table}

\section{Hyperparameters choice and training phase}\label{sec:hyperparam}
The choice of hyperparameters and settings in Section \ref{sec_background} was the result of a trial and error process to fit the neural networks that performed well over the three mentioned periods in Section \ref{sec_results_nn}. It is important to note that due to the time-consuming nature of testing all possible combinations of hyperparameters, a limited number of combinations were selected based on a sense of practice and intuition. 

The division of data into training and testing sets follows a random approach, with 75\% allocated to training and 25\% to testing. This choice is motivated by the goal of estimating count rates from historical data. However, it's important to note that in the context of online detection, the problem have to be framed into a count rate forecast. In this scenario, the dataset should be split based on time, with training data originating from the past and testing data covering future time periods.

During the development phase, a feature selection process was executed from the 60 input features. The objective was to retain only those features deemed essential and less correlated with one another. These features included Latitude, Longitude, Altitude, Sun equatorial position, Geocenter Earth equatorial position, Approximate McIlwain L value, and Fermi attitude quaternions. However, it was observed that the model's performance on the test set improved when all 60 features were employed. Consequently, the feature selection process was not incorporated into the pipeline.

The architecture was designed to generate 36 outputs, each of which corresponded to a different detector and energy range's count rate. The high correlation among these outputs motivated this choice, and a unified model capable of learning from all detectors' count rates was expected to generalize effectively. Training a separate model for each detector would have required 12 repetitions of the training process, which could have been computationally expensive. However, one challenge with this architecture was that some detector/energy range occasionally displayed significant errors, causing difficulties in neural network convergence. To address this issue, a learning rate decay strategy, as described below, was implemented.

For simplicity, here we report the final configuration settings: the first layers had 2048 neurons, the third layer had 1024 neurons, dropout was set to $0.02$, and the learning rate (denoted as $\eta$) was varied according to Equation \ref{lr_decay}, with $\beta_1 = 0.9$, $\beta_2 = 0.99$, and $\epsilon = 10^{-7}$. The models were trained for 64 epochs with a batch size of 2048, early stopping was applied after 32 epochs, the loss function used was MAE, and events known from the training set were removed.

The choice of a larger number of neurons (2048) for the first layers was made to ensure minimal residual in terms of MAE. Doubling the number of neurons did not lead to a significant increase in performance.

Initially, a constant learning rate of $0.0008$ was chosen. However, it was observed that especially in the first epochs, a higher learning rate was needed to reduce the loss function and avoid getting stuck in a local minimum. In the later steps, a learning rate of $0.0004$ was used to reach convergence and achieve a stable loss function. The values of $\beta_1 = 0.9$, $\beta_2 = 0.99$, and $\epsilon = 10^{-7}$ are similar to the default values for the optimizer of the Neural Networks.

Regarding the number of epochs, it was decided to stop training at 64 epochs because, along with a batch size of 2048, the neural network tended to achieve good and stable performance within this range.

To analyze the effects of dropout, removing events from the training set, and different loss functions, some experiments in the following sections are reported by tweaking these parameters.

\subsection{Dropout}
Dropout, introduced by Srivastava et al. (2014) \cite{JMLR:v15:srivastava14a}, is a technique that randomly deactivates neurons according to a user-defined probability. This regularization technique helps to smooth the loss function and can lead to faster convergence in certain cases. 

\begin{figure}[!htb]
\hspace*{-1cm} 
\centering
\begin{subfigure}{.52\linewidth}
    \centering
    \includegraphics[width=1.\textwidth]{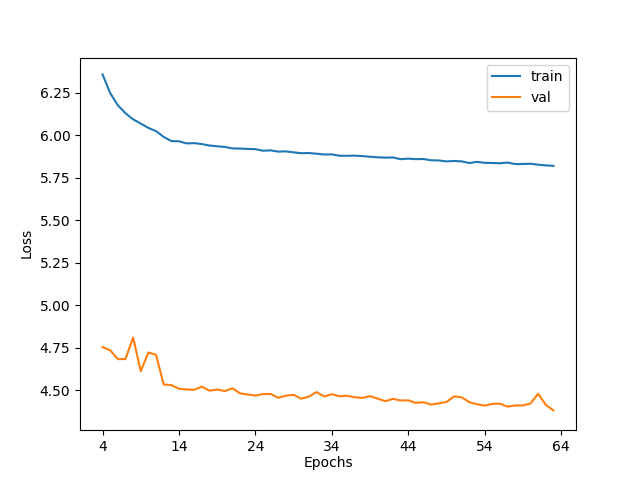}
    \caption{Dropout 0.2}
\end{subfigure}
\begin{subfigure}{.52\linewidth}
    \centering
    \includegraphics[width=1.\textwidth]{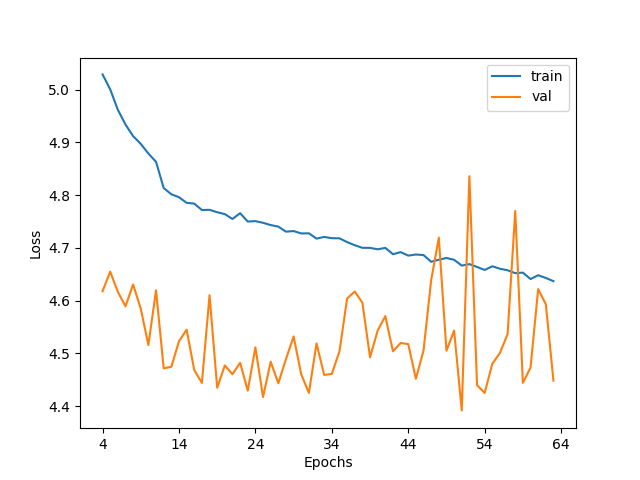}
    \caption{Dropout 0.02}
\end{subfigure}
    \hfill
\begin{subfigure}{.52\linewidth}
    \centering
    \includegraphics[width=1.\textwidth]{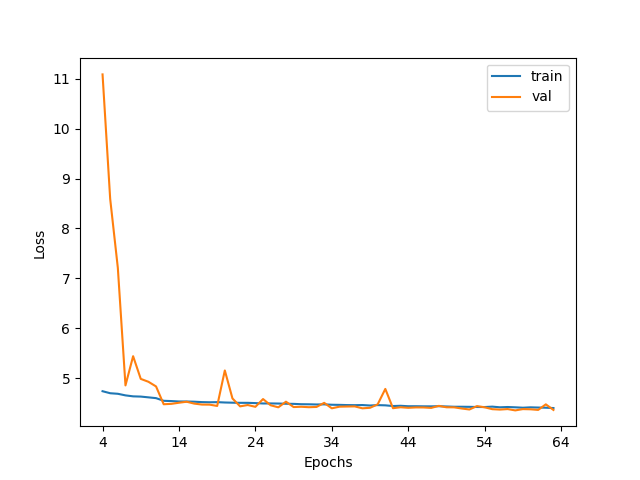}
    \caption{Dropout 0.0002}
\end{subfigure}
    \hfill
\caption{The figures display the MAE values on the training and validation sets as a function of the neural network's epoch for the chosen period of 2019. To improve readability, the figures start from the 4th epoch to avoid the initial high MAE values that would distort the y-axis scale. On the left side of the figures, we can observe the experiments conducted with dropout probabilities of 0.2, 0.02, and 0.0002.
\label{fig:dropout}}
\end{figure}

Figure  \ref{fig:dropout} illustrates the behaviour of the loss function (MAE) with different dropout probabilities: $0.2$, $0.02$, and $0.0002$. In all cases, the neural network converges, but it is noticeable that a lower dropout probability results in a higher initial MAE on the validation set. On the other hand, with a dropout probability of $0.02$, the model demonstrates stable and good performance even as early as the 4th epoch.

The presence of noise in the loss function behaviour during the final epochs may give the impression of a worse neural network. However, at the end of the 64 epochs, the neural network with the best MAE on the validation set across all epochs is chosen. The performance is similar in the two experiments, but it is worth emphasizing that the dropout parameter serves as a regularizer, preventing overfitting and ensuring that the loss function does not become stuck at a high MAE.

\subsection{MAE vs MSE - deletion/inclusion event in catalog}\label{sec:hyperparam_loss}
The choice of the appropriate loss function is indeed crucial in many applications of data science. In this case, we would like to emphasize the properties of MAE and MSE and provide a mathematical justification for their use.

\begin{theorem}\label{th:maemse}

Define a set of samples from a distribution denoted as $Y = {y_1, y_2, \ldots, y_n}$, a constant estimate $\hat{z}$.
To minimise the MSE $\frac{1}{n}\sum_{i=1}^{n}(y_i - \hat{z})^2$, $\hat{z}$ must be the mean $\sum_{i=1}^{n}(y_i)$. 
To minimise the MAE $\frac{1}{n}\sum_{i=1}^{n} \mid y_i - \hat{z} \mid$, $\hat{z}$ must be the median of $Y$.

\paragraph{MSE - mean}

$$\text{MSE}(\hat{z}) = \frac{1}{n}\sum_{i=1}^{n}(y_i - \hat{z})^2$$
$$\frac{d}{d\hat{z}}\text{MSE}(\hat{z}) = \frac{2}{n}\sum_{i=1}^{n}(\hat{z} - y_i)$$

Setting the derivative equal to zero to find the minimum:
$$\frac{2}{n}\sum_{i=1}^{n}(\hat{z} - y_i) = 0 $$
$$\sum_{i=1}^{n}(\hat{z} - y_i) = 0 $$
$$n\hat{z} - \sum_{i=1}^{n}y_i = 0 $$
$$n\hat{z} = \sum_{i=1}^{n}y_i $$
$$\hat{z} = \frac{1}{n}\sum_{i=1}^{n}y_i $$

Thus, the estimate $\hat{z}$ that minimizes the MSE is equal to the mean of the distribution, given by $\frac{1}{n}\sum_{i=1}^{n}y_i$.

\paragraph{MAE - median}

$$\text{MAE}(\hat{z}) = \frac{1}{n}\sum_{i=1}^{n}\mid y_i - \hat{z}\mid $$
$$\frac{d}{d\hat{z}}\text{MAE}(\hat{z}) = \frac{1}{n}\sum_{i=1}^{n}\text{sign}(y_i - \hat{z}) $$

Setting the derivative equal to zero to find the minimum:

$$\frac{1}{n}\sum_{i=1}^{n}\text{sign}(y_i - \hat{z}) = 0 $$
$$\sum_{i=1}^{n}\text{sign}(y_i - \hat{z}) = 0$$

Since the sign function returns $-1$ for negative values, $0$ for zero, and $1$ for positive values, the above equation indicates that the number of positive differences between $y_i$ and $\hat{z}$ is equal to the number of negative differences. In other words, $\hat{z}$ should be a value that divides the samples into two equal-sized groups. The median satisfy this condition because it is the value such that half of the samples are smaller or equal to it, and the other half are larger or equal to it. Therefore, by definition, the median minimizes the MAE. 
\end{theorem}

In the context of our background estimation, the estimated value $\hat{z}$ is not a simple constant because can depend by the previous count rates or other features. It can be demonstrated that a regressor using MSE and MAE approximates the conditional mean $E(Y \mid X = x)$ and the conditional median $\text{median}(Y \mid X = x)$, respectively, see Equations 2.13 and 2.18 in \cite{hastie2009elements} and the intuition is given by Theorem \ref{th:maemse}. Therefore, when employing MAE, the estimator's output behaves similar to the median, making it robust against outliers. Conversely, when using MSE, the output behaves more like the mean, which is not robust against outliers.

In the case where the training set contains significant event count rates, such as solar flares similar to Figure \ref{fig:bigsolarflare}, the estimator should treat these events as outliers since they do not belong to the common background dynamics. Even after normalizing the count rates by subtracting the mean and dividing by the standard deviation, the scaled dataset still retains the same outliers, preserving the proportion among the count rates.

\begin{figure}[H]
\centering
\begin{subfigure}{.62\linewidth}
    \centering
    \includegraphics[width=1.\textwidth]{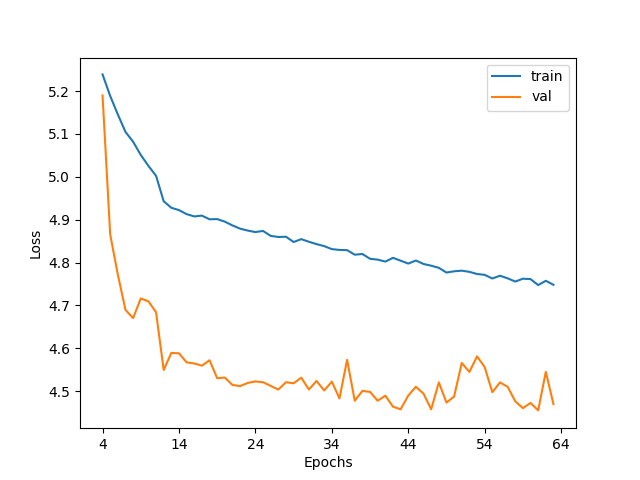}
    \caption{MAE including events in the training set \label{fig:maemse_loss_MAE_events}}
\end{subfigure}
    \hfill
\begin{subfigure}{.62\linewidth}
    \centering
    \includegraphics[width=1.\textwidth]{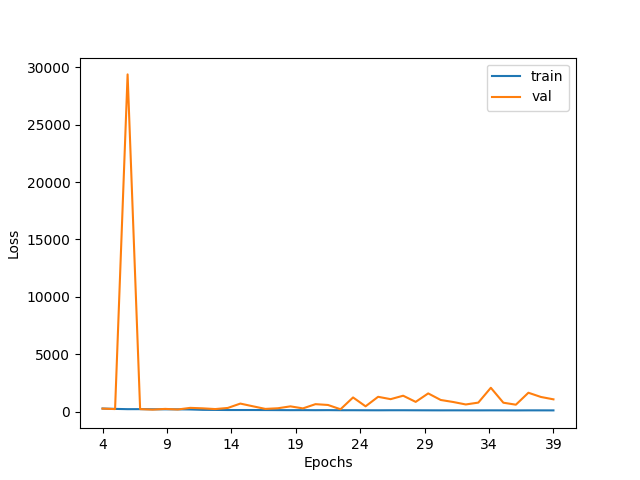}
    \caption{MSE excluding events in the training set \label{fig:maemse_loss_MSE}}
\end{subfigure}
    \hfill
\begin{subfigure}{.62\linewidth}
    \centering
    \includegraphics[width=1.\textwidth]{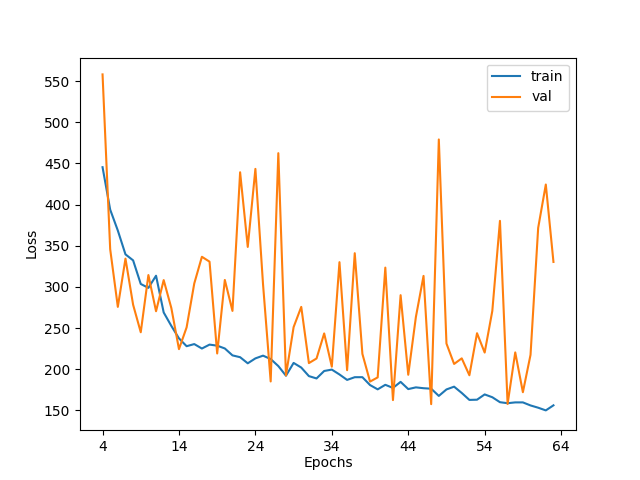}
    \caption{MSE including events in the training set \label{fig:maemse_loss_MSE_eventi}}
\end{subfigure}
\caption{Evaluation of the loss function on both the training and validation sets across different epochs of the neural networks. Source: Crupi et al. \cite{crupi2023searching}.
\label{fig:maemse_loss}}
\end{figure}

\begin{figure}[H]
\centering
\begin{subfigure}{1\linewidth}
    \centering
    \includegraphics[width=.9\textwidth]{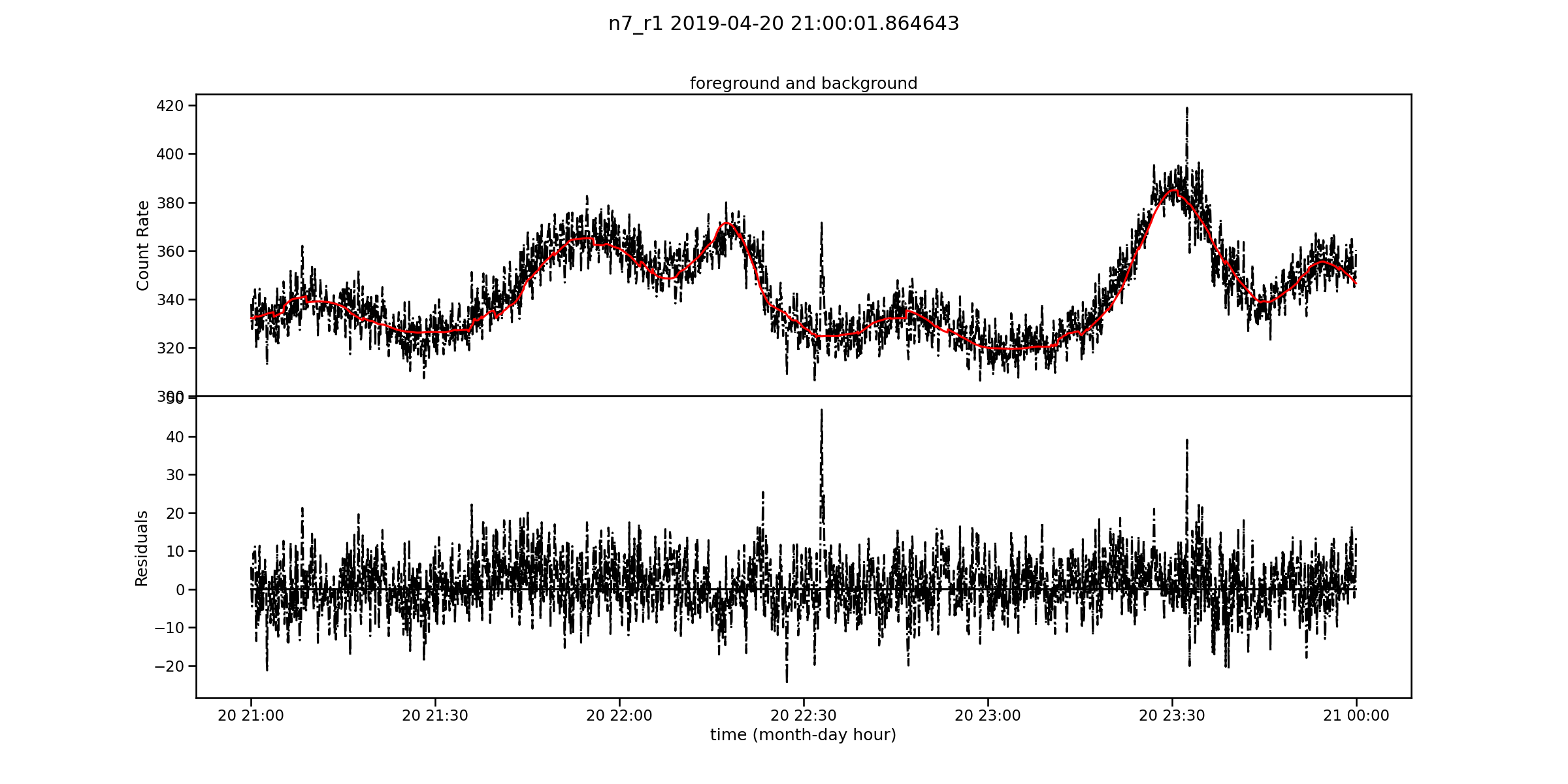}
    \caption{MAE including events in the training set \label{fig:maemse_day_MAE_events}}
\end{subfigure}
    \hfill
\begin{subfigure}{1\linewidth}
    \centering
    \includegraphics[width=.9\textwidth]{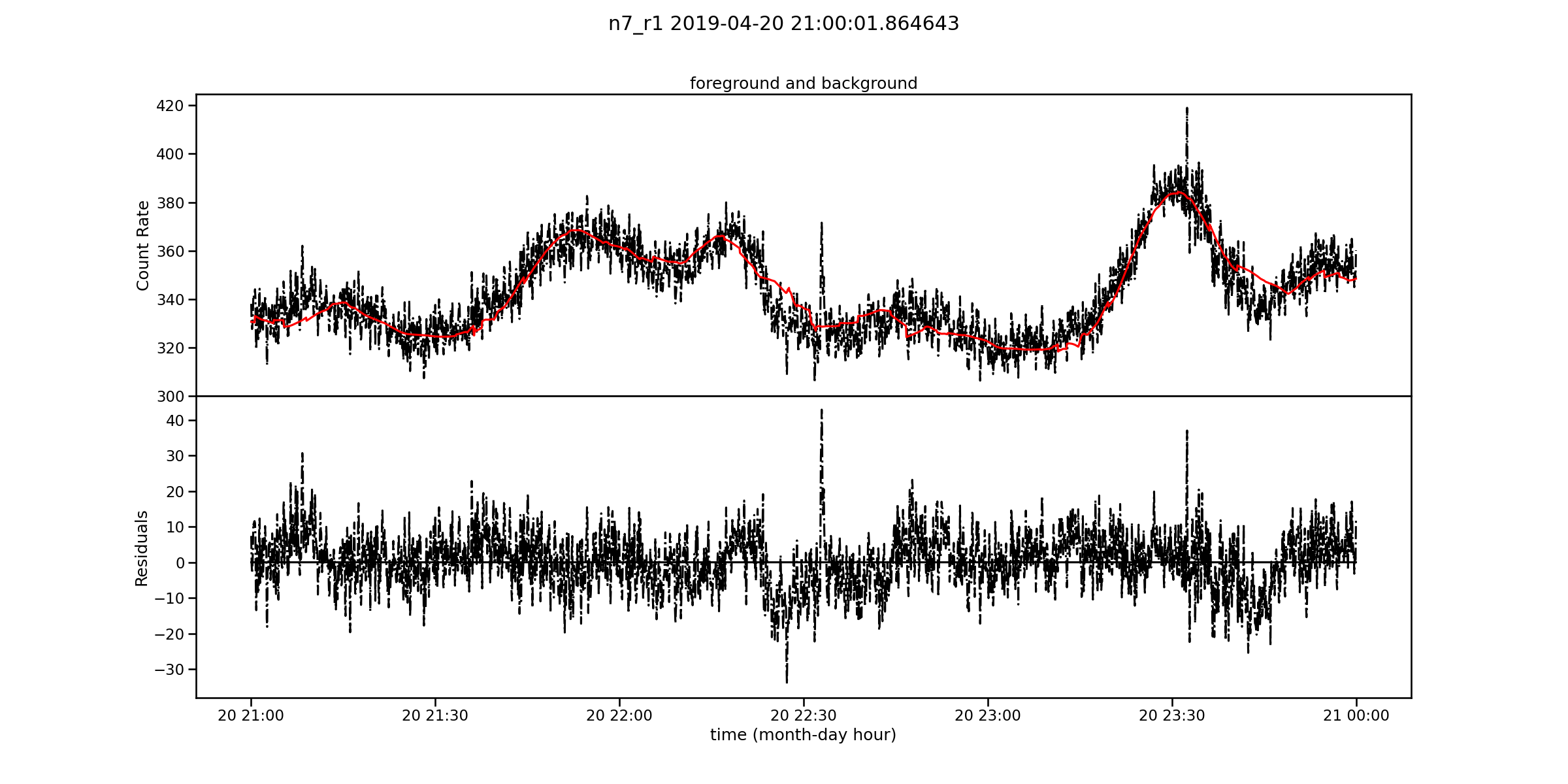}
    \caption{MSE excluding events in the training set \label{fig:maemse_day_MSE}}
\end{subfigure}
    \hfill
\begin{subfigure}{1\linewidth}
    \centering
    \includegraphics[width=.9\textwidth]{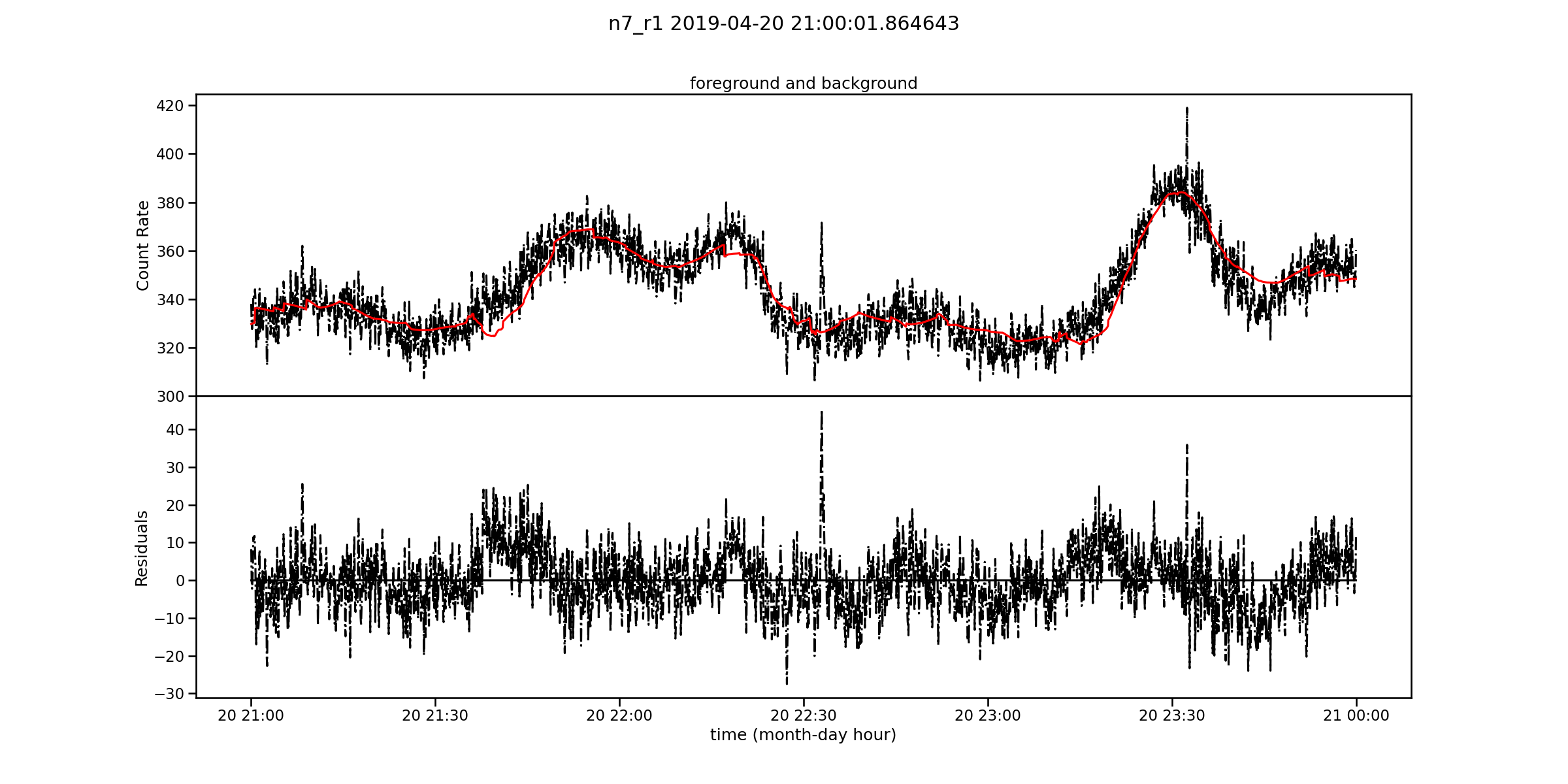}
    \caption{MSE including events in the training set \label{fig:maemse_day_MSE_eventi}}
\end{subfigure}
\caption{An excerpt of background prediction for a Neural Network (best one with loss on the validation set across all epochs) using: (a) MAE without deletion of events from the training set (b) MSE excluding events from the training set (c) MSE without deletion of events from the training set. Source: Crupi et al. \cite{crupi2023searching}.
\label{fig:maemse_day}}
\end{figure}

An empirical example showcasing the robustness of the method can be found in Section \ref{solarmaxmin}. Figure \ref{fig:maemse_loss} presents three training phases: a) MAE with inclusion of events from the Fermi/GBM catalog in the training set, b) MSE with exclusion of events from the Fermi/GBM catalog in the training set, and c) MSE with inclusion of events from the Fermi/GBM catalog in the training set. Only the run with MAE (a) exhibits good convergence, and Figure \ref{fig:maemse_day} showcases an example of prediction using MAE. For MSE (b), the validation loss initially peaks, then stabilizes but continues to increase, ultimately leading to an early stopping procedure (if the validation loss does not improve for 32 epochs, the training is interrupted). In the case of MSE with inclusion of events from the Fermi/GBM catalog (c), the neural network has converged, but the validation loss is higher and noisier than the training loss, indicating poor generalization. Figure \ref{fig:maemse_day} displays MSE predictions over a short period where the approximation of the background dynamics is not as accurate as with MAE.

Based on these plots, it becomes apparent why 64 epochs were sufficient for training the neural network, as the desired convergence and performance were achieved.

\subsection{Quantile regression}

By employing the quantile loss function defined in Equation \ref{eq:quantile_loss}, three neural networks are trained, each targeting a different quantile value: 0.1, 0.5, and 0.9. When $q=0.1$, the network estimates the 10th percentile of the distribution $\mathbb{P}(Y \mid X = x)$. For $q=0.5$, the network provides the median estimation, which is equivalent to minimizing Mean Absolute Error (MAE). Finally, when $q=0.9$, the network estimates the 90th percentile.

\begin{figure}[H]
\centering
\includegraphics[width=1.\textwidth]{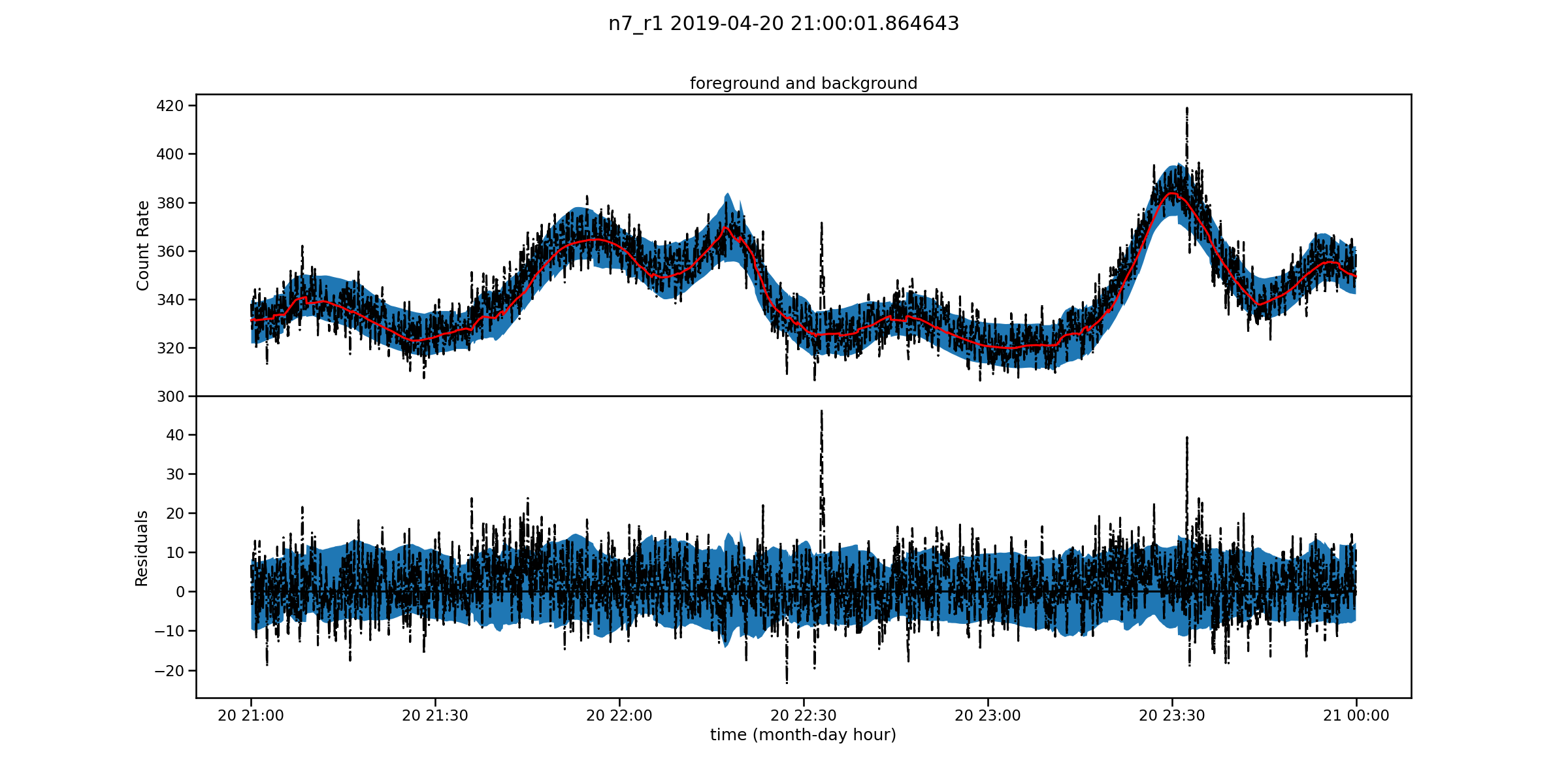}
\caption{Quantile regression for estimating the count rates of detector \texttt{n7} in energy range r1 over a three-hour period on April 20, 2019. The red line represents the $q=0.5$ (median) estimate, while the blue shaded region denotes the range between the upper value at $q=0.9$ and the lower value at $q=0.1$.}
\label{fig:quantile_regr_bkg}
\end{figure}

Quantile regression offers a range of confidence for count rate predictions, as illustrated in Figure \ref{fig:quantile_regr_bkg}. For the trigger algorithm, to enhance robustness in detection, the prediction with $q=0.9$ could serve as an estimate of the true count rates.

\subsection{Data quantity and out-of-time test set evaluation}

Once the model is trained, questions arise regarding 1) the amount of data required for the neural network to converge and achieve good performance. Additionally, it is interesting to understand 2) the network's behaviour when applied to data outside the training period.

To address the first question, thirteen experiments were conducted within the data period spanning 2019 to 2020, characterized by solar minima to minimize the impact of solar X/gamma rays and flares. Table \ref{tab:perf_out_of_time} displays results for training periods of 4, 7, 12, and 24 months, using 75\% of the dataset for training (30\% of which was further reserved as a validation set) and 25\% for testing. To facilitate convergence, a dropout of 0.002 was chosen, and the learning rate for the 24-month dataset was reduced by a factor of 10. The performance metrics, both MAE and MeAE, show no significant improvement with an increase in the dataset size. Performance remains consistently good and comparable beyond a 4-month training period. For reference, one month of data comprises approximately 500,000 data points. 

\begin{figure}[H]
	\centering
	\includegraphics[width=0.75\textwidth]{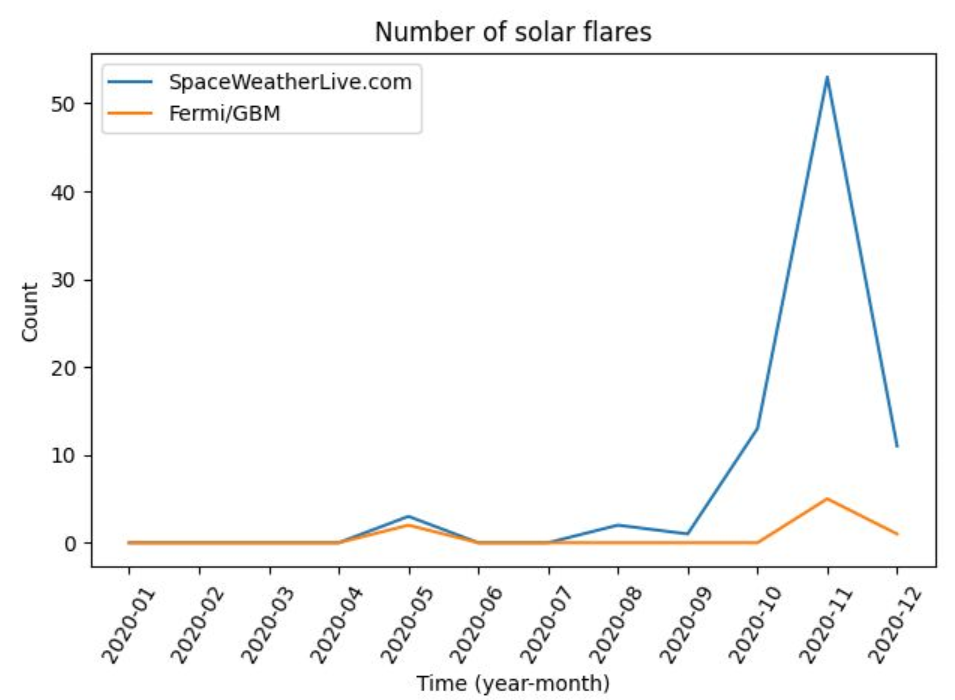}
	\caption{Solar flares counted by GOES-X satellite, source \href{https://www.spaceweatherlive.com/}{spaceweatherlive.com}, and by the Fermi/GBM, source Fermi/GBM trigger catalog.}
	\label{fig:solar_flare_in_2020}
\end{figure}

To address the second question, two group of experiments were conducted on data covering the period of 2019 and 2020. \# 6 to \# 9 reserve a "future" period as a test set with respect the train set,
 while \# 10 to \#13 display the performance of an NN trained on data "in the middle" with respect the train set. 
In particular, the \# 6 experiment reserves as test set July 2020 to January 2021 and \# 10 displays the performance of an NN trained on data excluding January 2020 to July 2020 (6 months or 25\% of the dataset). Training results align with previous experiments, as expected since the data is an extension of the previous dataset. However, applying the model to an out-of-time period results in slightly poorer performance compared to the in-time sets, with an MAE increase of approximately 2.5 counts/s and a MeAE increase of 1.0 counts/s. This indicates that while the NN did not perfectly generalize to a period outside the training set, it still yielded satisfactory results. \# 7, \# 8, \# 9 experiments are run over an increasing size of the train set but tested on the same 2 months data, Nov20 - Jan21. We can see the MAE even higher than the \# 6 and a closer examination of the MAE reveals a more pronounced impact on the r0 range, which is most susceptible to solar activity. This suggests that including a temporal feature to account for the varying effects of solar activity on the detectors would be beneficial over longer time periods. 
The number of solar flares, thus a proxy for solar activity, can be seen in Figure \ref{fig:solar_flare_in_2020}.
Finally, \# 11, \# 12, \# 13 experiments are the equivalent of the previous experiments but tested on a test set "in the middle", Jan20 - Mar20, which shows slightly better performance on the test set and not influenced by the train set size.

It is important to note that the actual NN's purpose is not to predict the future; thus, changes in architecture, such as LSTM, may be necessary for such predictions. The NN serves as a tool to interpolate expected count rates for chosen periods, demonstrating robustness against unexpected increases in count rates, such as those caused by solar flares or GRBs.

\begin{table}[H]
	\centering
	\hspace*{-2cm}
	\begin{tabular}{|p{1cm}|p{1.5cm} p{2.5cm} p{2.5cm}| p{1.5cm}p{1.5cm}p{1.6cm}p{1.6cm} ||} 
		\hline
		Exp \# & Total months & Train period & Test period & MAE train (counts/s) & MAE test (counts/s) & MeAE train (counts/s) & MeAE test (counts/s) \\ [0.5ex] 
		\hline\hline
		1 & 4 & Mar19 - Jul19 & in-time & 4.383 & 4.421 & 3.361 & 3.370 \\
		2 & 7 & Jan19 - Jul19 & in-time & 4.280 & 4.306 & 3.296 & 3.310 \\
		3 & 12 & Jan19 - Jan20 & in-time & 4.446 & 4.442 & 3.353 & 3.361 \\
		4 & 12 & Jan20 - Jan21 & in-time & 4.202 & 4.222 & 3.314 & 3.327 \\
		5 & 24 & Jan19 - Jan21 & in-time & 4.445 & 4.449 & 3.421 & 3.423 \\
		\hline
		6 & (13+5)+6 & Jan19 - Jul20 & Jul20 - Jan21 & 4.296 & 6.684 & 3.335 & 4.532 \\
		7 & (16+2)+2 & Jan19 - Jul20 & Nov20 - Jan21 & 4.888 & 7.518 & 3.744 & 4.543 \\
		8 & (18+2)+2 & Jan19 - Aug20 & Nov20 - Jan21 & 4.464 & 7.844 & 3.403 & 4.790 \\
		9 & (20+2)+2 & Jan19 - Nov20 & Nov20 - Jan21 & 4.256 & 7.790 & 3.321 & 4.681 \\
		\hline
		10 & (13+5)+6 & \text{Jan19 - Jan20} and \text{Jul20 - Jan21} & Jan20 - Jul20 & 4.390 & 6.980 & 3.365 & 4.626 \\
		11 & (16+2)+2 & \text{Jan19 - Jan20} and \text{May20 - Jan21} & Jan20 - Mar20 & 4.536 & 7.105 & 3.418 & 4.724 \\
		12 & (18+2)+2 & \text{Jan19 - Jan20} and \text{May20 - Jan21} & Jan20 - Mar20 & 4.787 & 6.667 & 3.631 & 4.778 \\
		13 & (20+2)+2 & \text{Jan19 - Jan20} and \text{Mar20 - Jan21} & Jan20 - Mar20 & 4.910 & 6.609 &  3.707 & 4.915 
		 \\ [1ex] 
		\hline
	\end{tabular}
	\caption{For each period, a neural network is trained, and MAE and MeAE metrics are presented, averaged per detector and range. For some experiments the total months od the dataset are described as: (train months + validation months) + test months. The first five rows compare the training and test results, demonstrating that the neural network does not suffer from overfitting when the test set is randomly chosen in the same period as the training set (in-time). However, from \# 6 to \# 13 experiments show that the NN did not perfectly generalize the background estimation when applied to a period not contained in the training set (out-of-time), but still yielded satisfactory results.}\label{tab:perf_out_of_time}
\end{table}

\section{XAI application}\label{sec:xai_nn}

In this section, XAI techniques such as Morris sensitivity analysis and Kernel SHAP are used to better understand and improve the NN. 

\subsection{Global}

\begin{figure}[H]
\centering
\hspace*{-1cm}   
\includegraphics[width=1.2\textwidth]{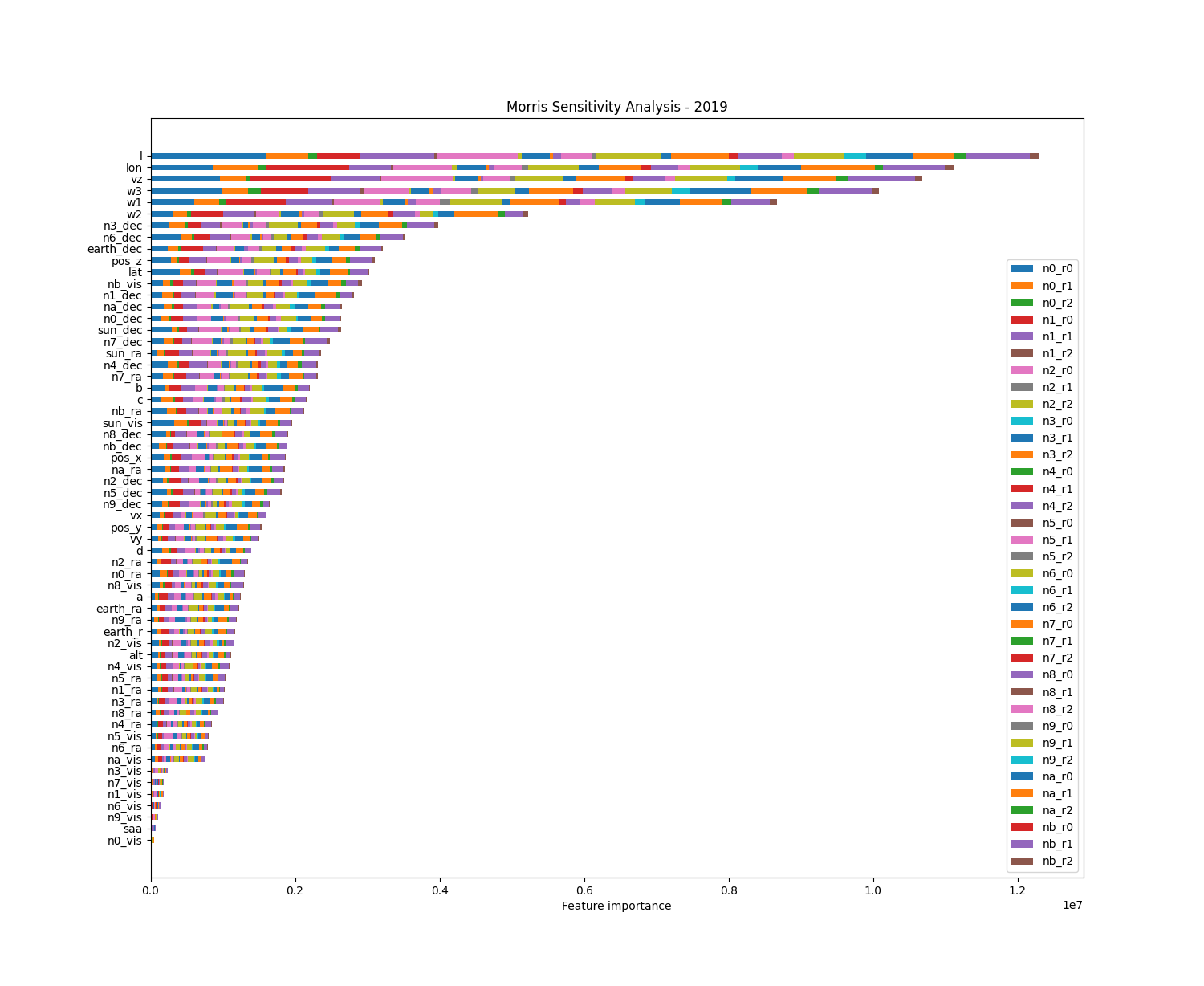}
\caption{Feature importance as determined by the Morris method for the 2019 dataset. Features are ranked by average importance, with the McIlwain L-parameter and longitude coordinates at the top. Conversely, certain detector flags indicating Earth's FoV occultation emerge as almost irrelevant factors in prediction. For a complete list of features and their descriptions, refer to Tables \ref{tab:feature} and \ref{tab:feature_det}.}
\label{fig:Morris_2019}
\end{figure}

\begin{figure}[H]
\centering
\hspace*{-1cm}   
\includegraphics[width=1.2\textwidth]{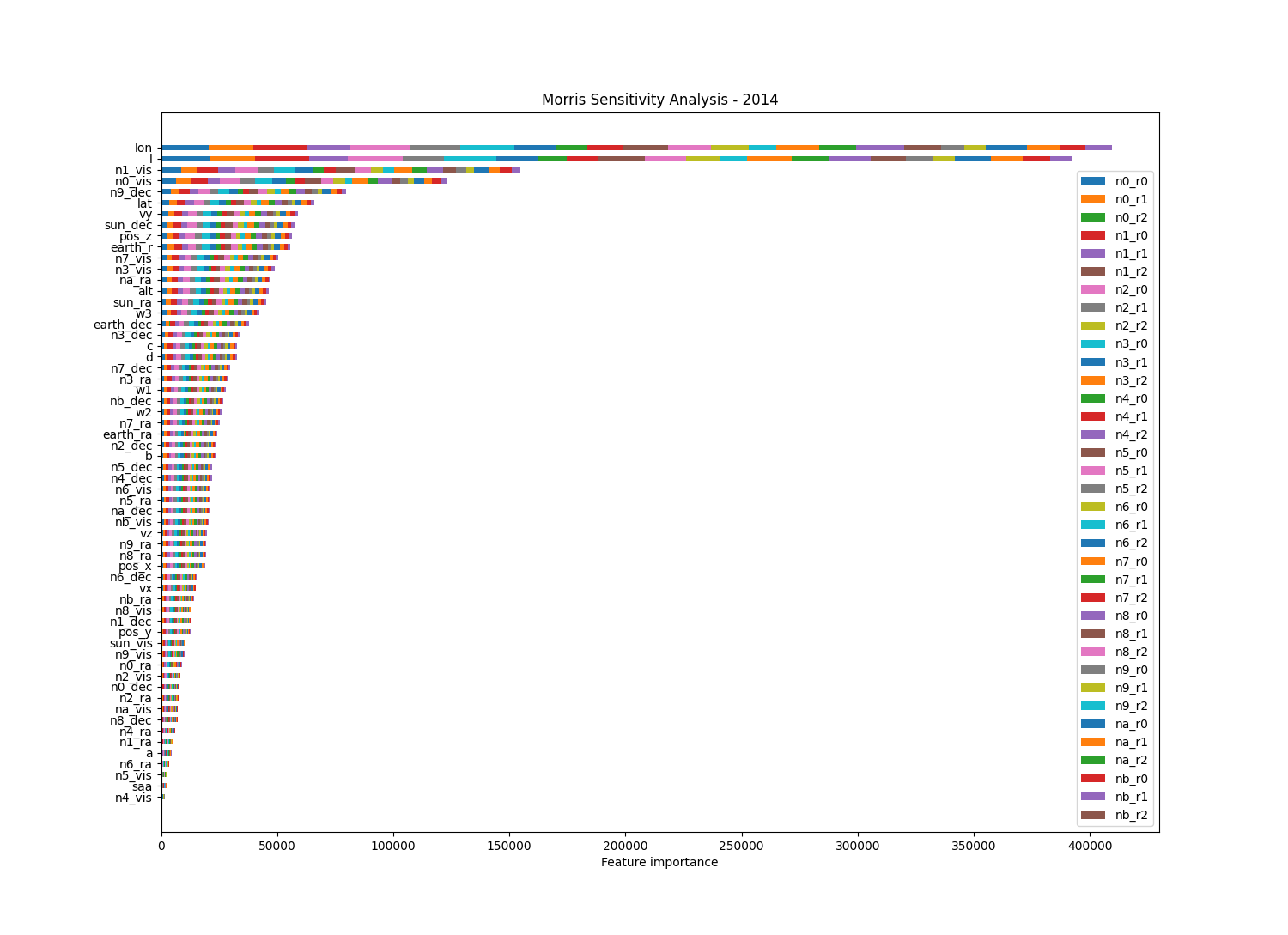}
\caption{Feature importance as determined by the Morris method for the 2014 dataset. Features are sorted by their average importance, with the McIlwain L-parameter and longitude coordinates as the most influential. In contrast to Figure \ref{fig:Morris_2019}, detector flags representing Earth's FoV occultation play a significant role in prediction. For a complete list of features and their descriptions, refer to Tables \ref{tab:feature} and \ref{tab:feature_det}.}
\label{fig:Morris_2014}
\end{figure}

To identify the most influential features and their impact on predictions, a Morris sensitivity analysis (refer to Section \ref{sec:xai}) is performed using the InterpretML library \cite{nori2019interpretml}. The choice for this method was due to its computational efficiency, as it perturbs one feature at a time. However, it doesn't consider non-linearities when estimating output, which it is addressed later in local explanations for individual data points.
Since the NN produces 36 outputs, feature importance is sorted based on the average importance. Figures \ref{fig:Morris_2019} and \ref{fig:Morris_2014} illustrate this for each detector/range combination. We applied this analysis to NNs trained on data from the 2019 and 2014 periods.
Figure \ref{fig:Morris_2014} reveals the significance of Earth's occultation of detectors, potentially due to high solar activity, which it is explored further in an example of the local analysis.

\subsection{Local}

\begin{figure}[H]
\centering
\includegraphics[width=.75\textwidth]{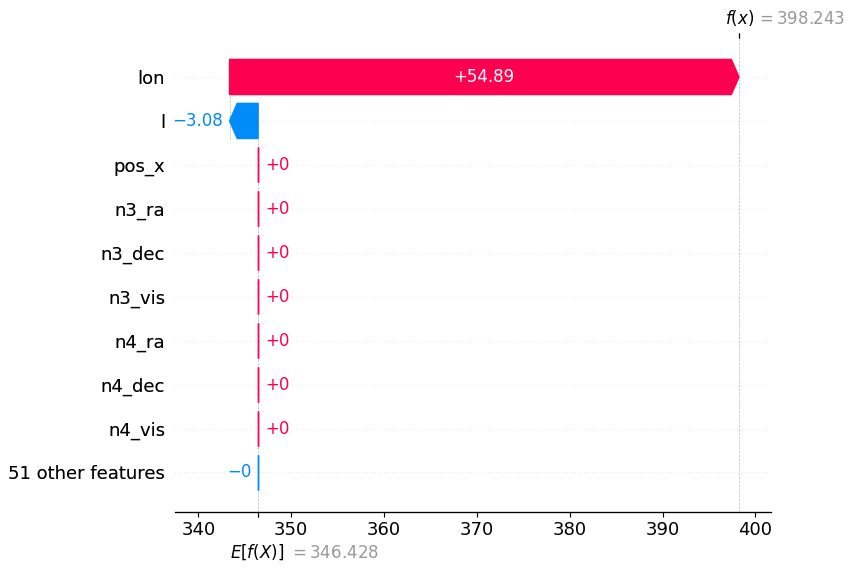}
\caption{Kernel SHAP feature importance for the peak datapoint. The spacecraft's geographical longitude is the most critical feature in determining the sudden count rate increase from a baseline count rate of 346 to 398 (the peak). This baseline is constructed using 15 data points before and after the peak, which are used for sampling values and perturbing the data point to provide an explanation.}
\label{fig:SHAP_expl_n0_r1}
\end{figure}

\begin{figure}[H]
\centering
\includegraphics[width=.85\textwidth]{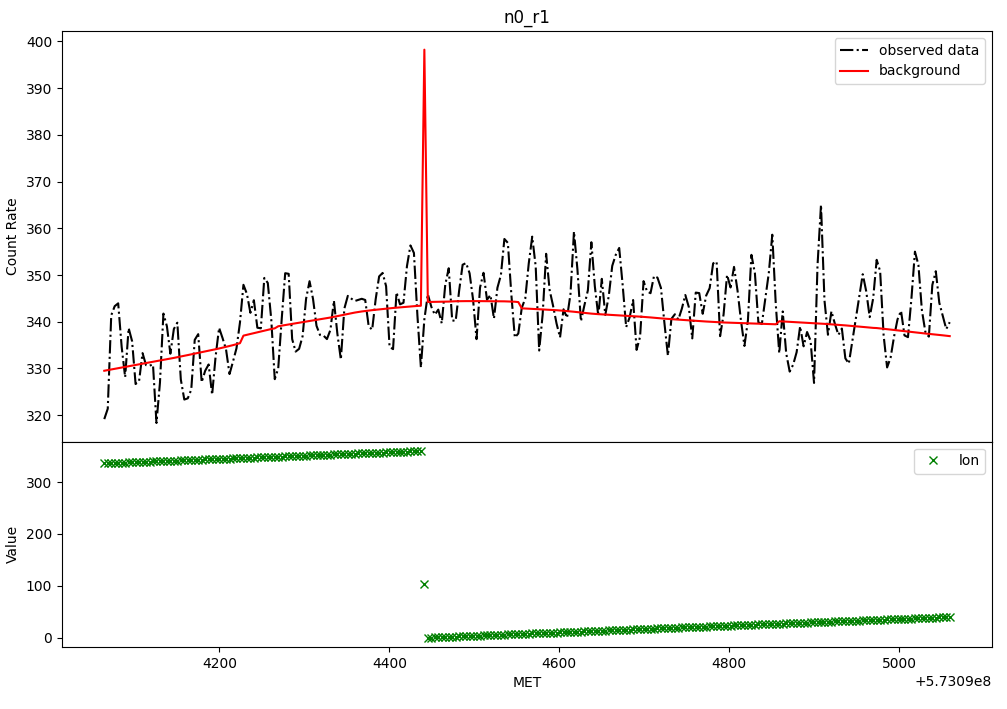}
\caption{Lightcurve of observed and estimated count rates for detector \texttt{n0} in energy range r1. An unexpected peak appers in the estimated count rates due to an error in the dataset for the longitude estimation.}
\label{fig:error_n0_r1_2019_lon}
\end{figure}

Local explanations focus on individual data points, and for this purpose, Kernel SHAP has been utilized. It allows for the generation of explanations that cover the entire NN, providing a feature importance list for each of the 36 outputs, aiding in debugging unusual predictions.
For instance, Figure \ref{fig:error_n0_r1_2019_lon} showcases an unexpected peak for detector \texttt{n8} in energy range r1. Kernel SHAP in Figure \ref{fig:SHAP_expl_n0_r1} highlights the importance of the longitude feature which is expressed in degrees, it gradually increases to $360^\circ$ and then returns to $0^\circ$ as expected. However, due to a dataset issue (an interpolation error in longitude estimation), the value jumps to around $100^\circ$, invalidating the result. The solution involves rectifying the dataset, as the NN strives to provide the best estimation possible. Nevertheless, this behaviour has minimal impact on the trigger algorithm's application.

\begin{figure}[!hbt]
	\centering
	\includegraphics[width=.75\textwidth]{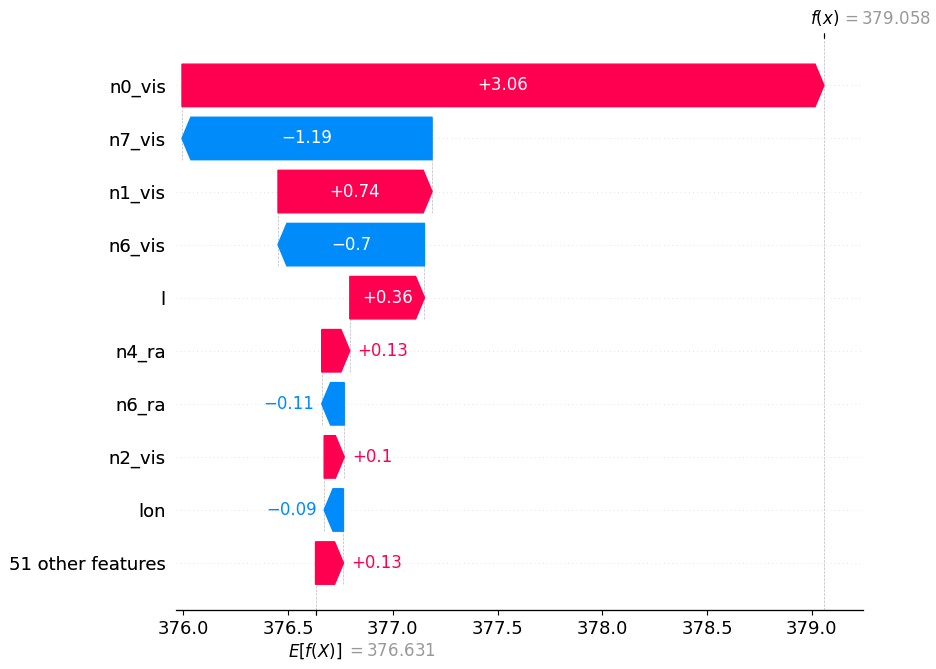}
	\caption{Kernel SHAP feature importance for a datapoint in a step of the estimated background in Figure \ref{fig:Bug_n8_r1_vis_2014}. Notably, the most influential features are those related to detector visibility, specifically whether the FoV is obscured by Earth or not. The baseline for this method is established by using 15 datapoints before and after the datapoint to be explained. These datapoints are used for value sampling and perturbation of the datapoint during the explanation process.}
	\label{fig:error_step_SHAP_n8_r1}
\end{figure}

\begin{figure}[!hbt]
	\centering
	\includegraphics[width=.85\textwidth]{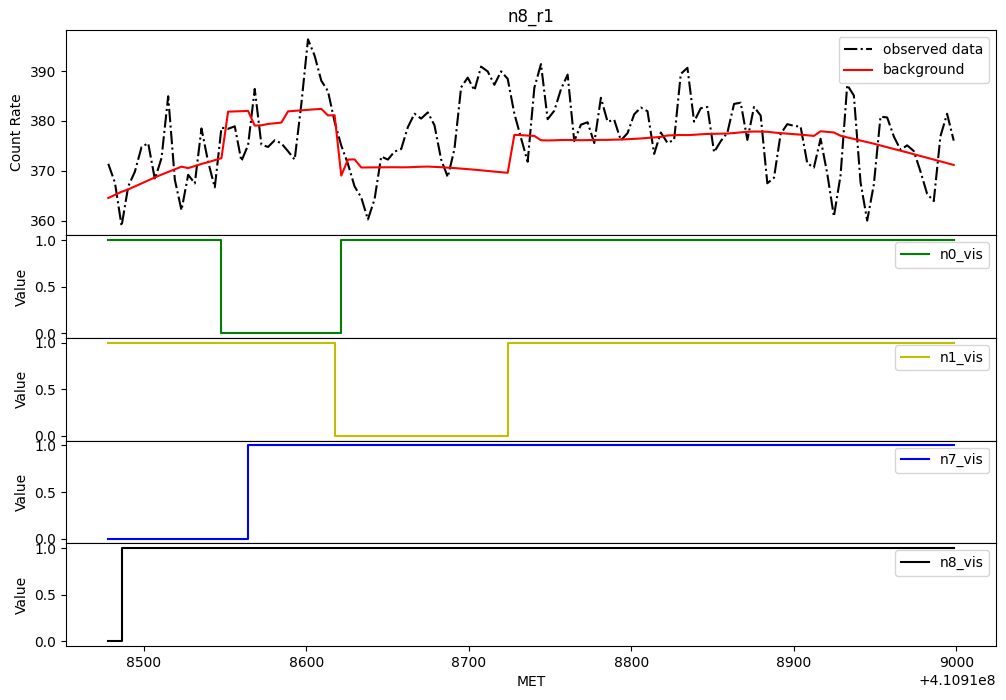}
	\caption{Observed and estimated count rate lightcurve for detector \texttt{n8} in energy range r1. The unexpected steps in estimated count rates are attributed to variations in features such as n0\_vis, n1\_vis, and n7\_vis, with n8\_vis remaining constant during this unusual estimation period.}
	\label{fig:Bug_n8_r1_vis_2014}
\end{figure}

Another significant error type, more severe due to its contribution to False Positive detections during the 2014 period (discussed in the subsequent chapter), involves sudden steps in the background estimation. Figure \ref{fig:Bug_n8_r1_vis_2014} illustrates this behaviour for detector \texttt{n8} in energy range r1. In the SHAP analysis (refer to Figure \ref{fig:error_step_SHAP_n8_r1}), the most influential features are related to a flag feature: 1 indicates the detector is visible, while 0 signifies that its FoV is obscured by the Earth.
Interestingly, what stands out is that detector \texttt{n8}'s behaviour seems influenced by the visibility of other detectors. This peculiar pattern cannot be attributed to data errors, as their correctness has been verified, but rather to a learned pattern within the neural network. Figure \ref{fig:Bug_n8_r1_vis_2014} presents the values of these features, demonstrating a clear correlation between the step behaviour and changes in these flags.

A potential simple solution in this scenario could involve omitting these features as inputs and evaluating whether model performance remains consistent and this behaviour subsides. However, it's worth noting that these features might serve a purpose in correcting for solar activity effects. This contrasts with the 2019 period, during which FoV occultation were deemed irrelevant in background estimation (see Figure \ref{fig:Morris_2019}). Moreover, it is noted that the impact of occultation on detectors facing the Sun (from \texttt{n0} to \texttt{n5}), cause a decrease in estimated count rates. In contrast, detectors from \texttt{n6} to \texttt{nb} exhibit behaviour similar to that depicted in Figure \ref{fig:Bug_n8_r1_vis_2014}. 

Another potential solution could be integrating these features in the last layer only if it associated to the detector (e.g., \text{n8}\_r1 count rate and n8\_vis).

\section{Background for GRB simulation}

In this last section of this chapter, we explore how the background estimation can be integrated into a simulation framework designed for modeling GRB events. For an in-depth understanding and detailed implementation, refer to the repository available at GitHub Repository \url{https://github.com/peppedilillo/synthburst}.
The primary objective of this simulation framework is to generate comprehensive lightcurves that faithfully capture the behaviour of GRBs under various background scenarios.

\begin{figure}[!hbt]
\centering
\includegraphics[width=1.\textwidth]{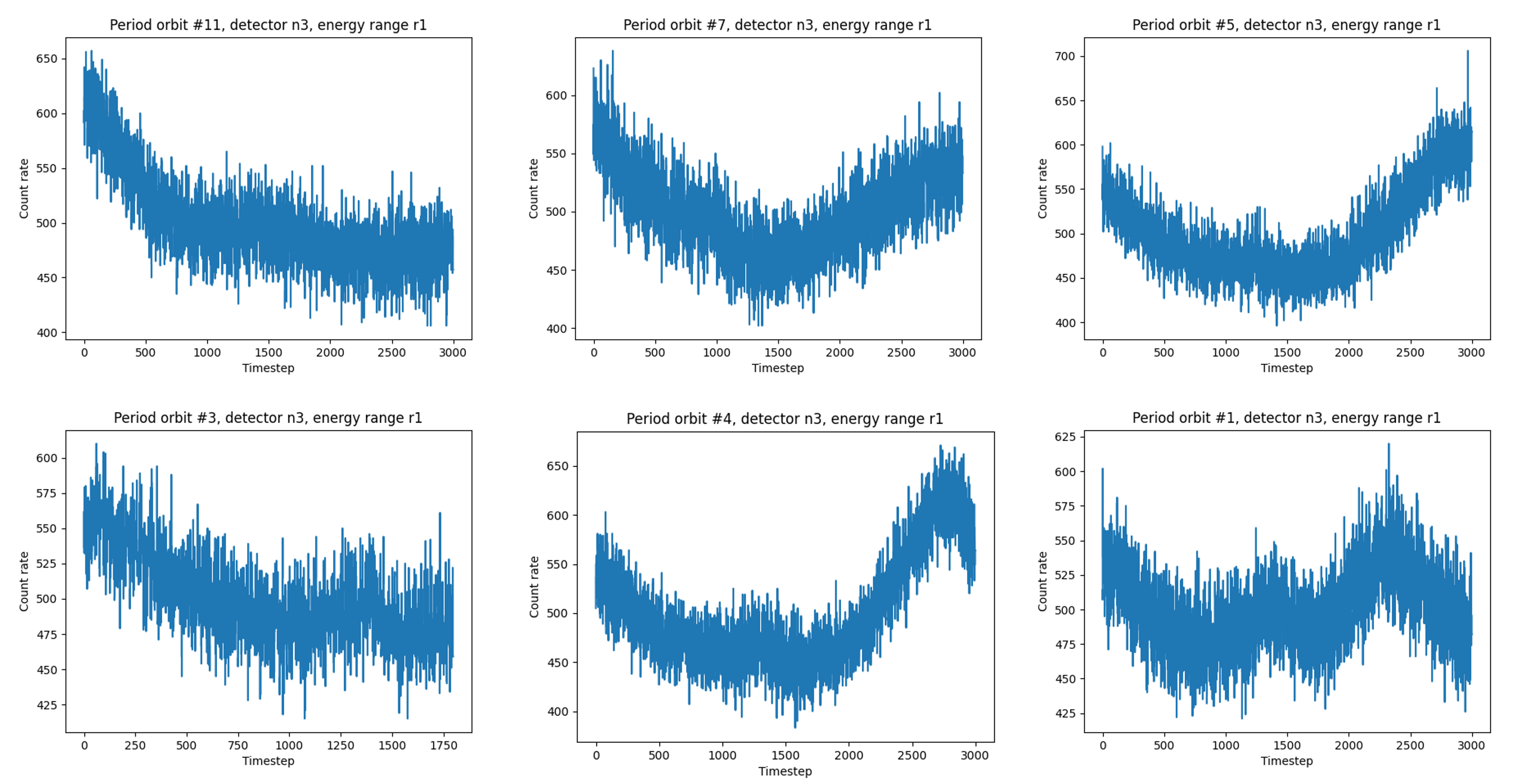}
\caption{Background templates for different parts of the orbit, corresponding to detector \texttt{n3} and energy range r1, estimated from data on 19 April 2019.}
\label{fig:bkg_templates}
\end{figure}

The framework introduces the concept of "background templates" which serve as representative models for background radiation. These templates are characterized by specific attributes, including the reference for a part of the satellite's orbit, the detector used, and the energy range considered. By varying these characteristics, a range of different environmental conditions can be simulated. These templates are then utilized as inputs to a trigger algorithm to assess its robustness under diverse background conditions. It's worth noting that these templates have a resolution of 4 s, as previously explained. However, users also have the flexibility to sample an Exponential distribution within their chosen bin size. This sampling enables the generation of photon arrival times, effectively creating an equivalent TTE object. Figure \ref{fig:bkg_templates} provides examples of background templates, each corresponding to different parts of the satellite's orbit and specific detector and energy range settings. These templates are drawn from data collected during the 2019 observation period.

\begin{figure}[!hbt]
\centering
\includegraphics[width=1.\textwidth]{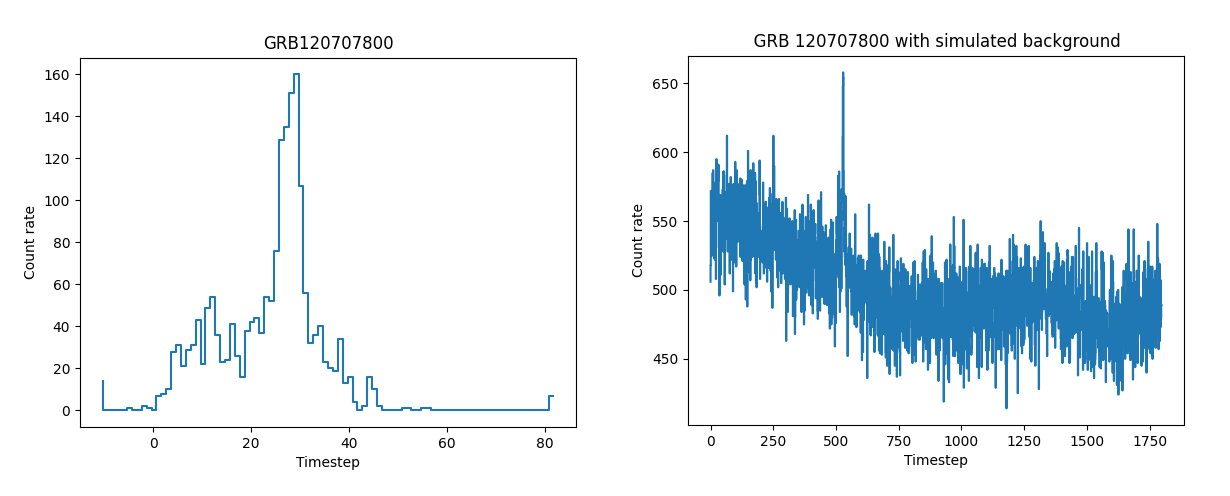}
\caption{Example of the simulation process, specifically showcasing the simulation of GRB120707 with background templates from a specific part of the satellite's orbit (as indicated by \#3 in Figure \ref{fig:bkg_templates}). This simulation is performed for detector \texttt{n3} in energy range r1, utilizing data from the 2019 observation period.}
\label{fig:GRB120707_sim}
\end{figure}

GRB event data can be downloaded from the Fermi GBM catalog in TTE format, allowing users to select from various types of GRBs, including short and long bursts, faint and bright, or other significant characteristics sensitive to the trigger algorithm. The simulation pipeline give in output the TTE data from GRB events are merged with the TTE background templates. By specifying a chosen temporal bin size, a simulation is generated, as illustrated in Figure \ref{fig:GRB120707_sim}.

\chapter{Framework for high-energy transient detection and event analysis}\label{chap:frm}

\section{FOCuS}
The background estimates so obtained in Chapter \ref{chap:bkg} are compared against the actual observations using Poisson-FOCuS. 
The significance of the excess in the count observations relative to the background model is quantified in units of standard deviations and recorded as a time series. Finally, these records are searched for intervals in the observation where the excess significance exceeds a threshold over one or more detector-energy band combinations.
\subsection{Description}\label{sec_trigger}
An efficient change-point and anomaly detection algorithm called FOCuS-Poisson (Functional Online CUSUM) \cite{ward2022poisson} is employed to find anomalous transients in Fermi GBM data relative to the NN estimates of the background. \\
The FOCuS-Poisson algorithm is executed sequentially over the time series of the observed count rate data and the background estimates, separately for each combination of detectors and energy range. For a given detector-energy range combination with label $i$ and a given time step $t$, FOCuS-Poisson outputs an estimate of the maximum significance in the observed count rate excess relative to the background, $m_t^{(i)}$. This value is computed over an optimal time interval ending at $t$ and starting at a past time-step $t - d$. Crucially, the interval length $d$ is not predetermined but rather assessed and optimized by the algorithm itself, conditionally on the observations.
The significance values $m_t^{(i)}$ are recorded, in units of standard deviations, in a table with dimensions $M \times N$, where $M$ equals the length of the input time series and $N$ equals the number of detector-energy range combination. From these table, candidate transients are extrapolated in two steps. The first step is to identify vertical table slices (time intervals, rows \textit{segments}) where a trigger condition is verified (e.g., the times when two detector-energy range combinations exceed a pre-set threshold). The second step is to cluster together segments whose start and end times are closer than a pre-defined value. This value is fiducially set to 600 s, a duration large enough to capture most long GRBs and equal to the duration of the Fermi-GBM TTE files \citep{von2014second}.\\
The user controls the search's output through three parameters. For the trigger condition to be verified it is required that the significance values exceed a threshold parameter $T$ over a minimum number detectors and energy ranges. Additionally, the user can limit the choice of the best interval to those whose length does not exceed a value $d_{\text{max}}$ or whose average intensity, given as a multiplicative factor of the observed count rates in relation to the integral of background values, is greater than a minimum $\mu_{\text{min}}$.

\begin{figure}[H]
\centering
\includegraphics[width=.58\textwidth]{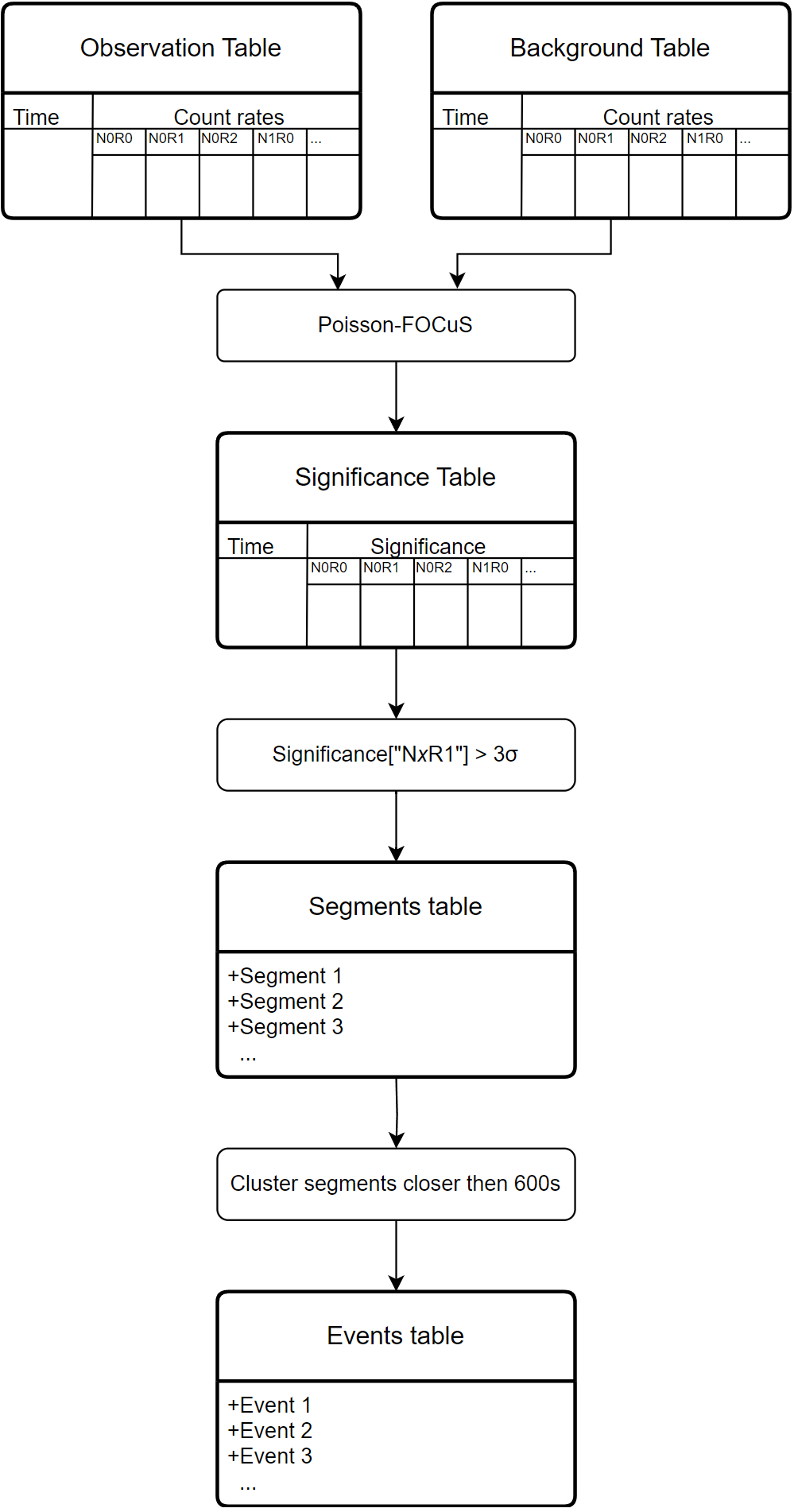}
\caption{A diagram representing the trigger algorithm pipeline's component. A Poisson-FOCuS container is given in input two tables. The first input table contains the NN's background count rate prediction, while the second reports the actual observations. The container outputs a table with same dimension as the inputs, and values representing statistical significance in unit of standard deviations. All tables share the same dimension and organization: columns are used to represent different combinations of detectors and energy ranges, while rows are used to represent different times. The output table is searched for time intervals in which statistical significance exceeds a threshold value over the energy range 50 - 300 keV (r1). Then, intervals close in time and exceeding the threshold are clustered together. Finally, clustered over-threshold intervals are reported in a list. Source: Crupi et al. \cite{crupi2023searching}.}
\end{figure}

\subsection{Application on background estimate}\label{sec:trigstat}

With reference to the technique described in Section \ref{sec_trigger}, the following detection parameters were used to obtain the results discussed in this section.
The trigger condition was defined to resolve whenever at least one detector observed enough count rates for the significance level to exceed a threshold $T = 3\sigma$ over the range of energy spanning $50$ keV and $300$ keV. This choice was made to ensure comparability with the approach used by the online search algorithm of Fermi-GBM and other major GRB monitoring experiments \cite{von2020fourth, kommers1999faint}, as well as to filter out softer events such as solar flares.

The FOCuS-Poisson algorithm was executed with the parameters $d_{\text{max}}$ and $\mu_{\text{min}}$  set to the values $120.4$ s and $1.2$ s, respectively. The choice of these parameters was driven by a trade-off between the need to find most astrophysical transients in our dataset both known and potentially unknown and the need to minimize the rate of false detection.

A filter was applied to exclude data points within 150 seconds before and after a SAA transit, specifically if the satellite remains in the SAA for at least 500s.
The purpose of filtering out data in proximity to the SAA is to reduce false detections. 
This is necessary due to various factors including the dynamic nature of the SAA environment, even on short time-scales \cite{zou2015short},  the spacecraft's apparent direct motion (Fermi enters the SAA at different geographic locations during each orbit), and the presence of a discontinuity in the observed data resulting from the instrument switch-off during the SAA transit.
These factors make estimating a reliable background count rate near the SAA challenging, often resulting in an underestimation of the background rate and, in turn, false detections by the trigger algorithm.
Through empirical analysis, we have determined that a filter duration of 150 seconds is the minimum required to ensure accurate estimation of background count rates. However, this precaution has the unfortunate consequence of preventing the detection of transients that occur during these filtered periods.

The transient search was performed over three distinct time periods, as defined in the previous section.
In the period spanning March 2019 and July 2019 a total of 100 events were identified. Of these, 74 events match the trigger time of events already in the Fermi/GBM Trigger Catalog \cite{von2020fourth}, one event is due to artifacts in the dataset, while the nature of the remaining 25 events is uncertain. These results, along other from the remaining test periods, have been summarized in Table \ref{tab:detections_stats}.
Over the same period, the Fermi/GBM Burst Catalog \cite{von2020fourth} reports on 96 known GRBs. 
Of these bursts, 15 are missing a counterpart in our dataset due to the clipping of data $150$ s before and after a SAA transit. 
Of the remaining 81 bursts (65 detected and 16 undetected), 68 have $T_{90}$ duration larger than the bin-length resolution of our dataset ($4.096$ s). 
We were able to correctly identify $60$ of these bursts ($88 \%$).
Finally, we detected 5 out of 13 ($34\%$) GRBs with $T_{90}$ duration inferior to the the bin-lenght resolution of our dataset.
These results are summarized in Figure \ref{fig:red_green} and Table \ref{tab:burstcat_stats}, the latter also reporting on results from other periods.

To measure the significance of the events in Table \ref{tab:burstcat_stats} the Standard Score $z$ is computed:
\begin{equation}
    z = \frac{x - \mu}{\sigma}
\end{equation}
where $\mu$ is the mean and $\sigma$ the standard deviation of the distribution $\mathcal{X}$.
Since we are dealing with count rates that follows the Poisson distribution, with sufficiently high count rates we can consider $\mu \approx \sigma^2$. Then the Standard Score can be approximated to:
\begin{equation}\label{eq:significance}
    S = \frac{N - B}{\sqrt{B}}
\end{equation}
where $N$ is the observed count rates integrated over an interval spanning the event's start time and end time\footnote{To avoid noise count rates and calculate the significance around the event's peak, only count rates greater than a quantile-based threshold were included in the integral.} and over each triggered detectors. $B$ is the total count rates comes from the background estimated by the NN, over the same event time. 
Standard Score is determined independently for each energy range $S_{r0}$, $S_{r1}$ and $S_{r2}$. The overall consistency for the event is defined as:
\begin{equation}
    C=max(S_{r0},\; S_{r1}, \; S_{r2}).
\end{equation}


\begin{table}[H]
\centering
\begin{tabular}{ |p{1cm}||p{1.1cm}|p{1.1cm}|p{1.8cm}|p{1.4cm}|p{1.8cm}|p{1.3cm}| }
 \hline
 \multicolumn{7}{|c|}{Transient detection statistics} \\
 \hline
 Period & Total events & Known & Consistency median known & Unknown & Consistency median unknown  & False Detections \\
 \hline
 2010 & 81   & 55  & $>10$ & 18  & 8.83 & 8 \\
 2014 & 195  & 81  & $>10$  & 71 & $>10$  & 43\\
 2019 & 100  & 74  & $>10$ & 25 & 8.71  & 1\\
 \hline
\end{tabular}
 \caption{\label{tab:detections_stats}  Total number of transients identified, number of transients with counterparts in the Fermi-GBM Trigger Catalog and its median consistency; number of transients of uncertain origin with no counterparts in the Fermi-GBM Trigger Catalog and its median consistency; false detection. Each table row corresponds to a different time period.}
\end{table}

\begin{table}[H]
\centering
\begin{tabular}{ |p{1cm}||p{1.4cm}|p{1.75cm}|p{2.2cm}|p{2cm}||p{1.4cm}|}
 \hline
 \multicolumn{6}{|c|}{Known GRBs detection statistic} \\
 \hline
 Period & Detected GRBs & Undetected GRBs & $T_{90} > 4.096$ s & $T_{90} < 4.096$ s & Missing (no data) \\
 \hline
 2010 & 41  & 25  & 39/52 (75 \%)  & 2/14 (14 \%) & 11\\
 2014 & 21  & 7 & 17/18 (94 \%) & 4/10 (40 \%) & 8\\
 2019 & 65  & 16 & 60/68 (88 \%) & 5/13 (34 \%) & 15 \\
 \hline
\end{tabular}
 \caption{\label{tab:burstcat_stats} Detected and undetected number of GRBs in the Fermi-GBM Burst Catalog, fraction of detected bursts with duration greater than the bin-length time resolution of the tested dataset, fraction of correctly detected bursts with duration smaller than the bin-length time resolution of the tested dataset. Each table row corresponds to a different time period. The last column of the table displays the number of missing events in our dataset, due to SAA data clipping (see Section \ref{sec:trigstat}). These missing events, along with the counts of detected and undetected GRBs, contribute to the total number of GRBs in the specific period as recorded in the Fermi-GBM Burst Catalog.
}
\end{table}


\section{Event analysis}
To further investigate the detected and undetected GRBs, we plot the flux (total fluence divided by T90) vs T90 for our triggered events in Figure \ref{fig:red_green}. The red points are events reported in the Fermi/GBM catalog but undetected by our method. The Fermi/GBM events with a duration less than our time binning (4.096s) are often undetected in our analysis because of the too coarse binning. We also miss a few longer events with low count rates. Reducing the time binning by using data with higher time resolution, such as \verb|CTIME| or \verb|TTE|, could be beneficial to capture shorter and fainter events. Despite the unfavorable adopted time binning of 4.096s, we recovered $\ge 75\%$ of the GRBs with $T_{90}$ greater than 4.096s, see Table \ref{tab:burstcat_stats}. 

\begin{figure}[H] 
	\hspace*{-0.5cm}
	\centering
	\subfloat[2019]{\includegraphics[width=1.1\textwidth]{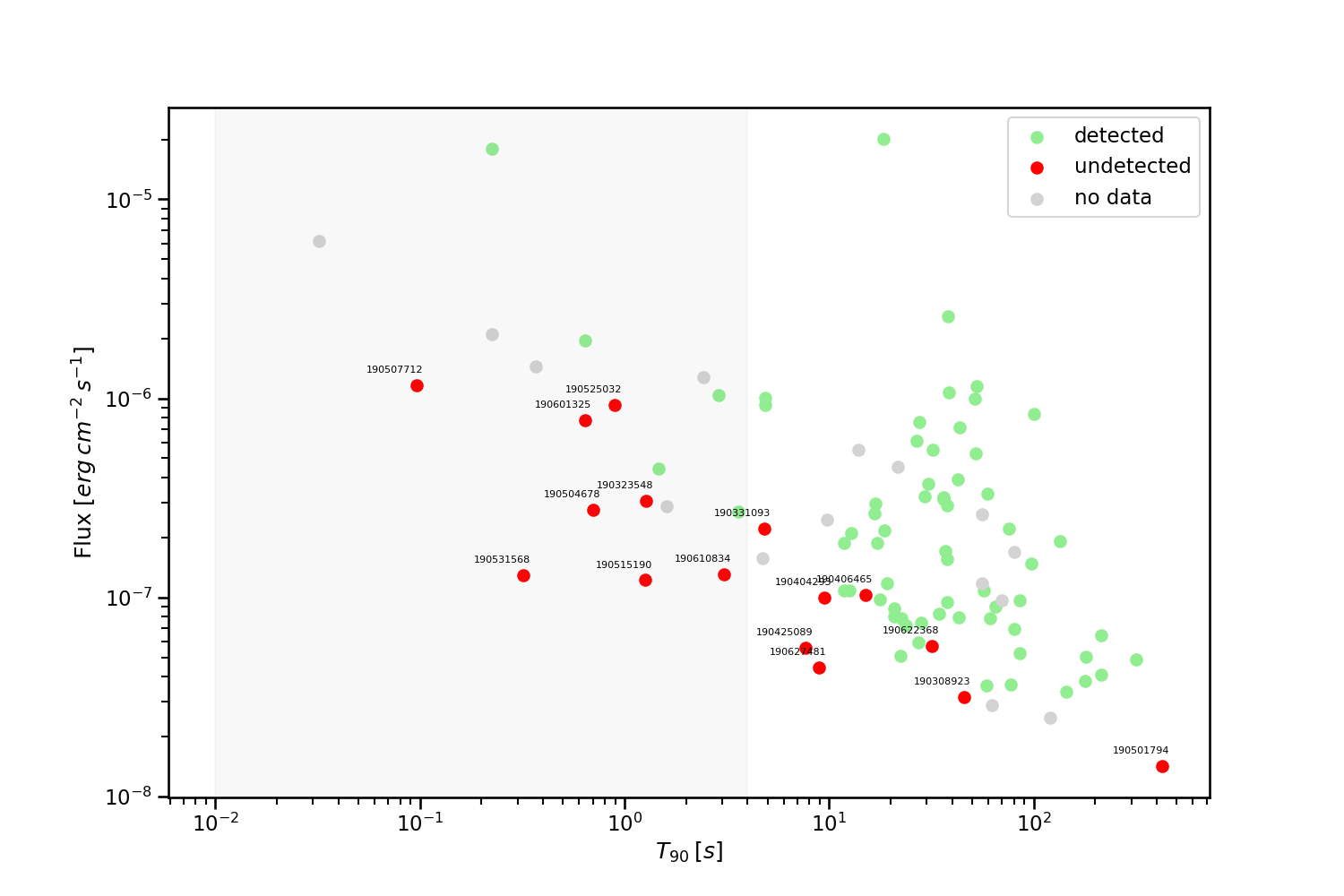}}
	\hfill
\end{figure}
\begin{figure}[H]\ContinuedFloat
	\hspace*{-0.5cm}
	\subfloat[2014]{\includegraphics[width=1.1\textwidth]{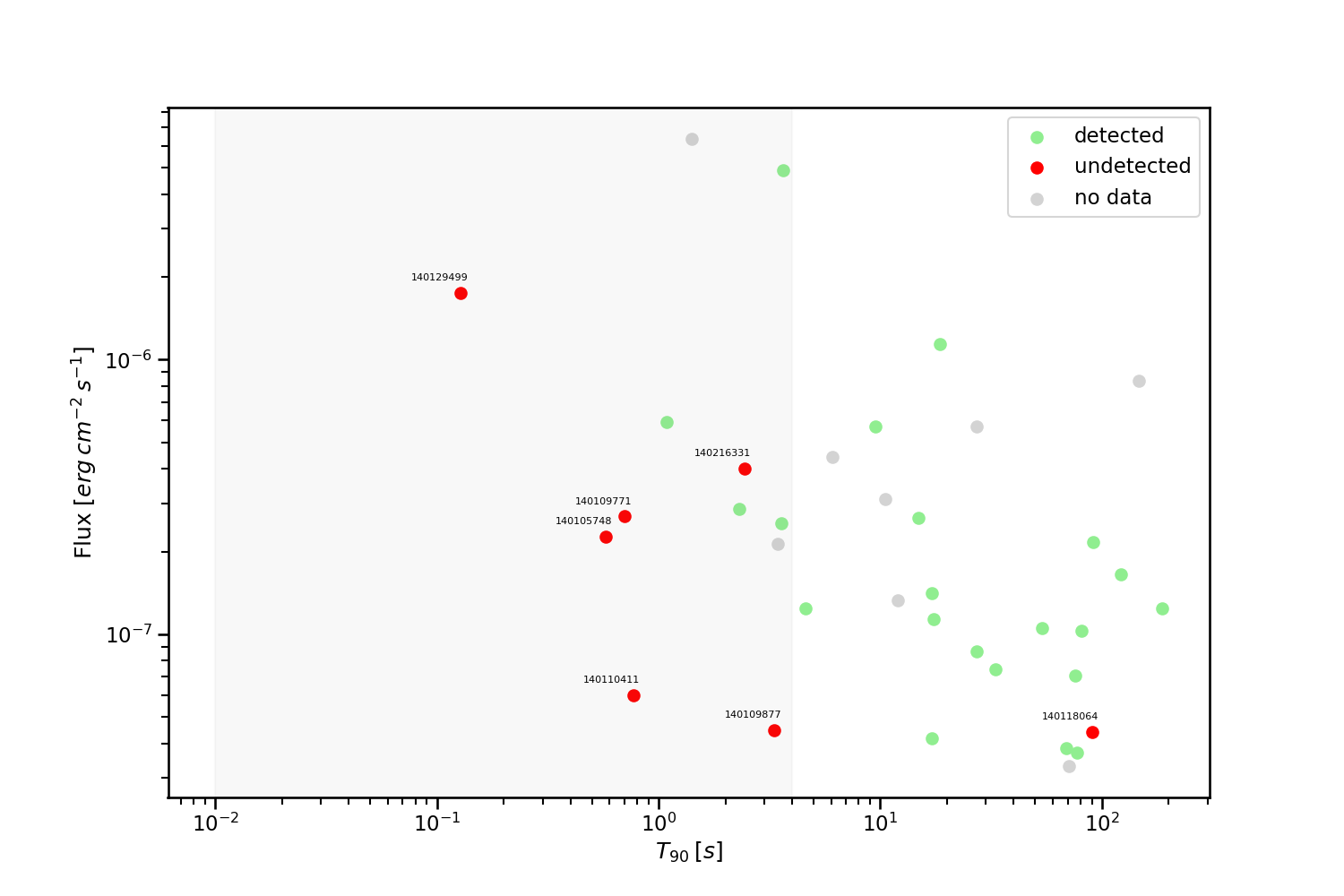}}
	\hfill
\end{figure}
\begin{figure}[H]\ContinuedFloat
	\hspace*{-0.5cm}
	\subfloat[2010]{\includegraphics[width=1.1\textwidth]{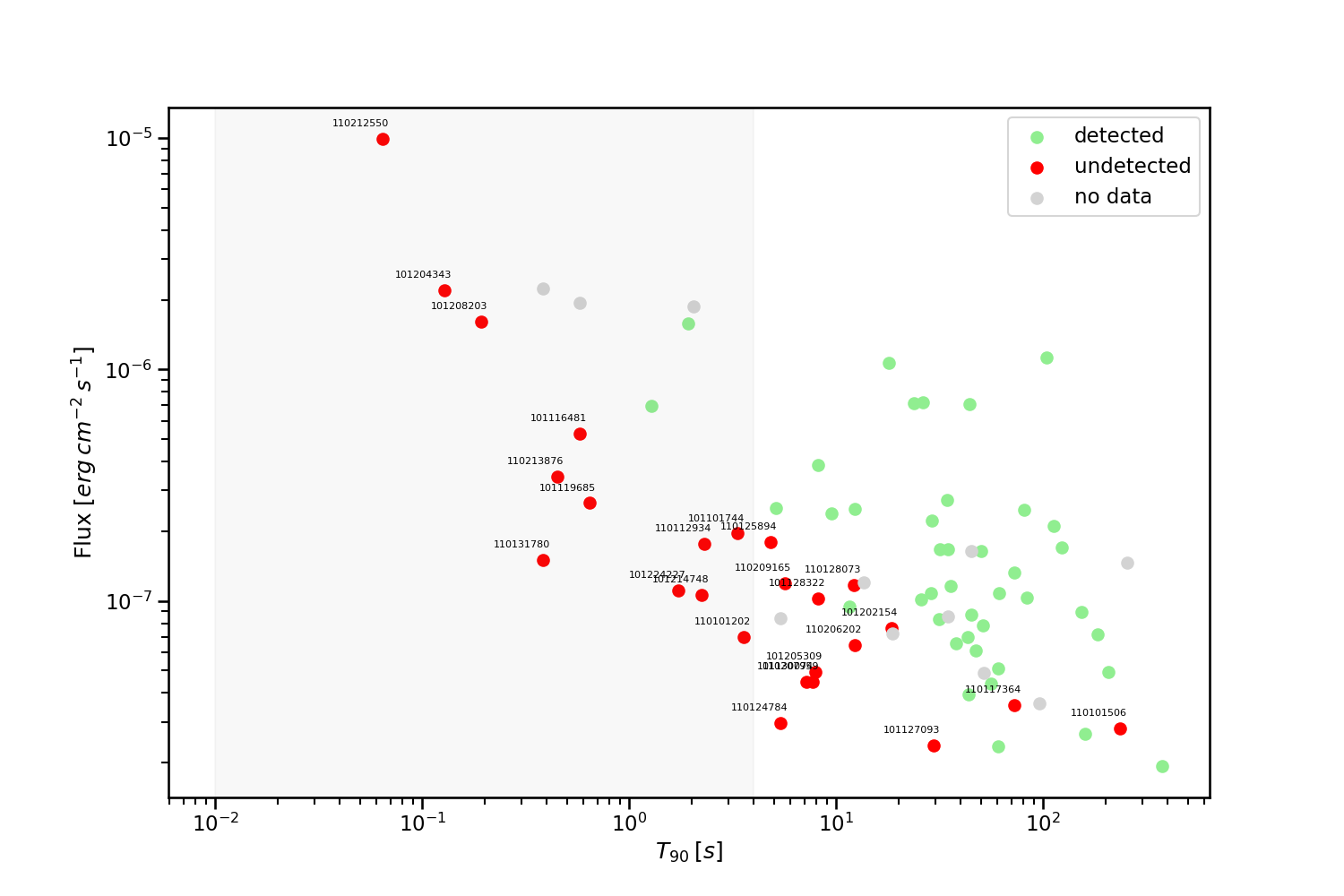}}
	\caption{\label{fig:red_green} GRB detection performances. Each dot represents a gamma-ray burst of the Fermi/GBM Burst Catalog discovered between March 1st and July 1st 2019 (a), 1 November 2010 to 19 February 2011 (b), 1 January 2014 to 28 February
		2014 (b) over the space spanned by the GRB's duration $T_{90}$ and flux, the latter computed as the ratio between the catalog's GRB fluence in band 10-1000 keV and $T_{90}$). Events in the shaded grey region have $T_{90}$ duration smaller than the bin-length time resolution of the dataset tested with the present framework ($4.096$ s, CSPEC data). Colors are used to identify the detection status within our search. In red the events unidentified with our method. Missing events (no data) are due to clipping of data 150 s before and after a SAA transit or portion of data that could not be preprocessed. Source: Crupi et al. \cite{crupi2023searching}.}
\end{figure}

\subsection{Rules for classification}\label{sec_rule}
We also detect many events not present in the Fermi/GBM catalog, 
and we use the methodology outlined in \cite{kommers1999faint} to characterize these transients. More specifically, we classify events as:
\begin{itemize}
\item Solar flare (SF) when the majority of the count rates are in the low-energy range and the Sun is in the field of view of the triggered detectors. 
\item Terrestrial Gamma-ray Flash (TGF) when most of the count rates are in the high-energy range and the event's  source reaches the detector from the Earth’s horizon.
\item Gamma-ray burst (GRB) when most of the count rates are in the $50-300$keV energy range, and the source direction is not occulted by the Earth and is distant from both the Sun and the galactic plane.
\item Galactic X-ray flash (GF) when the source direction is compatible with that of the galactic plane.
\item Uncertain (UNC) in all other cases. 
\end{itemize}
To determine the source direction, we employ a simple method based on the evaluation of the pointing and the relative photon count rate of the detectors. Further details can be found in Section \ref{localization}. 

Two classes of transient events are discussed further in this section: events already classified as GRBs in the Fermi/GBM trigger catalog; events not present in the Fermi/GBM catalog but classified by us as candidate GRBs. We report In Table \ref{tab:events} six more events that have no catalog counterpart, suggesting one or more of the previously mentioned categories. All these events are a cherry pick selection of the unknown events in Table \ref{tab:catalog_unk}.

\subsection{Localization}\label{localization}
For the standard reference for the localization of events found by GBM, look \cite{goldstein2020evaluation}.\\
In this work the localization is done by a simple geometric reasoning, but in future we hope to use more sophisticated algorithm of localization. To optimise the function loss it is employed a particle swarm optimiser\footnote{\href{https://github.com/tisimst/pyswarm}{https://github.com/tisimst/pyswarm}}.

Consider two vectors in the equatorial coordinates $\psi_d = (ra_d, dec_d)$ and $\psi_s = (ra_s, dec_s)$, respectively the pointing of a detector and the localization of the event source. The incidence intensity is modeled as the cosine between the angle $cos(\psi_d, \psi_s)$ is:
\begin{align*}
    cos(\psi_d, \psi_s) & = cos(\psi_{d,ra}) cos(\psi_{d,dec}) cos(\psi_{s,ra}) cos(\psi_{s,dec}) + \\
    & + sin(\psi_{d,ra}) cos(\psi_{d,dec}) sin(\psi_{s,ra}) cos(\psi_{s,dec}) +  \\
    & + sin(\psi_{d,dec}) sin(\psi_{s,dec}) 
\end{align*}
              
If the angle of incidence is grater than $\pi/2$ than the incident intensity must be set to 0. Finally we have \ref{ii}
\begin{equation}\label{ii}
\mathcal{I} (\psi_d, \psi_s) = max(cos(\psi_d, \psi_s), 0)
\end{equation}

The loss to optimise in Equation \ref{loss_pos}, where $i$ is a particular detector in $D$ detectors (in our case 12).
The energy range chosen is the one with the biggest residuals among detectors/energy ranges, then the count rates corresponding to the timestamp of the maximum value is given to the loss \ref{loss_pos} and minimized.

\begin{equation}\label{loss_pos}
    L_{\text{localization}} = \frac{\sum_{i=1}^{D} ( counts_s (\mathcal{I} (\psi_i, \psi_s) ) - counts_i )^2}{D}
\end{equation}
where $\psi_s$ and $counts_s$ are the unknown variables.

\section{Events highlight}\label{events}
We list in Table~\ref{tab:events} a selection of interesting events, including the one already discussed, which are not present in the GBM catalog and which deserve further analysis on lightcurves plot and localization. The Fermi satellite's position and event localization were straightforward to plot using the Fermi GBM data tools library \cite{GbmDataTools}.

\subsubsection*{GRB 190320A}
\begin{figure}[H]
\centering
\includegraphics[width=0.75\textwidth]{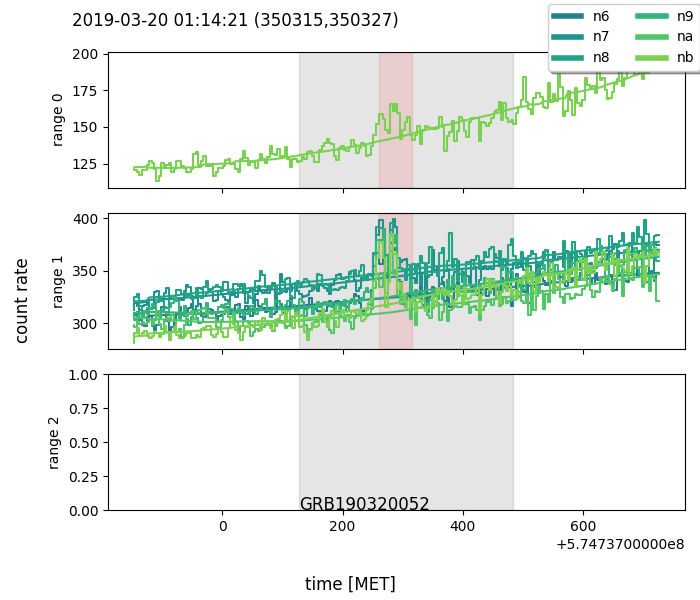}
\caption{\label{fig:grb20190320} The Fermi/GBM catalog GRB190320, as detected by our method.
Photon count rates from each triggered detector are plotted with step lines, across three energy bands spanning $28-50$ keV, $50-300$ keV and $300-500$ keV (Table \ref{tab:range}), with a resolution of $4.096$~s. 
The neural network's prediction of background count rates is represented by solid lines.
Different detectors are identified using different colors.
The GRB start and end MET time, as reported in the Fermi/GBM burst catalog, is represented by a grey shaded area. A red shaded area limits FOCuS-Poisson's best guess of the transient duration. Times are expressed in units of seconds according Fermi's standard mission elapsed time (MET). Source: Crupi et al. \cite{crupi2023searching}.
}
\end{figure}

At 01:14:16 UTC on March 20, 2019, the long GRB 190320052 triggered the Fermi/GBM on board trigger algorithm across detectors \texttt{n6} and \texttt{n9}. The estimated $T_{90}$ duration is $43$~s, with the highest emission component in the $50$-$300$ keV band.
In our analysis, the detectors \texttt{n6}, \texttt{n7}, \texttt{n8}, \texttt{n9} and \texttt{na} all exceeded a $3.0$~$\sigma$ significance threshold during the period event (Figure \ref{fig:grb20190320})  with a resulting consistency greater than $10$ on energy range r1 and $5.74$ on r2. The background estimate is comparable to a second order polynomial fitting in the soft energy range and first order polynomial fitting in the $50$-$300$ keV energy range.

\subsubsection*{Event 5 - 190420939}
\begin{figure}[H]
\centering
\includegraphics[width=0.75\textwidth]{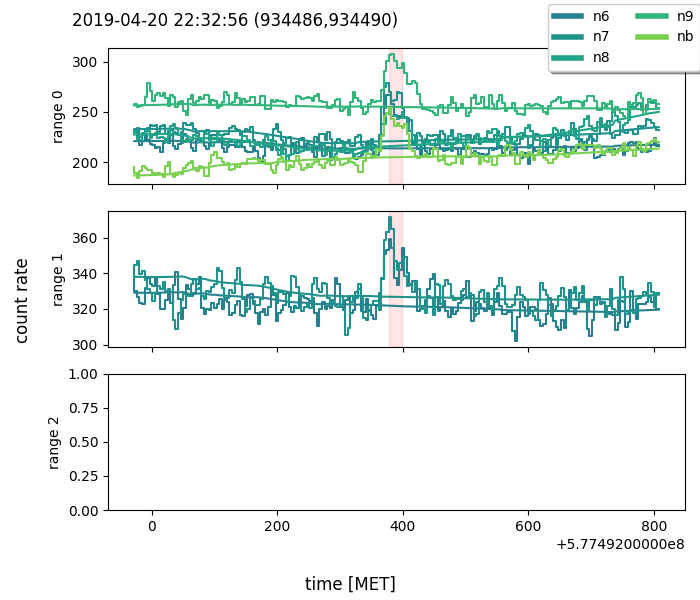}
\caption{\label{fig:grb20190420} The 190420939 transient event with no direct counter part in the Fermi/GBM trigger catalog. The event was classified as a candidate gamma-ray burst, according to the discussion presented in Section \ref{sec_rule}. For the corresponding localization see Figure \ref{fig:loc20190420}.
Photon count rates from each triggered detector are plotted with step lines, across three energy bands spanning $28-50$ keV, $50-300$ keV and $300-500$ keV (Table \ref{tab:range}), with a resolution of $4.096$~s. 
The neural network's prediction of background count rates is represented by solid lines.
Different detectors are identified using different colors. A red shaded area limits FOCuS-Poisson's best guess of the transient duration. Times are expressed in units of seconds according Fermi's standard mission elapsed time (MET). Source: Crupi et al. \cite{crupi2023searching}.}
\end{figure}

Figure \ref{fig:grb20190420} shows an event not present in the GBM trigger catalog, similar to GRB190320052 but with higher low-energy count rate. 
The event has been triggered by detectors \texttt{n6}, \texttt{n7}, \texttt{n8}, \texttt{na} and \texttt{nb} in the low energy band with a consistency greater than $10$. Two detectors provided a trigger in the $50$-$300$ keV energy band, with a consistency of $8.4$.

\begin{figure}[H]
\centering
\includegraphics[width=1.0\textwidth]{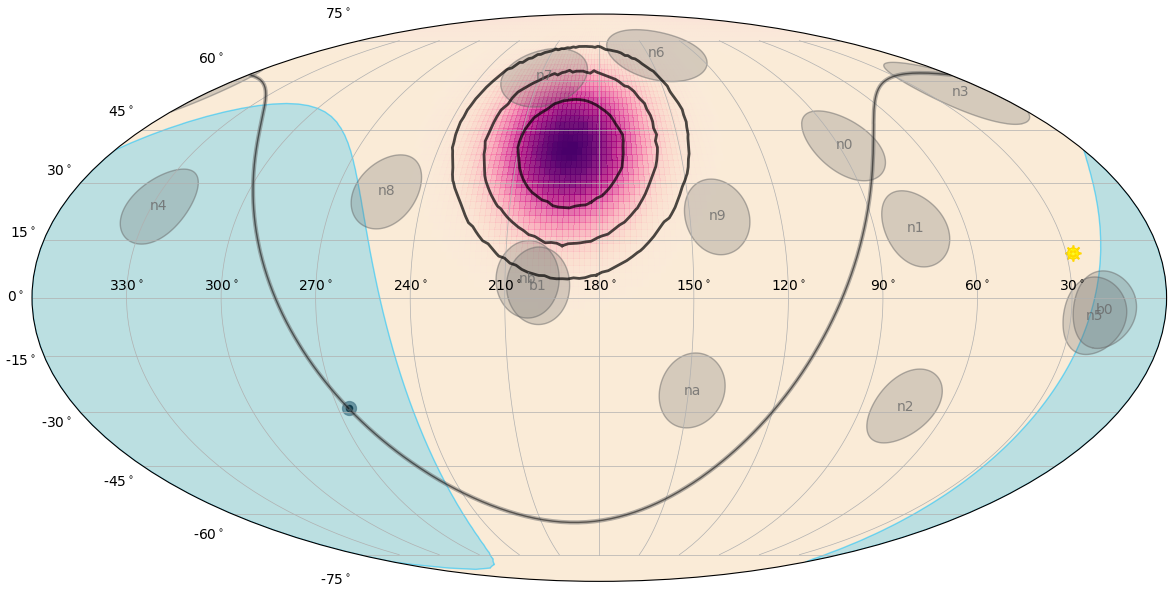}
\caption{\label{fig:loc20190420} Estimate of the candidate event's source localization over the celestial sphere at 2019-04-20 22:32:56 UTC. Source: Crupi et al. \cite{crupi2023searching}.}
\end{figure}

We can see from the localization estimate in Figure \ref{fig:loc20190420} that the event is far from the galactic plane, the Sun, and the Earth's horizon. With all of this information, this event could be a long soft GRB. 
The localization algorithm used is described in detail in Section \ref{localization}.

\subsubsection*{Event 1 - 140127222} 
\begin{figure}[H]
\hspace*{-1cm} 
\centering
\begin{subfigure}{.45\linewidth}
    \centering
    \includegraphics[width=1.25\textwidth]{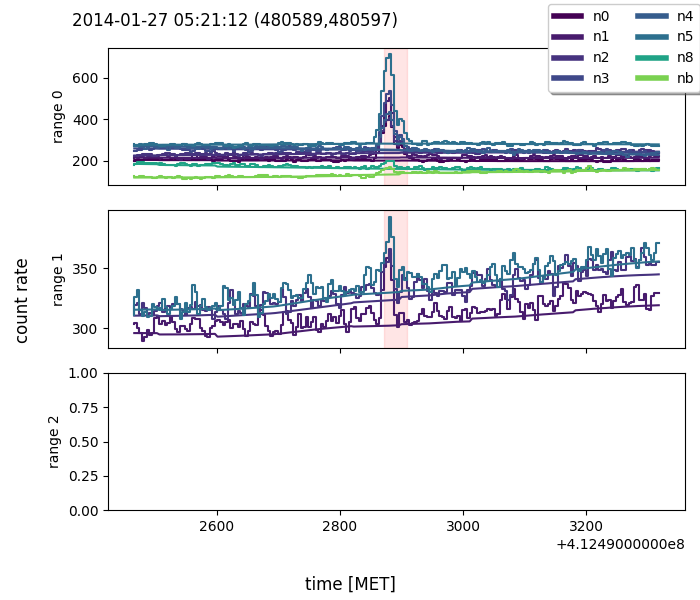}
\end{subfigure}
    \hfill
\begin{subfigure}{.5\linewidth}
    \centering
    \includegraphics[width=1.25\textwidth]{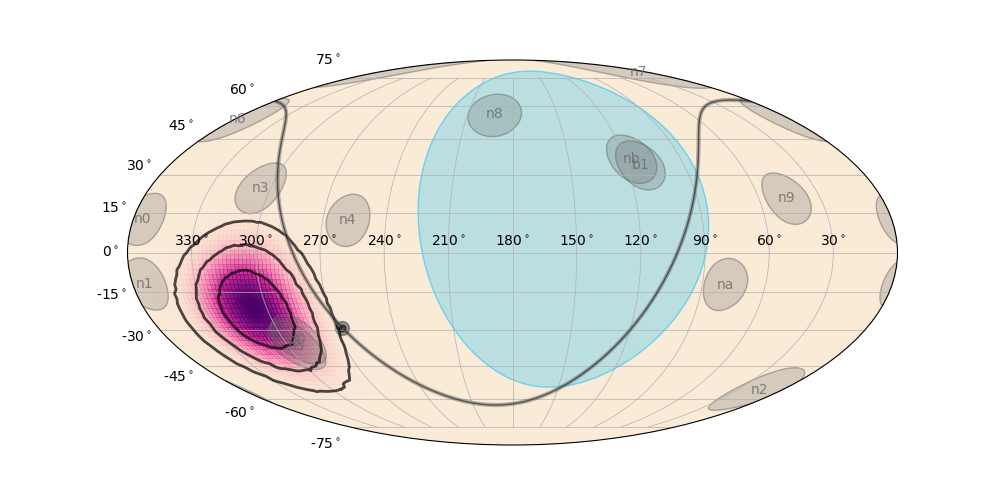}
\end{subfigure}

\caption{Lightcurve and localization for event id 1. The Sun is located under the purple spot. We classify this event as SF. Source: Crupi et al. \cite{crupi2023searching}.}
\label{fig:events1}
\end{figure}

This event has all the solar faced detectors triggered (\texttt{n0}, \texttt{n1}, \texttt{n2}, \texttt{n3}, \texttt{n4}  and \texttt{n5}) in the range r0 and its location is close to the Sun therefore is classified as Solar Flares, Figure \ref{fig:events1}.

\subsubsection*{Event 2 - 101111791} 
\begin{figure}[H]
\hspace*{-1cm} 
\centering
\begin{subfigure}{.45\linewidth}
    \centering
    \includegraphics[width=1.25\textwidth]{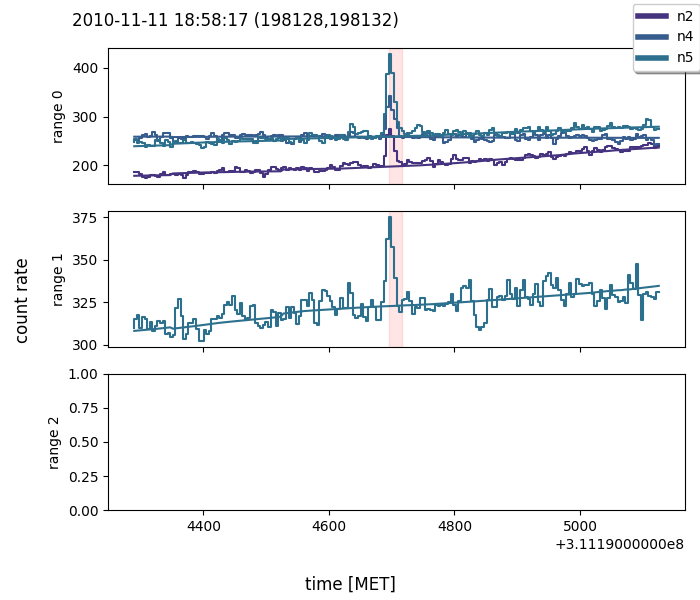}
\end{subfigure}
    \hfill
\begin{subfigure}{.5\linewidth}
    \centering
    \includegraphics[width=1.25\textwidth]{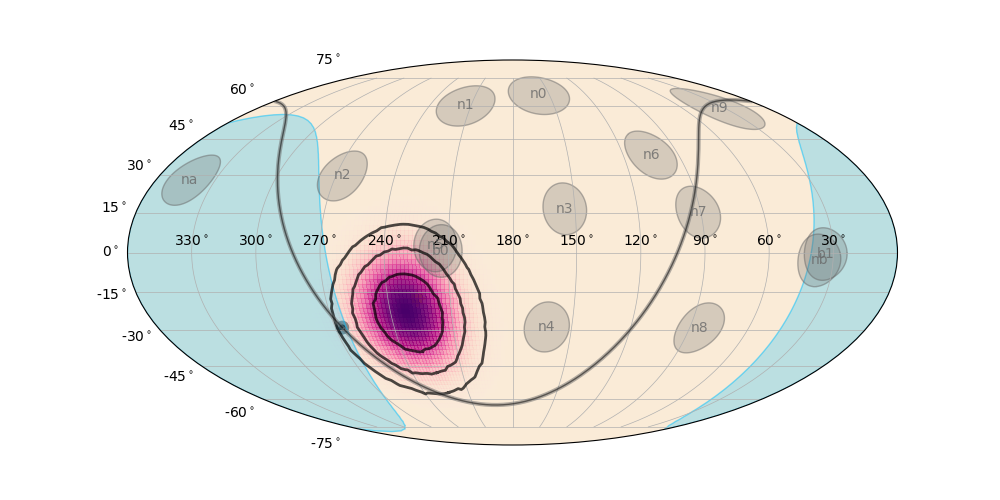}
\end{subfigure}

\caption{Lightcurve and localization for event id 2. The Sun is located under the purple spot. We classify this event as SF. Source: Crupi et al. \cite{crupi2023searching}.}
\label{fig:events2}
\end{figure}

Events 101111791 is classified as Solar Flares because three of the solar faced detector are triggered (\texttt{n2}, \texttt{n4}  and \texttt{n5}) their location is close to the Sun and the count rates are quite significant in the energy range r0, Figure \ref{fig:events2}.

\subsubsection*{Event 3 - 140112583} 
\begin{figure}[H]
\hspace*{-1cm} 
\centering
\begin{subfigure}{.45\linewidth}
    \centering
    \includegraphics[width=1.25\textwidth]{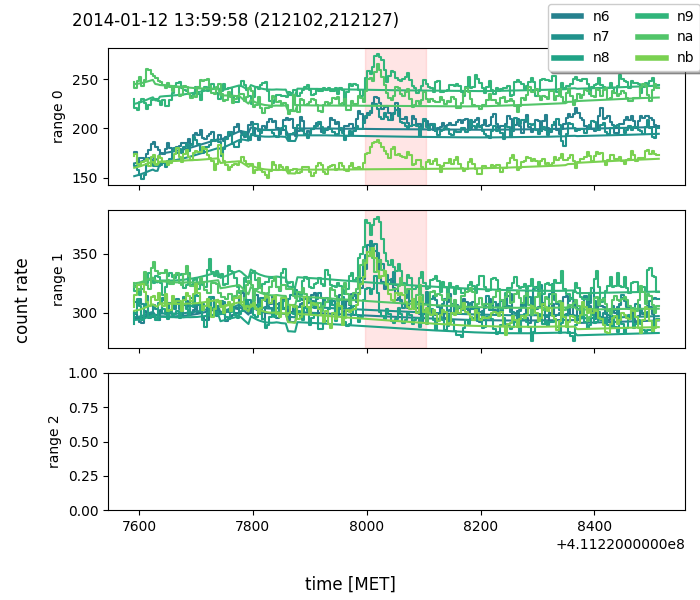}
\end{subfigure}
    \hfill
\begin{subfigure}{.5\linewidth}
    \centering
    \includegraphics[width=1.25\textwidth]{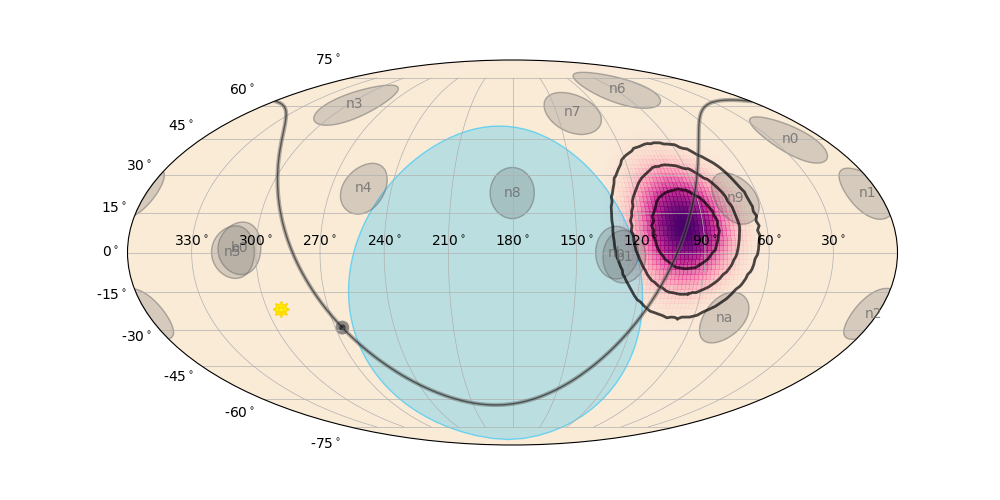}
\end{subfigure}
\caption{Lightcurve and localization for event id 3. Because of its location near the galactic plane and the Earth's horizon, we could classify this event as TGF or GF. Source: Crupi et al. \cite{crupi2023searching}.}
\label{fig:events3}
\end{figure}

Because event 3 is far from the Sun yet close to the galactic plane and the Earth's horizon, it might be a Galactic X-ray flash or a Terrestrial Gamma-ray Flash, Figure \ref{fig:events3}.

\subsubsection*{Event 4 - 190404542} 
\begin{figure}[H]
\hspace*{-1cm} 
\centering
\begin{subfigure}{.45\linewidth}
    \centering
    \includegraphics[width=1.25\textwidth]{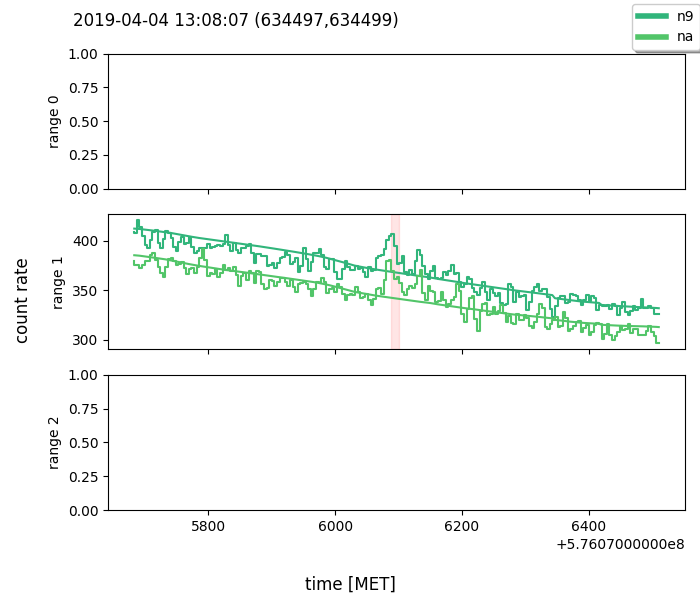}
\end{subfigure}
    \hfill
\begin{subfigure}{.5\linewidth}
    \centering
    \includegraphics[width=1.25\textwidth]{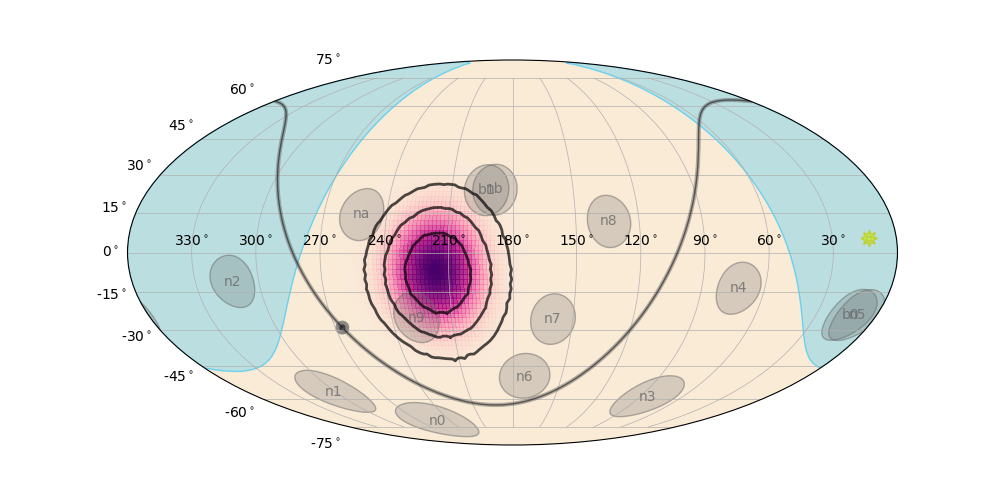}
\end{subfigure}
    \hfill
\begin{subfigure}{1\linewidth}
	\centering
	\includegraphics[width=0.8\textwidth]{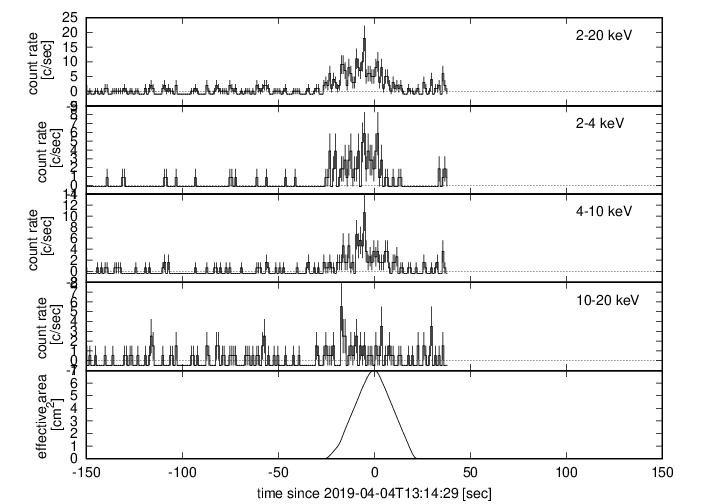}
\end{subfigure}
\caption{Lightcurve (top left) and localization (top right) for event id 4. We classify this event as a GRB. Source: Crupi et al. \cite{crupi2023searching}. The bottom panel shows the lightcurve in four energy bands, all below the energy range involved in this work, for GRB 190404B \href{http://maxi.riken.jp/grbs/190404b/}{http://maxi.riken.jp/grbs/190404b/}, detected by the MAXI satellite, \copyright 2009 JAXA/Riken/MAXI-team.}
\label{fig:events4}
\end{figure}

Event 6 is categorized as GRBs for the same reasons as event 5, Figure \ref{fig:events4}. According to the  \href{https://www.mpe.mpg.de/~jcg/grb190404B.html}{GCN Circular notice}, GRB 190404B was detected by the Monitor of all-sky X-ray image (MAXI) satellite \cite{matsuoka2009maxi} in an energy range below 20 keV, thus outside the lower energy band r0. It was estimated to be located at $(\text{RA} = 221^{\circ}, \text{Dec} = -22^{\circ})$, which is near the localization of event 4. This event occurred at 13:14:34.00 UTC on April 4, 2019, approximately six minutes after event 4. 

\newpage

\subsubsection*{Event 6 - 190606555} 
\begin{figure}[H]
\hspace*{-1cm} 
\centering
\begin{subfigure}{.45\linewidth}
    \centering
    \includegraphics[width=1.25\textwidth]{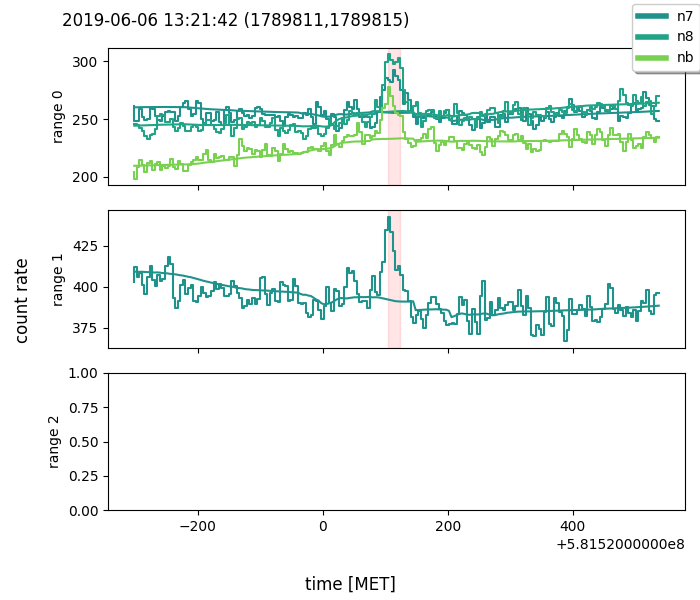}
\end{subfigure}
    \hfill
\begin{subfigure}{.5\linewidth}
    \centering
    \includegraphics[width=1.25\textwidth]{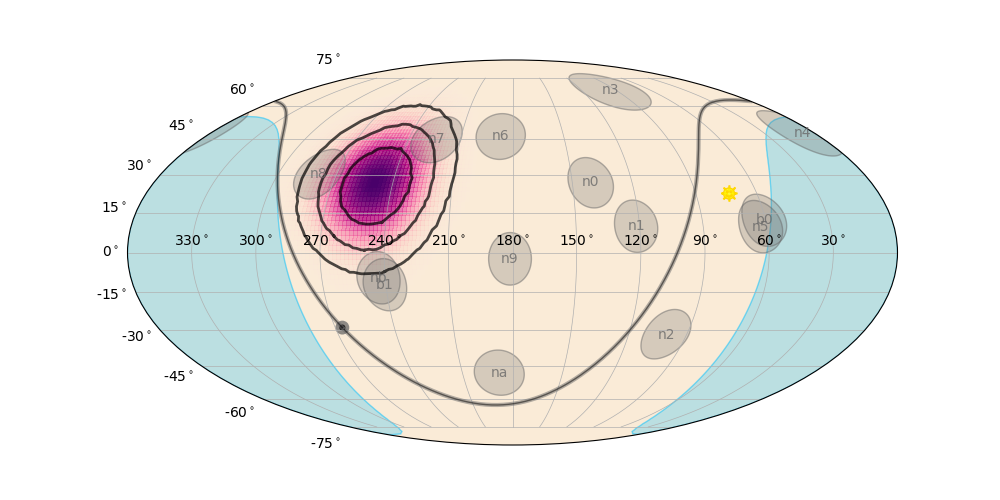}
\end{subfigure}
\caption{Lightcurve and localization for event id 6. We could classify this event as a GRB or, because of its proximity to the galactic plane, a GF. Source: Crupi et al. \cite{crupi2023searching}.}
\label{fig:events6}
\end{figure}

Event 6 is categorized as GRBs for the same reasons as event 5, however because it is near the galactic plane, event 6 might be a Galactic X-ray burst, Figure \ref{fig:events6}.

\subsubsection*{Event 7 - 110215667} 
\begin{figure}[H]
\hspace*{-1cm} 
\centering
\begin{subfigure}{.45\linewidth}
    \centering
    \includegraphics[width=1.25\textwidth]{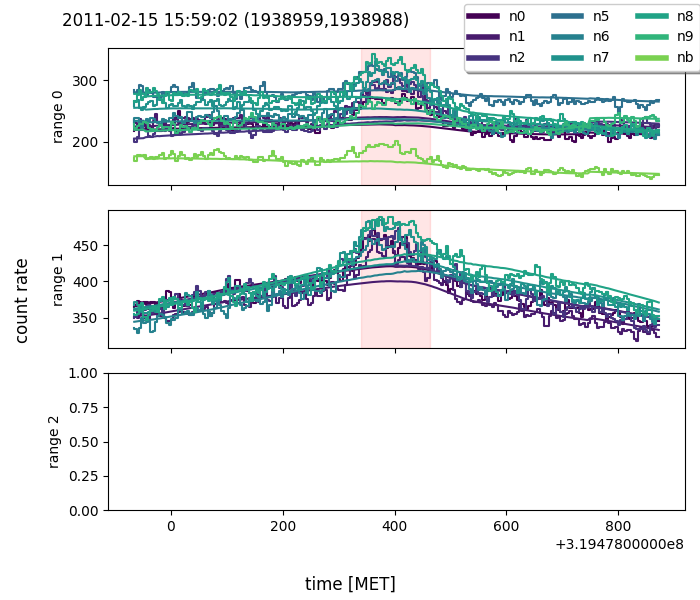}
\end{subfigure}
    \hfill
\begin{subfigure}{.5\linewidth}
    \centering
    \includegraphics[width=1.25\textwidth]{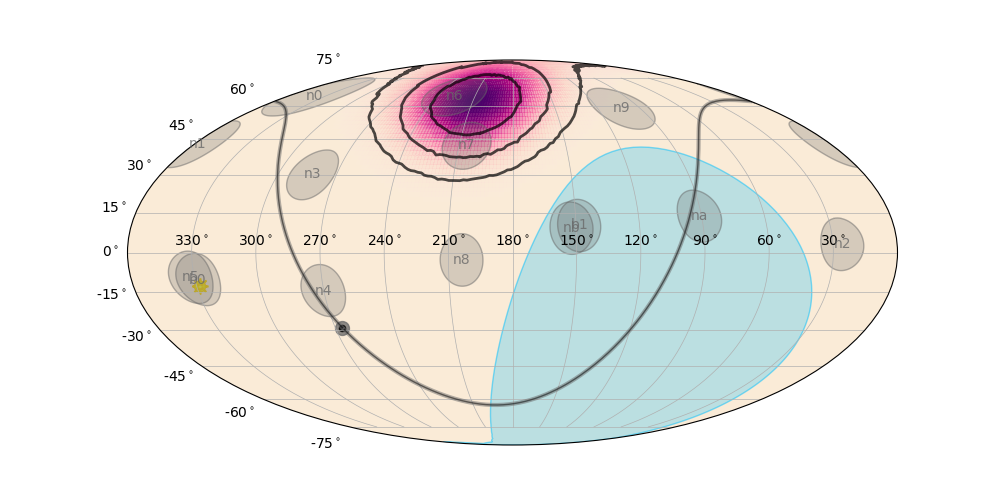}
\end{subfigure}
    \hfill
\begin{subfigure}{1.\linewidth}
    \centering
    \includegraphics[width=1\textwidth]{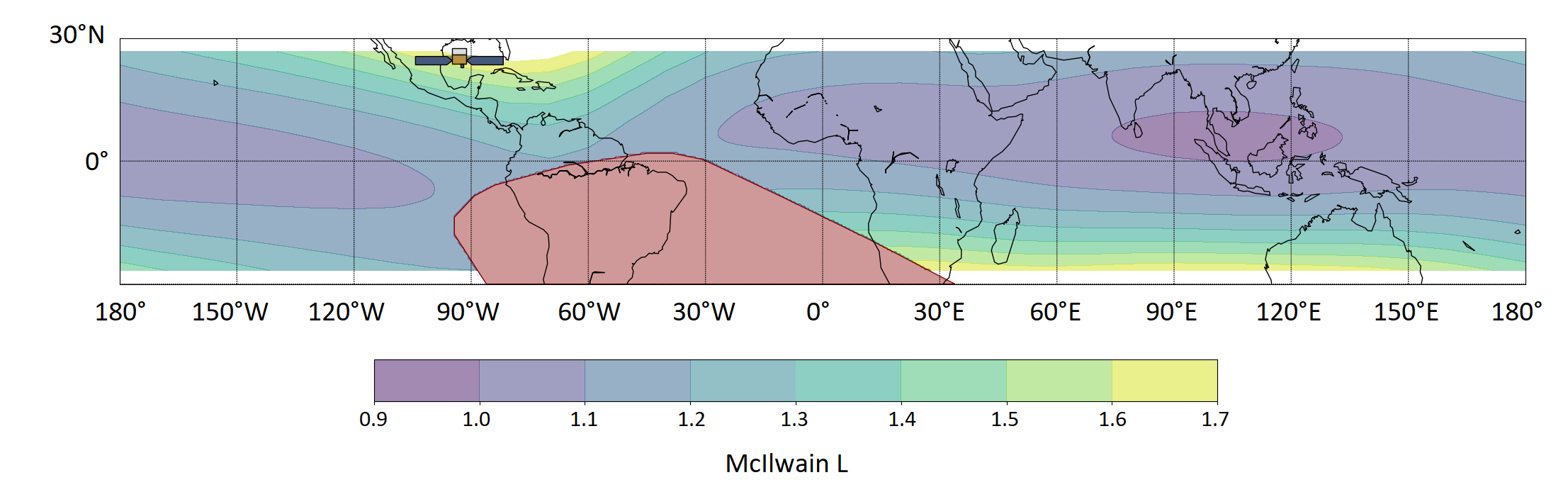}
\end{subfigure}
\caption{In the first two figures (top), the lightcurve and localization for event id 7 are shown. The third one (bottom) shows where the GBM satellite is located on Earth. Local Particles events like LOCLPAR1905205 and LOCLPAR1904085 have occurred in this region. We classify this event as uncertain. Source: Crupi et al. \cite{crupi2023searching}.
\label{fig:events7}}
\end{figure}

\begin{figure}[H]
    \centering
    \includegraphics[width=1.\textwidth]{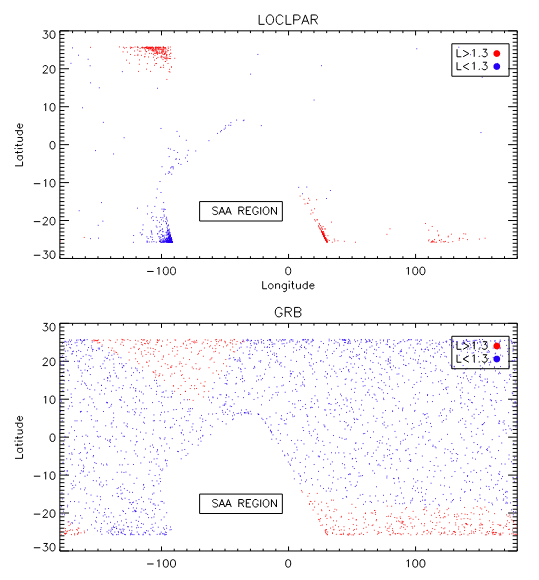}
    \caption{
     Earth's latitude and longitude coordinates of Fermi. Each point on the map represents the occurrence of a transient event, with red points indicating an L McIlwain parameter greater than 1.3, and blue points indicating otherwise. The upper figure displays Fermi's positions at the time of Local Particles (LOCLPAR) occurrences, while the lower figure shows Fermi's positions at the time of GRB occurrences. \cite{keskin2021comparative} \copyright Keskin (2021). Reproduced with permission.}
    \label{fig:LP-map}
\end{figure}

Finally, in event 7, nine detectors with roughly equal intensities are triggered, suggesting that this event is likely due to Local Particles. This is further validated by the satellite's position at high geomagnetic latitude (Figure \ref{fig:events7}), which is highly correlated with the localization of charged particle events as discussed in \cite{keskin2021comparative, von2020fourth} and illustrated in Figure \ref{fig:LP-map}; as a result, the event is classified as uncertain.
While these events are not the primary focus of this research, it should be noted as potential precursors to earthquakes in previous studies such as \cite{picozza2021looking, anagnostopoulos2010radiation}.

The complete catalog of unknown and known events for the three time periods analyzed can be found in Appendix in Tables \ref{tab:catalog_unk} and \ref{tab:catalog_kn}, respectively. The events are reported with the trigger time, duration, the triggered detectors, the Standard Score for each energy range, and a significance classification. Unknown events were assigned tentative transient classes using the methodology described in this section.

\begin{table}[H]
\hspace*{-0.5cm}
\centering 
\begin{tabular}{c p{0.15\textwidth} c p{0.15\textwidth} c c p{0.15\textwidth} c}
 \multicolumn{8}{c}{} \\
ID & Trigger time & T (s) & Detectors triggered & RA ($^\circ$) & Dec ($^\circ$) & Transient class & C \\\hline
1 & 2014-01-27 05:21:12 &	32.77 &	n0 n1 n2 n3 n4 n5 n8 nb & 306 & -22 & SF & $>10$ \\
2 & 2010-11-11 18:58:17 &   16.38 &	n2 n4 n5 & 230 & -20 & SF & $>10$ \\
3 & 2014-01-12 13:59:58 &	102.40 &	n6 n7 n8 n9 na nb & 105 & 10 & GF/TGF/ GRB &  $>10$ \\
4 & 2019-04-04 13:08:07 &   8.19 &	n9 na & 220 & -10 & GRB & 4.93 \\
5 & 2019-04-20 22:32:56 &  16.38 &	n6 n7 n8 n9 nb & 187 & 40 & GRB & $>10$ \\
6 & 2019-06-06 13:21:42 &   16.38 &	n7 n8 nb & 250 & 25 & GRB/GF & 9.12 \\
7 & 2011-02-15 15:59:02 &  118.79 & n0 n1 n2 n5 n6 n7 n8 n9 nb & 208 & 62 & UNC & $>10$
\end{tabular}
\caption{\label{tab:events} List of interesting events. We report the ID, the start time in MET and UTC, the end time in MET, the detectors triggered during the event, the localisation expressed in right ascension and declination, the proposed transient class  and the consistency of the event.}
\end{table}

\section{Automatic classification}

While the initial classification was conducted manually, as indicated by the transient count in Table \ref{tab:event_stats}, within the HERMES mission and several years of data collection it is particularly useful an automated classification system. Several strategies can be considered for this purpose.
The chosen approach focuses on addressing a relatively small database of around 400 events sourced from the catalog in Appendix \ref{appendix}. The primary goal is to produce interpretable results, similar to those obtained from rule-based methods. The goal of this method is to provide a identify potential biases in the classification process. To accomplish this, a feature engineering procedure was used to help to reveal underlying patterns through rule extraction.
The goal of this automated classification process is to categorize events quickly while also providing insights, primarily through rule-based explanations. These findings are meant to aid future investigations into the true nature of these transient events.

\begin{table}[H]
    \centering
    \begin{tabular}{c|cc|c}
        type of transient & num. known & num. unidentified & total \\ \hline
        GRB     & 127 & 30 & 157\\
        SF      &  78 & 24 & 102\\
        UNC(LP) &   3 & 57 &  60\\
        GF      &   0 & 10 &  10\\
        TGF     &   1 &  8 &   9\\
        UNC     &   1 &  3 &   4\\
        FP      &   0 & 52 &  52\\
    \end{tabular}
    \caption{Number of transients classes according to the GBM classification \ref{tab:catalog_kn} and manual classification in Tables \ref{tab:catalog_unk}. FP refers to False Positive events.}
    \label{tab:event_stats}
\end{table}

The primary focus of this section will be on the three majority classes: GRB, SF, UNC(LP). It's important to note that the "FP" category (False Positives) is included in the analysis to explore the possibility of excluding events during the classification phase.

Firstly, features have been defined based on the resulting catalog, derived from event significance and detector triggers, as presented in Table \ref{tab:catalog_feature}. These features include: 
\begin{enumerate}
	\item Parameters like significance in various energy ranges and statistics related to the triggered detectors. Additionally, features have been extracted based on satellite information and event localization, as outlined in Table \ref{tab:satellite_feature}.
	\item Data such as right ascension, declination, solar and Earth coordinates, localization in galactic coordinates, and flags indicating Earth and Sun visibility.
	\item Feature engineering involving statistics and wavelet transforms applied to the event's lightcurve over energy ranges with higher significance. These features are detailed in Table \ref{tab:feature_extracted} and include metrics such as peak counts, peak-to-peak distances, kurtosis, skewness, maxima, minima, medians, means, standard deviations, and various wavelet-based features. The computations have been performed using the TSFEL software \citep{barandas2020tsfel}.
\end{enumerate}

Furthermore, 
These comprehensive sets of features provide a rich dataset for further analysis and classification of transient events.

\begin{table}[H]
    \centering
    \begin{tabular}{p{0.3\columnwidth}|p{0.7\columnwidth}}
    feature name & description  \\ \hline
      $S_{r0}$, $S_{r1}$, $S_{r2}$  & Significance of the event per each energy range defined as in Equation \ref{eq:significance} \\
       duration  & Duration of the event defined by the trigger algorithm \\
       qtl\_cut\_r0, qtl\_cut\_r1, qtl\_cut\_r2 & Quantile-based threshold chosen to in event significance calculation per each energy range \\
       sigma\_r0\_ratio, sigma\_r1\_ratio, sigma\_r2\_ratio & $S_{r0}/\text{duration}$, $S_{r1}/\text{duration}$, $S_{r2}/\text{duration}$\\
       HR10, HR21 & $S_{r0} / S_{r1}$ and $S_{r1} / S_{r2}$ \\
    num\_det\_rng, num\_det & Number of detector/range and detector triggered, number of detector triggered per energy range \\
    $\text{flg}_{r_j}$ & Flag, equal to 1 if range $r_j$ has been triggered  \\
    $\text{flg}_{n_i}$ & Flag, equal to 1 if detector $n_i$ has been triggered  \\
    avg\_det\_sun & Average value of the $\text{flg}_{n_i}$, only for the detectors solar faced \\
    avg\_det\_noSun &  Average value of the $\text{flg}_{n_i}$, only for the detectors not solar faced \\
    num\_anti\_coincidence &  Number of opposite detectors triggered for an event\\
    \end{tabular}
    \caption{List of features included in the complete catalog version. These features refer to the significance of the event in various energy ranges as well as statistics on the detectors that were triggered. A higher significance in energy range r1 is expected for GRB, making these features critical for event characterization.}
    \label{tab:catalog_feature}
\end{table}

\begin{table}[H]
    \centering
    \begin{tabular}{p{0.3\columnwidth}|p{0.7\columnwidth}}
    feature name & description  \\ \hline
    ra, dec & Right ascension and declination of the event localization estimate \\
    ra\_montecarlo, dec\_montecarlo &  Right ascension and declination average over the localizations estimated over Poisson perturbed count rates \\
    ra\_std, dec\_std &  Right ascension and declination standard deviation over the localizations estimated over Poisson perturbed count rates \\
     ra\_earth, dec\_earth & Right ascension and declination of the center of Earth\\
    ra\_sun, dec\_sun & Right ascension and declination of the Sun\\
    diff\_sun & Distance of the Sun respect the event defined as $\mid ra - ra\_sun \mid + \mid dec - dec\_sun \mid$\\
    diff\_earth & Distance of the Earth respect the event defined as $\mid ra - ra\_earth \mid + \mid dec - dec\_earth \mid$\\
    l\_galactic, b\_galactic & Localization of the event in galactic coordinates \\
    earth\_vis & Flag indicator if the localization is in the direction of the Earth \\
    sun\_vis & Flag indicator if the Sun is visible at the event time \\
    l & L McIlwain parameter \\
    lat\_fermi, lon\_fermi, alt\_fermi & Position of Fermi in latitude, longitude and altitude \\
    dist\_saa\_lon   &  Distance of Fermi longitude to the nearest point of SAA in longitude \\
    dist\_saa\_lat   &  Distance of Fermi latitude to the nearest point of SAA in latitude \\
    dist\_saa   &  Sum of Distance of Fermi to SAA in longitude and latitude \\
    dist\_pole\_nord\_lon  &  Fermi's longitude distance to the northernmost zone with a high L McIlwain value \\
    dist\_pole\_nord\_lat   & Fermi's latitude distance to the northernmost zone with a high L McIlwain value \\
    dist\_pole\_sud\_lon   &  Fermi's longitude distance to the southernmost zone with a high L McIlwain value \\
    dist\_pole\_sud\_lat   &  Fermi's latitude distance to the southernmost zone with a high L McIlwain value \\
    \end{tabular}
    \caption{List of features related to satellite positioning and the localization of the Sun, Earth, and event estimation. These features play a crucial role in discerning the nature of the event, whether its localization is in proximity to the Sun (potential SF), the SAA (potential LP), or other conditions. They include right ascension, declination, distance calculations, galactic coordinates, and more, contributing valuable insights into event characteristics.}
    \label{tab:satellite_feature}
\end{table}

\begin{table}[H]
    \centering
    \begin{tabular}{p{0.25\columnwidth}|p{0.75\columnwidth}}
    feature name & description  \\ \hline
    fe\_np, fe\_bkg\_np & Number of peaks from a defined neighbourhood of the lightcurve and background estimation \\
    fe\_pp, fe\_bkg\_pp & Peak to peak distance in the lightcurve and background estimation \\
    fe\_kur, fe\_bkg\_kur & Kurtosis of the lightcurve and background estimation \\
    fe\_skw, fe\_bkg\_skw & Skewness of the lightcurve and background estimation\\
    fe\_max, fe\_bkg\_max & Max value of the lightcurve and background estimation \\
    fe\_min, fe\_bkg\_min & Min value of the lightcurve and background estimation \\
    fe\_med, fe\_bkg\_med & Median value of the lightcurve and background estimation \\
    fe\_mea,  fe\_bkg\_mea & Average value of the lightcurve and background estimation \\
    fe\_std, fe\_bkg\_std & Standard deviation of the lightcurve and background estimation \\
    fe\_wam & CWT absolute mean value of each wavelet scale \\
    fe\_wen & CWT energy of each wavelet scale \\
    fe\_wstd & CWT std value of each wavelet scale\\
    fe\_wet & CWT entropy \\
    fe\_bkg\_step\_max & Maximum of the differences of values of the lightcurve with respect to the previous timestep\\
    fe\_bkg\_step\_min & Minimum of the differences of values of the lightcurve with respect to the previous timestep\\
    fe\_bkg\_step & $max(\mid fe\_bkg\_step\_min \mid, \mid fe\_bkg\_step\_max \mid) $\\
    fe\_bkg\_step\_med & Median of the differences of values of the lightcurve with respect to the previous timestep\\
    *\_ratio\_* & list of features determined by the ratio of the previous ones over fe\_med, fe\_std, fe\_min, fe\_max, fe\_bkg\_max, fe\_bkg\_min or a combination of those \\
    \end{tabular}
    \caption{List of features extracted from the event's lightcurve and background estimation. These features encompass statistical measures and attributes derived from the Continuous Wavelet Transform (CWT), employing the Ricker wavelet with scale parameters ranging from 1 to 10. The features offer insights into the event's temporal characteristics and statistical properties.}
\label{tab:feature_extracted}
\end{table}

\newpage
Each event class is trained using a one-versus-all machine learning classifier. Alternatively, a multiclassification model could be employed, but this approach tends to be more complex to interpret. Another possibility involves training one class against a different class (one-versus-one), but this entails managing a greater number of models with less data for each model. Decision Trees, Random Forest, and RuleFit\footnote{Implemented via the GitHub library \href{https://github.com/csinva/imodels}{imodels} based on \cite{Singh_imodels_a_python_2021}.} are employed for this classification task, with the latter employing a L1-regularized linear model (Lasso) and boolean output rules derived from decision trees within a Random Forest, enhancing the interpretability of the model \citep{friedman2008predictive}.

The inclusion of Decision Trees serves multiple purposes: firstly, with a relatively small dataset of approximately 400 events, a generic machine learning model might encounter issues with overfitting or insufficient data points; secondly, employing a logically explainable model aids in uncovering potential biases; thirdly, the decision process underlying this kind of model is more likely to be human interpretable compared to a model which operates through summations like regression; and finally, the identified rules can incorporate domain knowledge, enhancing the robustness and physical interpretability of the classification model.

On the other hand, Random Forest, an ensemble of Decision Trees, is known as a "black box" model, as its predictions are more challenging to interpret but typically result in improved classification performance. To provide interpretability, additional explainers such as \texttt{SHAP} for feature importance-based explanations or \texttt{Anchor} for rule-based explanations can be applied. RuleFit, in addiction, offers an output of coefficients corresponding to the most significant rules extracted from the Decision Trees within the Random Forest, providing a comprehensive global explanation of the classification problem.

The majority of events in the catalog fall into the categories of GRBs (157) and SF (102), with a third class LP (60), frequently occurring in the catalog of unknown events. Furthermore, FP (52) are introduced to be distinguished during the classification process.

The dataset is constructed with a list of features $X$ for each event, as previously discussed, and a target variable $y \in \{0,1\}$, where 1 indicates membership in the considered event class, while 0 indicates otherwise. It is important to note that because of the inherent uncertainty in unknown events, an event may be associated with multiple classes at the same time. This classification flexibility accounts for the ambiguity that frequently occurs when dealing with unknown or uncharacterized events.

The dataset is divided into training (80\%) and testing (20\%) subsets, maintaining proportional labeling between the sets. Feature selection involves utilizing a Lasso logistic regressor, including only features with coefficients distinct from 0. The models and their respective hyperparameters are as follows:
\begin{enumerate}
    \item \textbf{Decision Tree} classifier (Scikit-Learn): 
    
     \texttt{max\_depth=3, class\_weight="balanced", criterion="gini", \\
       min\_impurity\_decrease=0.01, min\_samples\_leaf=2, \\
       splitter="best"}.
    \item \textbf{Random Forest} classifier (Scikit-Learn): 
    
     \texttt{n\_estimators=200, max\_depth=4, class\_weight="balanced", \\
     	min\_impurity\_decrease=0.01}. 
    \item \textbf{RuleFit} (imodels): 
    
     \texttt{n\_estimators=200, tree\_size=4, max\_rules=30}.
     The dataset was normalized, only for this model, with a quantile transformer in order to facilitate the converge of the lasso linear regression.
    \item The \textbf{Manual rules} are hand-crafted considering the DT, the most important rules by RuleFit and guided by the classification methodology described in Section \ref{sec_rule}:
    \begin{enumerate}
        \item Rule for SF: \texttt{HR10 $\le$ 0.392 AND diff\_sun < 63.49}
        \item Rule for GRB: \texttt{HR10 > 0.449 AND HR21 $\le$ 0.375 AND fe\_wet > 2.054}
        \item Rule for LP: \texttt{((dist\_saa\_lon <= 9 AND dist\_saa\_lat <= 3.6) 
        					 OR 
                             (dist\_polo\_nord\_lon <= 19 AND dist\_polo\_nord\_lat <= 7.6) 
                             OR 
                             (dist\_polo\_sud\_lon <= 19 AND dist\_polo\_sud\_lat <= 7.6))
                             AND 
                             (num\_det >= 9 OR fe\_skw <= 0.345)
                              AND 
                             (diff\_sun > 35 OR $max($ra\_std, dec\_std$) > 100$)}
        \item Rule for FP: \texttt{$\frac{fe\_bkg\_step\_min}{max(\mid fe\_bkg\_max \mid, \mid fe\_bkg\_min \mid))}$ $\le -0.028$ OR \\
                            $\frac{fe\_bkg\_step\_min}{max(\mid fe\_bkg\_max \mid, \mid fe\_bkg\_min \mid))} > 0.03$ OR fe\_std $< 4.8$}.
    \end{enumerate}
\end{enumerate}

\begin{figure}[!hbt]
    \centering
    \includegraphics[width=1.\textwidth]{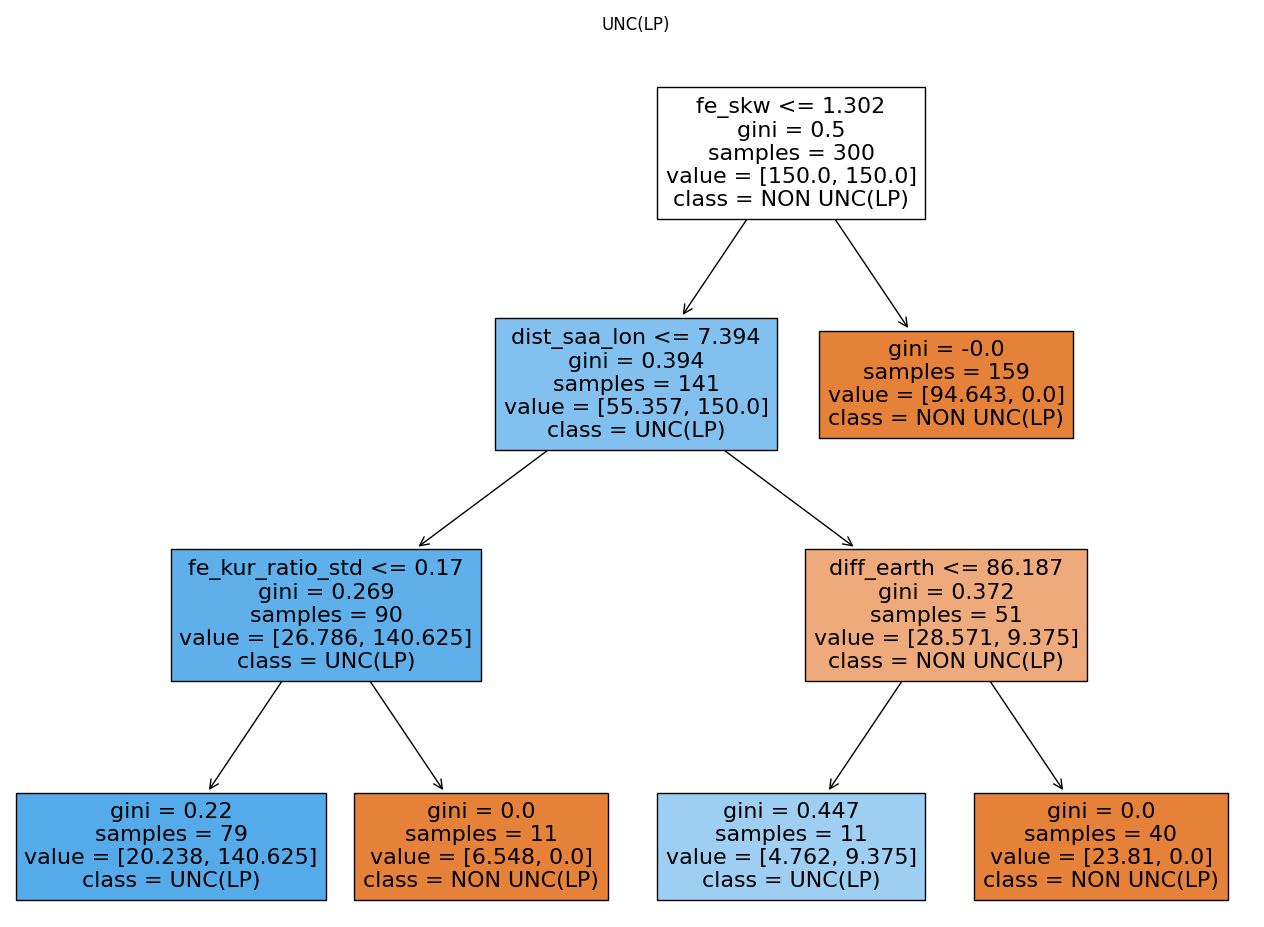}
    \caption{Decision Tree model for LP class classification. While it correctly incorporates a feature related to the distance from the SAA, this alone is insufficient for building an effective model. Notably, the model erroneously splits on the right branch over the feature \texttt{diff\_earth}, which lacks physical relevance and introduces bias. This is because LP should not exhibit a clear localization, and LP classification should not be influenced by the Earth's position.}
    \label{fig:dt_unc(LP)}
\end{figure}

Table \ref{tab:classification_result} presents the performance metrics of four distinct models on the test dataset. The Random Forest (RF) and RuleFit models outperformed the simple rule-based method in three out of four classes, with LP being the exception. This discrepancy can be attributed to the fact that in the Manual Rules are included rules based on the distance from the SAA and the Earth's poles, particularly in regions with higher McIlwain parameter values (see Figure \ref{fig:LP-map}). Figure \ref{fig:dt_unc(LP)} depicts the Decision Tree (DT) paths that employ distance from the SAA. However, the model does not account for the higher L McIlwain value zone (colored in yellow in Figure \ref{fig:events7}), which is a crucial aspect of LP classification.
RuleFit demonstrated superior performance to RF in SF classification, indicating that manipulation of important rules (in this case, only 30 rules) can lead to improved generalization. 

However, while simplifying the rules and integrating domain knowledge, Manual rules outperform the greedy DT and are comparable to RF, except in the case of GRB classification. This finding suggests the existence of a more complex underlying pattern for GRB events.

\begin{table}[!hbt]
\centering
  \begin{tabular}{|c|c|c|c|c|c|}
    \hline
    \multirow{2}{*}{Class} &
    \multicolumn{1}{c|}{} &
      \multicolumn{1}{c}{DT} &
      \multicolumn{1}{|c}{RF} &
      \multicolumn{1}{|c}{FuleFit} &
      \multicolumn{1}{|c|}{Manual rule} \\
    & Metric &  &  &  &  \\
    \hline
    GRB & Precision  & 81\% & \textbf{93}\% & 87\% & 87\% \\
        & Recall   & 84\% & \textbf{90}\% & 87\% &  84\% \\
        & F1-score   & 83\% & \textbf{92}\% & 87\% &   85\% \\
    \hline  
    SF & Precision  & 80\% & 80\% & \textbf{87}\% & 83\% \\
        & Recall   & \textbf{95}\% & \textbf{95}\% & \textbf{95}\% &  \textbf{95}\% \\
        & F1-score   & 87\% & 87\% & \textbf{91}\% &   89\% \\
    \hline  
    LP & Precision  & 43\% & 73\% & \textbf{100}\% & 73\% \\
        & Recall   & 83\% & 67\% & 58\% &  \textbf{92}\% \\
        & F1-score   & 57\% & 70\% & 74\% &   \textbf{81}\% \\
    \hline
    FP & Precision  & 50\% & 82\% & \textbf{89}\% & 71\% \\
        & Recall   & \textbf{100}\% & 90\% & 80\% &  \textbf{100}\% \\
        & F1-score   & 67\% & \textbf{86}\% & 84\% &   83\% \\
    \hline
  \end{tabular}
  \caption{Classification performance over four different models, Decision Tree (DT), Random Forest (RF), RuleFit and Manual rule, one per each of the four classes defining a binary classification task. In bold are highlighted the highest metrics among the models.}
  \label{tab:classification_result}
\end{table}

Indeed, relying heavily on HR10 and HR21 in both DT and manual rules for determining the GRB class can introduce bias by favoring events within energy range r1.  These rules may lead to selection bias by favoring GRB events within energy range r1. To improve the generalization of the GRB class, it's essential to include a more extensive dataset with a greater representation of GRB events. In addiction, a careful analysis of classification model errors, including both false positives and false negatives, could help understanding the logic behind the classification (see Section \ref{sec:error_analysis}).

It was run a training on the shape of the lightcurve using wavelets highlights an important aspect. While using all features from Table \ref{tab:feature_extracted} yields comparable performance to Table \ref{tab:classification_result}, focusing solely on wavelet-transformed features results in a drop in the F1-score to around 70\%. This result emphasizes the importance of conducting further experiments with lightcurves with higher temporal resolution (e.g., TTE files) to assess the significance of these features. Additionally, ML algorithms designed for time series analysis, such as RNN or CNN, as demonstrated in the \cite{zhang2023application}, shows promise as the dataset expands to include more events. However, these models' interpretability remains a challenge, and more research is needed to elucidate their decision-making processes and improve their transparency and explainability.

The confusion matrices for the multiclassification task on the train and test sets using the Random Forest algorithm are shown in table \ref{tab:CM}. The Balanced Accuracy for the train set is 97\%, while it is 86\% for the test set, indicating possible overfitting. To address this issue and improve generalization, cross-validation with hyperparameter tuning on the train set could be used. Before fine-tuning the models, an error analysis should be performed to determine whether the problems are due to the algorithm or the dataset.

For the evaluation phase, unknown event labels were assigned to a single class, with priority given to SF, followed by LP and GRB. This prioritization helps prevent potential confusion between the major classes, being more precise in GRB labelling.

\begin{table}[!ht]
    \centering
    \renewcommand{\arraystretch}{1.5}
    \begin{tabular}[t]{ll|c|c|c|c|}
        \multicolumn{2}{c}{} & \multicolumn{4}{c}{Predicted Classes} \\
        \multicolumn{2}{c}{} & \multicolumn{1}{c}{GRB} & \multicolumn{1}{c}{LP} & \multicolumn{1}{c}{SF} & \multicolumn{1}{c}{FP} \\
        \cline{3-6}
        \multirow{4}{*}{\rotatebox[origin=c]{90}{Actual Classes}}
        & GRB & 118 & 4 & 0 & 2 \\ \cline{3-6}
        & LP  & 0 & 47 & 0 & 1 \\ \cline{3-6}
        & SF  & 1 & 2 & 78 & 0 \\ \cline{3-6}
        & FP  & 0 & 1 & 0 & 41 \\ \cline{3-6}        
    \end{tabular} 
    \begin{tabular}[t]{ll|c|c|c|c|}
        \multicolumn{2}{c}{} & \multicolumn{4}{c}{Predicted Classes} \\
        \multicolumn{2}{c}{} & \multicolumn{1}{c}{GRB} & \multicolumn{1}{c}{LP} & \multicolumn{1}{c}{SF} & \multicolumn{1}{c}{FP} \\
        \cline{3-6}
        \multirow{4}{*}{\rotatebox[origin=c]{90}{Actual Classes}}
        & GRB & 28 & 1 & 1 & 1 \\ \cline{3-6}
        & LP  & 0 & 9 & 2 & 1 \\ \cline{3-6}
        & SF  & 2 & 0 & 19 & 0 \\ \cline{3-6}
        & FP  & 0 & 0 & 1 & 9 \\ \cline{3-6}
    \end{tabular}
    \caption{Confusion Matrices (CM) for the multiclassification task. On the left the CM on the train set with a Balanced Accuracy of 97\%. On the right the CM table with a Balanced Accuracy of 86\%.}
    \label{tab:CM}
\end{table}

\subsection{Error analysis}\label{sec:error_analysis}

This section investigates false negative and false positive classifications in order to identify potential model biases or labeling inaccuracies within the GRB class. We look specifically at Random Forest (RF) false negatives, such as events GRB 140113624, GRB 140218427, and 2014\_193, as well as two Manual rule false negatives, GRB 110119931 and GRB 190531840. Furthermore, are investigated two cases of RF false positives among unknown events, denoted 2011\_6 and 2011\_7.

\paragraph{GRB 140113624 - false negative for RF}

\begin{figure}[!hbt]
\hspace*{-1cm} 
\centering
\begin{subfigure}{.45\linewidth}
    \centering
    \includegraphics[width=1.25\textwidth]{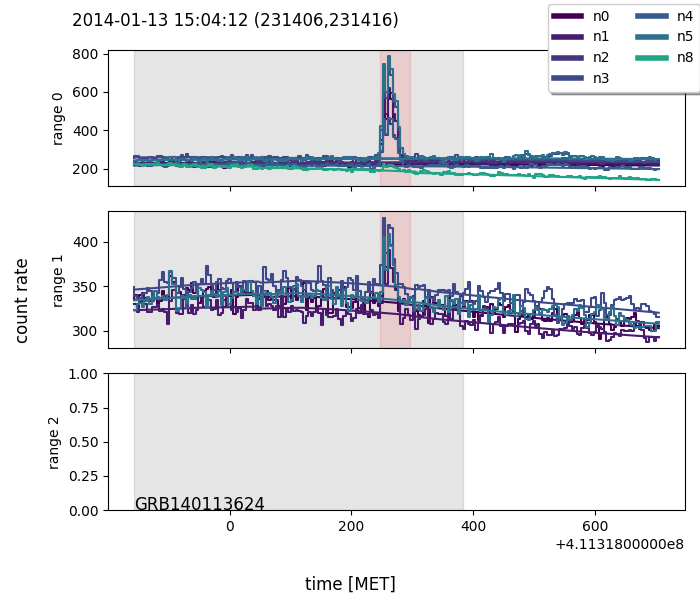}
\end{subfigure}
    \hfill
\begin{subfigure}{.5\linewidth}
    \centering
    \includegraphics[width=1.2\textwidth]{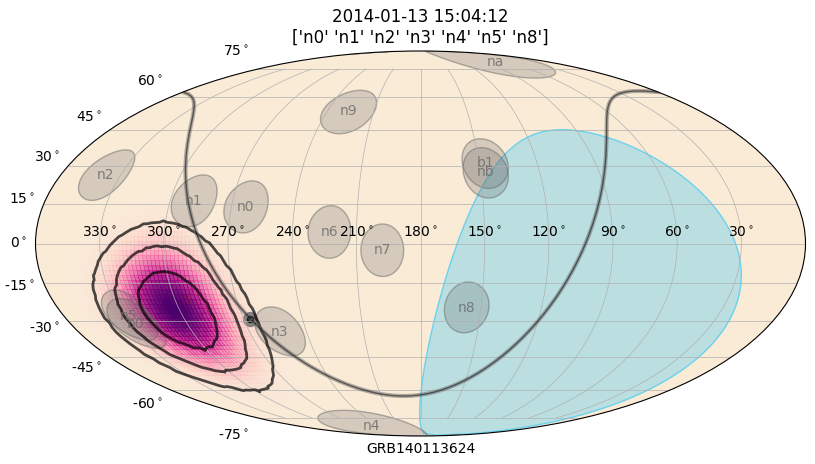}
\end{subfigure}
    \hfill
\begin{subfigure}{1\linewidth}
    \centering
    \includegraphics[width=1.\textwidth]{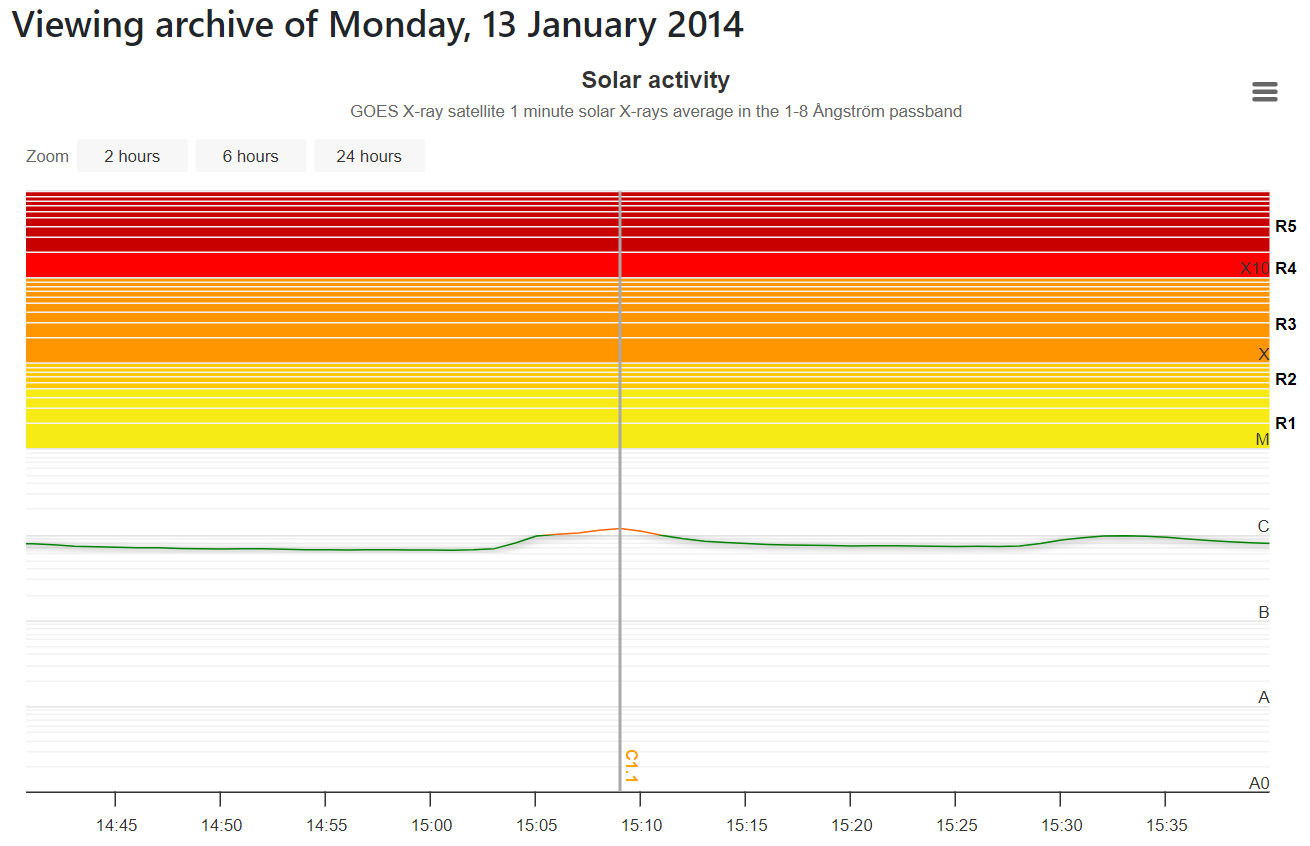}
\end{subfigure}
\caption{
The first two figures (top) depict GRB 140113624 and its estimated localization coincident with the Sun. The third image (bottom) displays solar activity around the event's trigger time (15:04:12), as observed by the \href{https://www.spaceweatherlive.com/}{GOES-X} ray satellite, showing a peak in solar activity (15:09:00). Credits for the latter image goes to SpaceWeatherLive.com.
\label{fig:events60}}
\end{figure}

Figure \ref{fig:events60} depicts GRB 140113624, which was initially classified as SF rather than a GRB. Notably, the localization estimate on the lightcurve peak corresponds to the Sun's position. However, it's important to consider that this event lasts approximately 300 seconds, making it invisible with 4-second binning (T90 is 4 seconds). As a result, the event displayed is more likely to have the correct classification, whereas the initial trigger have missed it.

Even the Random Forest for SF classifies this event as a SF with high confidence (97\%) and provides the following Anchor rule: \texttt{diff\_sun $\le$ 32.58 AND HR10 $\le$ 0.27}. Conversely, the Random Forest for GRB assigns a low probability (10\%) and generates an Anchor Rule as follows: \texttt{sigma\_r0\_ratio $\le$ 0.06 AND $\frac{fe\_kur}{fe\_med}$ > 1.13 AND fe\_bkg\_min > 253.10}. The first Anchor rule offers a straightforward criterion for classifying as SF, while the second rule excludes GRB classification based on certain statistical properties.

\newpage

\paragraph{GRB 140218427 - false negative for RF}

\begin{figure}
    \centering
    \includegraphics[width=.8\textwidth]{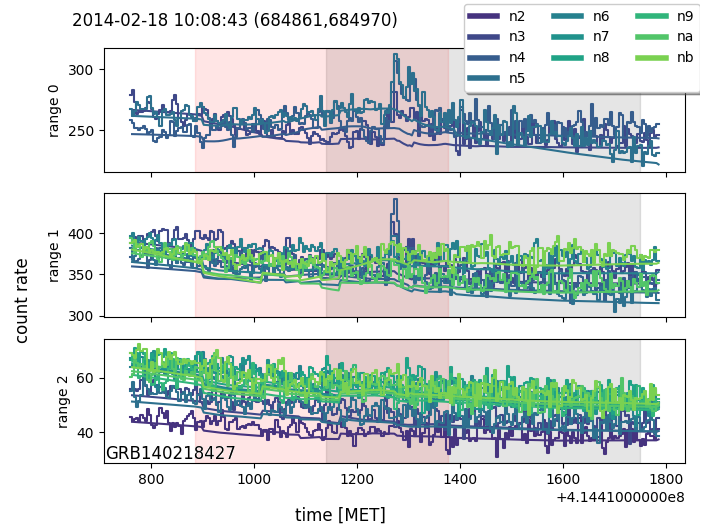}
    \caption{GRB 140218427 observed across three energy ranges, with the softer energy range being the most significant. The background estimation has non-smooth characteristics and visible steps.}
    \label{fig:event136}
\end{figure}

Figure \ref{fig:event136} showcases GRB 140218427, which the RF misclassifies as a FP class with a probability of 84.7\%. The associated Anchor rule is quite extensive:
\texttt{qtl\_cut\_r1 $\le$ 0.35 \\ AND $\frac{fe\_bkg\_step\_max}{fe\_std}$ > 0.07 AND fe\_max $\le$ 78.94 AND \\ $\frac{fe\_bkg\_step\_max}{fe\_bkg\_std}$ > 0.07 AND dec\_std > 153.00 AND \\ $\frac{fe\_bkg\_med}{fe\_bkg\_max}$ $\le$ 0.98 AND $\frac{fe\_bkg\_step\_max}{\mid fe\_bkg\_max \mid - \mid fe\_bkg\_min \mid}$ > 0.20}. Interestingly, the features used in this rule closely resemble those utilized for FP classification. Specifically, the background estimation exhibits noticeable steps rather than smooth behaviour.

\paragraph{2014\_193 - false negative for RF}

\begin{figure}[!hbt]
\hspace*{-1cm} 
\centering
\begin{subfigure}{.5\linewidth}
    \centering
    \includegraphics[width=1.1\textwidth]{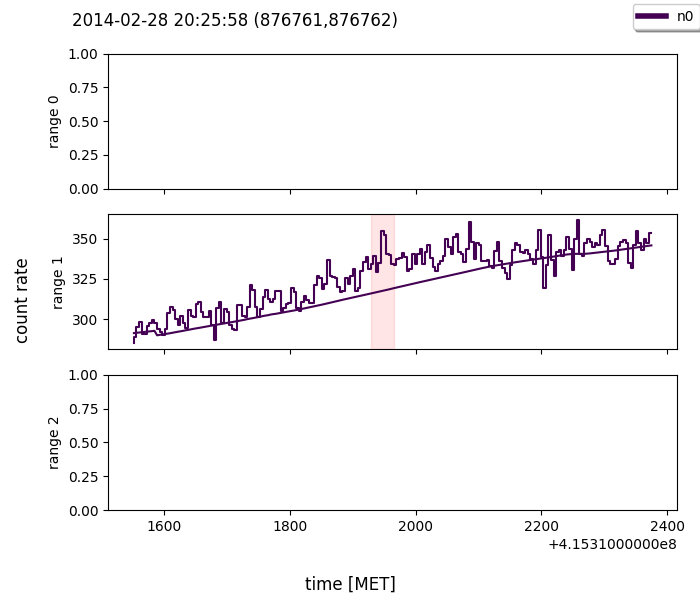}
\end{subfigure}
    \hfill
\begin{subfigure}{.5\linewidth}
    \centering
    \includegraphics[width=1.1\textwidth]{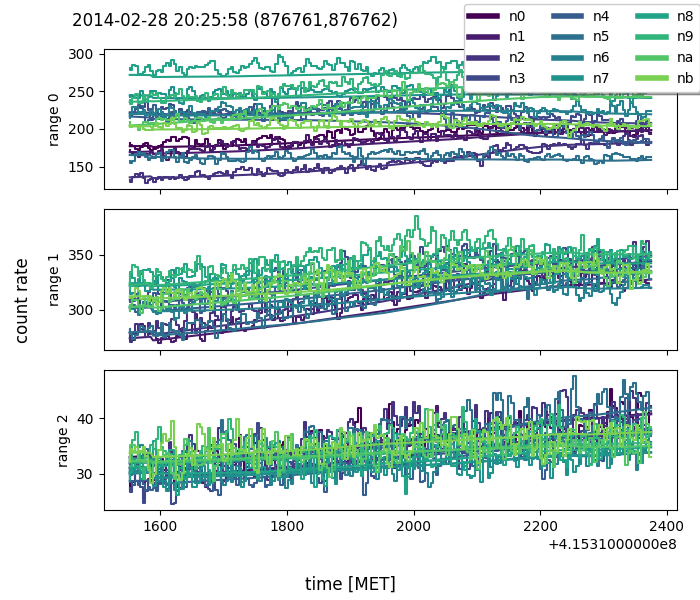}
\end{subfigure}
    \hfill
\begin{subfigure}{1\linewidth}
    \centering
    \includegraphics[width=1\textwidth]{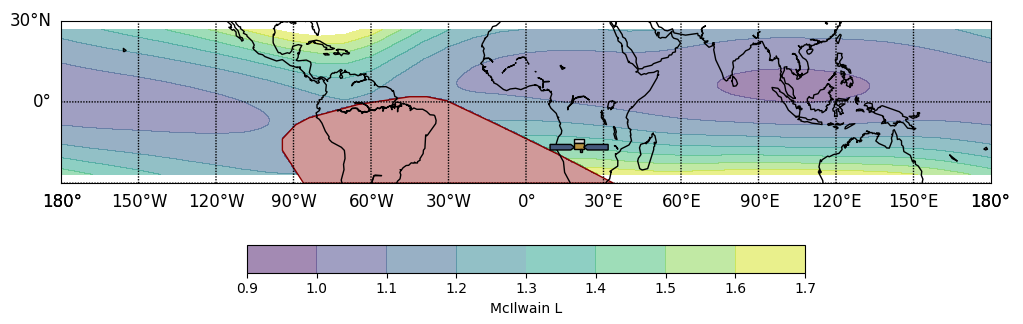}
\end{subfigure}
\caption{The first two figures (top) display an unknown event initially labelled as a GRB. The left image shows the triggered detectors, while the right shows the non-triggered detectors. The bottom image illustrates Fermi's location during the trigger phase.
\label{fig:events193}}
\end{figure}

Figure \ref{fig:events193} depicts an unknown event (2014\_193) initially classified as a GRB, but upon closer examination, it can be classified as a LP event, especially when considering the location of Fermi and other detector count rate residuals. The RF for LP assigns a prediction probability of 54\% with the following Anchor rule: \texttt{ra\_sun > 292.93 AND $\frac{fe\_med}{fe\_std}$ > 0.43 AND dist\_saa\_lon $\le$ 24.27 AND $\frac{fe\_mea}{fe\_std}$ > 2.00 AND dec\_sun > -13.01 \\ AND $\frac{fe\_bkg\_step\_med}{fe\_std}$ > 0.00}.

This rule incorporates features related to the distance from the SAA, statistical feature extracted and Sun declination, although it's not entirely clear how the Sun's position influences the classification. On the other hand, the RF for GRB predicts a low probability (29\%) with the following Anchor rule: \texttt{$\frac{fe\_min}{fe\_std}$ > -0.06 AND $\frac{fe\_med}{\mid fe\_max \mid - \mid fe\_min \mid}$ > 0.54}.

\newpage
\paragraph{GRB 110119931 - false negative for Manual Rules}

\begin{figure}[!hbt]
\hspace*{-1cm} 
\centering
\begin{subfigure}{.5\linewidth}
	\hspace*{-1.5cm} 
    \centering
    \includegraphics[width=1.5\textwidth]{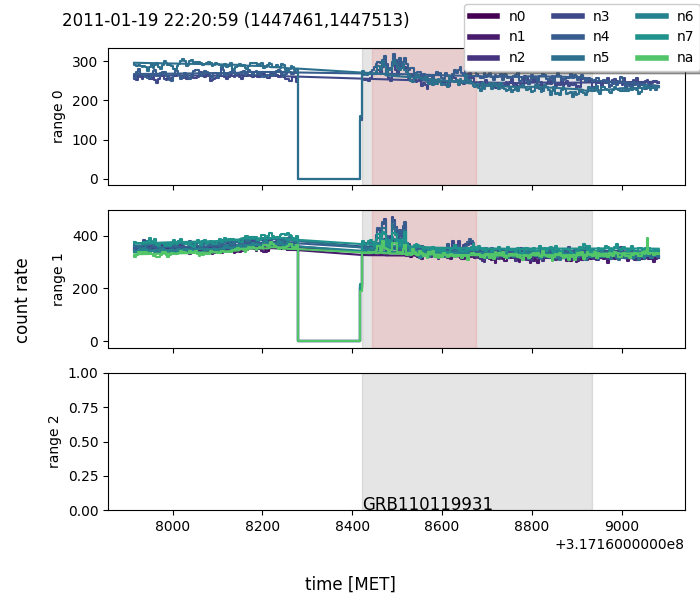}
\end{subfigure}
    \hfill
\begin{subfigure}{1\linewidth}
    \centering
    \includegraphics[width=1\textwidth]{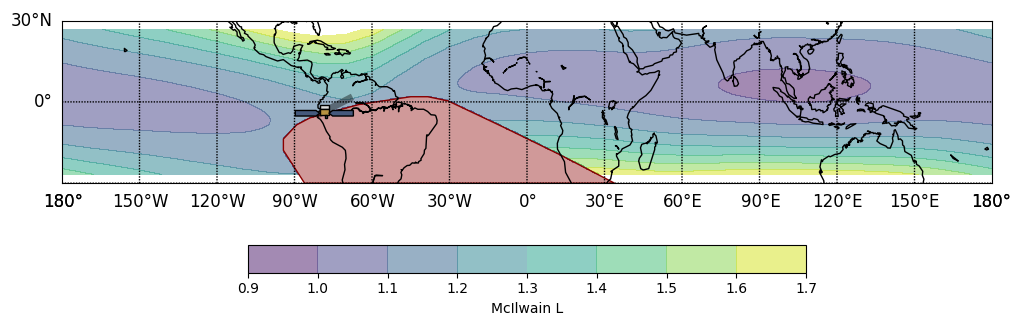}
\end{subfigure}
    \hfill
\caption{On top is GRB 110119931, which was initially misclassified as a non-GRB and as a LP event using Manual rules. The bottom figure illustrates the equatorial coordinates of the Fermi satellite's location, highlighting its proximity to the SAA.
\label{fig:events42}}
\end{figure}

Figure \ref{fig:events42} illustrates the event GRB 110119931, which was initially classified as a non-GRB and as a LP event according to the Manual rules. This misclassification can be attributed to the condition $\text{dist\_saa} > 10.225$, which tends to exclude many LP events but occasionally results in the misclassification of genuine GRBs. Further analysis involving shape features is required to address this issue.

In contrast, the RF correctly classifies the event with a probability of 72\%. Its Anchor rule is as follows: \texttt{$\frac{fe\_bkg\_skw}{\mid fe\_bkg\_max \mid - \mid fe\_bkg\_min \mid}$ > -0.06 AND $\frac{fe\_kur}{fe\_med}$ > 1.13 AND fe\_bkg\_min > 253.10}. Notably, this rule relies on statistics derived from the lightcurve distribution, rather than the hardness ratio proxy used in the Manual rules.

\paragraph{GRB 190531840 - false negative for Manual Rules}

\begin{figure}[!hbt]
\hspace*{-1cm} 
\centering
\begin{subfigure}{1\linewidth}
    \centering
    \includegraphics[width=0.75\textwidth]{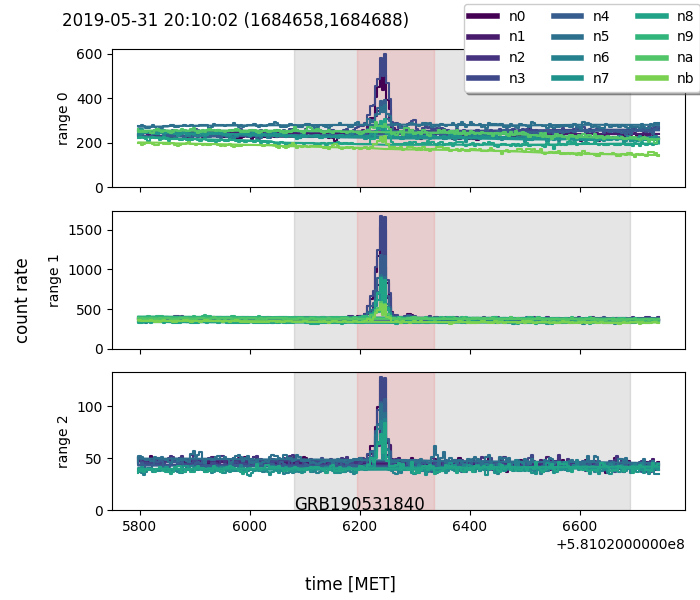}
\end{subfigure}
\caption{GRB event exhibiting high energy levels across all energy bands. This type of event may be underrepresented in the catalog, potentially leading to bias in the ML algorithm.}
\label{fig:events72}
\end{figure}

The manual rule for classifying GRBs has an energy range bias, which results in performance that is approximately 6\% lower than that of the RF. Due to high count rates in all energy bands, Event 3 is misclassified as a non-GRB, similar to GRB 190531840, as shown in Figure \ref{fig:events72}. This bias is undesirable, and it highlights the need for a second analysis that includes spectral, statistical, and temporal factors to investigate such events further.

\paragraph{2011\_6 - false positive for RF}

There are only two false positive events identified by the RF in the test set, meaning they are labeled as non-GRBs but classified as GRBs. For event 2011\_6, the probability of being a GRB is 78\%, with an Anchor rule of \texttt{HR21 $\le$ 0.27 AND $\frac{fe\_min}{fe\_std}$ $\le$ -0.40 AND $\frac{fe\_skw}{\mid fe\_bkg\_max \mid - \mid fe\_bkg\_min \mid}$ > 0.64}. While the RF for SF gives a probability of 0.05 with Anchor rule: \texttt{diff\_sun > 32.58 AND HR10 > 1.29}. Meanwhile, the RF for SF gives a low probability of 5\%, with an Anchor rule of \texttt{diff\_sun > 32.58 AND HR10 > 1.29}.

\begin{figure}[!hbt]
\hspace*{-1cm} 
\centering
\begin{subfigure}{.5\linewidth}
    \centering
    \includegraphics[width=1.\textwidth]{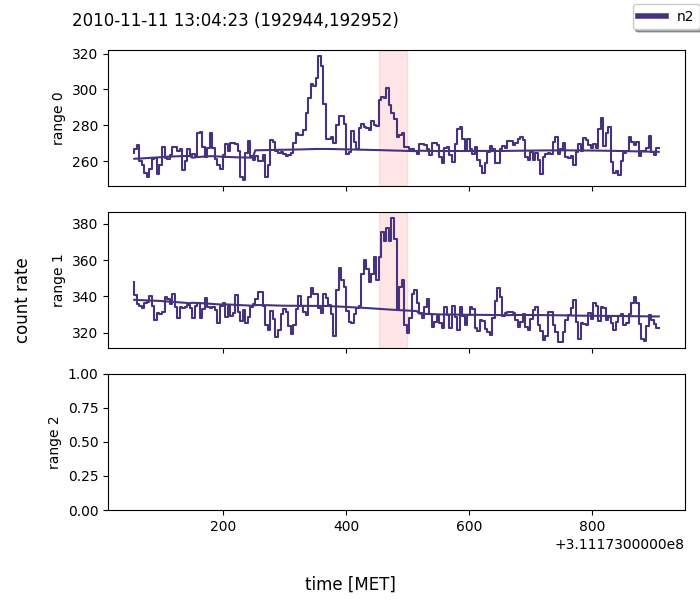}
\end{subfigure}
    \hfill
\begin{subfigure}{.55\linewidth}
    \centering
    \includegraphics[width=1\textwidth]{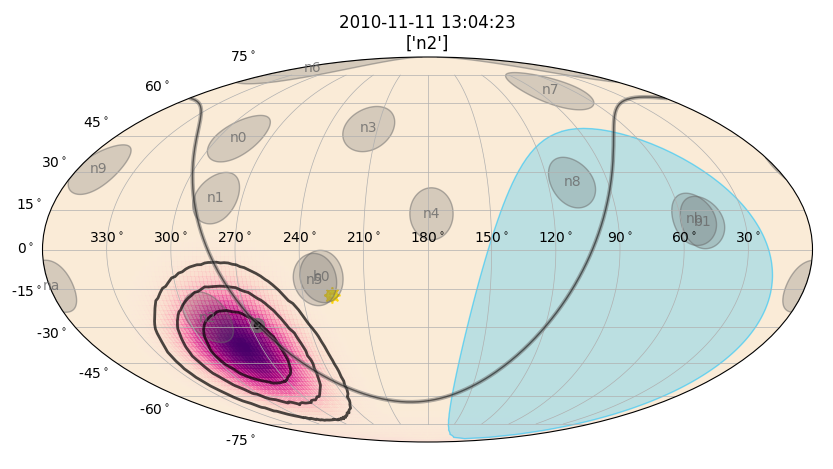}
\end{subfigure}
\hspace*{-1cm}
\begin{subfigure}{.5\linewidth}
    \centering
    \includegraphics[width=1.\textwidth]{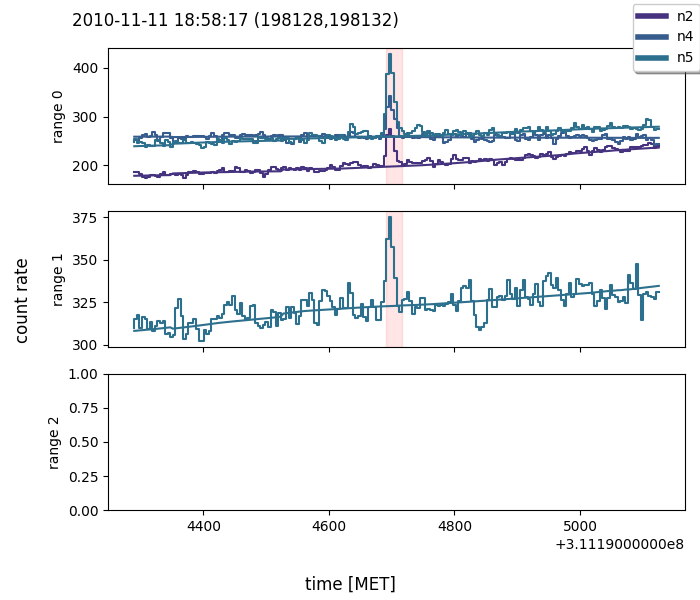}
\end{subfigure}
    \hfill
\begin{subfigure}{.55\linewidth}
    \centering
    \includegraphics[width=1\textwidth]{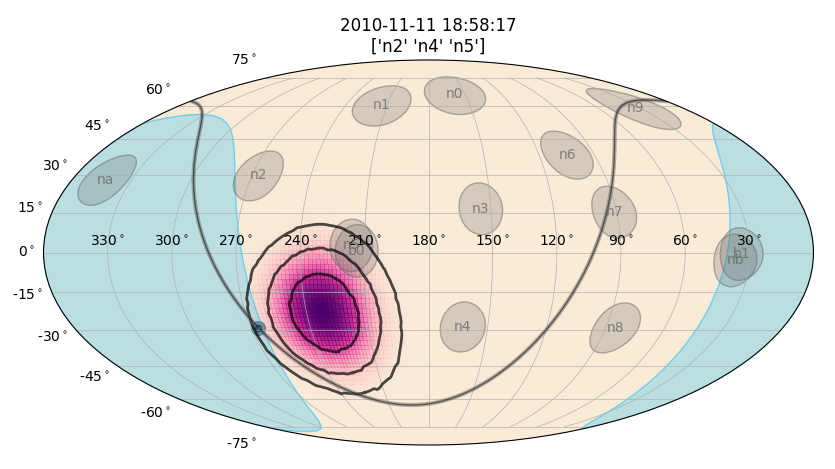}
\end{subfigure}
    \hfill
\caption{Lightcurves of events 2011\_6 (top) and 2011\_7 (bottom) in the triggered detectors and their corresponding localization estimates. Event 2011\_6 is notably close to the galactic center, potentially be classified as GF. However, the GF class is not included in the classification task.
\label{fig:events6and7}}
\end{figure}

Figure \ref{fig:events6and7} shows that the event's localization is not directly over the sun, indicating that could not be a SF event. Instead, because the event is so close to the galactic center, it was manually labeled as GF. To improve the classification algorithm, it may be beneficial to extend the classification to GF, incorporating more of those in the training dataset.

\paragraph{2011\_7 - false positive for RF}

Event 2011\_7 (Figure \ref{fig:events6and7}) was classified as SF primarily due to a peak in activity observed by the \href{https://www.spaceweatherlive.com/}{GOES-X} ray satellite. The RF for GRB assigned a probability of 63\%, driven by the Anchor rule: \texttt{fe\_wstd3 > 0.21 AND fe\_wet > 2.12 AND $\frac{fe\_min}{fe\_std}$ $\le$ -0.06}.

Conversely, the RF for SF, while slightly lower with a probability of 61\%, aligned closely with the Manual rule, reflected in the Anchor rule:  \texttt{diff\_sun $\le$ 32.58 AND HR10 $\le$ 0.27}. These probabilities are quite close, explaining the classification error.

\newpage

\paragraph{Summary}

The output results aim to provide possible event classifications, but it's important to note that further comprehensive analysis will be necessary to validate these classifications. 
This could involve using a larger dataset of events as input to the ML algorithms to obtain more precise outputs. Additionally, subsequent analysis could involve tasks such as spectral examinations or cross-referencing with additional data sources related to pulsars, galactic objects, and solar activity, potentially with higher temporal resolution using data like TTE files.

In summary, the Random Forest (RF) classification on the test dataset shows promise, and the error analysis often provides insights into understandable misclassifications, which, in some instances, actually confirm the logic behind the initial classifications. For instance, in the case of event 2011\_7, the close probabilities of GRB and SF classifications highlight the challenge of distinguishing between them. The misclassification of event 2011\_6 underscores the importance of including GF in the classification task, given the event's proximity to the galactic center. Furthermore, GRB 140113624's misclassification likely arises because it is more likely an SF, as suggested by the Manual rules. Similarly, event 2014\_193 could potentially be reclassified as LP in the dataset. Additionally, the misclassification of GRB 140218427 can be attributed to the non-smooth background estimation and the event's soft characteristics, indicating areas for background NN estimator refinement.

While the Manual rules offer a straightforward and transparent classification approach, they may introduce bias related to energy range (being either too soft or too hard) or misclassify GRB events as LP or SF, especially when these events are detected in a zone where Fermi is close proximity to the SAA (e.g., GRB 110119931) or the location estimated is near the Sun.

\chapter*{Conclusion and future perspective}
\addcontentsline{toc}{chapter}{Conclusion}
\label{chap:concl}
\markboth{CONCLUSION}{CONCLUSION}
\section*{Overview}

In this thesis it is reported an overview of GRBs, their observational properties, and possible progenitors, as well as the instruments on board satellites that detect them. 

The Chapter \ref{chp:grb_ml} serves has an overview of the applications and opportunities that AI has in the GRB context. To the best of my knowledge, this was the first comprehensive review of its kind in the literature. Clustering of GRBs was investigated in order to categorize them and gain insight into their possible progenitors. The step of the detection and classification of transients is critical because it is responsible for creating a catalog that should represent the larger population of events. The quality and quantity of this catalog are essential factors that influence the outcomes of subsequent research and analyses. Further research focus on using GRBs as standard candles in cosmological distance, and finally, GRB reconstruction and/or simulation may be used to complete some analyses that are difficult to complete because of missing data, or to test detection algorithms with selected GRB or background conditions.

The new generation of satellites (see Section \ref{sec:HERMES_mission}), particularly those operating with limited resources such as CubeSats, necessitates innovative algorithms to maximize data utility. My contribution is in this vein; I developed a novel framework for detecting high energy transients using a data-driven approach, which is described in Chapters \ref{chap:bkg} and \ref{chap:frm}. This approach may be included into an AI application for the detection of long and faint GRBs, that could have been missed by other techniques.

This novel method for high-energy transient detection \cite{crupi2023searching} integrate the precise estimation from a NN \cite{crupi2022background} with an efficient trigger algorithm \cite{ward2023poisson, dilillo2023gamma}. The method has been designed to be applied to HERMES Pathfinder data, but it can be extended to analyze data from other space-based high-energy missions. In this work I have presented an application using Fermi/GBM data, which served as the training dataset for the framework. 

The first step is to estimate the background count rate with a NN using satellite data that may be used to build a physical background model. The accuracy of the background estimate is measured using Mean Absolute Error and Median Absolute Error. An experiment is carried out to assess the robustness of the background estimator during the periods of solar maxima (2014) and solar minima (2020), demonstrating that the background estimation is stable enough to have comparable performance in both periods. Because HERMES Pathfinder will be deployed near the next solar maximum, a scenario of expected count rates can be seen in the 2014 Fermi/GBM data observation.
It has also been integrated explainability methods into the framework. SHAP allows users to understand and interpret specific predictions made by the NN by revealing the key contributing factors and features.
This method, as a debugging process, allow me to identify a problem in the feature creation of the input dataset and biases within the NN, particularly in the 2014 period.

The background is then used by Poisson-FOCuS, an evolution iteration of the CUSUM algorithm, to detect transient events more efficiently. This method is tested on three distinct periods of Fermi/GBM data, which are temporally binned into 4-second intervals. The method successfully extract the majority of known GRBs lasting more than 4 seconds. It also identifies transients that are not present in the Fermi/GBM catalog, which are detailed in the appendix. A thorough examination of seven such previously unknown events is provided, along with the information needed to classify them. 

In the final phase of this work, I describe the implementation of a supervised ML algorithm for automatic classification. Furthermore, using transparent ML algorithm allows for the examination of the prediction logic, making it easier to identify potential biases within the model.

\section*{Future perspectives}

There were no ultra-long GRBs detected in the 9 months of analyzed data. However, candidate long GRBs were identified, some of which had softer spectra than typical GRBs. Based on these promising results, the natural progression could be to extend the work the entire 15-year Fermi dataset.

The focus of future work will be on improving the neural network's prediction capabilities. It is worth to investigate the use of Recurrent Neural Networks and expand the training dataset in order to achieve a more precise estimation, particularly in areas near the South Atlantic Anomaly. I hope to improve detection of shorter events with higher precision by reducing time binning, for example, from 4 s to 1 s. To accomplish this, \verb|CTIME| or \verb|TTE| files would be required. Actually, the Neural Network described in Chapter \ref{chap:bkg} must be applied in the same time period as the training data and does not generalize perfectly outside of that time period.

Accordingly to the explainability analysis the longitude feature has to be fixed in the dataset, whereas a the detector FoV occultation features should have been treated differently (e.g., removed) and retrain the NN, in order to reduced the biased induced in the NN, which are causes of unexpected steps in the background estimation. It should be noted that solar activity alters the background significantly, so an information of time as a feature may be required to extend the period of training.

In terms of the trigger algorithm, hyperparameter tuning may be required to optimize the detection of cataloged transients (e.g., Fermi/GBM) while keeping false positives to a minimum. 

Regarding other AI applications discussed in Chapter \ref{chp:grb_ml}, beyond of the scope of this thesis but worth to mention, it is suggested to leverage on the most comprehensive GRB information by representing a GRB as an image (Figure \ref{fig:grb_image}), taking into account both temporal and spectral aspects. The handling of various detector instruments should also be carefully considered. I believe that it is valuable to investigate novel AI architectures such as Transformers and the Diffusion model for a variety of tasks. One particularly appealing direction is the generation of GRBs while taking into account specific physical properties (e.g., T90, HR, power-law index of Band function, etc).

\section*{Contributions}

In conclusion, there are many exciting opportunities for future work in this field. A well-prepared dataset is essential for inspiring computer scientists to apply AI techniques to GRBs. My major contribution, DeepGRB \href{https://github.com/rcrupi/DeepGRB}{github.com/rcrupi/DeepGRB}, provides an end-to-end solution for detecting and classifying transients, with modularity for future enhancements. The research contributions related to the background estimator and the application of the trigger algorithm can be found in the following references: \cite{crupi2022background, crupi2023searching, dilillo2023gamma}.
Furthermore, I contributed to SynthBurst \href{https://github.com/peppedilillo/synthburst}{github.com/peppedilillo/synthburst} provides functionality to leverage background estimation for simulating GRBs in various contexts for testing new trigger algorithms. Finally, ImageGRB \href{https://github.com/rcrupi/ImageGRB}{github.com/rcrupi/ImageGRB} describes a method for representing Fermi/ GBM GRBs as images, allowing them to be used in clustering, image generation or in semi-supervised regression tasks such as redshift estimation for GRBs, which could leverage both those with known and unknown redshift values.

\pagestyle{fancy}
\renewcommand{\chaptermark}[1]{\markboth{\MakeUppercase{APPENDICE\ \thechapter.\ #1}}{}}

\appendix
\label{appendix}

\chapter{Catalog tables}\label{chp:appendix}

\begin{footnotesize}
\setlength\LTleft{-1.5cm}
\begin{longtable}{llp{0.15\textwidth}rp{0.15\textwidth}p{0.2\textwidth}llll}
\toprule
{\#} & ID & Trigger time &  T (s) & Detectors triggered & Catalog trigger name & S r0 & S r1 & S r2 & CE \\
\midrule
\endhead
1   &    2010\_1 &  2010-11-02 14:16:36 &      8.19 &                                   n2 &       UNKNOWN: GRB/GF &     3.99 &     3.38 &        0 &      P \\
6   &    2010\_6 &  2010-11-11 13:04:23 &     32.77 &                                   n2 &        UNKNOWN: SF/GF &     4.29 &     5.81 &        0 &      P \\
7   &   2010\_7* &  2010-11-11 18:58:17 &     16.38 &                             n2 n4 n5 &           UNKNOWN: SF &    $>10$ &     4.57 &        0 &      R \\
10  &   2010\_10 &  2010-11-12 23:46:52 &     19.73 &                                   n1 &          UNKNOWN: UNC &        0 &     9.18 &     5.31 &      P \\
14  &   2010\_14 &  2010-11-22 13:23:34 &      8.19 &                                n5 nb &      UNKNOWN: UNC(LP) &        0 &     7.45 &     5.41 &      R \\
33  &   2010\_34 &  2010-12-28 18:22:27 &    106.50 &                                   na &      UNKNOWN: UNC(LP) &        0 &     8.08 &        0 &      P \\
38  &   2010\_40 &  2011-01-16 11:40:21 &     20.48 &                                n2 n5 &          UNKNOWN: GRB &        0 &     5.36 &        0 &      S \\
44  &   2010\_50 &  2011-01-31 15:52:43 &      8.19 &                             n5 n6 nb &      UNKNOWN: UNC(LP) &        0 &     7.08 &     4.27 &      R \\
\end{longtable}

\begin{longtable}{llp{0.15\textwidth}rp{0.15\textwidth}p{0.2\textwidth}llll}
\toprule
{\#} & ID & Trigger time &  T (s) & Detectors triggered & Catalog trigger name & S r0 & S r1 & S r2 & CE \\
\midrule
\endhead
0   &    2010\_0 &  2010-11-01 21:34:05 &     34.81 &                             n3 n4 n5 &     GRB101101899 &     5.54 &     8.51 &        0 &      R \\
2   &    2010\_2 &  2010-11-02 20:10:09 &     41.47 &                       n6 n7 n8 n9 nb &     GRB101102840 &      5.9 &    $>10$ &        0 &      R \\
3   &    2010\_3 &  2010-11-03 12:13:28 &    114.69 &           n0 n1 n2 n3 n4 n5 n9 na nb &   SFLARE10110350 &    $>10$ &    $>10$ &        0 &      R \\
4   &    2010\_4 &  2010-11-04 19:26:16 &      4.43 &                                   n2 &     GRB101104810 &        0 &     3.22 &        0 &      P \\
5   &    2010\_5 &  2010-11-07 00:16:24 &    341.51 &                       n0 n1 n3 n4 n5 &     GRB101107011 &     8.84 &    $>10$ &        0 &      R \\
8   &    2010\_8 &  2010-11-12 22:10:31 &     17.92 &                                n2 n5 &     GRB101112924 &     5.36 &     7.77 &        0 &      R \\
9   &    2010\_9 &  2010-11-12 23:36:52 &     96.26 &                          n3 n6 n7 n8 &     GRB101112984 &    $>10$ &    $>10$ &        0 &      R \\
11  &   2010\_11 &  2010-11-13 11:35:33 &    202.24 &           n0 n1 n2 n6 n7 n8 n9 na nb &     GRB101113483 &    $>10$ &    $>10$ &        0 &      R \\
\end{longtable}
\end{footnotesize}
\clearpage

\backmatter

\clearpage 
\addcontentsline{toc}{chapter}{Bibliography}
\nocite{*}
\bibliography{bib/PhD_Thesis}
\bibliographystyle{plainnat}

\chapter*{}
\thispagestyle{empty}
\vspace*{3cm}

\begin{center}
\hfill Che sia d'orgoglio ai miei genitori e d'ispirazione per mio fratello. \\
\end{center}

\end{document}